%% file: DustPolarization_Astrophysics_ArXiV_March7.tex
\documentclass[longauth,traditabstract]{aa}

\usepackage[nonamebreak]{natbib}
\usepackage[stable]{footmisc}
\usepackage{comment}
\usepackage{amsmath}
\usepackage{txfonts}
\usepackage{placeins}
\usepackage{natbib}
\bibpunct{(}{)}{;}{a}{}{,} 
\usepackage{graphicx}
\usepackage{epstopdf}
\usepackage{ifthen}
\usepackage[table,usenames,dvipsnames]{xcolor}
\definecolor{linkcolor}{rgb}{0.6,0,0}
\definecolor{citecolor}{rgb}{0,0,0.75}
\definecolor{urlcolor}{rgb}{0.12,0.46,0.7}
\usepackage[breaklinks, colorlinks, urlcolor=urlcolor,
linkcolor=linkcolor,citecolor=citecolor,pdfencoding=auto]{hyperref}
\hypersetup{linktocpage}
\usepackage{overpic}
\usepackage{fixltx2e}
\usepackage{rotating}
\usepackage{ulem}

\input{Planck}
\input{Macros}

\begin{document}

\title{\textit{Planck} 2018 results. XII. Galactic astrophysics using polarized dust emission}
\authorrunning{Planck Collaboration}
\titlerunning{Polarized thermal emission from Galactic dust}

\input{L11_Dust_B_authors_and_institutes.tex}

\date{\vglue -1.5mm \today\vglue -5mm}
\abstract{\vglue -3mm 
Observations of the submillimetre emission from Galactic dust, in both total intensity $I$ and polarization, have received tremendous interest thanks to the \Planck\ full-sky maps.
In this paper we make use of such full-sky maps of dust polarized emission produced from the third public release of \Planck\ data.
%
As the basis for expanding on astrophysical studies of the polarized thermal emission from Galactic dust, we present full-sky maps of the dust polarization fraction $\PsI$, polarization angle $\psi$, and dispersion function of polarization angles $\mathcal{S}$. 
The joint distribution (one-point statistics) of $\PsI$ and $\NH$ confirms that the mean and maximum polarization fractions decrease with increasing $\NH$. The uncertainty on the maximum observed polarization fraction, $\pmax=22.0^{+3.5}_{-1.4}$\,\% at 353\GHz\ and 80\arcmin\ resolution, is dominated by the uncertainty on the Galactic emission zero level in total intensity, in particular towards diffuse lines of sight at high Galactic latitudes.
Furthermore, the inverse behaviour between $\PsI$ and $\mathcal{S}$ found earlier is seen to be present at high latitudes. This follows the $\mathcal{S} \propto p^{-1}$ relationship expected from models of the polarized sky (including numerical simulations of magnetohydrodynamical turbulence) that include effects from only the topology of the turbulent magnetic field, but otherwise have uniform alignment and dust properties. Thus, the statistical properties of $\PsI$, $\psi$, and $\S$ for the most part reflect the structure of the Galactic magnetic field. 
Nevertheless, we search for potential signatures of varying grain alignment and dust properties.
First, we analyse the product map $\StimesPsI$, looking for residual trends. While the polarization fraction $\PsI$ decreases by a factor of \maxdropp\ between $\NH=10^{20}$\,\cmsq\ and $\NH=\uptoNH$\,\cmsq, out of the Galactic plane, this product $\StimesPsI$ only decreases by about $\maxdropSp\,\%$. 
Because $\S$ is independent of the grain alignment efficiency, this demonstrates that the systematic decrease in $\PsI$ with $\NH$ is determined mostly by the magnetic-field structure and not by a drop in grain alignment. This systematic trend is observed both in the diffuse interstellar medium (ISM) and in molecular clouds of the Gould Belt. Second, we look for a dependence of polarization properties on the dust temperature, as we would expect from the radiative alignment torque (RAT) theory. We find no systematic trend of $\StimesPsI$ with the dust temperature $\Td$, whether in the diffuse ISM or in the molecular clouds of the Gould Belt. In the diffuse ISM, lines of sight with high polarization fraction $\PsI$ and low polarization angle dispersion $\S$ tend, on the contrary, to have colder dust than lines of sight with low $p$ and high $\S$. We also compare the \Planck\ thermal dust polarization with starlight polarization data in the visible at high Galactic latitudes. The agreement in polarization angles is remarkable, and is consistent with what we expect from the noise and the observed dispersion of polarization angles in the visible on the scale of the \Planck\ beam. The two polarization emission-to-extinction ratios, $\RPp$ and $\Rsv$, which primarily characterize dust optical properties, have only a weak dependence on the column density, and converge towards the values previously determined for translucent lines of sight. We also determine an upper limit for the polarization fraction in extinction, $\pv/\ebv$, of 13\,\% at high Galactic latitude, compatible with the polarization fraction $\PsI\approx20\,\%$ observed at 353\GHz. Taken together, these results provide strong constraints for models of Galactic dust in diffuse gas.
}


\maketitle

\section{Introduction}

Interstellar dust grains are heated by absorption of the interstellar radiation field (ISRF), the ambient ultraviolet (UV), visible, and near-infrared radiation produced by the ensemble of stars in the Galaxy.  The grains cool via thermal emission, which is in the far-infrared/submillimetre, as determined by the equilibrium temperature corresponding to a balance between absorbed and emitted power. Thermal emission from the larger grains that dominate the mass in the grain size distribution can be modelled as that of a modified blackbody (MBB) with emissivity $\epsilon_\nu=\kappa_\nu B_\nu(\Td)$, where the absorption coefficient $\kappa_\nu$ depends on the dust properties~\citep{Krugel2003}.  The equilibrium temperature is observed to be of order $20\,\mathrm{K}$ \citep{planck2013-p06b} for the ISRF found in the bulk of the interstellar medium (ISM).

Starlight polarization, discovered by \citet{Hall1949} and \citet{Hiltner1949}, was quickly ascribed to differential extinction by aspherical dust grains with a preferential alignment related to the configuration of the interstellar magnetic field~\citep{Davis-Greenstein-1949,Davis1951}. Over the years, a number of theories have been put forward to explain how this alignment occurs and is sustained, despite gas collisions~\citep[see the review by][]{Andersson15}. The mechanism favoured currently involves radiative torques acting on grains subject to anisotropic illumination (RAT; see, e.g., \citealp{Hoang2016}). 

For thermal processes, Kirchhoff's law states that differential extinction implies differential emission and so the submillimetre thermal emission from dust grains is also polarized, orthogonally to that of extinction.  Thus, for dust grains aligned with respect to the Galactic magnetic field (GMF), the observed emission is also partially linearly polarized~\citep{Stein66,Hildebrand1988}. Because the spin axis of a dust particle is perpendicular to its long axis and alignment is statistically parallel to the local orientation of the magnetic field, the polarization of starlight transmitted through interstellar dust reveals the average orientation of the magnetic field projected on the plane of the sky, whereas the direction of polarized emission is rotated by 90$^\circ$ with respect to the magnetic field.

Observations of this submillimetre emission from Galactic dust, in both total intensity and polarization, have drawn strong attention, thanks to the \Planck\footnote{\Planck\ (\url{http://www.esa.int/Planck}) is a project of the European Space Agency (ESA) with instruments provided by two scientific consortia funded by ESA member states and led by Principal Investigators from France and Italy, telescope reflectors provided through a collaboration between ESA and a scientific consortium led and funded by Denmark, and additional contributions from NASA (USA).} full-sky maps, whose sensitivity and sky-coverage largely supersede the previously-available data from ground-based, balloon-borne~\citep[e.g.,][]{deBernardis1999,Benoit04}, and space observations~\citep[e.g.,][]{gold2010}.

Over the course of four years (2009--2013), \Planck\ surveyed the entire sky in nine frequency bands, from 30\,GHz to 857\,GHz, providing the best maps to date of the cosmic microwave emission, with unprecedented sensitivity, and angular resolutions varying from 30\arcmin\ at 30\,GHz to 4\parcm8 at 857\,GHz. All but the two highest-frequency channels (545\,GHz and 857\,GHz) were sensitive to linear polarization of the observed radiation. In these seven bands, most of the polarized signal is of Galactic origin, with polarized synchrotron emission dominating at the low-frequency end of the spectrum, and polarized thermal emission from Galactic dust dominating at the high-frequency end. At 353\,GHz, which is therefore the highest-frequency polarization-sensitive channel of {\Planck}, polarized thermal dust emission is about two orders of magnitude stronger than the polarized cosmic microwave background~\citep{planck2014-a01}. It is therefore the channel we use to study this Galactic emission, and several \Planck\ papers have already provided analyses of earlier releases of this data to investigate the link between dust polarization and physical properties of the ISM, most notably the structure of the Galactic magnetic field, properties of dust grains, and interstellar turbulence. In Appendix~\ref{sec:appendix:PlanckPapers}, we provide a summary of the main results of these {\Planck} papers, to serve as a useful reference.

In this paper, one in a series associated with the 2018 release of data from the \Planck\ mission \citep{planck2016-l01}, we use all-sky maps of dust polarized emission produced from this third public release of \Planck\ data (hereafter the {\DRThree} data release or PR3) to expand on some of these studies of the polarized thermal emission from Galactic dust. More specifically, our analysis first focuses on a refined statistical analysis of the dust emission's polarization fraction and polarization angle over the full sky, in the fashion of \cite{planck2014-XIX} but based on a post-processing of the {\DRThree} data that minimizes the contamination from components other than dust. One of the results from that paper, confirmed by a comparison with numerical simulations of magnetohydrodynamical (MHD) interstellar turbulence~\citep{planck2014-XX}, is the nearly inverse proportionality of the polarization fraction $\PsI$ and the local dispersion of polarization angles $\mathcal{S}$. Here we propose an interpretation of this relationship, showing that it is a generic result of the turbulent nature of interstellar magnetic fields. We therefore further analyse the \Planck\ data by considering the product $\StimesPsI$, which allows us to search for deviations from this first-order relationship. Deviations might be related to changes in the properties of the dust or of its alignment with respect to the magnetic field. In the final part of the paper, we present an updated comparison of the dust polarized emission with stellar polarization data in the visible, following \cite{planck2014-XXI}, but with a much larger sample of stellar polarization data. For aspects of the analysis of polarized thermal dust emission related to component-separation, i.e., the angular power spectra and spectral energy distributions (SEDs) of the $E$ and $B$ modes, we refer the reader to \cite{planck2016-l11A}.

The paper is organized as follows. In Sect.~\ref{sec:PlanckMaps}, we present the \planck\ maps of Stokes parameters that are used in the subsequent analysis. In Sect.~\ref{sec:full-sky-polarization-maps}, we present the full-sky maps of thermal dust polarization derived from these Stokes maps. In Sect.~\ref{sec:full-sky-analysis}, we present a statistical overview of these dust polarization maps over the full sky, using the tools and analysis introduced in~\cite{planck2014-XIX}. In Sect.~\ref{sec:NewResults} we expand on this statistical analysis, looking for trends beyond the first-order correlations exhibited by the data. In Sect.~\ref{sec:stars}, we update our comparison with the stellar polarization data, greatly expanding on the sample presented initially in~\cite{planck2014-XXI}. Finally, Sect.~\ref{sec:conclusions} presents our conclusions. Seven appendices complete the paper. In Appendix~\ref{sec:appendix:PlanckPapers}, as already mentioned, we offer a summary of the main results of earlier {\Planck} papers dealing with the polarized thermal emission from Galactic dust. In Appendix~\ref{sec:appendix:GNILC}, we show complementary, variable resolution Stokes maps at 353\,GHz and present the Stokes covariance maps that are used to assess the statistical uncertainties affecting \Planck\ polarization data presented in this work. Appendix~\ref{sec:appendix:E2E} describes our approach to estimating the systematic uncertainties in the data, based on a set of end-to-end (E2E) simulations. Appendix~\ref{sec:appendix:SdP} explains the relationship of the polarization angle dispersion function $\S$ to the polarization gradients commonly used in polarization studies at lower frequencies. Appendix~\ref{sec:appendix:PsI-other-offsets} provides supplementary figures showing how the behaviour of polarization fraction with total gas column density is affected by the uncertainty on the Galactic zero level. Appendix~\ref{sec:appendix:Sp} provides a demonstration of the inverse relationship between the polarization fraction $\PsI$ and the polarization angle dispersion function $\S$, based on a phenomenological model of magnetized interstellar turbulence. 
Finally, Appendix~\ref{sec:appendix:nands} assesses the noise and systematics that affect the data used in the comparison of visible and submillimetre polarization properties (Sect.~\ref{sec:stars}).

\section{Processing \Planck\ maps for Galactic science}
\label{sec:PlanckMaps}

The Stokes $I$, $Q$, and $U$ maps at 353\,GHz that we use in this paper are based on products from the {\DRThree} data release. The processing steps applied to the data to compute the \DRThree\ frequency maps are presented in~\cite{planck2016-l02} and~\cite{planck2016-l03} for the Low Frequency Instrument (LFI) and High Frequency Instrument (HFI), respectively. For HFI, the $Q$ and $U$ products used at 353\,GHz make use of the polarization-sensitive bolometers (PSBs) only, ignoring the spider-web bolometer (SWB) data, as recommended in~\cite{planck2016-l03}, while the rest, including $I$ at 353\,GHz, make use of the complete data set (PSB+SWB).

For our Galactic science applications, we use maps that result from post-processing with the Generalized Needlet Internal Linear Combination (\GNILC) algorithm, developed by~\cite{Remazeilles2011b}; this filters out the cosmic infrared background (CIB) anisotropies, a key feature for Galactic science. These \GNILC\ maps, derived from the {\DRThree} maps, are presented and characterized in~\cite{planck2016-l04}, and so we simply recall a few key properties of this post-processing step in the next subsection (Sect.~\ref{sec:GNILCprocessing}). The \GNILC\ maps used here have a uniform resolution of 80\arcmin.

In Sects.~\ref{subsec:gbregions},~\ref{view}, and~\ref{sec:stars}, where we require data at a uniform resolution that is finer than 80\arcmin, and where we are less concerned by the presence of CIB anisotropies, we use maps derived more directly from the {\DRThree} 353\,GHz Stokes maps and their covariance maps. The required post-processing to produce these alternative Stokes maps (\asmaps) is also described below.

As another important post-processing step, we need to establish the desired zero level in the total intensity maps for Galactic dust emission, as described in Sect.~\ref{sec:I-offset}.

\subsection{\GNILC\ and \asmap\ post-processing}
\label{sec:GNILCprocessing}

\GNILC\ is a wavelet-based component-separation method that makes use of both spectral and spatial information to disentangle multidimensional components of the sky emission. In practice it combines data from the different \Planck\ bands and outputs maps at any desired frequency.

In \citet{planck2016-XLVIII}, \GNILC\ was applied to \Planck\ 2015 total intensity data, effectively separating Galactic thermal dust emission and CIB anisotropies over the entire sky, while simultaneously filtering out noise and cosmic microwave background (CMB) contributions. In regions of low dust column density, it was found that the CIB anisotropies are well above the noise, correlated spatially, and provide a significant contribution to the emission power spectrum.  We are interested in polarization properties for Galactic dust emission over the full sky, including high-latitude diffuse lines of sight, for which {\GNILC}-processing significantly reduces contamination of the $I$ map by CIB anisotropies. 

For the {\DRThree} data release we go further, applying \GNILC\ not only in total intensity, but also in polarization, thus providing maps of polarized Galactic thermal dust emission in which the contamination by polarized CMB emission and instrumental noise has been reduced.

The \GNILC\ algorithm optimizes the component separation given the local variations of the contamination. At high Galactic latitudes and small angular scales, the local dimension of the Galactic signal subspace estimated by \GNILC, i.e., the number of components in the Galactic signal, can be null because in this regime the data become compatible with a mixture of CIB, CMB, and noise.\footnote{In the case of polarized intensity, the CIB is assumed not to contribute to the signal.} Therefore, the effective resolution of the \GNILC\ dust maps is not uniform but variable over the sky, with an effective beam whose full-width at half-maximum (FWHM) increases from the Galactic plane towards high latitudes. The local resolution depends on the local signal-to-nuisance ratio, which varies differently for intensity and for $E$- and $B$-mode polarization.\footnote{In practice, \GNILC\ ingests full-sky $Q$ and $U$ maps, converts these to $E$ and $B$ maps for component separation, and then converts back to $Q$ and $U$ for the output maps.} 
Therefore, the optimal \GNILC\ resolution should, by design, be different for total intensity and for polarization maps. However, for consistency in the astrophysical study of dust intensity and polarization, where the polarization fraction $\PsI = P/I$ is of interest, we adopt a common resolution by imposing that the variable resolution of the \GNILC\ dust maps should be driven by the more stringent signal-to-nuisance ratio of the $B$-mode data. In practice, in the Galactic plane the signal-to-noise ratio (S/N) in polarization is sufficiently large to allow for the use of the nominal \Planck\ resolution at 353\,GHz, while for high Galactic latitude regions data smoothed to 80\arcmin\ are required.

The \GNILC\ method is also able to provide Stokes maps at a uniform resolution of 80\arcmin\ over the entire sky, enabling the analysis of polarization properties over the entire sky at a common resolution.  It should be noted that in this case, and to avoid oversampling, the output maps are subsequently downgraded from the original {\healpix}\footnote{{\tt https://healpix.jpl.nasa.gov}}~\citep{Gorski05} resolution $N_\mathrm{side}=2048$ to $N_\mathrm{side}=128$.

The equivalent \asmap\ post-processing step is to subtract the total intensity CMB {\tt SMICA} map~\citep{cardoso2008,planck2016-l04} from the \DRThree\ total intensity map at 353\,GHz. No subtraction of CIB anisotropies is performed. Compared to the dust signal at 353\,GHz, the CMB polarized signal is small, less than 1\,\%~\citep{planck2016-l04}, and subtracting that would add noise unnecessarily.

\subsection{Zero level for total intensity of Galactic thermal dust emission}
\label{sec:I-offset}

We recall that \Planck\ had very little sensitivity to the absolute level of emission and so the zero level of the maps of $I$ must be set using ancillary data. This is of central interest for our study, because for the most diffuse lines of sight it directly impacts polarization fractions through $\PsI = P/I$.

\DRThree\ HFI frequency maps, as delivered \citep{planck2016-l03}, deliberately include a model of the CIB monopole.  As a first step towards maps suitable for Galactic science, this needs to be subtracted. \GNILC\ post-processing does not adjust the monopoles contained in the input maps and so the CIB monopole needs to be subtracted explicitly, frequency by frequency, as for \asmaps.  At 353\,GHz the intensity of the model CIB monopole is $0.13\,\mathrm{MJy\,sr^{-1}}$, or $452\,\mu\mathrm{K_{CMB}}$ using the unit conversion $287.5\,\mathrm{MJy\,sr^{-1}\,K_{CMB}^{-1}}$ given in~\cite{planck2016-l03}.

This CIB-subtracted total intensity map has a zero level that by construction is based on a correlation of the emission at high Galactic latitudes with the column density of the ISM traced by the 21-cm emission of {\sc Hi} at low column densities.  Nevertheless, this Galactic offset needs to be refined. A favoured method is again based on a correlation of dust emission with {\sc Hi}, as described in~\cite{planck2013-p03f},~\cite{planck2013-p06b},~\cite{planck2016-XLVIII}, and \cite{planck2016-l03}. After the \GNILC\ processing, we apply the same {\sc Hi} correlation procedure to the output maps of $I$, in particular finding that a Galactic {\sc Hi} offset of $36\,\mu\mathrm{K_{CMB}}$ should be added to the 353-GHz \GNILC\ total intensity map used for polarization at the uniform 80\arcmin\ resolution.  The statistical error of about $2\,\mu\mathrm{K_{CMB}}$ is small compared to the systematic uncertainties that we now discuss.

Because the dust total intensity versus {\sc Hi} correlation has an upward curvature, the estimates of the offset and slope are dependent on the column density range used for the fit. Furthermore, there is an additional source of uncertainty, related to the possibly significant emission from dust that is in the warm ionized medium (WIM), and therefore associated with {\sc Hii} rather than with neutral hydrogen {\sc Hi}.  The fractional contribution might be most important at low {\sc Hi} column densities, i.e., in the diffuse ISM.

To assess the systematic effect related to the WIM-associated dust, we rely on an estimate of the total column density of the WIM towards high Galactic latitudes by~\cite{gaensler-et-al-2008}, $N_\mathrm{H,{WIM}}=8 \times 10^{19}\,\mathrm{cm^{-2}}$. Assuming the same SED in the submillimetre per proton as per H atom, and using the results of~\cite{planck2013-XVII}, this translates to 54\,$\mu\mathrm{K_{CMB}}$ at 353\,GHz. If all of the dust emission associated with the WIM were uncorrelated with the {\sc Hi}-associated dust, then this value would need to be added to the Galactic {\sc Hi} offset. On the other hand, part of any dust emission associated with the WIM is probably correlated with {\sc Hi} as well, and in the extreme case of 100\,\% correlation, there would be no correction due to the WIM. 

To account for this effect, we adopt a central value of $27\,\mu\mathrm{K_{CMB}}$ which, when added to the Galactic {\sc Hi} offset, gives a fiducial total Galactic offset of $\goff\,\mu\mathrm{K_{CMB}}$ (corresponding to $0.0181\,\mathrm{MJy\,sr^{-1}}$ at 353\,GHz), to be added back to the \GNILC\ total intensity map at 353\,GHz, after the CIB monopole subtraction.
This fiducial value will be used in the rest of our analysis.
It has an uncertainty that we estimate to be $\goffu\,\mu\mathrm{K_{CMB}}$ (corresponding to $\pm0.0115\,\mathrm{MJy\,sr^{-1}}$). As mentioned above, the offset affects the statistics of the polarization fraction of dust polarized emission. To quantify the effect of an offset uncertainty in the range estimated, we also use intensity maps resulting from the addition of a total Galactic offset of $\goffl\,\mu\mathrm{K_{CMB}}$ ($0.0066\,\mathrm{MJy\,sr^{-1}}$) and $\goffh\,\mu\mathrm{K_{CMB}}$ ($0.0296\,\mathrm{MJy\,sr^{-1}}$), referred to as {\it low} and {\it high}, respectively. Note, however, that these correspond to fainter and brighter intensity maps, leading to higher and lower polarization fractions, respectively.

The procedure to adjust the \asmap\ intensity map at 353\,GHz after CIB-monopole subtraction is the same. In this case the fiducial Galactic offset is $\goffasm\,\mu\mathrm{K_{CMB}}$, a value that is, not surprisingly, very close to that for {\GNILC}.

\subsection{\GNILC\ Stokes maps}
\label{sec:gsm}

\begin{figure}
\includegraphics[width=0.5\textwidth]{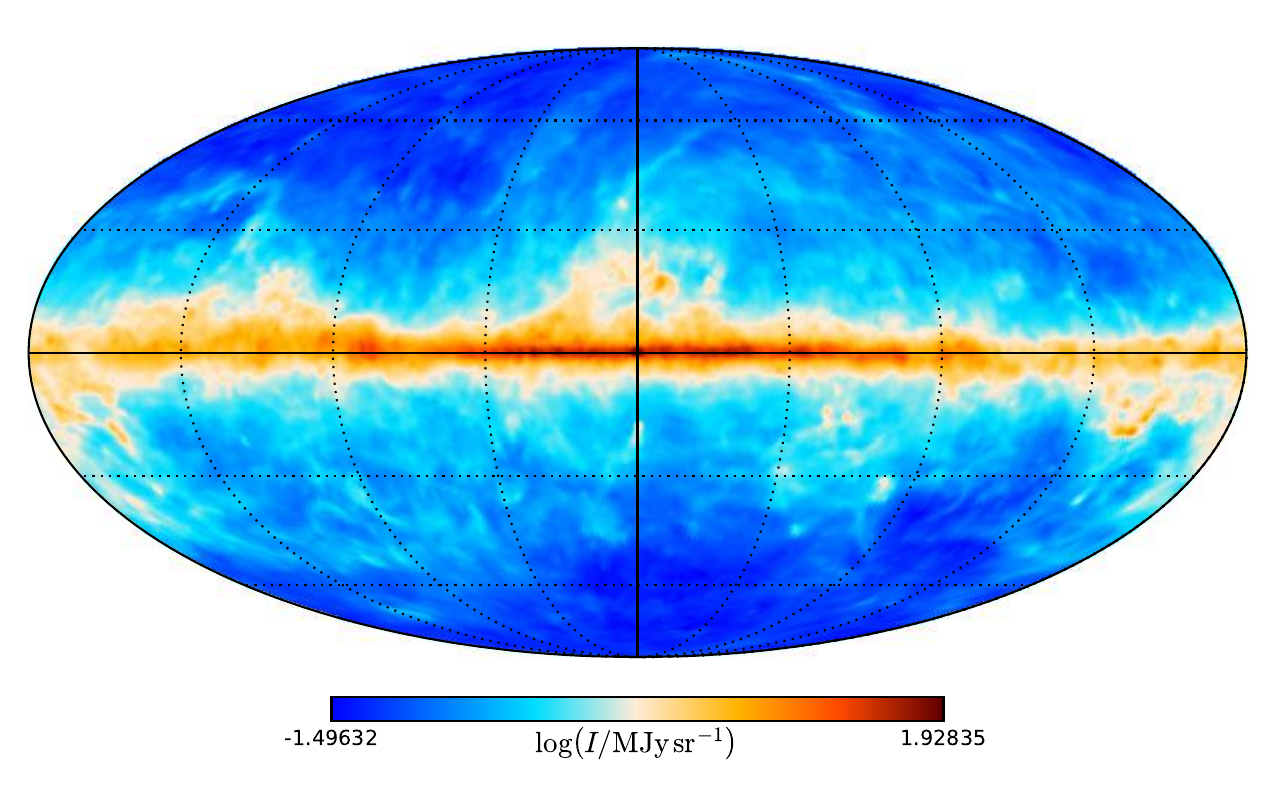}
\includegraphics[width=0.5\textwidth,trim={0 0 0 0.5cm},clip]{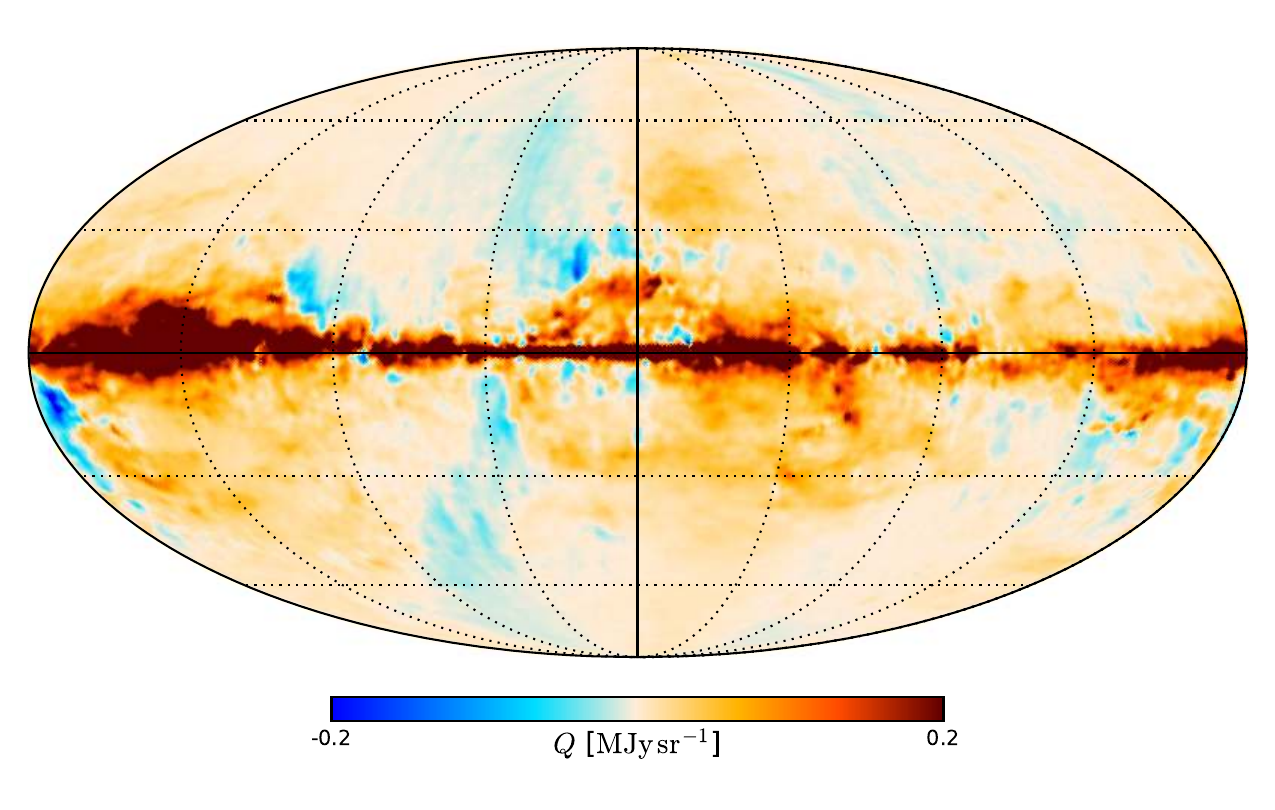}
\includegraphics[width=0.5\textwidth,trim={0 0 0 0.5cm},clip]{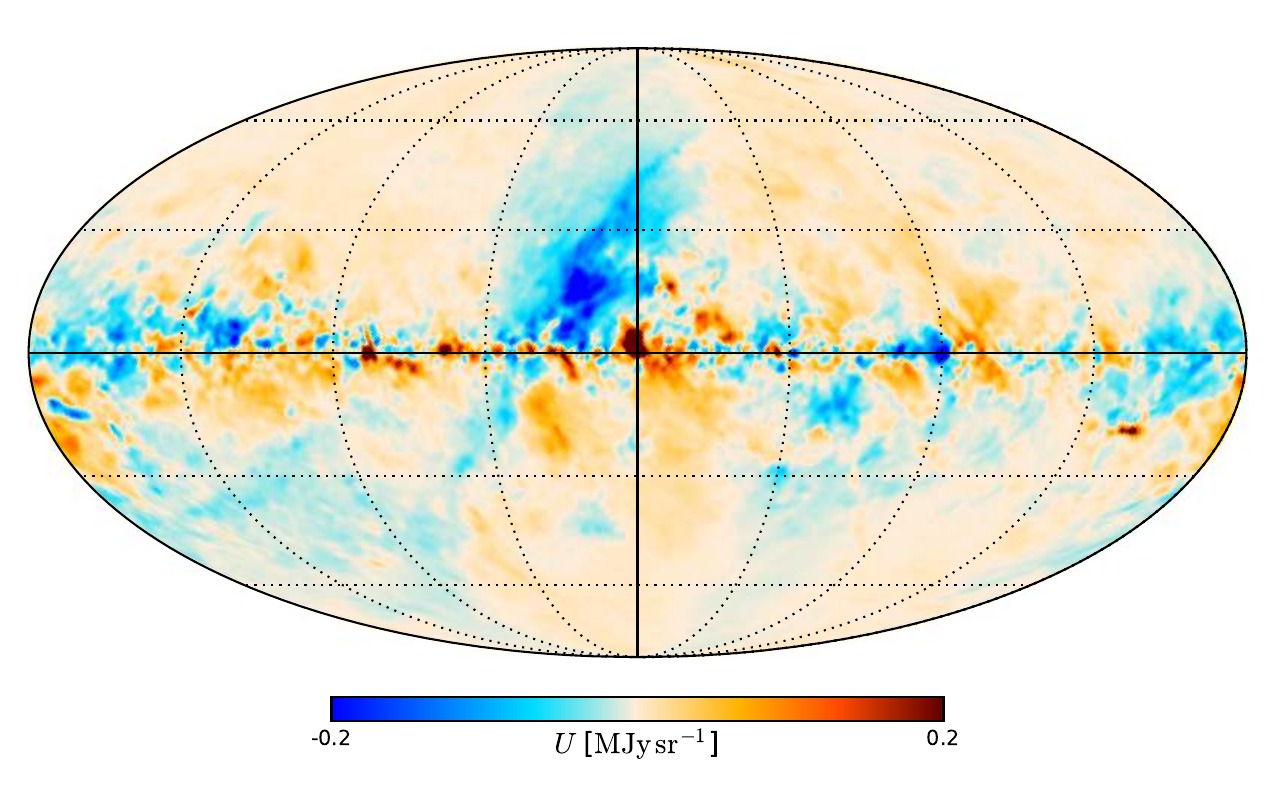}
\includegraphics[width=0.5\textwidth,trim={0 0 0 0.5cm},clip]{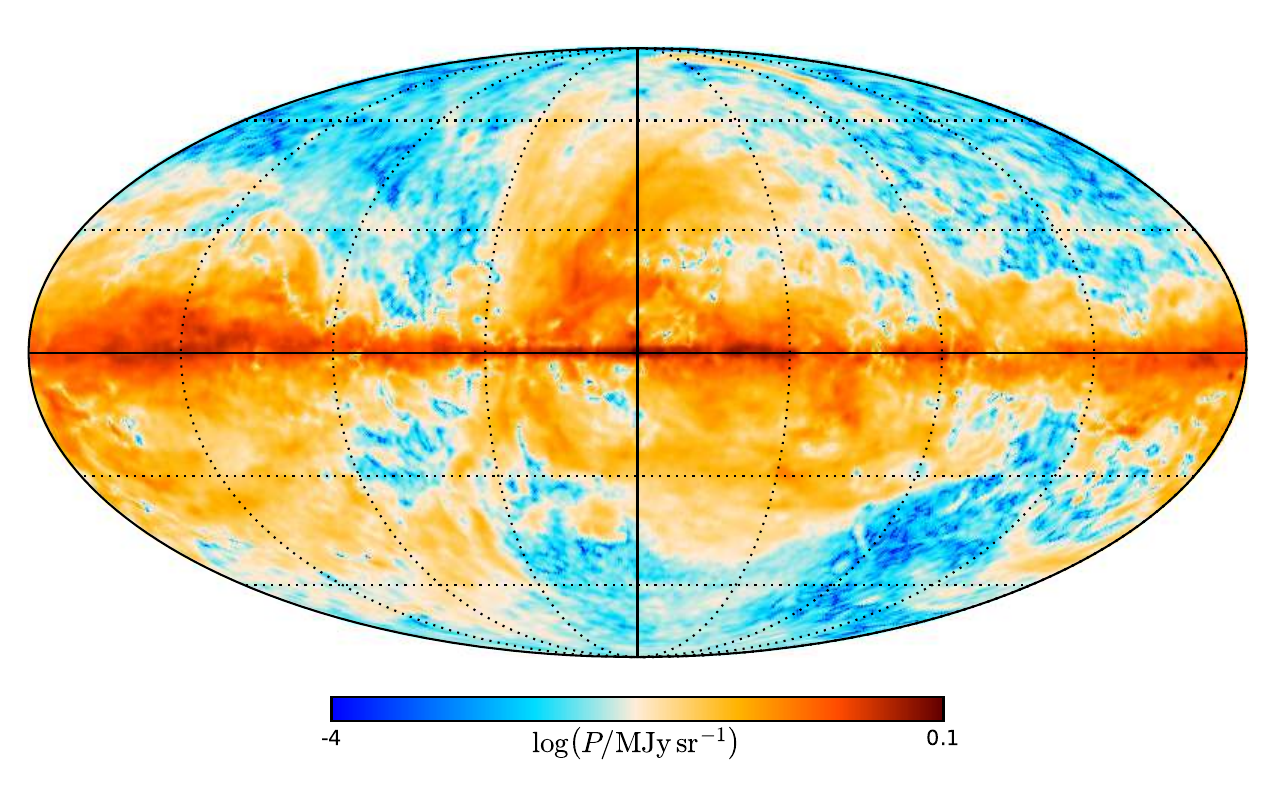}
\caption[]{From top to bottom: \GNILC\ maps of Stokes $I$, $Q$, and $U$, and polarized intensity $P$ at 353\,GHz and uniform 80\arcmin\ resolution in Galactic coordinates, centred on the Galactic centre (GC). The Galactic plane (GP) appears clearly in all maps. The scales for $I$ and $P$ are logarithmic, while those for $Q$ and $U$ are linear.}
\label{fig:IQU_353_GNILC-uniform}
\end{figure}

For the 353-GHz data used here, after the adjustments of the zero level of $I$ just discussed, the \GNILC\ Stokes $I$, $Q$, and $U$ maps are converted to astrophysical units using the already mentioned conversion factor $287.5\,\mathrm{MJy\,sr^{-1}\,K_{CMB}^{-1}}$. The resulting \GNILC\ Stokes maps at 353\,GHz and uniform 80\arcmin\ resolution are shown\footnote{In this paper, all maps are shown either with a Mollweide projection of the full sky, in Galactic coordinates centred on the Galactic centre (GC), or with an orthographic projection of both hemispheres. In this latter case, the northern Galactic hemisphere is always on the left and the southern Galactic hemisphere on the right, with the rotation of each hemisphere such that the Galactic centre ($l=0^\circ$) is towards the top.} in Fig.~\ref{fig:IQU_353_GNILC-uniform}. The total intensity map corresponds to the fiducial offset value. The \GNILC\ Stokes maps at 353\,GHz and variable resolution over the sky are shown in Appendix~\ref{sec:appendix:GNILC}, alongside the \GNILC-processed covariance maps $\sII$, $\sIQ$, $\sIU$, $\sQQ$, $\sQU$, and $\sUU$ that are used in Sect.~\ref{sec:pol_uncertainties} to estimate the statistical uncertainties on the dust polarization properties.

We note that for studies involving the polarization angle dispersion function $\mathcal{S}$ (Sect.~\ref{sec:S_maps_sec3}), we use Stokes maps and covariance maps that are further smoothed to a 160\arcmin\ FWHM uniform resolution, and downgraded to $N_\mathrm{side}=64$.

\subsection{Alternative Stokes maps (\asmaps)}
\label{sec:asm}

For \asmaps, as a final step after converting to astrophysical units, we smooth the Stokes $I$, $Q$, and $U$ maps uniformly to 10\arcmin, 20\arcmin, 40\arcmin, 60\arcmin, 80\arcmin, and 160\arcmin, downgrading the \healpix\ resolution to $N_\mathrm{side}=1024$, $N_\mathrm{side}=512$, $N_\mathrm{side}=256$, $N_\mathrm{side}=128$, and $N_\mathrm{side}=64$, respectively. The covariance matrix maps $\sII$, $\sIQ$, $\sIU$, $\sQQ$, $\sQU$, and $\sUU$ are consistently smoothed from \DRThree\ data to the same resolutions using the procedure described in Appendix A of~\cite{planck2014-XIX}.

\section{Full-sky thermal dust polarization maps}
\label{sec:full-sky-polarization-maps}
In this section, we present the maps of Galactic thermal dust polarization over the full sky, derived from the \GNILC-processed Stokes $I$, $Q$, and $U$ maps at uniform 80\arcmin\ resolution. 

\subsection{Polarization fraction and angle maps}
\label{sec:p_maps}

From the \GNILC\ maps of Stokes parameters $I$, $Q$, and $U$ at 353\,GHz, we build maps of the polarized intensity $P$, polarization fraction $\PsI$, and polarization angle $\psi$. The convention used for the Stokes parameters in the {\DRThree} data release is to measure polarization angles from the direction of the Galactic north and positively towards Galactic west in accordance with the {\healpix} convention used in cosmology~\citep[see][for further discussion]{planck2016-ES}. However, as in 
\citet{planck2014-XIX}, we conform here to the IAU convention, polarization angles $\psi$ being counted positively towards Galactic east, and so they are computed simply by changing the sign of Stokes $U$ in the {\DRThree} data. Thus
\begin{equation}
\label{eq:p_psi_def}
P=\sqrt{Q^2+U^2} \qquad p=\frac{P}{I} \qquad \psi=\frac{1}{2}\mathrm{atan2}(-U,Q) \, ,
\end{equation}
where the two-argument function $\mathrm{atan2}(-U,Q)$ is used in place of $\mathrm{atan}(-U/Q)$ to avoid the $\pi$-ambiguity. Conversely, the Stokes parameters can be recovered from the total intensity, the polarization fraction, and the polarization angle via
\begin{equation}
Q=p\,I\cos\left(2\psi\right) \qquad U=-p\,I\sin\left(2\psi\right)\, .
\end{equation}

The presence of noise in the Stokes maps can bias the estimates of $P$, $\PsI$, and $\psi$ \citep{PMA1,PMA2}, so that naive estimators $\hat{P}$, $\hat{p}$, and $\hat{\psi}$ computed using Eq.~\eqref{eq:p_psi_def} directly on the noisy data do not adequately represent the true values at low S/N. Alternative estimators have been developed, most notably for the polarized intensity and the polarization fraction (the bias on the polarization angle is usually negligible). For the polarization fraction, we use the modified asymptotic (MAS) estimator introduced by~\cite{plaszczynski14} and defined through
\begin{equation}
p_\mathrm{MAS}=\hat{p}-\varsigma^2\frac{1-e^{-\hat{p}^2/\varsigma^2}}{2\hat{p}} \, ,
\end{equation}
where $\varsigma$ is a noise-bias parameter that depends on the geometrical properties of the (assumed Gaussian) 2-dimensional distribution of the noise in $(Q,U)$ space, assuming a noise-free total intensity $I$. From the 353-GHz \GNILC\ covariance matrices at the uniform 80\arcmin\ resolution, we can compute this noise-bias parameter and find that $\varsigma^2<10^{-3}$ over the full sky, which shows that the debiasing performed by the MAS estimator is small.

\begin{figure*}[htbp]
\centerline{\includegraphics[width=0.5\textwidth]{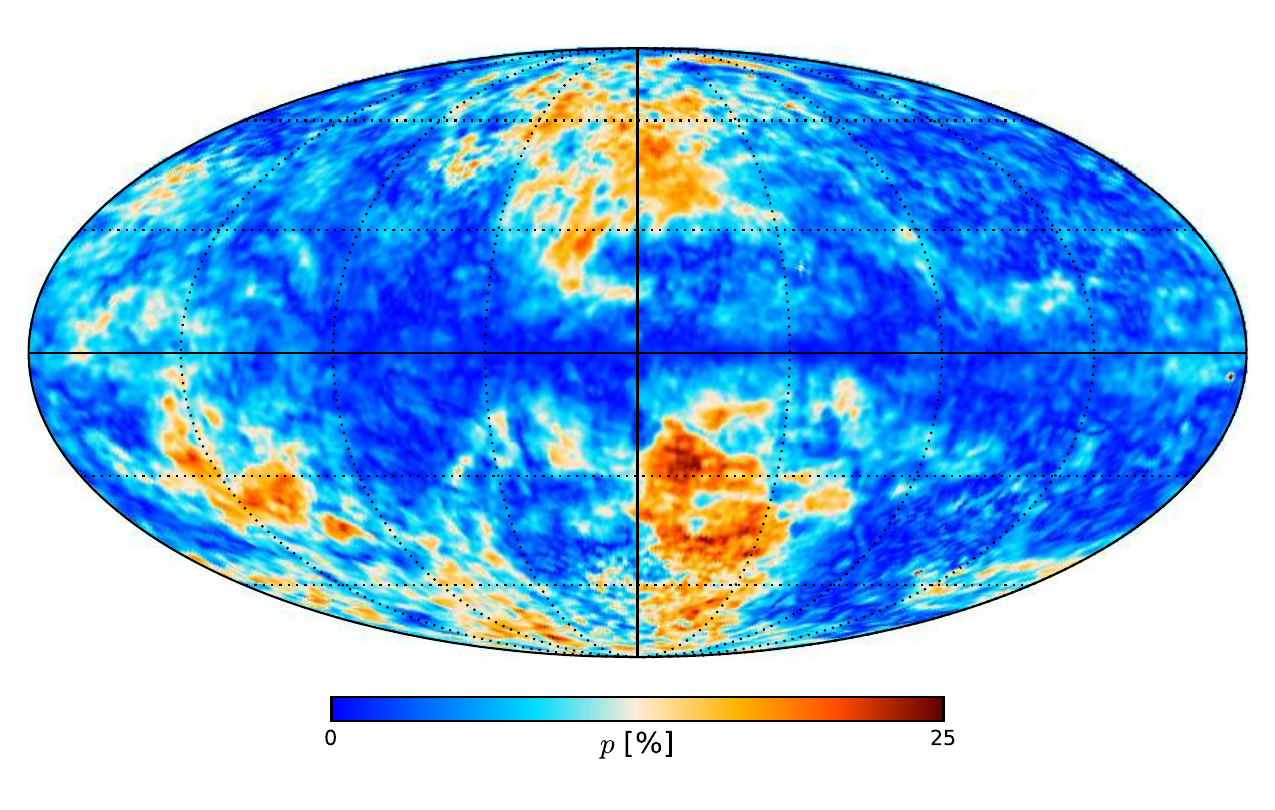}
\includegraphics[width=0.5\textwidth]{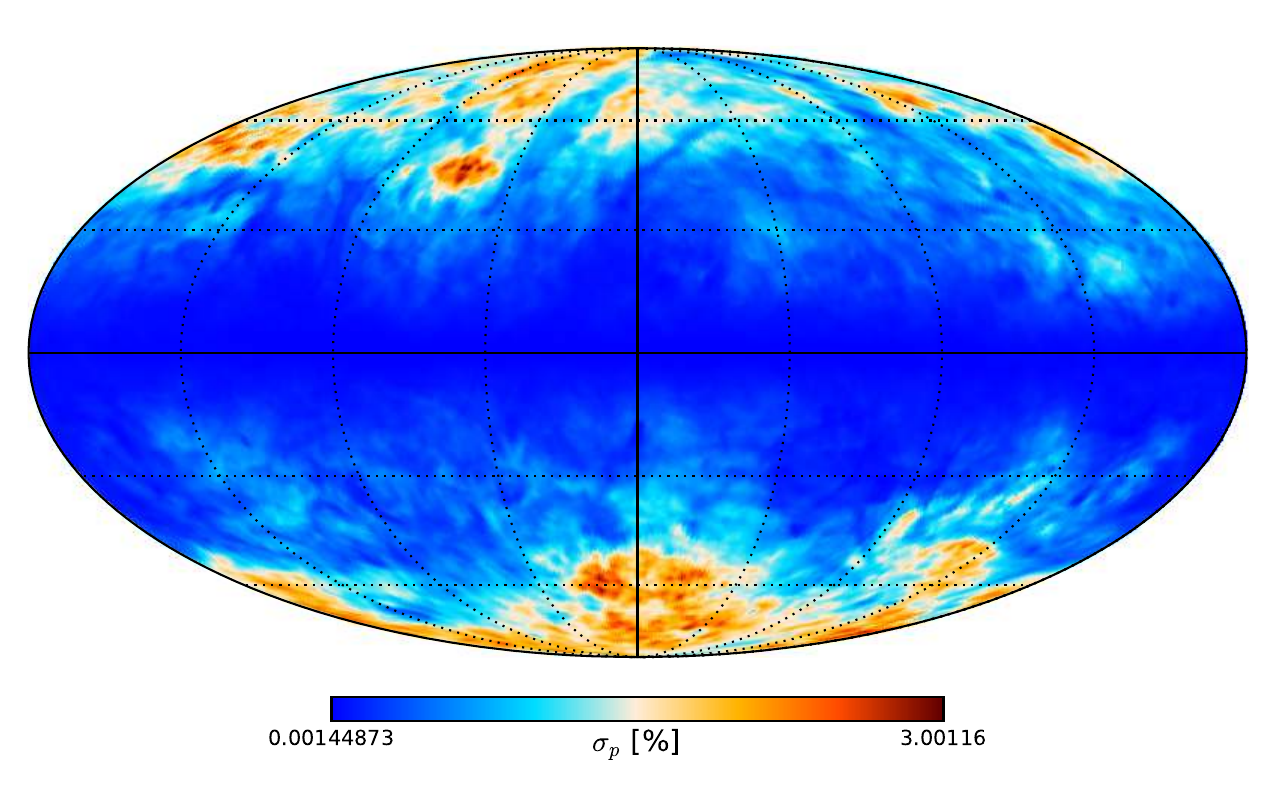}
}
\centerline{\includegraphics[width=0.5\textwidth]{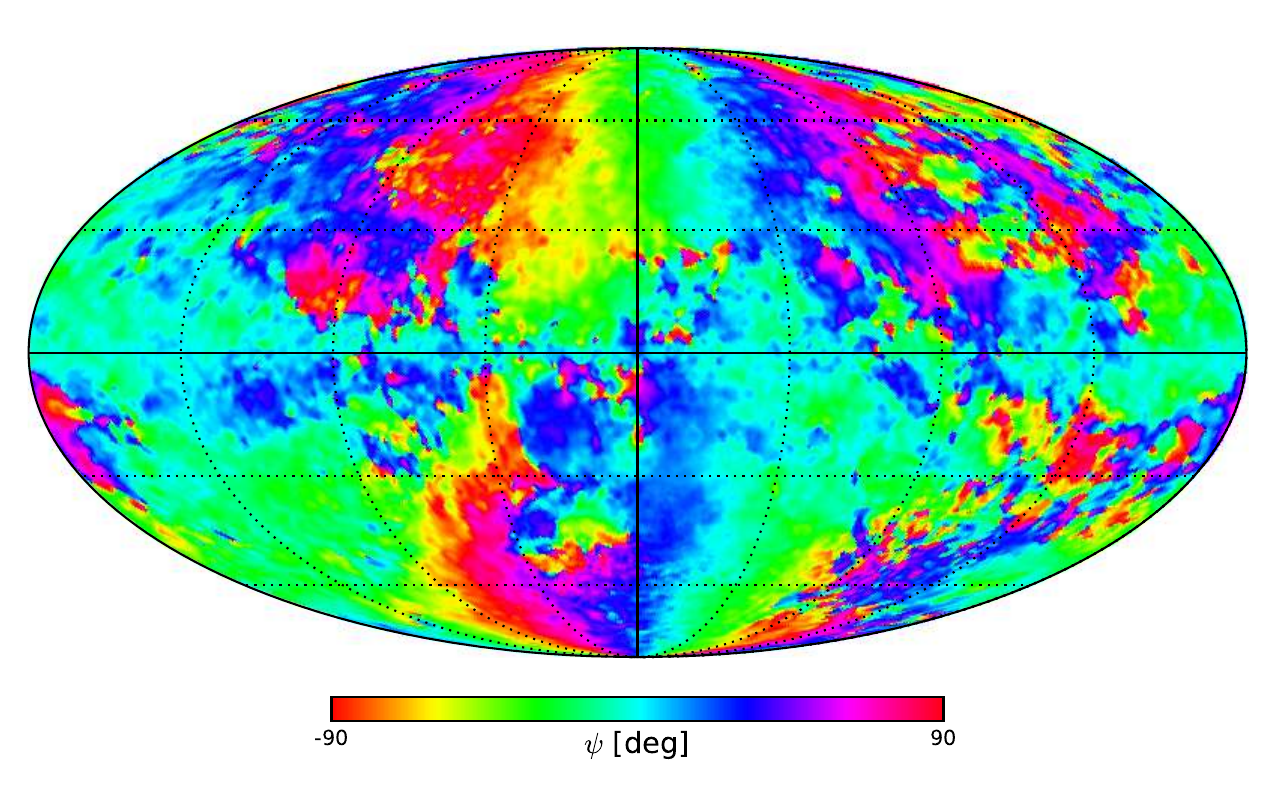}
\includegraphics[width=0.5\textwidth]{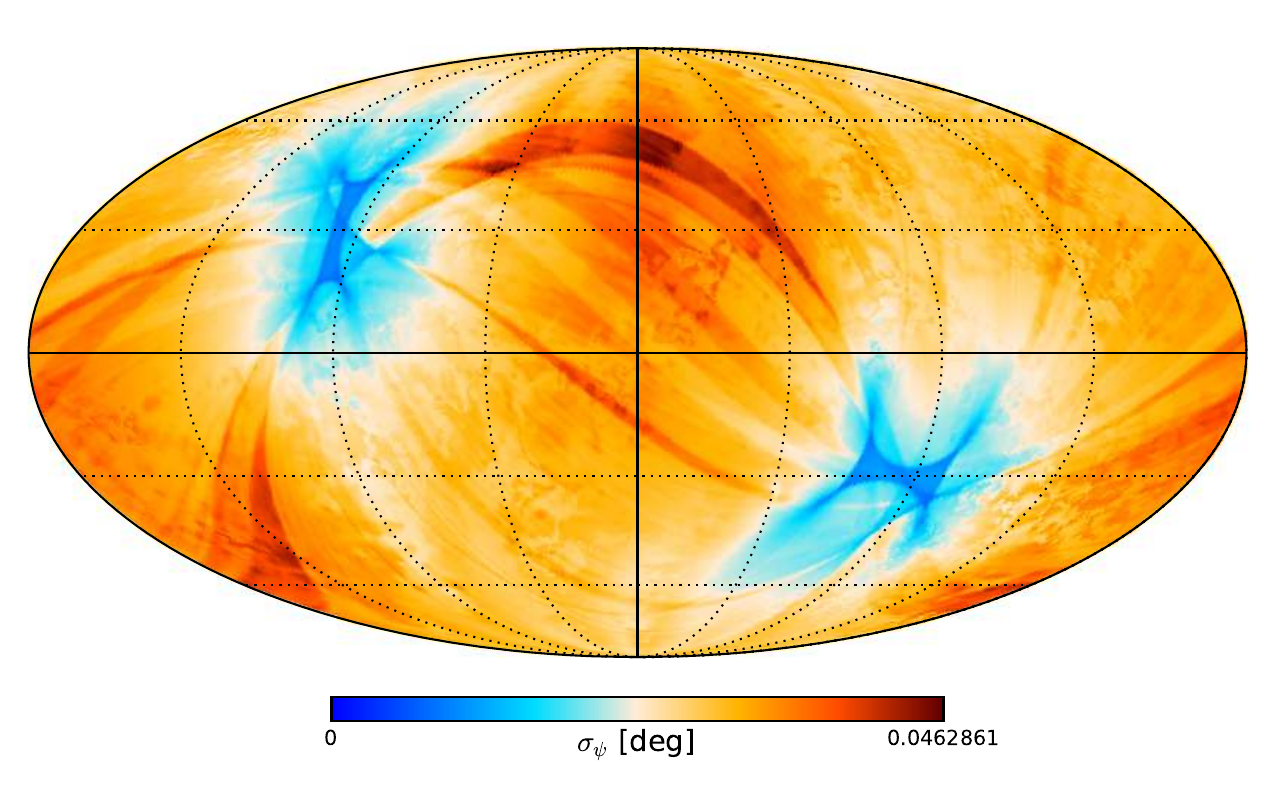}
}
 \caption[]{
Polarization maps for the \GNILC-processed data at 353\,GHz and uniform 80\arcmin\ resolution: polarization fraction $\PsI$ (top left) and associated statistical uncertainty $\sigma_p$ (top right), polarization angle $\psi$ (bottom left) and associated statistical uncertainty $\sigma_\psi$ (bottom right). The pattern in the $\sigma_\psi$ map arises from the \Planck\ scanning strategy.
}
\label{fig:P_p_psi_GNILC-80arcmin}
\end{figure*}

Because the noise in total intensity is small, this is a reasonable approach that can also be used to provide a MAS estimate of the polarized intensity, $P_\mathrm{MAS}$. For notational simplicity, we hereafter drop the subscript `MAS' and write $\PsI$ to mean $\PsI_\mathrm{MAS}$ and $P$ to mean $P_\mathrm{MAS}$. For the \GNILC-processed 353-GHz data at the uniform 80\arcmin\ resolution, the resulting polarized intensity $P$ map is shown in Fig.~\ref{fig:IQU_353_GNILC-uniform} {(bottom row)}, while the polarization fraction $\PsI$ and the polarization angle $\psi$ maps are shown in Fig.~\ref{fig:P_p_psi_GNILC-80arcmin}. We note that the total intensity offset used for these maps is the fiducial one. The choice of offset has an impact on $\PsI$ (as we will discuss in Sect.~\ref{sec:DFs}) but not on $\psi$ or $P$.

The overall structure of the polarization fraction and angle maps is consistent with that found over a smaller fraction of the sky in~\cite{planck2014-XIX}. We note in particular that the Galactic plane exhibits low polarization fractions, except towards the `Fan' region, near the anticentre, and that the structures seen in $\PsI$ do not generally correspond to structures in total intensity. The polarization angle map $\psi$ shows that the magnetic field is essentially parallel to the Galactic plane at low Galactic latitudes $|b|$, and the large-scale patterns at higher latitudes can be broadly interpreted as arising from the projection of the local magnetic field in the Solar neighbourhood~\citep{planck2016-XLIV,alves-et-al-2018}.

\subsection{Estimation of uncertainties}
\label{sec:pol_uncertainties}

There are several types of uncertainties that need to be taken into account in our analysis of the dust polarization maps. 

First, there is statistical noise, whose contribution to the uncertainties can be estimated using the covariance maps of the \GNILC-processed data. This was evaluated by performing a set of Monte Carlo simulations of Stokes $I$, $Q$, and $U$ maps, taking the \GNILC\ maps as means of a multivariate normal distribution with covariances given by the \GNILC\ covariance maps.\footnote{This procedure results in simulations containing twice as much noise as the original data; however, our main purpose is not to estimate the statistical noise precisely, but rather to assess whether bias is significant.} A set of 1000 simulations was computed; results for a set half this size do not change significantly, confirming that 1000 is sufficient. From these simulations we computed 1000 maps of not only $\PsI$ and $\psi$, but also other derived quantities, such as the polarization angle dispersion function (Sect.~\ref{sec:S_maps_sec3}). As discussed in Sect.~\ref{sec:full-sky-analysis}, these are instrumental in detecting any remaining bias (after using the MAS estimator), by investigating whether statistical properties (e.g., the histogram of $\PsI$) computed on the \GNILC\ maps shown in Fig.~\ref{fig:P_p_psi_GNILC-80arcmin}, are compatible with the ensemble average of the same properties computed on the Monte Carlo simulations. When, as expected, the quantities in the polarization maps are unbiased, the standard deviations of these maps, and of any derived quantity that we compute using the simulations, yield reliable statistical uncertainties. 

\begin{figure}[htbp]
\includegraphics[width=0.5\textwidth]{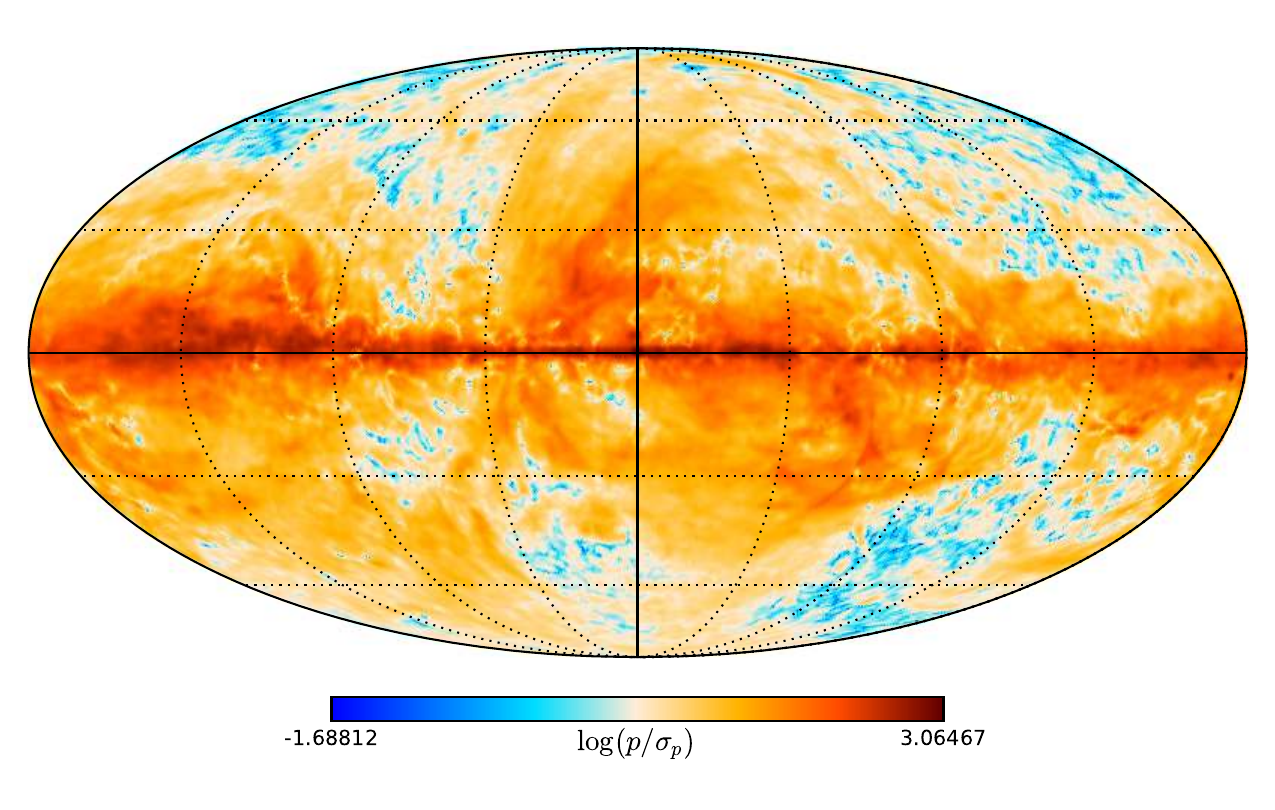}
\includegraphics[width=0.5\textwidth]{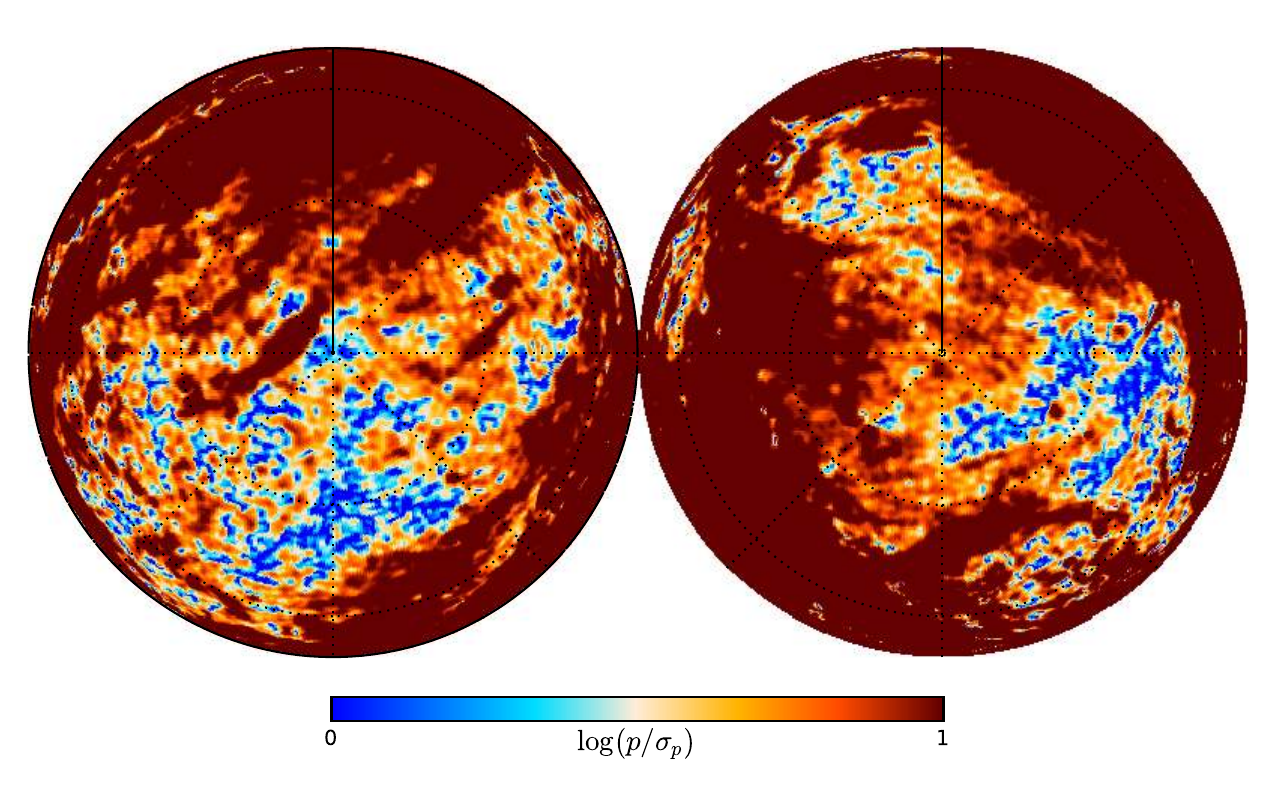}
 \caption[]{
Signal-to-noise ratio (S/N) $\PsI/\sigma_p$ for the polarization fraction in the \GNILC-processed data at 353\,GHz and uniform 80\arcmin\ resolution. The polar view (bottom) uses a range $1\leqslant p/\sigma_p\leqslant 10$ to bring out low S/N regions.
}
\label{fig:sigma_p_psi_GNILC_80acm}
\end{figure}

Using this approach, we computed polarization fraction and polarization angle uncertainty maps $\sigma_p$ and $\sigma_\psi$ (shown in Fig.~\ref{fig:P_p_psi_GNILC-80arcmin}). These are actually very close to the ones obtained using equations B.2 and B.3 of \citet{planck2014-XIX}, which are valid at sufficiently high S/N in polarization $\PsI/\sigma_p$.\footnote{We note a typo in equation B.3 of \citet{planck2014-XIX}, which should include a factor of $1/P$ on the right-hand side.} Fig.~\ref{fig:sigma_p_psi_GNILC_80acm} shows the polarization S/N map for the \GNILC-processed data at 353\,GHz and uniform 80\arcmin\ resolution. At this resolution, $\PsI/\sigma_p>3$ over most of the sky and thus the estimate of the S/N is robust~\citep{PMA1}.\footnote{No polarization S/N cut is applied in the following analyses of distribution functions and correlations.}

The statistical absolute uncertainty on polarization fractions is at most 3\,\%, and the statistical uncertainty on polarization angles is completely negligible, at less than $0.1^\circ$. Furthermore, based on the results of~\cite{PMA1}, we are confident that the polarization angle bias is less than 10\,\% of this value. Indeed, at 80\arcmin\ resolution, 99.9\,\% of the sky pixels have an effective ellipticity below 1.25. This quantity characterizes the asymmetry between the noise distributions on $Q$ and $U$ maps in a rotated reference frame that cancels correlated noise between the two. \citet{PMA1} show that in this case the bias on the polarization angle is at most of order 7--8\,\% of the statistical uncertainty $\sigma_\psi$.

Second, we need to estimate the impact of residual systematics arising from the \Planck\ data processing. This is accomplished via a set of 100 end-to-end (E2E) simulations that take a model sky as input, simulate the timelines of the instrument taking into account all known systematics, and process these simulated timelines with the mapmaking pipeline described in~\cite{planck2016-l03}. These E2E simulations are described in detail in appendix~A of~\cite{planck2016-l11A}. The statistical comparison between the input and output polarization maps, which we discuss in Appendix~\ref{sec:appendix:E2E}, shows that the absolute uncertainties from residual systematics are estimated to be $\pm0.5\,\%$ on $\PsI$ and $\pm8\deg$ on $\psi$. 

We note that these E2E simulations include realizations of random data noise and so already include part of the statistical uncertainty that is addressed by the Monte Carlo simulations based on the covariance matrices. 

Finally, as already mentioned in Sect.~\ref{sec:I-offset}, the quantitative analysis of $\PsI$ towards diffuse lines of sight depends strongly on the value of the Galactic offset used to set the zero level of total intensity for Galactic dust emission. To take this source of uncertainty into account, following the discussion in Sect.~\ref{sec:I-offset} we consider a fiducial case in which the Galactic offset is $\goff\,\mu\mathrm{K_{CMB}}$ and also consider a range of $\pm \goffu\,\mu\mathrm{K_{CMB}}$ about this central value.

\subsection{Polarization angle dispersion function}
\label{sec:S_maps_sec3}

The polarization angle dispersion function $\mathcal{S}$, introduced in~\citet{planck2014-XIX} quantifies the local (non-)uniformity of the polarization angle patterns on the sky by means of the local variance of the polarization angle map at a certain scale parameterized by a lag $\delta$. It is defined as
\begin{equation}
\label{eq:defS}
\mathcal{S}\left(\boldsymbol{r},\delta\right)=\sqrt{\frac{1}{N}\sum_{i=1}^N\left[\psi(\boldsymbol{r}+\boldsymbol{\delta}_i)-\psi(\boldsymbol{r})\right]^2} \, ,
\end{equation}
where the sum extends over the $N$ pixels, indexed by $i$ and located at positions $\boldsymbol{r}+\boldsymbol{\delta}_i$, within an annulus centred on $\boldsymbol{r}$ and having inner and outer radii $\delta/2$ and $3\delta/2$, respectively. Regions where the polarization angle tends to be uniform exhibit low values of $\mathcal{S}$, while regions where the polarization patterns are more chaotic exhibit larger values, with $\mathcal{S}=\pi/\sqrt{12}\approx52^\circ$ when the polarization angles are completely uncorrelated spatially.

\begin{figure}
\includegraphics[width=0.5\textwidth]{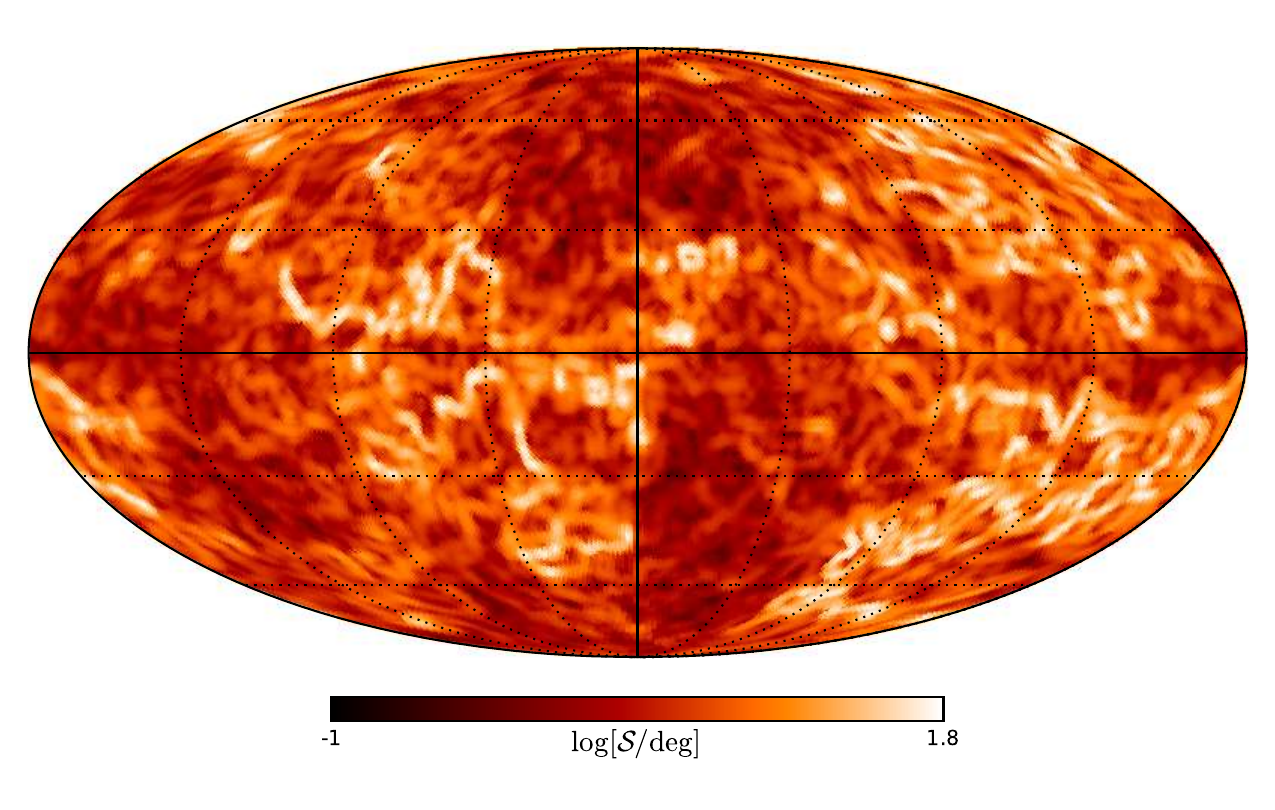}
\includegraphics[width=0.5\textwidth]{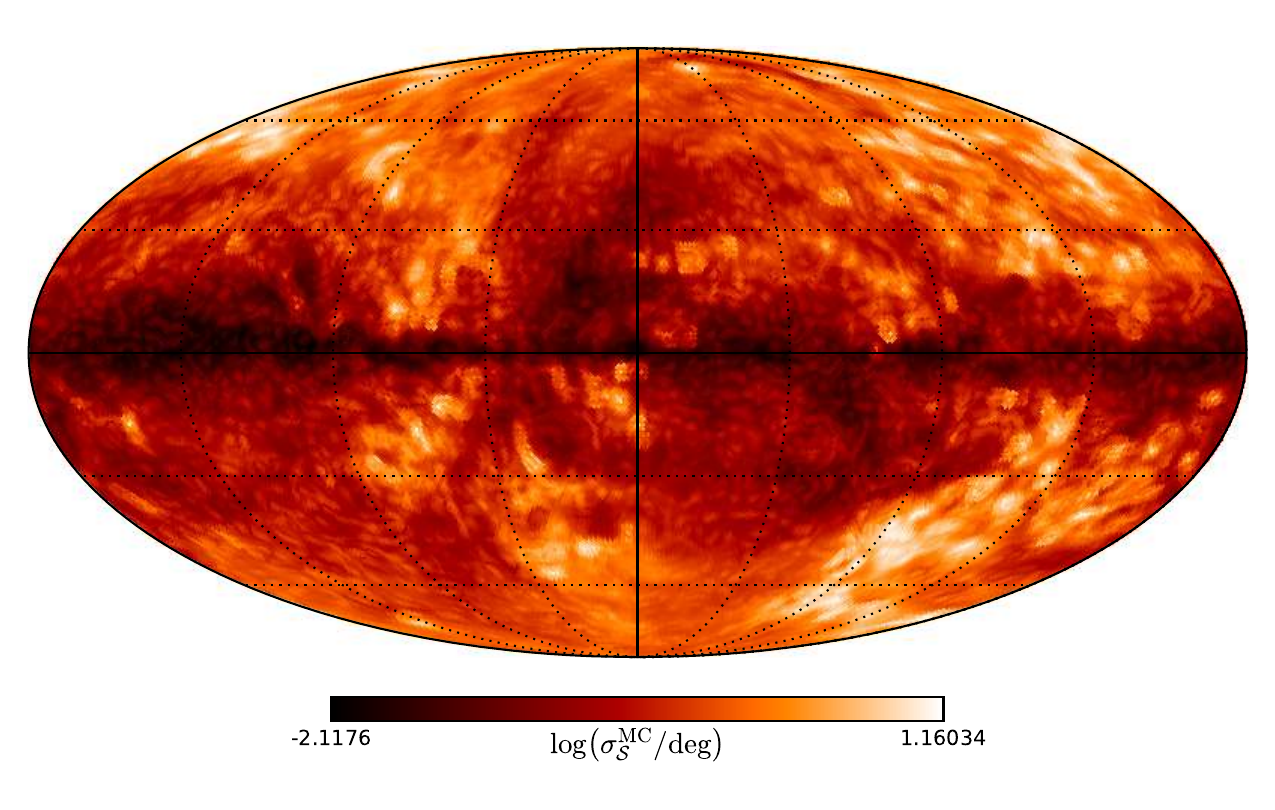}
\caption[]{{\it Top:} Polarization angle dispersion function $\mathcal{S}$ computed from the \GNILC-processed data at 353\,GHz and uniform 160\arcmin\ resolution, using a lag $\delta=80\arcmin$. {\it Bottom:} Statistical uncertainty $\sigma^\mathrm{MC}_\mathcal{S}$ computed from the Monte Carlo simulations on maps with the same 160\arcmin\ resolution and $\delta=80\arcmin$ lag.
}
\label{fig:S_GNILC_160arcmin}
\end{figure}

A map of $\mathcal{S}$ at 60\arcmin\ resolution and using a lag of 30\arcmin, based on \Planck\ 2013 data, was shown over a restricted region of the sky in~\citet{planck2014-XIX}. We can now present the $\mathcal{S}$ map over the full sky, based on the \GNILC-processed {\DRThree} data release at 353\,GHz. Because $\mathcal{S}$ is built from the polarization angle $\psi$, it is independent of the value chosen for the total intensity offset. However, when computed at uniform 80\arcmin\ resolution and using a lag $\delta=40\arcmin$, $\mathcal{S}$ is still significantly biased (see Sect.~\ref{sec:DFs}). For this reason, we use maps smoothed to 160\arcmin\ and adopt a correspondingly larger lag $\delta=80\arcmin$.\footnote{When considering the Monte Carlo simulations discussed in the previous subsection, we find that the ratio of the ensemble average map $\langle\mathcal{S}\rangle$ to the map $\mathcal{S}$ computed from the smoothed \GNILC\ data have a mean of 0.90 and a median value of 0.97, with a standard deviation of 0.14. For comparison, when working at 80\arcmin\ resolution and a lag of $\delta=40\arcmin$, these values shift to 0.81, 0.87, and 0.19, respectively, which quantifies the bias that remains when working at 80\arcmin\ resolution.} This map is shown in the top panel of Fig.~\ref{fig:S_GNILC_160arcmin}. We computed the statistical uncertainty $\sigma^\mathrm{MC}_\mathcal{S}$ using the Monte Carlo approach discussed in Sect.~\ref{sec:pol_uncertainties}, but based on the Stokes maps smoothed to 160\arcmin\ resolution. The map of $\sigma^\mathrm{MC}_\mathcal{S}$ is shown in the bottom panel of Fig.~\ref{fig:S_GNILC_160arcmin}. Quite large values, up to $14\deg$, are reached in some regions, but we will see in Sect.~\ref{sec:DFs} that this is compatible with the noise in the data (see also Sect.~\ref{sec:noisebiasS}). 

\subsection{Relationship of $\mathcal{S}$ to alternative estimators}
\label{subsec:alternative_S}

Synchrotron studies in the radio domain frequently use another estimator of the uniformity of polarization patterns, the polarization gradient introduced by~\cite{Gaensler11} and defined as
\begin{equation}
|\nabla{P}|=\sqrt{\left(\frac{\partial Q}{\partial y}\right)^2+\left(\frac{\partial Q}{\partial z}\right)^2+\left(\frac{\partial U}{\partial y}\right)^2+\left(\frac{\partial U}{\partial z}\right)^2} \, ,
\label{eq:polgrad}
\end{equation}
where $y$ and $z$ refer to an orthogonal coordinate system on the plane of the sky. We show in Appendix~\ref{sec:appendix:SdP} that, as far as the \Planck\ thermal dust polarization data are concerned, $|\nabla{P}|$ is strongly correlated with $\S$, though not perfectly because of the contribution from the polarized intensity in $|\nabla{P}|$. This can be mitigated by considering an angular version of the polarization gradient defined as~\citep{Burkhart12}
\begin{equation}
|\nabla{\psi}|=\sqrt{\left[\frac{\partial (Q/P)}{\partial y}\right]^2+\left[\frac{\partial (Q/P)}{\partial z}\right]^2+\left[\frac{\partial (U/P)}{\partial y}\right]^2+\left[\frac{\partial (U/P)}{\partial z}\right]^2} \, ,
\label{eq:unitpolgrad}
\end{equation}
which encodes only the angular content of the polarization.\footnote{Other advanced diagnostics from polarization gradients are discussed in \citet{Herron18a}, but further discussion of these is beyond the scope of this paper.} In Appendix~\ref{sec:appendix:SdP}, we show not only that $|\nabla{\psi}|$ is better correlated with $\S$ than $|\nabla{P}|$ is, but also that this can be demonstrated analytically, with
\begin{equation}
\S(\boldsymbol{r},\delta)\approx\frac{\delta}{2\sqrt{2}}|\nabla{\psi}| \, ,
\end{equation}
the linear dependence of $\S$ on the lag being revealed simply through a first-order Taylor expansion. We do not use this estimator $|\nabla{\psi}|$ in the rest of this paper, but note that in practice it might be easier to compute than $\mathcal{S}$.

\subsection{Noise and bias in $\S$}
\label{sec:noisebiasS}

An estimate of the variance of $\S$ due to noise is \citep{planck2014-XIX,PMA3}:
\begin{eqnarray}\label{Eq-sigmaS}
\sigma^2_{\mathcal{S}}\left(\boldsymbol{r},\delta\right) & = & 
\frac{\sigma^2_\psi(\boldsymbol{r})}{N^2\mathcal{S}^2}
\left(\sum_{i=1}^N\psi(\boldsymbol{r}+\boldsymbol{\delta}_i)-\psi(\boldsymbol{r})\right)^2 \nonumber\ \\
&+& \frac{1}{N^2\mathcal{S}^2}
\sum_{i=1}^N
\sigma^2_\psi(\boldsymbol{r}+\boldsymbol{\delta}_i)
\left(\psi(\boldsymbol{r}+\boldsymbol{\delta}_i)-\psi(\boldsymbol{r})\right)^2 
\, .
\end{eqnarray}
Just like for $\PsI$, noise on Stokes parameters $Q$ and $U$ induces a bias on $\S$. Unlike for $\PsI$, however, this bias can be positive or negative, depending on whether the true value is, respectively, smaller or larger than the value $\pi/\sqrt{12}\approx 52\deg$ obtained for fully random polarization angles~\citep{PMA3}.
As prescribed by \cite{Hildebrand09} and \cite{planck2014-XIX}, we use the following debiasing scheme
\begin{equation}
  \S_{\rm db}=\begin{cases}
    \sqrt{\S^2-\sigma^2_\S} & \text{if $\S >\sigma_\S$},\\
    0 & \text{otherwise}.
  \end{cases}
\end{equation} 
This expression works well for S/N on $\S$ larger than 3, which we ensure by smoothing the Stokes maps. For notational simplicity, in the rest of this paper, we write $\S$ to mean the debiased value $ \S_{\rm db}$ of the polarization angle dispersion function.

\section{Statistics of thermal dust polarization maps}
\label{sec:full-sky-analysis}

In this section, we provide a statistical analysis of the quantities represented in the Galactic thermal dust polarization maps derived above. We start by discussing the distribution functions of $p$, $\psi$, and $\mathcal{S}$. We then examine the joint distributions of $p$ and total gas column density on the one hand, and of $\S$ and $p$ on the other hand. Finally, we look into how one striking feature of these maps, i.e., the inverse relationship between $\S$ and $p$, is well reproduced by relatively simple Gaussian models of the \Planck\ polarized sky.

\subsection{Distribution functions for $p$, $\psi$, and $\mathcal{S}$ }
\label{sec:DFs}

\subsubsection{Polarization fraction}
\label{subsec:polfrac}

\begin{figure}[htbp]
\includegraphics[width=0.5\textwidth]{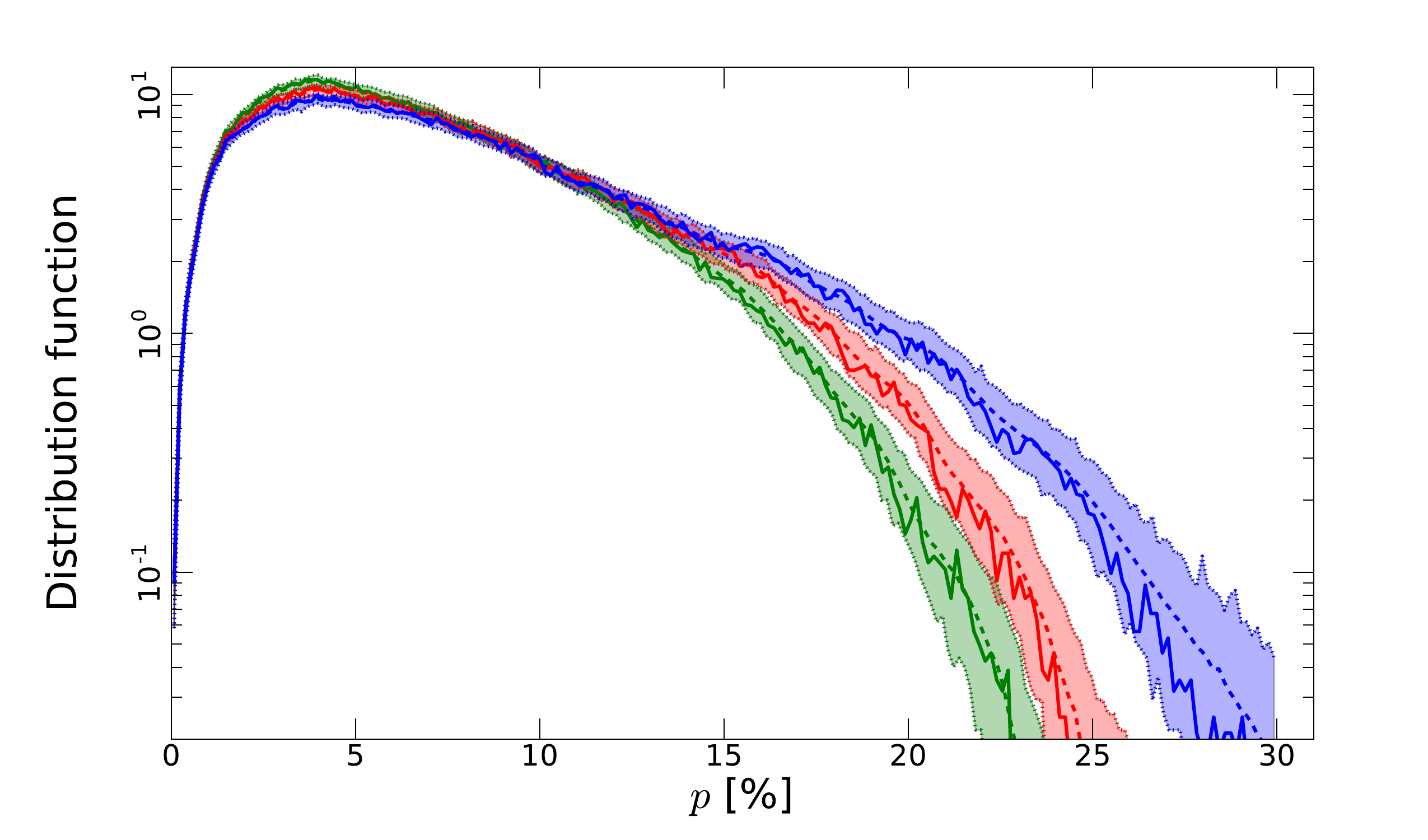}
\caption[]{Distribution functions of the polarization fraction $\PsI$ in the \GNILC\ data at 353\,GHz and uniform 80\arcmin\ resolution. The solid red curve corresponds to the fiducial Galactic offset for the total intensity, whereas blue and green correspond to the cases of low and high offset, respectively. The dashed curves show the mean over the 1000 Monte Carlo histograms, and the envelopes shown as coloured regions span the range of the 1000 histograms.
}
\label{fig:PI_MC_histograms}
\end{figure}

The distribution function (DF, or histogram) for $\PsI$ over the full sky is shown in Fig.~\ref{fig:PI_MC_histograms} The solid red curve is the histogram for the \GNILC\ map of $\PsI$ for the fiducial offset in $I$, while the solid blue and green curves are the corresponding histograms for the low and high offsets, respectively. These clearly show the significant effect induced by the uncertainty on the total intensity offset. We note, however, that the polarization fractions observed reach at least 20\,\% for any choice of the total intensity offset, setting strong constraints for dust models.

For comparison, the corresponding dashed coloured curves are the means of the DFs from the 1000 Monte Carlo simulations.  Compared to the solid curves, there is only a small bias, shifting the DF towards higher $\PsI$ in the tail of the distribution.  This is less pronounced for the green curves (high total intensity offset) because for this case the statistical changes in $I$ are less important.\footnote{Corresponding DFs and values for the naive estimator $\PsI$ (not shown) are very similar, underlining that the bias is quite small already at 80\arcmin\ resolution.}

The coloured regions encompassing the mean histograms show the minimum and maximum values of the histogram for any given bin of $\PsI$ over the 1000 samples, i.e., the envelope within which all 1000 histograms lie.  Lines defining the edges of the envelope would themselves not be distribution functions; however, they give an idea of the possible spread of the $\PsI$ histograms with varying noise realizations.

It is of interest, for dust models in particular, to estimate $\PsI_{\rm max}$, the largest value of $\PsI$ over the full sky. To estimate this and provide further quantification, we compute, for each of the total intensity offset values, the 90th, 95th, 98th, 99th, and 99.9th percentiles for the \GNILC\ histogram from the data, which we write as $h(p)$, and for each of the 1000 Monte Carlo histograms, which we write as $h(p_i)$, with $1\leqslant i\leqslant 1000$.  From the latter we calculate the mean and the standard deviation, which gives an estimate of the statistical uncertainty of $\PsI_{\rm max}$ in a single realization, such as the data.

These numbers are given in Table~\ref{table-PI_stats}, alongside the corresponding values for the average $\PsI$ map over the 1000 Monte Carlo realizations, which we write $h(\overline{p_i})$, and those for the mean histogram over the 1000 realizations,\footnote{Those are shown as dashed curves in Fig.~\ref{fig:PI_MC_histograms}.} which we write $\overline{h(p_i)}$. The percentiles for the average $\PsI$ map are always very close to those for the data, which is not surprising because the data were taken as the mean for the Monte Carlo realizations. More interestingly, the percentiles $h(p_i)$ are systematically larger than the corresponding values for the data, with very low statistical uncertainties. We note that this discrepancy is significantly smaller for the high total intensity offset than for the low total intensity offset, at least for the highest percentiles. This shows that $\PsI_{\rm max}$ from the data is likely biased by a similar amount and is to be adjusted accordingly. We also point out that the percentiles for the mean histogram $\overline{h(p_i)}$ are larger still, by about 0.1--0.3\,\%. Consequently, we give a conservative estimate of the bias on the polarization fraction percentiles  (and therefore on $\PsI_\mathrm{max}$) by considering the difference $\overline{h(p_i)}-h(p)$. A rough debiasing of the data percentiles by this quantity is achieved by subtracting this value from $h(p)$. For instance, the estimated bias for the 99.9th percentile at 80\arcmin\ resolution in the fiducial offset case is about 0.66\,\%. Subtracting this from the data percentile, we obtain a debiased value of 22.00\,\%.

Finally, we emphasize that the truly dominant source of uncertainty in the determination of characteristic values of the $\PsI$ distribution is the offset in $I$. It is larger than the statistical uncertainty, which is of order 0.01--0.10\,\%, or the impact of the residual systematics that has been estimated in Appendix~\ref{sec:appendix:E2E} to be typically 0.5\,\%.

Performing the same debiasing for the low and high offset values, and gathering these results for the 99.9th percentile, we obtain a debiased value of $22.0^{+{3.5}}_{-{1.4}}\pm0.1\,\%$ for the maximum dust polarization fraction observed at 80\arcmin\ resolution and 353\,GHz over the full sky, where the first uncertainty relates to the systematic effect of the total intensity offset and the effects of residual systematics, and the second covers the statistical uncertainty estimated from the 1000 Monte Carlo realizations.

For completeness, Table~\ref{table-PI_stats} also gives the same percentiles for the maps smoothed to 160\arcmin\ resolution, showing a further reduction of the bias $\overline{h(p_i)}-h(p)$. In that case, we find that the maximum dust polarization fraction observed is $21.4{^{+2.2}_{-1.2}}\pm0.1\,\%$.  This value of $\PsI_{\rm max}$ and the debiased value at 80\arcmin\ agree quite well. This shows that smoothing has little effect on the polarization fraction. Of course, the amount of smoothing applied should not be excessive, because of the potential impact of beam depolarization at higher FWHM. In Appendix~\ref{A-depolar}, we quantify the effect of smoothing on $\PsI$ and $\pmax$ in the framework of the analytical model presented in Sect.~\ref{sec:rmodels}. It is found that smoothing from one resolution to another leads to a decrease in $\PsI^2$ by an amount that is
statistically independent of the value of the polarization fraction. Considering $\PsI$ itself, this means that the effect of smoothing is very small if $\PsI$ is large, e.g., $\PsI\approx\pmax$ (Appendix~\ref{sec:cmp_analytic_numeric}). We conclude that our derivation of $\pmax$ is not so much affected by the resolution and much more so by the offset in $I$.

These results are consistent with the finding of~\cite{planck2014-XIX} that $\PsI_\mathrm{max}>19.8\,\%$ at 60\arcmin\ resolution over a smaller fraction of the sky. We have also checked that they are not significantly affected when selecting only those pixels on the sky for which the S/N in polarization is $\PsI/\sigma_p>3$.

As was pointed out in \cite{planck2014-XIX} and \cite{planck2014-XX}, the level of observed polarization fractions is strongly dependent on the angle $\Gamma$ of the mean magnetic field with respect to the plane of the sky (see Appendix~\ref{sec:appendix:Sp} and Fig.~\ref{ref-frame}). The distribution function of $\PsI$ must depend on this mean orientation of the Galactic magnetic field. Compared to what would be obtained for a mean field that is everywhere in the plane of the sky, the distribution should be more peaked towards lower values, as we do observe, but the value of $\PsI_{\rm max}$ might still be high, reflecting those parts of the sky with a favourable orientation, i.e., in the plane of the sky. Although the estimate of $\PsI_{\rm max}$ based on percentiles would be impacted, such an analysis (requiring a model of the large-scale GMF) is beyond the scope of this paper.

\begin{table}[htbp!] 
\begingroup
\newdimen\tblskip \tblskip=5pt
\caption{Statistics from the distribution functions of $p$, given as percentages. 
}
\label{table-PI_stats}                           
\nointerlineskip
\vskip -3mm
\footnotesize
\setbox\tablebox=\vbox{
   \newdimen\digitwidth 
   \setbox0=\hbox{\rm 0} 
   \digitwidth=\wd0 
   \catcode`*=\active 
   \def*{\kern\digitwidth}
   \newdimen\signwidth 
   \setbox0=\hbox{+} 
   \signwidth=\wd0 
   \catcode`!=\active 
   \def!{\kern\signwidth}
\halign{\hbox to 1.0in{#\leaderfil}\tabskip 1.0em&
\hfil#\hfil\tabskip 2.0em& \hfil#\hfil& \hfil#\hfil& \hfil#\hfil\tabskip 0pt\cr
\noalign{\doubleline}
\omit \hfil Percentile\hfil& $h\left(p\right)$& $h\left(\overline{p_i}\right)$&
 $h\left(p_i\right)$& $\overline{h\left(p_i\right)}$\cr
\noalign{\vskip 3pt\hrule\vskip 5pt}
\omit& \multispan4\hfil Resolution 80\arcm, intensity offset low\hfil\cr
\noalign{\vskip -5pt}
\omit&\multispan4\hrulefill\cr
\noalign{\vskip 2pt}
90& 15.01 & 14.82 & $15.14\pm0.01$& 15.67\cr
95& 17.67 & 17.63 & $17.92\pm0.02$& 18.37\cr
98& 20.53 & 20.55 & $20.88\pm0.02$& 21.22\cr
99& 22.24 & 22.29 & $22.76\pm0.03$& 23.17\cr
99.9& 26.43 & 26.50 & $27.64\pm0.10$& 27.37\cr
\noalign{\vskip 4pt\hrule\vskip 5pt}
\omit& \multispan4\hfil Resolution 80\arcm, intensity offset fiducial\hfil\cr
\noalign{\vskip -5pt}
\omit&\multispan4\hrulefill\cr
\noalign{\vskip 2pt}
90& 13.35 & 13.35 & $13.48\pm0.01$& 14.02\cr
95& 15.62 & 15.65 & $15.81\pm0.01$& 16.27\cr
98& 17.90 & 17.93 & $18.16\pm0.02$& 18.52\cr
99& 19.36 & 19.39 & $19.63\pm0.02$& 20.02\cr
99.9& 22.66 & 22.68 & $23.01\pm0.05$& 23.32\cr
\noalign{\vskip 4pt\hrule\vskip 5pt}
\omit& \multispan4\hfil Resolution 80\arcm, intensity offset high\hfil\cr
\noalign{\vskip -5pt}
\omit&\multispan4\hrulefill\cr
\noalign{\vskip 2pt}
90& 12.20 & 12.23 & $12.32\pm0.01$& 12.82\cr
95& 14.25 & 14.27 & $14.41\pm0.01$& 14.77\cr
98& 16.42 & 16.44 & $16.56\pm0.01$& 16.87\cr
99& 17.72 & 17.74 & $17.90\pm0.02$& 18.22\cr
99.9& 21.08 & 21.10 & $21.21\pm0.04$& 21.52\cr
\noalign{\vskip 4pt\hrule\vskip 5pt}
\omit& \multispan4\hfil Resolution 160\arcm, intensity offset low\hfil\cr
\noalign{\vskip -5pt}
\omit&\multispan4\hrulefill\cr
\noalign{\vskip 2pt}
90&\hfil$14.39$\hfil&\hfil$14.41$\hfil&\hfil$14.43\pm0.02$\hfil&\hfil$14.77$\hfil\cr
95&\hfil$16.99$\hfil&\hfil$17.01$\hfil&\hfil$17.05\pm0.02$\hfil&\hfil$17.32$\hfil\cr
98&\hfil$19.59$\hfil&\hfil$19.59$\hfil&\hfil$19.65\pm0.03$\hfil&\hfil$19.87$\hfil\cr
99&\hfil$21.11$\hfil&\hfil$21.12$\hfil&\hfil$21.23\pm0.04$\hfil&\hfil$21.52$\hfil\cr
99.9&\hfil$24.07$\hfil&\hfil$24.07$\hfil&\hfil$24.38\pm0.08$\hfil&\hfil$24.52$\hfil\cr
\noalign{\vskip 4pt\hrule\vskip 5pt}
\omit& \multispan4\hfil Resolution 160\arcm, intensity offset fiducial\hfil\cr
\noalign{\vskip -5pt}
\omit&\multispan4\hrulefill\cr
\noalign{\vskip 2pt}
90&\hfil$12.77$\hfil&\hfil$12.78$\hfil&\hfil$12.82\pm0.01$\hfil&\hfil$13.12$\hfil\cr
95&\hfil$15.05$\hfil&\hfil$15.05$\hfil&\hfil$15.09\pm0.02$\hfil&\hfil$15.37$\hfil\cr
98&\hfil$17.18$\hfil&\hfil$17.18$\hfil&\hfil$17.23\pm0.02$\hfil&\hfil$17.47$\hfil\cr
99&\hfil$18.52$\hfil&\hfil$18.51$\hfil&\hfil$18.55\pm0.03$\hfil&\hfil$18.82$\hfil\cr
99.9&\hfil$21.70$\hfil&\hfil$21.70$\hfil&\hfil$21.76\pm0.06$\hfil&\hfil$21.97$\hfil\cr
\noalign{\vskip 4pt\hrule\vskip 5pt}
\omit& \multispan4\hfil Resolution 160\arcm, intensity offset high\hfil\cr
\noalign{\vskip -5pt}
\omit&\multispan4\hrulefill\cr
\noalign{\vskip 2pt}
90&\hfil$11.67$\hfil&\hfil$11.68$\hfil&\hfil$11.70\pm0.01$\hfil&\hfil$12.07$\hfil\cr
95&\hfil$13.71$\hfil&\hfil$13.72$\hfil&\hfil$13.75\pm0.02$\hfil&\hfil$14.02$\hfil\cr
98&\hfil$15.86$\hfil&\hfil$15.86$\hfil&\hfil$15.90\pm0.02$\hfil&\hfil$16.12$\hfil\cr
99&\hfil$17.05$\hfil&\hfil$17.06$\hfil&\hfil$17.07\pm0.02$\hfil&\hfil$17.32$\hfil\cr
99.9&\hfil$20.41$\hfil&\hfil$20.40$\hfil&\hfil$20.40\pm0.06$\hfil&\hfil$20.62$\hfil\cr
\noalign{\vskip 4pt\hrule\vskip 5pt}
}}
\endPlancktable                    
\tablenote{{\rm a}} The columns are the following, from left to right: $h(p)$ refers to the DF of the data; $h(\overline{p_i})$ refers to the DF of the average $p$ map over the 1000 Monte Carlo realizations; $h(p_i)$ refers to the individual Monte Carlo realizations of the $p$ maps (the values listed give the mean and standard deviation over the 1000 realizations); and $\overline{h(p_i)}$ refers to the average DF over the 1000 realizations, as shown in Fig.~\ref{fig:PI_MC_histograms}.\par
\endgroup
\end{table}

\subsubsection{Polarization angle}
\label{subsec:polangle}

\begin{figure}[htbp]
\includegraphics[width=0.5\textwidth]{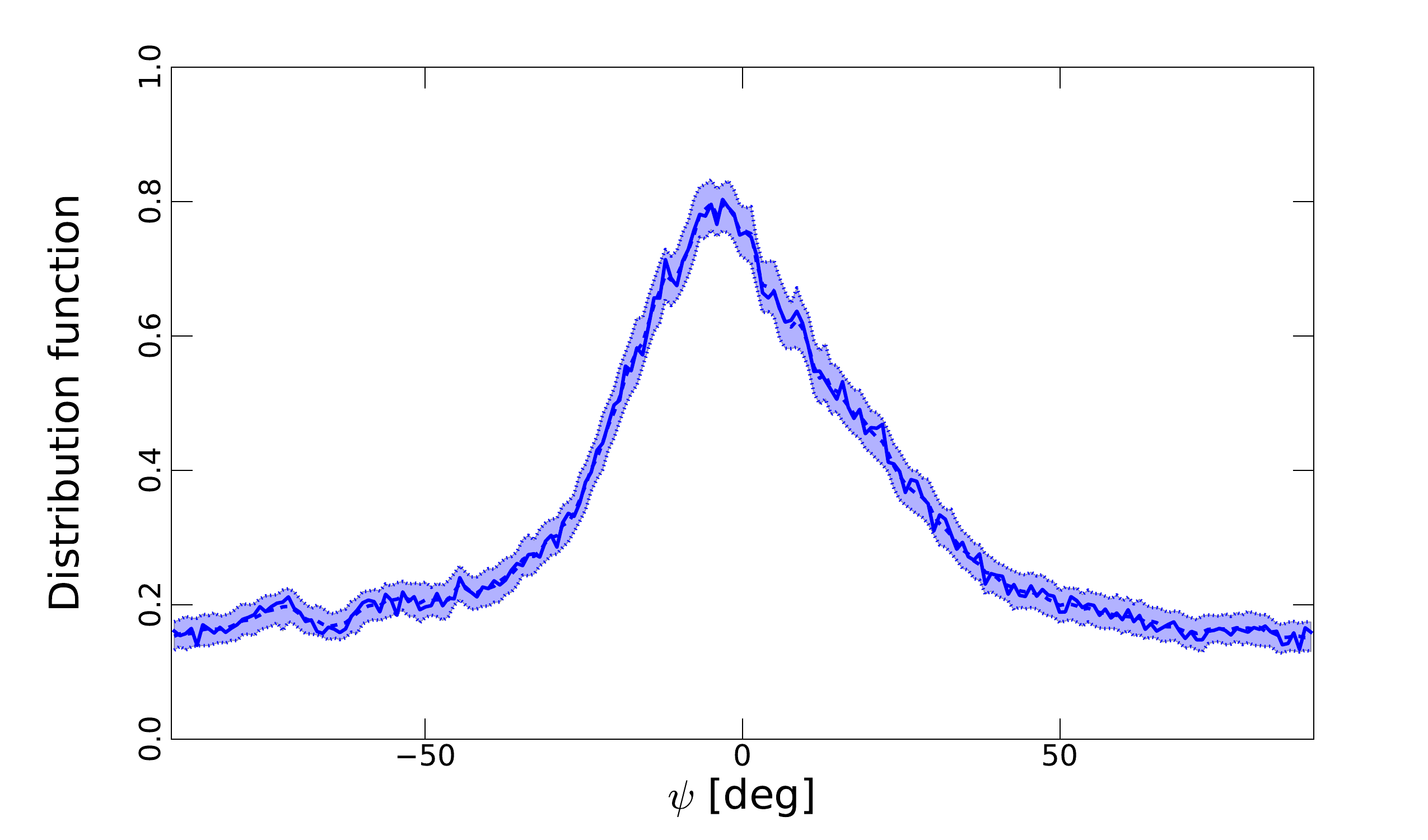}
 \caption[]{Distribution function for the polarization angle $\psi$ in Galactic coordinates for the \GNILC\ data at 353\,GHz and uniform 80\arcmin\ resolution. The solid curve shows the histogram of the polarization angles computed directly from the \GNILC\ data, the dashed curve gives the mean of the 1000 Monte Carlo histograms, and the blue region shows the envelope spanned by the 1000 histograms.
}
\label{fig:psi_MC_histograms}
\end{figure}

Figure~\ref{fig:psi_MC_histograms} shows the distribution function for the polarization angle $\psi$, for which the value of the total intensity offset is unimportant. 
The comparison between the histogram for the \GNILC\ map of $\psi$ and the mean histogram over the Monte Carlo realizations shows that there is virtually no noise bias.
The histograms peak around 0\deg, which corresponds to an orientation of the GMF parallel to the Galactic plane. Quantitatively, over the 1000 Monte Carlo samples, the ensemble average of the mean polarization angle is $-0\pdeg64\pm0\pdeg03$. This value is compatible with the earlier measurement in~\cite{planck2014-XIX} (see their figure 3).

\subsubsection{Polarization angle dispersion function}
\label{subsec:polangledisp}

\begin{figure}
\includegraphics[width=0.5\textwidth]{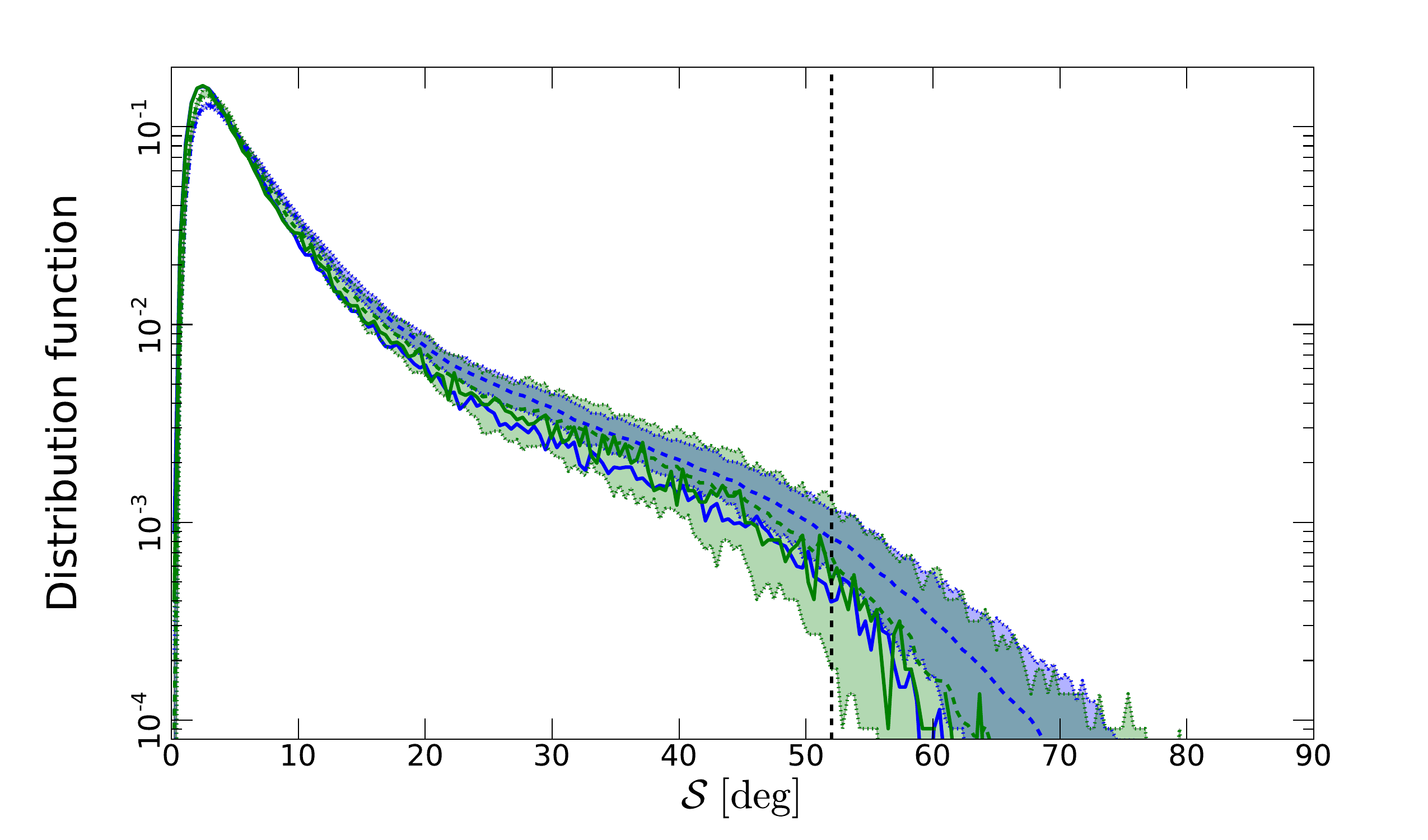}
\caption[]{Distribution functions of the polarization angle dispersion function $\mathcal{S}$ in the \GNILC\ data at 353\,GHz.  The cases shown are for the 160\arcmin\ resolution using a lag $\delta=80\arcmin$ (in green), and for the 80\arcmin\ resolution using a lag $\delta=40\arcmin$ (in blue). The solid curves show the histograms computed directly from the \GNILC\ maps, the dashed curves give the mean histogram from the 1000 Monte Carlo realizations for each case, and the coloured regions show the envelope. The dashed vertical line indicates the value $\pi/\sqrt{12}\approx 52\deg$ corresponding to a completely random polarization pattern.
}
\label{fig:PDF_S_GNILC_160arcmin}
\end{figure}

Finally, the distribution function of $\mathcal{S}$ is shown in Fig.~\ref{fig:PDF_S_GNILC_160arcmin}. Results for the case of a 160\arcmin\ FWHM and lag $\delta=80\arcmin$ are shown in green, and for the case of a 80\arcmin\ FWHM and lag $\delta=40\arcmin$ in blue. As above for $\PsI$ and for $\psi$, the solid lines are for the \GNILC\ maps, the dashed lines are the Monte Carlo means, and the coloured regions show the span of histograms for the 1000 Monte Carlo realizations. 

\begin{figure}
\includegraphics[width=0.5\textwidth]{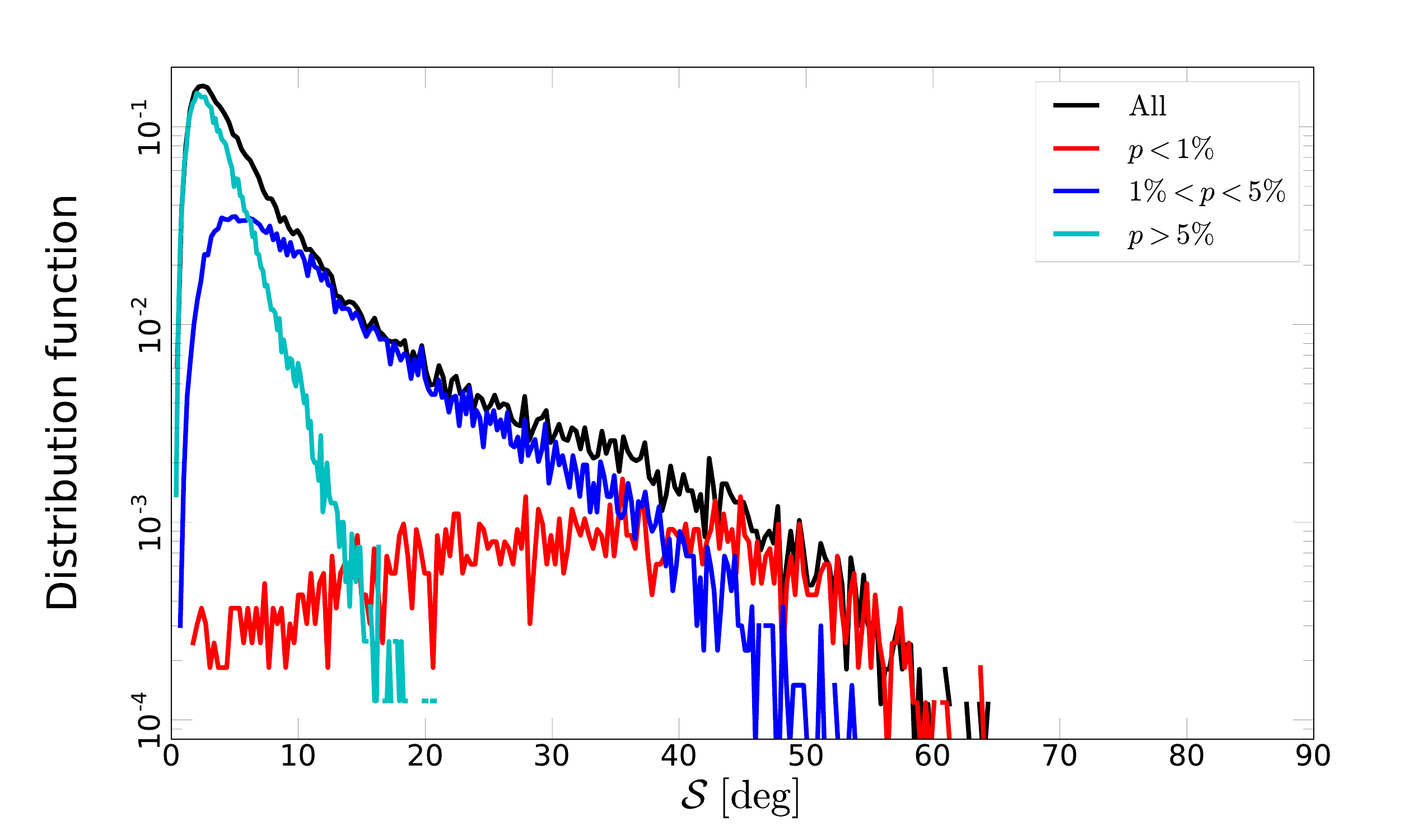}
\caption[]{Distribution functions of $\mathcal{S}$ at 160\arcmin\ resolution and using a lag of $\delta=80\arcmin$, for different ranges of $\PsI$, using the fiducial total intensity offset. The distribution function for all points is shown in black and for different ranges of $\PsI$ in separate colours. The distribution functions for the different subsets are scaled to the fractional number of points contained in each range. 
}
\label{fig:DF_S_different_p}
\end{figure}

It is interesting that these distributions have a tail passing through 52\deg, the value of $\S$ for randomly oriented polarization. As noted by~\cite{PMA3}, if an orientation distribution produces a value of $\S$ that is somewhat lower (higher) than this, then the addition of noise tends to make $\S$ larger (smaller), towards 52\deg. This tail in the full DF in Fig.~\ref{fig:PDF_S_GNILC_160arcmin} is strongly associated with regions where $\PsI$ is small and more susceptible to the influence of noise, as is apparent in Fig.~\ref{fig:DF_S_different_p}, which shows the distribution function of $\mathcal{S}$ for different ranges in polarization fraction ($\PsI<1\,\%$, $1\,\%<p<5\,\%$, and $\PsI>5\,\%$) for the \GNILC\ data at 160\arcmin\ resolution and with a lag $\delta=80\arcmin$. The large values of $\S$ are also associated with large values of the scatter $\sigma_\S^\mathrm{MC}$, as shown by the widening of the envelope at high values of $\S$ in Fig.~\ref{fig:PDF_S_GNILC_160arcmin}. The width of the envelope at 160\arcmin\ resolution is compatible with the largest values found in the map of $\sigma_\S^\mathrm{MC}$ (Fig.~\ref{fig:S_GNILC_160arcmin}).

Fig.~\ref{fig:PDF_S_GNILC_160arcmin} shows that, for the case of an 80\arcmin\ FWHM and lag $\delta=40\arcmin$, at large values of $\mathcal{S}$ the mean DF of the Monte Carlo realizations is clearly biased with respect to the distribution function of the data, which does not even fit within the region spanned by the 1000 Monte Carlo simulations. On the other hand, for 160\arcmin\ FWHM and lag $\delta=80\arcmin$, the bias is much less apparent and so we focus on the results for this case.
Despite the long tail at large $\S$, most of the points in this tail have low occurrence rates, underlining the regularity of the polarization angle on large scales. At 160\arcmin\ resolution and a lag of $\delta=80\arcmin$, the distribution of values in the $\mathcal{S}$ map for the data peaks at $1\pdeg7$, with mean and median values of $7\pdeg6$ and $4\pdeg6$, respectively. The same characteristic values over the 1000 Monte Carlo simulations are, respectively, $1\pdeg9\pm0\pdeg6$, $8\pdeg29\pm0\pdeg01$, and $5\pdeg12\pm0\pdeg01$. Using the 99th percentile, most of the points in the data have $\S\leqslant43\pdeg6$, while the Monte Carlo simulations give an estimate of $45\pdeg3\pm0\pdeg2$. We give these values for reference in the future, for instance in work comparing \Planck\ data with MHD simulations and analytical models.

We stress that while the smoothing to 160\arcmin\ is warranted here for studies including the high-latitude sky, this requirement for smoothing should not be generalized.  Indeed, when the analysis is restricted to the approximately 42\,\% of the sky considered in~\cite{planck2014-XIX}, we find that no such bias exists when working at 80\arcmin\ FWHM and lag $\delta=40\arcmin$. Incidentally, this confirms the results shown in~\cite{planck2014-XIX} at 60\arcmin\ resolution and $\delta=30\arcmin$.

\subsection{Two-dimensional distribution functions}
\label{sec:twod}

In this section we investigate the 2-dimensional joint distribution functions of polarization fraction $p$ and another variable. Therefore, instead of simply presenting a scatter plot, we display a 2-dimensional histogram made by binning in the two dimensions and encoding the number in each bin by colour.

\subsubsection{Polarization fraction versus total gas column density}
\label{sec:pandI}

\begin{figure}[htbp]
\includegraphics[width=0.55\textwidth,trim=50 0 0 50,clip=true]{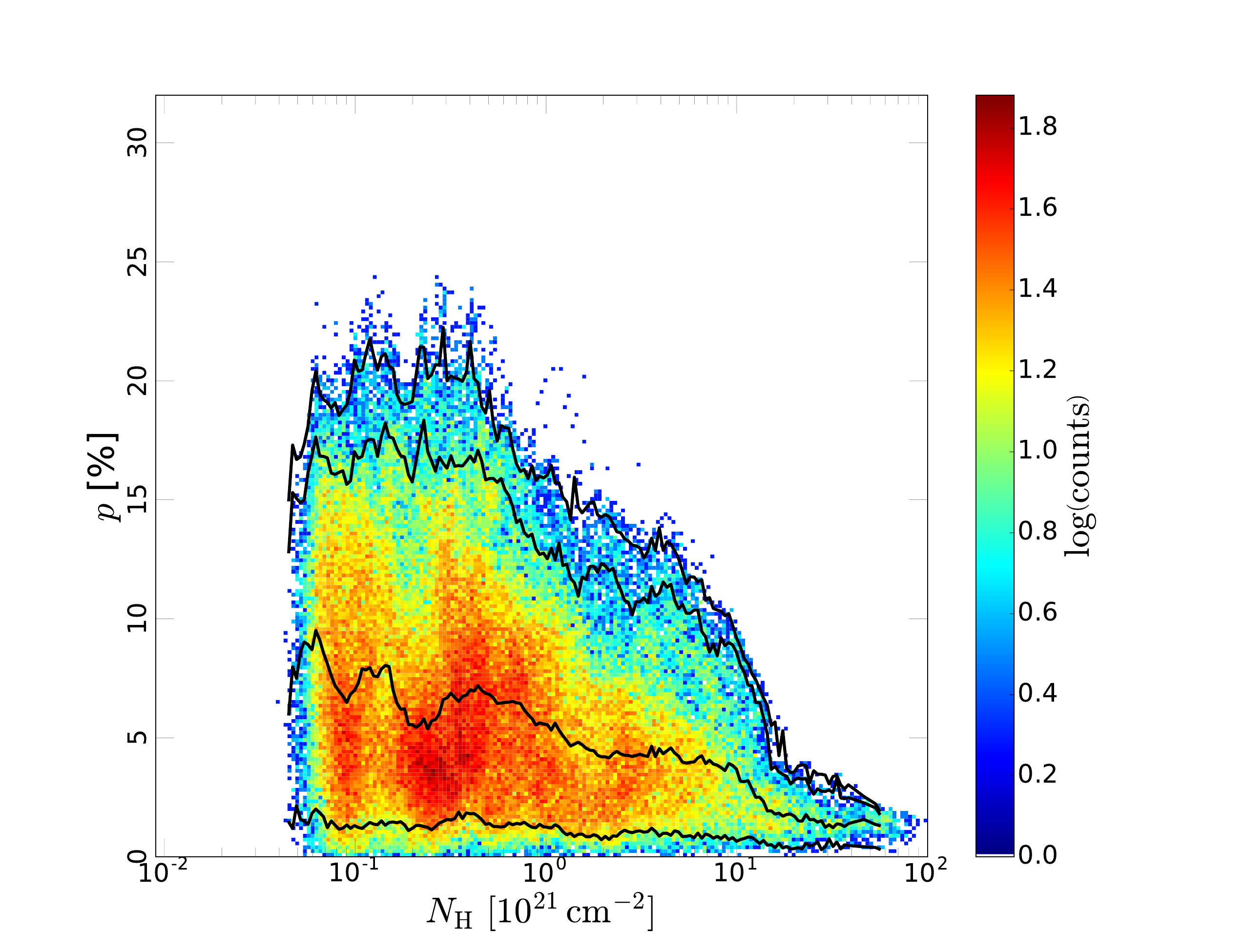}
\caption[]{
Two-dimensional histogram showing the joint distribution function of the polarization fraction $p$ from the \GNILC\ data, at 353\,GHz and uniform 80\arcmin\ resolution, and the total gas column density $\NH$. This plot uses the fiducial total intensity offset. The black lines show the 5th, 95th, and 99th percentiles of the $p$ distribution in each $\NH$ bin, as well as the median $p$ in each $\NH$ bin. 
}
\label{fig:PI_I_GNILC}
\end{figure}

In Fig.~\ref{fig:PI_I_GNILC} we display the 2-dimensional histogram of $p$ and total gas column density $\NH$, using the \GNILC\ data, at 353\,GHz and uniform 80\arcmin\ resolution, with the fiducial total intensity offset, over the full sky. We discuss the determination of $\NH$ at the beginning of Sect.~\ref{sec:NewResults}. No cut in either S/N or Galactic latitude has been performed here. The colour scale encodes the logarithm of counts in each bin, while the black curves show the 5th, 95th, and 99th percentiles of the $p$ distribution in each $\NH$ bin, as well as the median polarization fraction in each $\NH$ bin.

To explore the sensitivity of this distribution and characteristic curves to statistical noise, we use the Monte Carlo approach described above. We first compute the 2-dimensional distribution function of $p$ and $\NH$ for each of the 1000 simulations, along with the curves giving the median and the 5th, 95th, and 99th percentiles of $p$ within each bin of $\NH$. We then compute the average curve for each of these four quantities, as well as their dispersions within each $\NH$ bin. 

We find that these exhibit small statistical dispersions, but that towards the most diffuse lines of sight ($\NH<10^{20}\,\mathrm{cm^{-2}}$), the maximum polarization fractions (measured for instance by the 99th percentile curve) for the Monte Carlo simulations are slightly higher than the corresponding values from the data. As expected, this bias is in the same sense as discussed in Sect.~\ref{subsec:polfrac} for the distribution function of $p$. Recall that this is for 80\arcmin\ resolution; when working at 160\arcmin\ resolution this bias disappears.

The joint $\left(\NH,p\right)$ distribution has qualitatively the same behaviour as that found over a smaller fraction of the sky in~\cite{planck2014-XIX}: a large scatter of $p$ towards diffuse lines of sight and a decrease in the maximum $p$ as $\NH$ increases. 

For completeness, we show in Appendix~\ref{sec:appendix:PsI-other-offsets} the effect of the total intensity offset. It is negligible at the high intensity end, where the histograms are similar whether we use the fiducial, high, or low offset values. At the low intensity end, on the other hand, the effect of the offset is more marked. There is a significant increase in characteristic values (highest percentiles) of $p$ for decreasing $\NH$ when taking the low offset, and conversely a marked decrease in the maximum $p$ when taking the high offset. 

One might wonder if it would be possible to constrain the offset by assuming that $p$ should reflect dust properties at low column densities, and therefore that the offset should be such that the maximum $p$ is approximately constant at low $\NH$. In this respect, the fiducial offset seems more adequate than either the high or low cases, as can be seen by comparing Fig.~\ref{fig:PI_I_GNILC} with Fig.~\ref{fig:PI_I_GNILC_otheroffsets}.

The sharp downturn of the maximum polarization fraction observed near $\NH\approx10^{22}\,\mathrm{cm^{-2}}$ corresponds to the strong depolarization occurring on lines of sight that probe high column density structures that are not resolved at 80\arcmin.

\subsubsection{Polarization angle dispersion versus polarization fraction}
\label{sec:S_maps}

In \cite{planck2014-XIX}, we discovered an inverse relationship between the polarization fraction $p$ and the polarization angle dispersion function $\S$, working with data over approximately 42\,\% of the sky, at a resolution of 60\arcmin\ and a lag of $\delta=30\arcmin$.
We have verified quantitatively on the same sky region and using the same methodology that the same inverse relationship holds with the {\DRThree} data release; the maps of polarization are very similar where the S/N is high, as expected. In this limited sky region, we also find that the results are only slightly dependent on the adopted Galactic offset.

\begin{figure}
\includegraphics[width=0.5\textwidth]{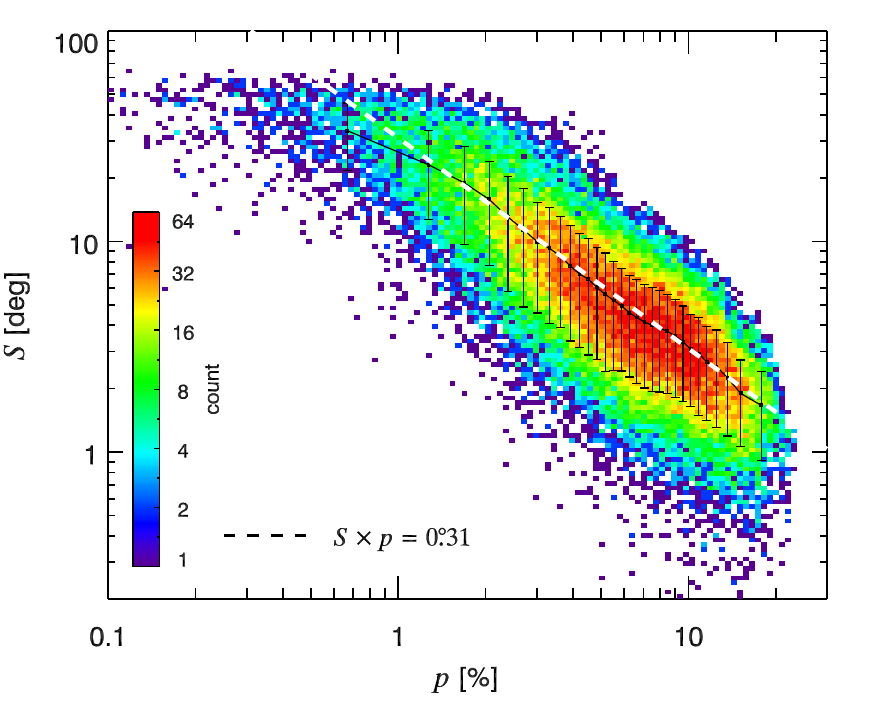}
\caption[]{
Two-dimensional histogram showing the joint distribution function of $\mathcal{S}$ and $p$ at 160\arcmin\ resolution, using a lag $\delta=80\arcmin$. The black curve is the running mean of $\S$ as a function of the mean $p$, in bins of ordered $p$, with each bin containing the same number of pixels. The error bars represent the standard deviation of $\S$ in each bin of $p$. The dashed white line shows our fit $\S = 0\pdeg31/\PsI$ to this running mean.
}
\label{fig:S_p_GNILC_160arcmin_delta80}
\end{figure}

Extending to the full sky at 160\arcmin\ resolution and a lag of $\delta=80\arcmin$, we present the 2-dimensional histogram of the joint distribution function of $\S$ and $p$ in Fig.~\ref{fig:S_p_GNILC_160arcmin_delta80}. The data clearly show that the inverse relationship seen at low and intermediate Galactic latitudes in~\cite{planck2014-XIX} persists in the high-latitude sky. In Fig.~\ref{fig:S_p_GNILC_160arcmin_delta80} we also
display the running mean of $\mathcal{S}$ in each bin of $p$ for the data.\footnote{We note that the linear fitting of the mean $\log\S$ per bin of $\log{p}$ that was originally used in~\cite{planck2014-XIX} and~\cite{planck2014-XX} is not the optimal procedure to quantify the inverse relationship between $\S$ and $p$.} We show in the next section that simple analytical models suggest that the relationship is indeed $\left\langle\S\right\rangle_p\propto 1/p$. Such a trend is shown in Fig.~\ref{fig:S_p_GNILC_160arcmin_delta80} as the dashed white line.

\subsection{Relationship to models}
\label{sec:rmodels}

All of the properties discussed so far, namely the distribution functions of $p$, $\psi$, and $\mathcal{S}$, the decrease in the maximum polarization fraction with increasing column density, and the inverse relationship between $\mathcal{S}$ and $p$, are consistent with the analysis first presented in~\cite{planck2014-XIX}.
Subsequently, phenomenological models of the polarized sky incorporating Gaussian fluctuations of the magnetic field have been developed \citep{planck2016-XLIV,ghosh-et-al-2017,vansyngel-et-al-2017,levrier-et-al-2018}. 
Interestingly, although these models were built to reproduce some 1- and 2-point statistics of polarized emission maps, they were not tailored to reproduce the inverse relationship between $\mathcal{S}$ and $p$ that is evident in the {\Planck} data, and yet they are able to do so very readily and robustly. 
A similar inverse relationship between $\S$ and $p$ was also observed in synthetic polarization maps built from numerical simulations of MHD turbulence~\citep{planck2014-XX}.

In Appendix~\ref{sec:appendix:Sp} we present an analytical derivation of this property, based on the most basic version of these phenomenological models. In that framework, the emission is assumed to arise from a small number $N$ of layers, each emitting a fraction $1/N$ of the total intensity, and harbouring a magnetic field that is the sum of a uniform component and a turbulent Gaussian component. The main parameters of the model, besides $N$, are the intrinsic polarization fraction $p_0$,\footnote{This parameter is related to $p_\mathrm{max}$, the maximum polarization fraction observed, by
$\pmax=p_0/(1-p_0/3)$ \citep{planck2014-XX}.}
the level of the turbulent magnetic field $\fM$ relative to the magnitude of the uniform component, and the spectral index $\alpha_{\rm M}$ of the spatial distribution of this turbulent component. 
\begin{figure}
\includegraphics[width=0.5\textwidth]{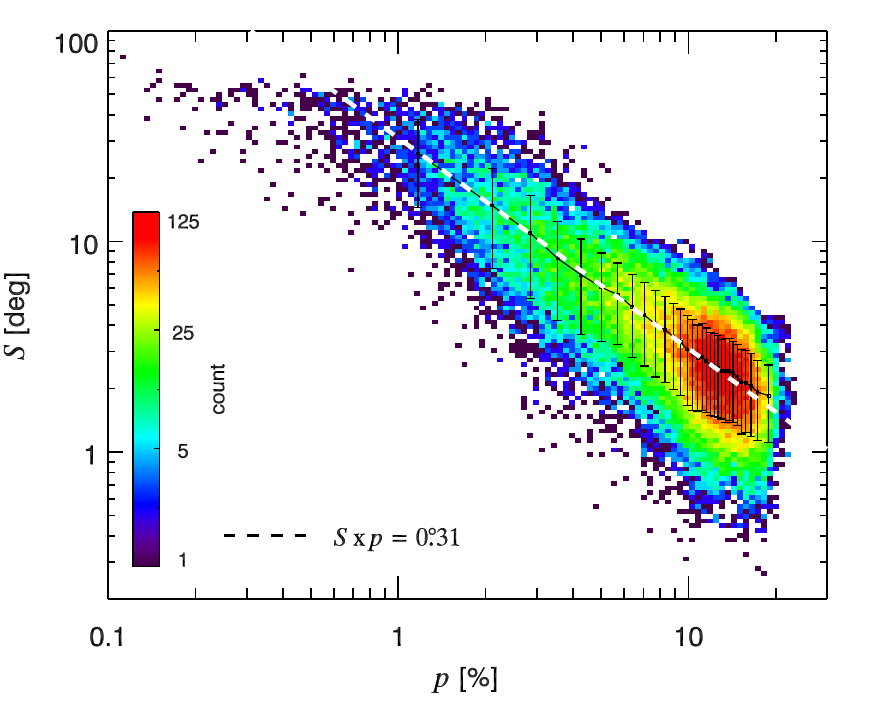}
\caption[]{
Same as Fig.~\ref{fig:S_p_GNILC_160arcmin_delta80}, but for a phenomenological model of the polarized sky, as described in the text. The dashed white line is the same as in Fig.~\ref{fig:S_p_GNILC_160arcmin_delta80}.
}
\label{fig:S-vs-p-Andrea}
\end{figure}
In Fig.~\ref{fig:S-vs-p-Andrea} we show the 2-dimensional distribution function of $\mathcal{S}$ and $p$ at 160\arcmin\ resolution, using a lag $\delta=80\arcmin$, for a polarized sky built from such a Gaussian phenomenological model, with $\alpha_{\rm M}=-2.5$, $f_{\rm M}=0.9$, $N=7$, and $p_0=26\,\%$. This choice of parameters, within the range of good fits in~\cite{planck2016-XLIV}, leads to the mean analytical relation (see Appendix~\ref{sec:cmp_analytic_numeric}) $\left\langle \S\right\rangle_p~=~0\pdeg34/p$, which is very close to a fit to the observational trend, $\left\langle \S\right\rangle_p~=~0\pdeg31/p$, overplotted in Figs.~\ref{fig:S_p_GNILC_160arcmin_delta80} and~\ref{fig:S-vs-p-Andrea}. We show in Appendix~\ref{sec:cmp_analytic_numeric} that this relationship depends weakly on the resolution $\omega$ according to
\begin{equation}
\label{Eq-Sp_lag}
\left\langle \S\right\rangle_p=\frac{0\textrm{\pdeg}31}{p}\left(\frac{\omega}{160\arcmin}\right)^{0.18}\,.
\end{equation}

Because changes of dust properties or dust alignment are not included in these phenomenological models nor in the synthetic observations from MHD simulations, we conclude that the inverse relationship between $\mathcal{S}$ and $p$ is a generic statistical property that results primarily from the topology of the magnetic field.

We also note that neither the phenomenological model of~\cite{planck2016-XLIV}, nor the MHD simulation in~\cite{planck2014-XX}, account for the 3D structure of the ordered (mean) component of the GMF on large scales. The imprint of this ordered component on the dust polarization can be readily identified in the map of the dust polarization angle (Fig.~\ref{fig:P_p_psi_GNILC-80arcmin}). It also impacts the polarization fraction map on large angular scales and thereby the dependency of $p$ on Galactic latitude. For synchrotron polarization, this has been quantified by~\cite{page2007} and~\cite{miville-deschenes-et-al-2008} using Galaxy-wide models of the GMF. As discussed in~\cite{alves-et-al-2018}, a comprehensive model of dust polarization would also need to take into account the structure of the GMF on the scale of the Local Bubble (100--200\,pc).

\section{Insight from interrelationships and Galactic context}
\label{sec:NewResults}

Further insight into statistical measures of the polarization can be gained not only by considering them in relation to one another, but also by studying how they jointly vary with other physical parameters such as dust temperature $\Td$ or column density, and how these relationships evolve from the diffuse ISM to molecular clouds.

An important parameter in this study is the total amount of dust along the line of sight, or dust column density, which is usually quantified by the dust optical depth $\tau$ (at 353\,GHz). Because dust emission is optically thin at this frequency, this relates the modified blackbody (MBB) model of the emission to the total intensity via
\begin{equation}
\label{eq:Inu}
I_\nu=\tau\,B_\nu(\Td)\left(\frac{\nu}{353\,\mathrm{GHz}}\right)^\beta \, ,
\end{equation}
where $\beta$ is the observed dust spectral index~\citep{planck2013-p06b}. It is also common to rescale from dust optical depth to entirely different units like colour excess in the optical, $\ebvsub$,\footnote{We write $\ebvsub$ instead of simply $\ebv$ to emphasize that this colour excess is computed from the dust optical depth derived from \Planck\ data, and to distinguish it from other estimates used in Sect.~\ref{sec:stars}.} 
or total column density of hydrogen $\NH$. The calibrations of such rescalings are uncertain and possibly dependent on the environment. This is not important for our results below and we use the MBB parameters $\tau$ and $\Td$ from \cite{planck2016-XLVIII}, the calibration from \cite{planck2013-p06b} at 353\,GHz,
\begin{equation}
\label{eq:ebvsub}
\ebvsub=\left(1.49\pm0.03\right)\times 10^4\,\tau \, ,
\end{equation}
and the relation $\NH =5.87\times10^{21}\,$cm$^{-2}\times \ebvsub$ \citep{Bo78,R09} to estimate $\NH$. It is preferable to use $\tau$ converted to $\NH$ rather than an estimate of the gas column density derived from {\hi} because of the presence of dust in the WIM that is sampled by all of our polarized and unpolarized observables.

We note that in this section we use not only the \GNILC\ maps at 80\arcmin\ and 160\arcmin\ resolution, but also the alternative Stokes maps (Sect.~\ref{sec:asm}) at finer resolutions of 40\arcmin, 20\arcmin, and 10\arcmin.

\subsection{Origin of the observed variations of the polarization fraction $\PsI$}
\label{subsec:originpvar}

\begin{figure}
\includegraphics[width=0.5\textwidth]{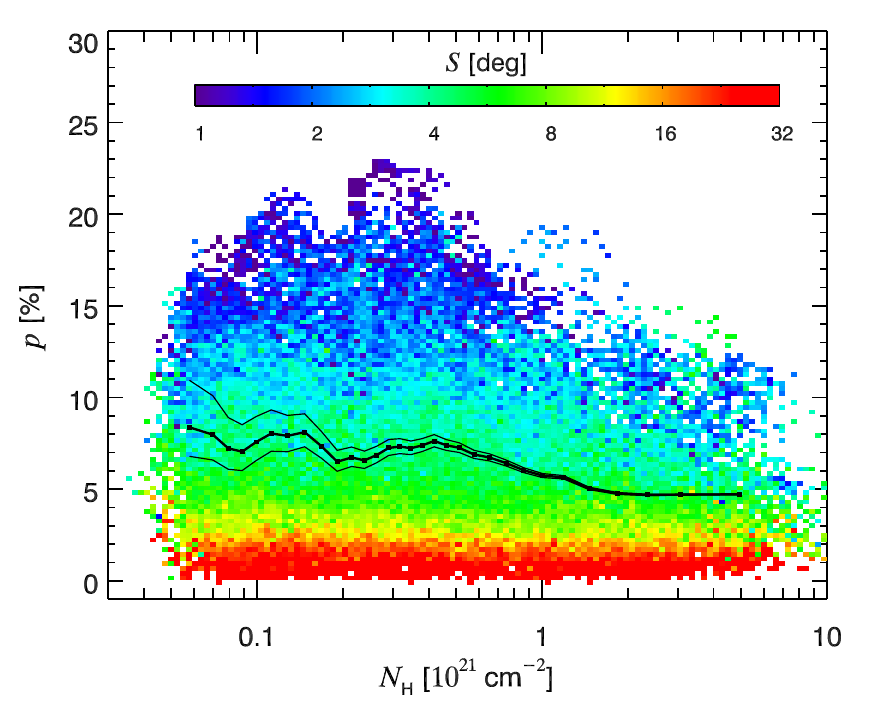}
\caption{
Polarization fraction $\PsI$ as a function of the total gas column density, $\NH$, coloured by the polarization angle dispersion function $\S$ (on a logarithmic scale). The resolution of the data is 160\arcmin, in order to limit the bias in $\S$. The black curve is the running mean of $\PsI$ as a function of the mean $\NH$, in bins of $\NH$ that contain the same number of pixels. The top, middle, and bottom running means are calculated for the low, fiducial, and high intensity offsets, respectively.
} 
\label{Fig_strat_DISM}
\end{figure}

The mutual correlations between $\PsI$, $\S$, and the column density $\NH$ were studied in detail for the particular case of the Vela C molecular cloud by \cite{fissel-et-al-2016} using BLASTpol data.
From the present \DRThree\ data, Fig.~\ref{Fig_strat_DISM} shows how these correlations appear for the more diffuse ISM ($4\times10^{19}$\,\cmsq$<\NH<10^{22}$\,\cmsq) over the whole sky, excluding only the latitudes close to the Galactic plane ($|b| < 5 \deg$).
Significant variations about the trend of $\PsI$ with $\NH$ prevent modelling it by a simple relationship. For $\NH < 5\times10^{20}$\,\cmsq, the mean value is compatible with a constant, then decreases over the range 0.5--$2\times10^{21}$\,\cmsq, and eventually becomes rather flat again. 
Colouring\footnote{This is done in practice to represent the mean value of $\S$ over points that fall within a given bin in $\PsI$ and $\NH$.} the data with $\S$ (on a logarithmic scale)
we see from the stratification of the data in Fig.~\ref{Fig_strat_DISM} that there is a gradient of $\S$ mainly perpendicular to the observed trend of $\PsI$ with respect to $\NH$. This analysis indicates that the decreases in $\PsI$ with $\S$ and with $\NH$ are mostly independent of each other.

\begin{figure}[htbp]
\includegraphics[width=\hhsize,trim=-20 0 0 0, clip=true]{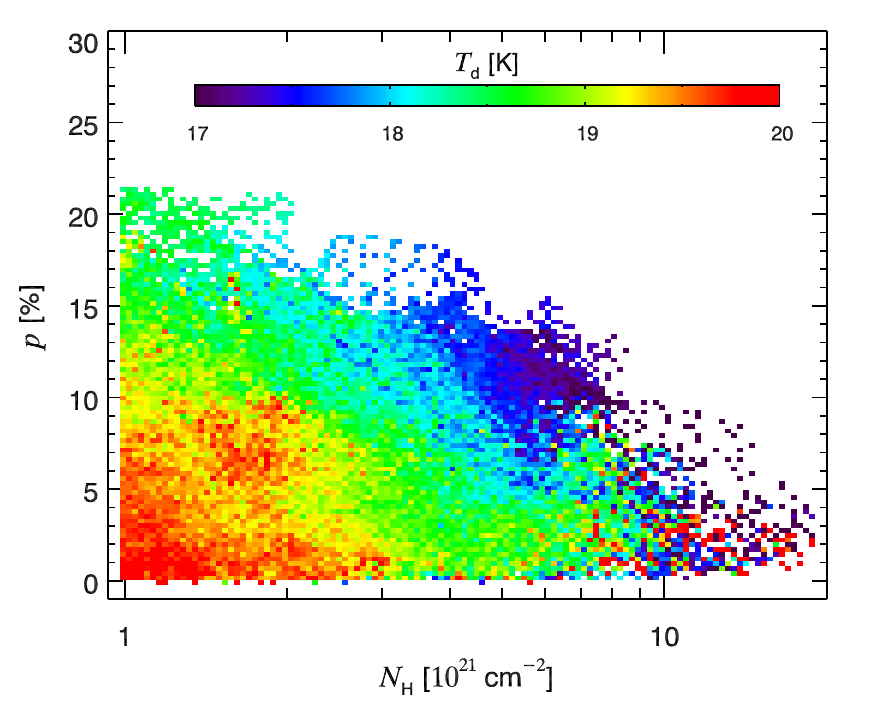}
\includegraphics[width=\hhsize,trim=-20 0 0 0, clip=true]{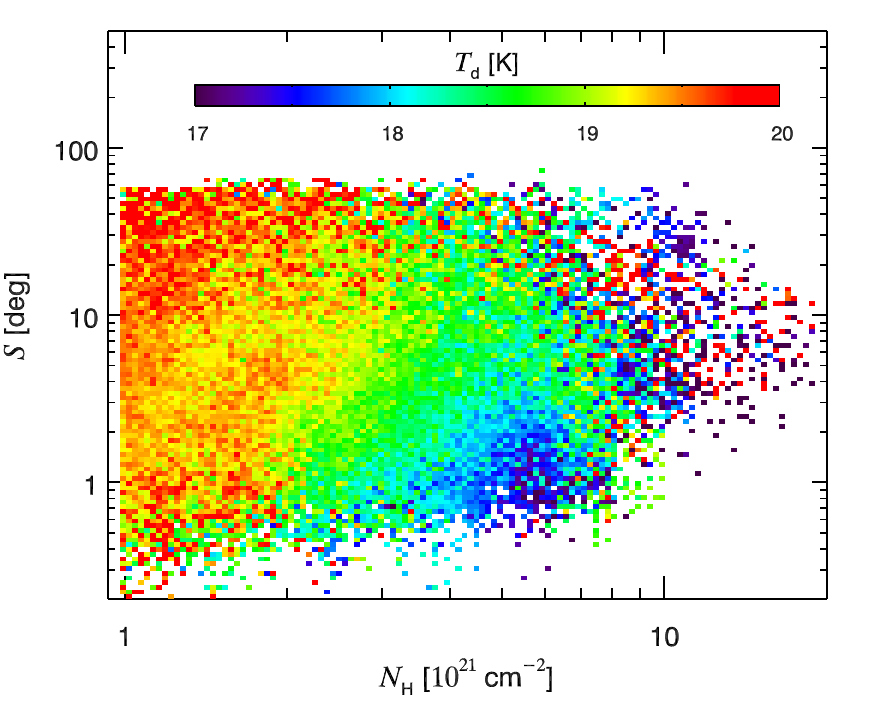}
\caption{Polarization fraction $\PsI$ (top panel), and polarization angle dispersion function $\S$ (bottom panel) coloured by the dust temperature, $\Td$, as a function of the column density $\NH$. We note that the resolution of the data used here is $40\arcmin$. Estimates of $\S$ are nevertheless not biased, because only lines of sight with $\NH\geqslant 10^{21}\,\mathrm{cm^{-2}}$ are considered.
}
\label{PsI_S_NH_byT}
\end{figure}

Figure~\ref{PsI_S_NH_byT} shows how the variations of $\PsI$ and $\S$ at a given column density are related to the dust temperature $\Td$.
Dust tends to be systematically cooler when $\PsI$ is high and warmer when $\PsI$ is low (top panel). This is observed at all but the highest column densities ($\NH > \maxNHSpT$\,\cmsq). On the other hand, the opposite is seen in $\S$ (bottom panel). The mirror symmetry between the two panels of Fig.~\ref{PsI_S_NH_byT} shows convincingly that there is in fact no physical relation between the polarization fraction $\PsI$ and the dust temperature $\Td$ in the diffuse ISM. Even if it seems that, at any column density, high $\PsI$ corresponds to colder dust and low $\PsI$ to warmer dust, the bottom panel demonstrates that the value of $\PsI$ is actually driven by $\S$, i.e., by the magnetic field structure and the depolarization produced by its variations along the line of sight and within the beam.

\subsection{Exploring beyond first-order trends using $\StimesPsI$}
\label{subsec:stimesp}

In Sect.~\ref{sec:rmodels}, we concluded that the inverse relationship between $\mathcal{S}$ and $\PsI$ is a generic statistical property that results from the topology of the magnetic field alone, and that a trend $\mathcal{S} \propto 1/p$, close to that observed, is expected on the basis of simple analytical models~(Appendix~\ref{sec:appendix:Sp}).
It is therefore interesting to explore beyond this underlying cause for the inverse relationship, in search of evidence for the impact of other physical factors, such as dust alignment efficiency, elongation, and composition.  For this we can use the product $\StimesPsI$, which removes the impact of the magnetic field structure \textit{statistically}. This does not mean that the product depends only on properties of the dust, e.g., the maximum polarization fraction $\pmax$. As explained in Appendix~\ref{sec:appendix:Sp}, the product $\StimesPsI$ also depends on the length over which dust structures are probed along the line of sight, as well as on the ratio of the turbulent to ordered magnetic field. Nevertheless, it is interesting to try this approach, as also emphasized by the mirror symmetry seen between the two panels of Fig.~\ref{PsI_S_NH_byT}.

\begin{figure*}
\includegraphics[scale=0.44,trim={0 40 5 10},clip]{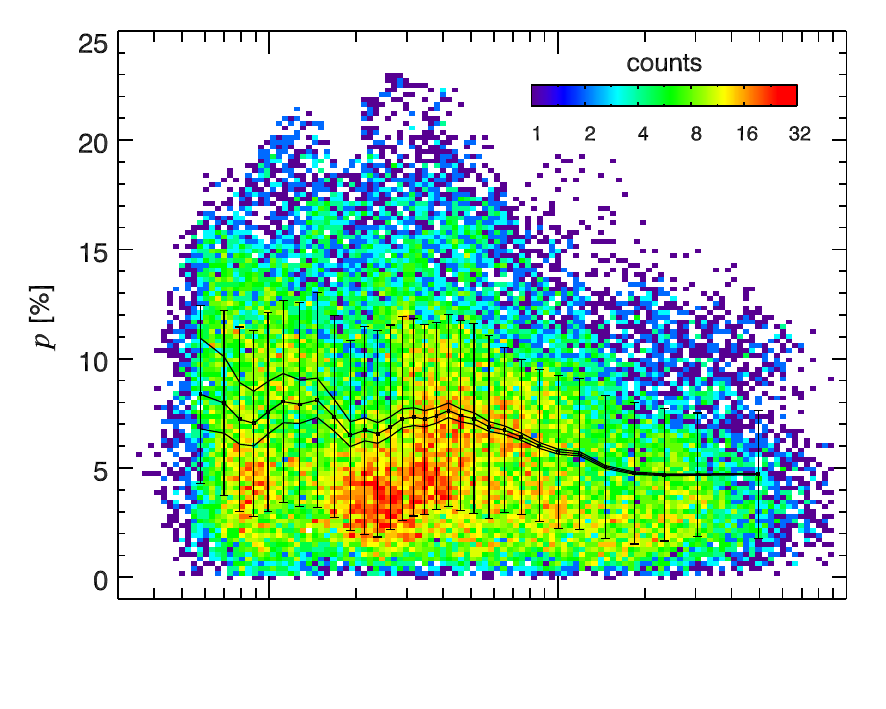}
\includegraphics[scale=0.44,trim={45 40 5 10},clip]{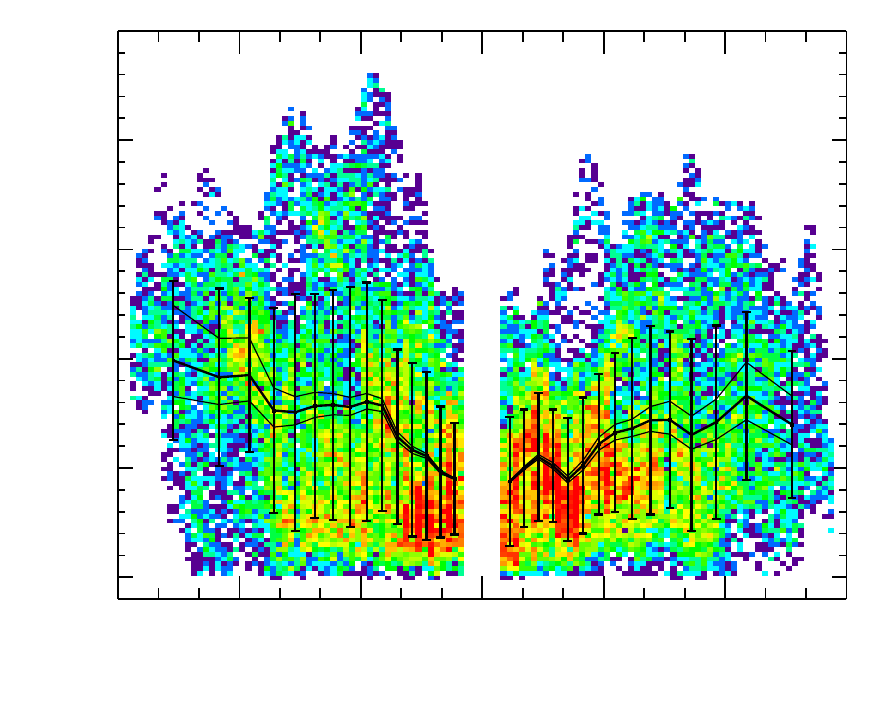}
\includegraphics[scale=0.44,trim={45 40 5 10},clip]{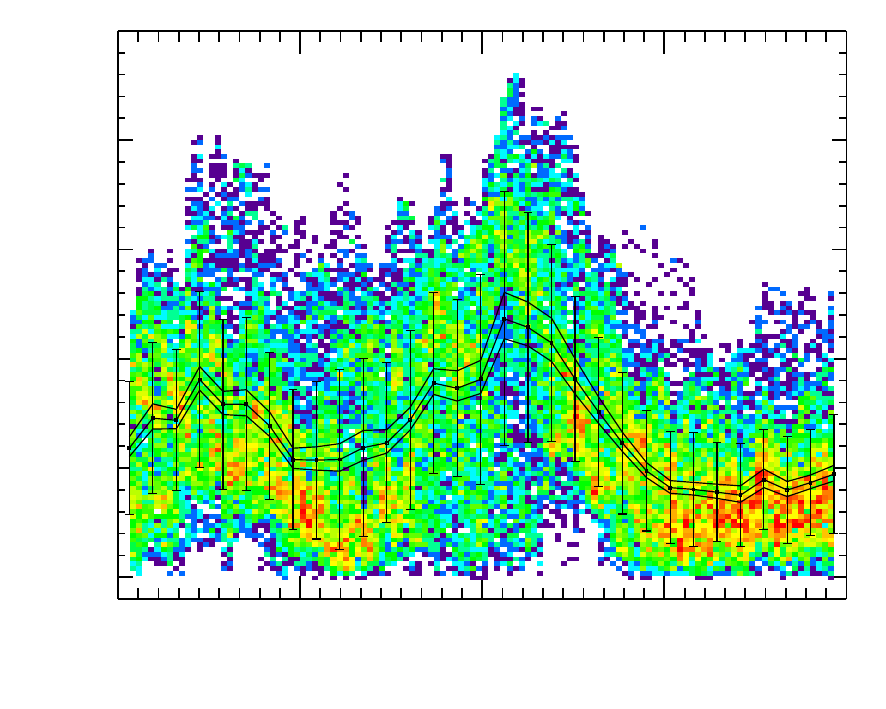}
\includegraphics[scale=0.44,trim={0 50 5 10},clip]{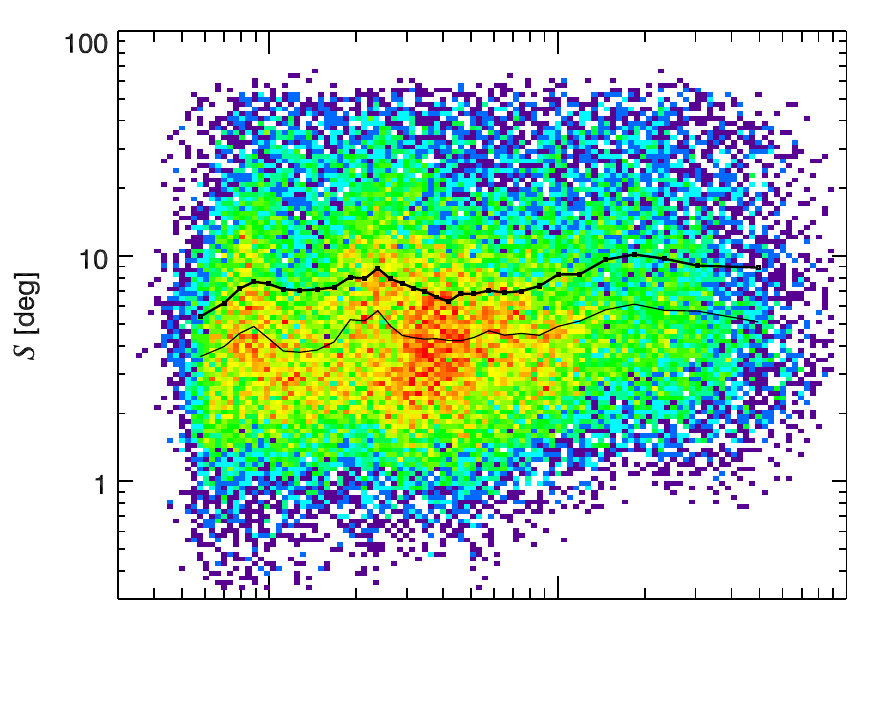}
\includegraphics[scale=0.44,trim={45 50 5 10},clip]{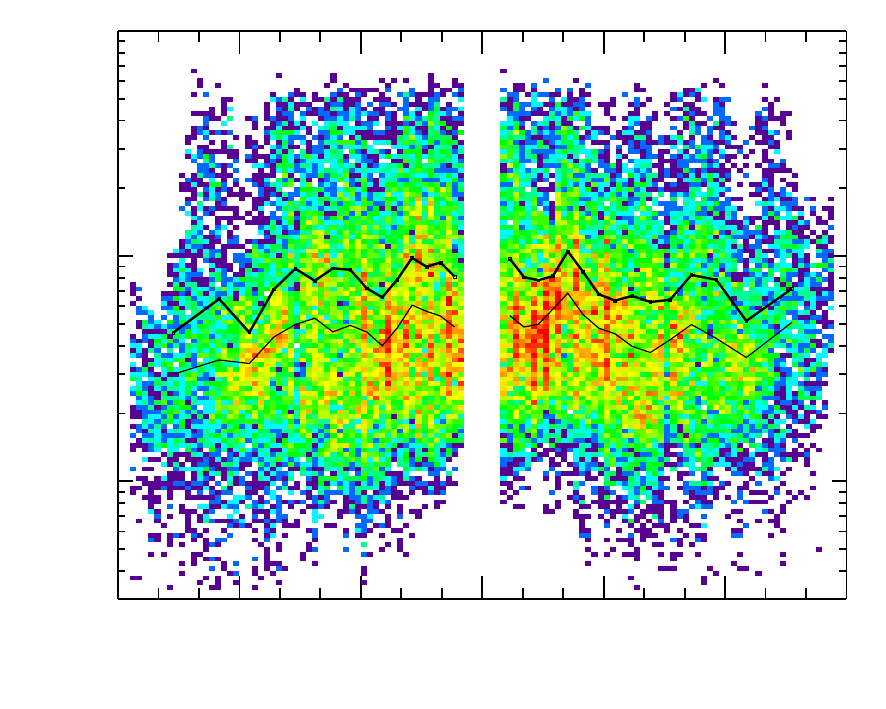}
\includegraphics[scale=0.44,trim={45 50 5 10},clip]{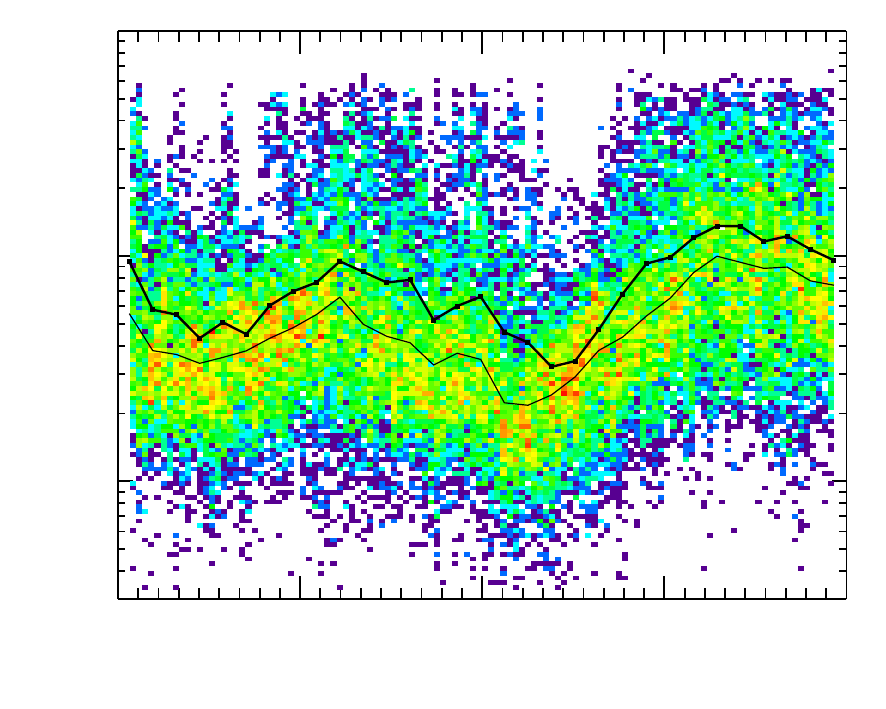}
\includegraphics[scale=0.442,trim={0 0 5 10},clip]{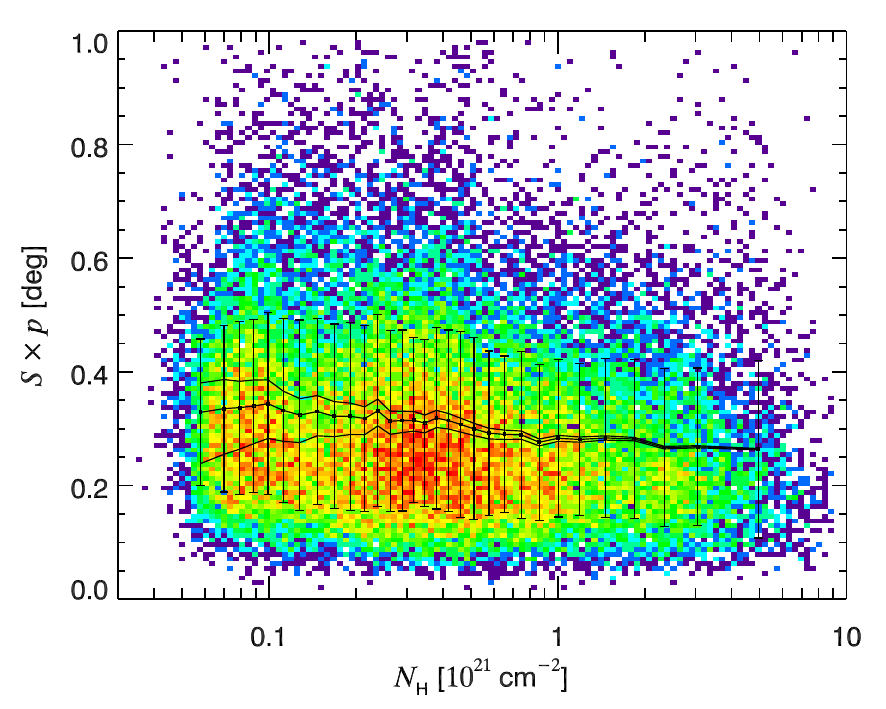}
\includegraphics[scale=0.442,trim={45 0 5 10},clip]{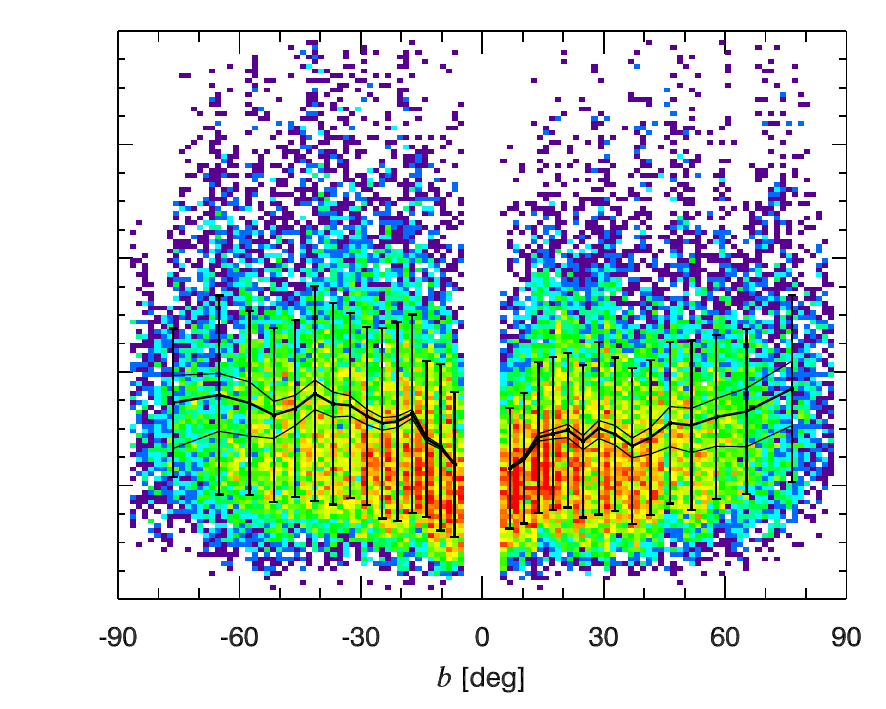}
\includegraphics[scale=0.442,trim={45 0 5 0},clip]{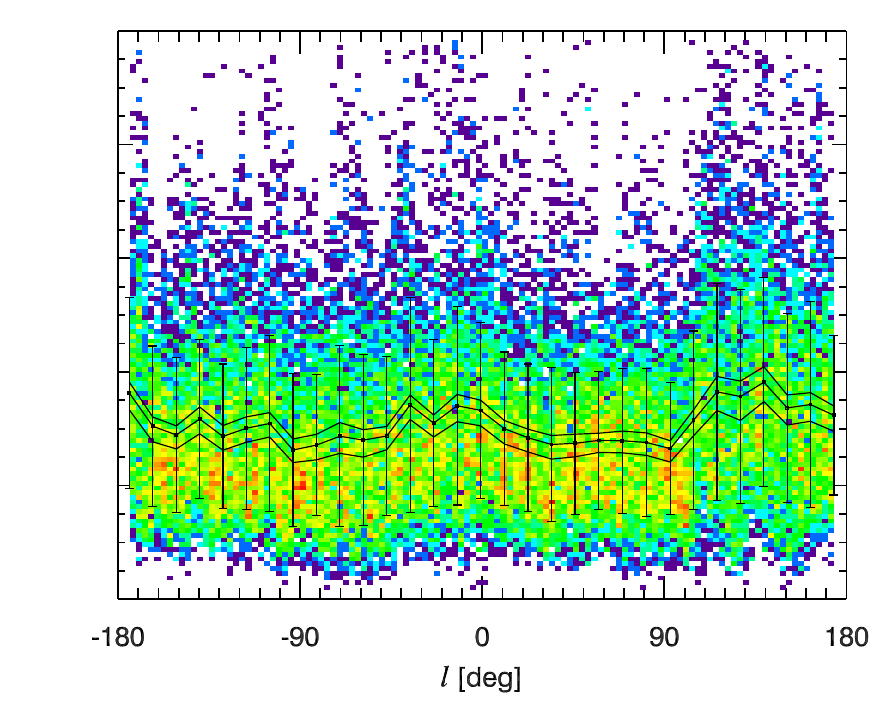}
\caption{
Two-dimensional histograms with background colours encoding the density of points on a logarithmic scale, showing
$\PsI$ (top), $\S$ (middle), and $\StimesPsI$ (bottom) as a function of the column density $\NH$ (left), Galactic latitude $b$ (middle), and Galactic longitude $l$ (right). The resolution is 160\arcmin. The colour bar shown in the top left panel is common to all plots. Black curves show the running means calculated as in Fig.~\ref{Fig_strat_DISM}, with error bars representing the scatter in each bin. For $\S$, which is on a logarithmic scale, the median trend shown (thin black line) follows the density of points more faithfully than does the mean (thicker black line).}
\label{Fig_PsIxS}
\end{figure*}

Accordingly, Fig.~\ref{Fig_PsIxS} compares the variations of not only $\PsI$ and $\S$, but also $\StimesPsI$ with $\NH$, Galactic latitude $b$, and Galactic longitude $l$. It should be noted that throughout this entire analysis lines of sight close to the Galactic plane ($|b| < 5\deg$) are excluded.

As expected, the product $\StimesPsI$ has smaller variations with $\NH$, $b$, and $l$ than exhibited by $\PsI$ and $\S$ separately, and the decrease in $\StimesPsI$ with $\NH$ is systematic, without significant departures. 

Away from the Galactic plane, the dependence of $\StimesPsI$ on $b$ is less pronounced than it is for $\PsI$ and also more symmetric between positive and negative latitudes. The strong dependence of $\PsI$ on $b$ that can be attributed to the systematic change in the orientation of the mean magnetic field with respect to the line of sight is mitigated in $\StimesPsI$, confirming our interpretation. However, there are still small variations of $\StimesPsI$ over a large spatial scale that remain to be interpreted. Towards the Galactic plane there is a pronounced dip, that is probably due to the accumulation of variously polarized structures along the line of sight at these low latitudes \citep{JKD92}. This dependence on the latitude will be further discussed in Sect.~\ref{view}.

As with the dependence on $\NH$, the variations of the product $\StimesPsI$ with Galactic longitude $l$ are much less pronounced (of the order of 30\,\%) than those of $\PsI$ and $\S$ independently (a factor 3 or so in each case).

\subsection{Dedicated study for six molecular regions in the Gould Belt}
\label{subsec:gbregions}

The radiative torques theory \citep[RAT;][]{Lazarian2007,Hoang2016} makes strong predictions for the dependence of the dust alignment on local physical conditions, namely the intensity and anisotropy of the radiation field, and the angle between the magnetic field and the anisotropic radiation field. Dense regions, screened from the interstellar radiation field and possibly with embedded sources, should be promising regions in which to identify evidence for RATs \citep{VD15,W17}. 

To probe this possibility, we have selected a set of six 12\deg$\times$12\deg\ molecular regions in the Gould Belt~\citep{dame-et-al-2001}. These are listed in Table~\ref{table-GB}. All but one (Aquila Rift) were already studied using \Planck\ 2013 data in~\cite{planck2014-XX}. The higher S/N in these bright, high column density regions enables an analysis at a higher resolution (40\arcmin, $\nside=256$), and the uncertainty on the offset in total intensity $I$ can be safely ignored. For this study, we therefore make use of the \asmaps~(Sect.~\ref{sec:asm}) with the fiducial total intensity offset.

\begin{table}[tb]                 
\begingroup
\newdimen\tblskip \tblskip=5pt
\caption{Selected molecular regions in the Gould Belt.}
\label{table-GB}                            
\nointerlineskip
\vskip -3mm
\footnotesize
\setbox\tablebox=\vbox{
   \newdimen\digitwidth 
   \setbox0=\hbox{\rm 0} 
   \digitwidth=\wd0 
   \catcode`*=\active 
   \def*{\kern\digitwidth}
   \newdimen\signwidth 
   \setbox0=\hbox{+} 
   \signwidth=\wd0 
   \catcode`!=\active 
   \def!{\kern\signwidth}
\halign{\hbox to 1.5in{#\leaderfil}\tabskip 0em&
\hfil#\hfil\tabskip 1em& \hfil#\hfil\tabskip 0pt\cr
\noalign{\doubleline}
\omit\hfil Region\hfil& Longitude $l$ [deg]& Latitude $b$ [deg]\cr
\noalign{\vskip 3pt\hrule\vskip 5pt}
Taurus&           173& $-15$\cr
Orion&            211& $-16$\cr
Chamaeleon-Musca& 300& $-13$\cr
Ophiuchus&        354& $!15$\cr
Polaris&          120& $!27$\cr
Aquila Rift&      *18& $!24$\cr
\noalign{\vskip 3pt\hrule\vskip 5pt}
}}
\endPlancktable                    
\endgroup
\end{table}

\begin{figure}[htbp!]
\includegraphics[width=\hhsize,trim=-20 0 0 0, clip=true]{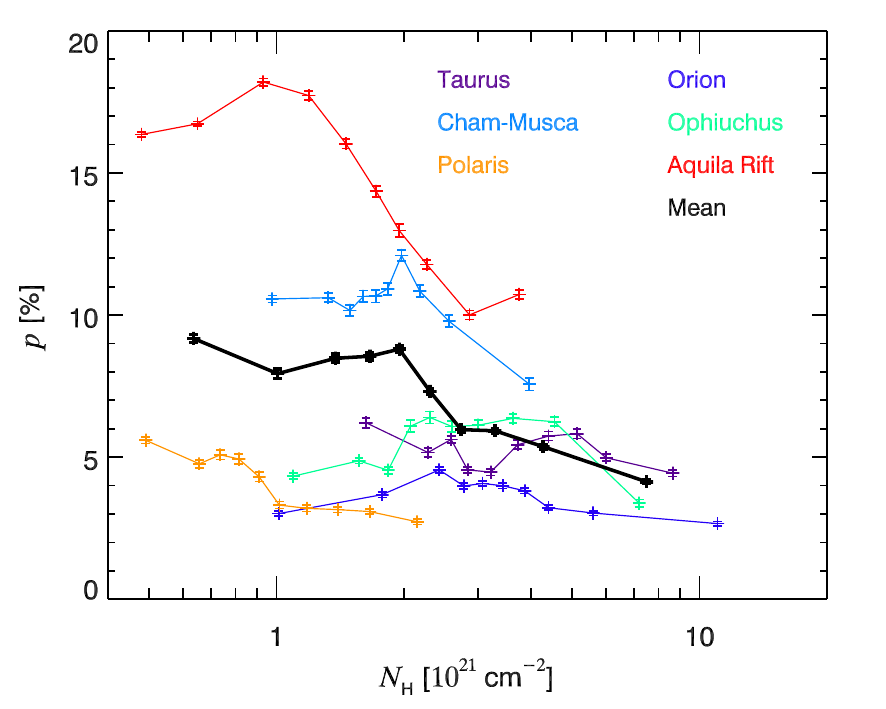}
\includegraphics[width=\hhsize,trim=-20 0 0 0, clip=true]{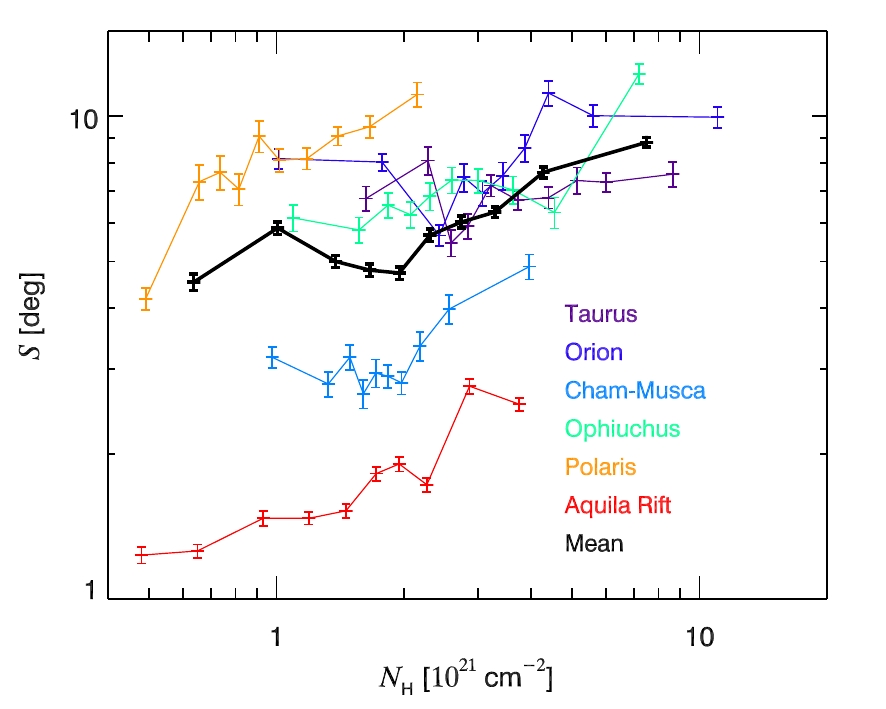}
\includegraphics[width=\hhsize,trim=-20 0 0 0, clip=true]{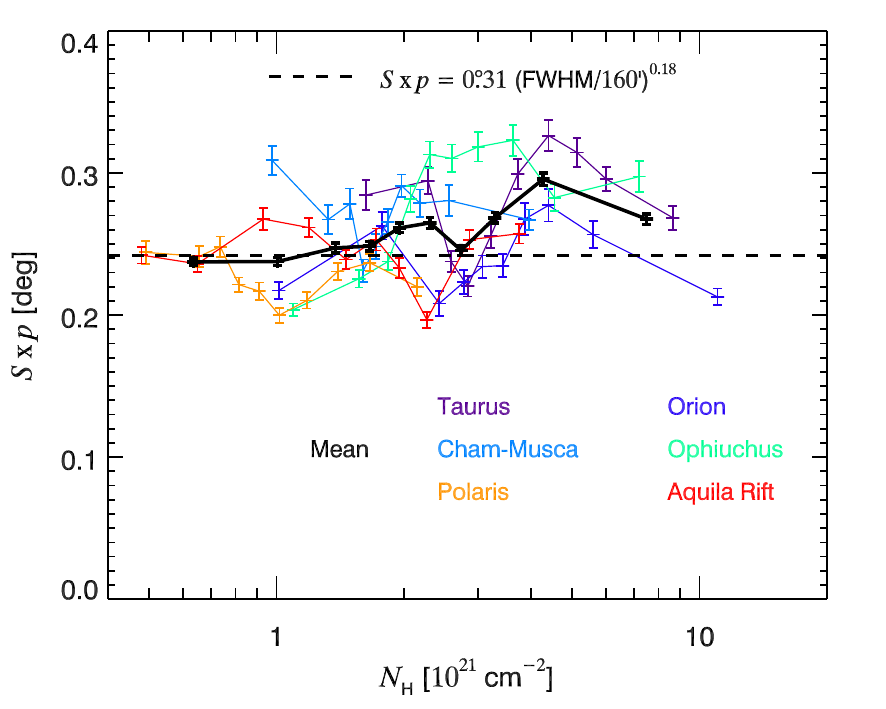}
\caption{Means from 2-dimensional distributions of polarization properties and column density $\NH$, for selected regions in the Gould Belt, at a resolution of 40\arcmin: $\PsI$ (top); $\S$ (middle); and $\StimesPsI$ (bottom). All bins in $\NH$ contain the same number of pixels $n$, approximately 250. Error bars correspond to the uncertainty on the mean, i.e., $\sigma/\sqrt{n}$, where $\sigma$ is the statistical dispersion in the corresponding bin. The dashed horizontal line in the bottom panel is the mean value of $\PsIxS$ at 160\arcmin\ (Fig.~\ref{fig:S_p_GNILC_160arcmin_delta80}), corrected for its dependence on the resolution, as per Eq.~\eqref{Eq-Sp_lag}.
} 
\label{PsI_S_NH_regions}
\end{figure}

Figure~\ref{PsI_S_NH_regions} (top panel) shows that the variation of $\PsI$ with $\NH$ is very diverse in these molecular clouds, as was already observed in~\cite{planck2014-XX}. For some (e.g., Aquila Rift and Chamaeleon) $\PsI$ is fairly high in the more diffuse envelope but progressively decreases towards denser parts. Others (e.g., Polaris and Orion) have low $\PsI$ at all column densities, while for one (Ophiuchus) $\PsI$ increases, then decreases. In each region, the corresponding variations of $\S$ with $\NH$ (Fig.~\ref{PsI_S_NH_regions}, {middle panel}) are clearly inversely related to those of $\PsI$, so that the product is by contrast almost constant and uniform across the sample of clouds, as can be seen in Fig.~\ref{PsI_S_NH_regions} ({bottom panel}).

\begin{figure}[htbp]
\includegraphics[width=\hhsize]{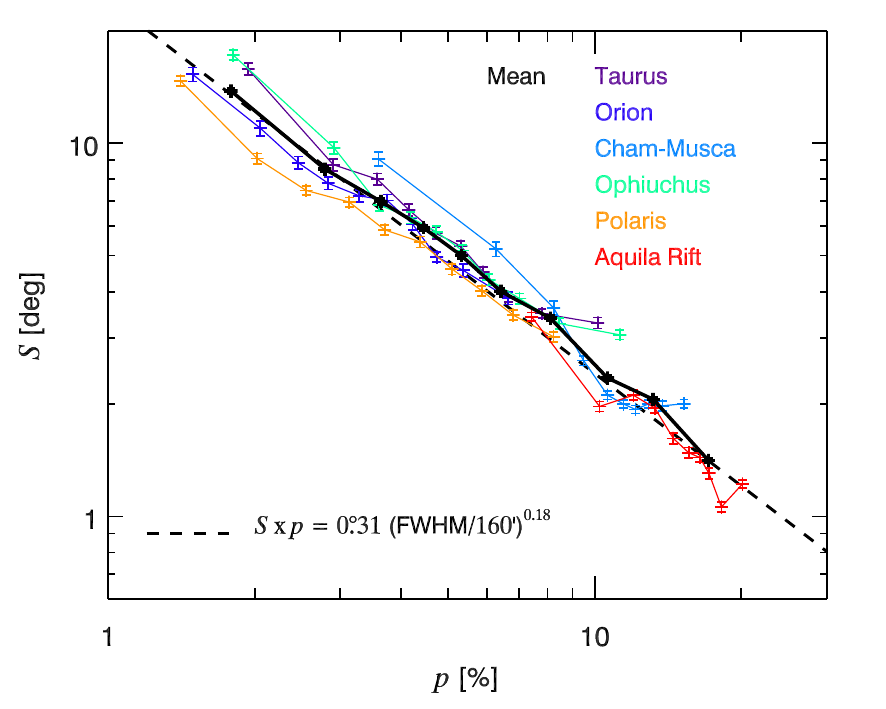}
\includegraphics[width=\hhsize]{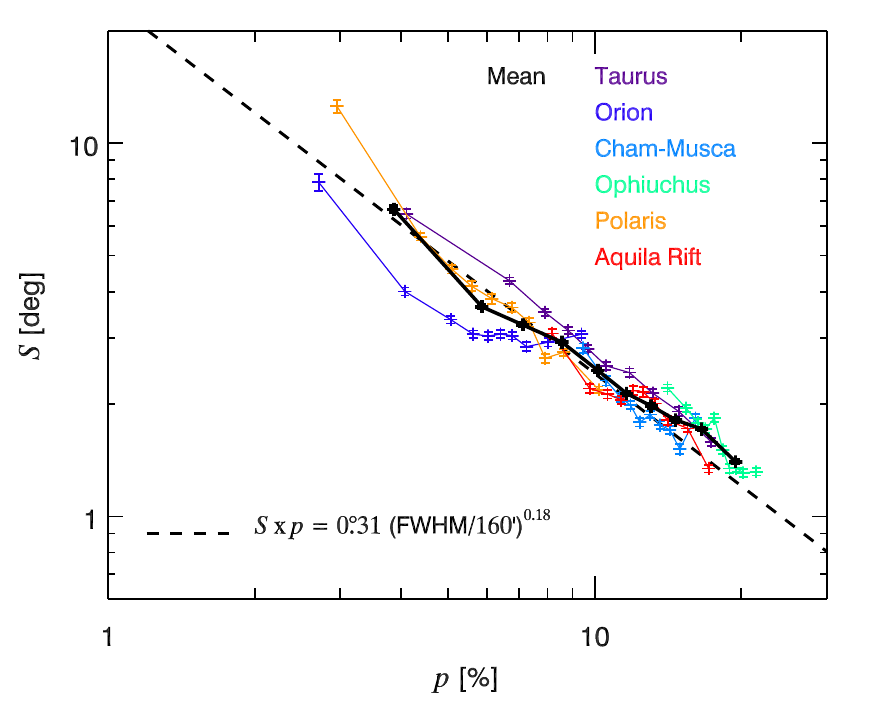}
\caption{Mean $\S$ as a function of $\PsI$ in selected regions in the Gould Belt for the \Planck\ data (top) and for our phenomenological model (bottom, see text), at a resolution of 40\arcmin. The black curve indicates the mean trend averaged over all regions. The dashed line is the fit to the mean $\S=f(\PsI)$ trend at 160\arcmin\ (Fig.~\ref{fig:S_p_GNILC_160arcmin_delta80}), corrected for its dependence on the resolution, as per Eq.~\eqref{Eq-Sp_lag}.
All bins in $\PsI$ contain the same number of pixels, $n \approx 250$. Error bars correspond to the uncertainty on the mean, i.e., $\sigma/\sqrt{n}$, where $\sigma$ is the statistical dispersion in the corresponding bin.
}
\label{S_PsI_regions}
\end{figure}

Figure~\ref{S_PsI_regions} (top panel) directly shows this inverse trend between $\S$ and $\PsI$ in the selected molecular clouds. The various curves show the mean $\S$ in each bin of $\PsI$, all bins containing the same number of pixels. Also plotted is the average curve over all the different regions. Despite differences in mean column densities, regions as different as Polaris and Orion all fall on the same correlation line. This figure demonstrates that most of the variations of $\PsI$ with $\NH$ in Fig.~\ref{PsI_S_NH_regions} can be attributed to variations of $\S$ alone, and eventually to the variation of the magnetic field structure along the line of sight and in the plane of the sky.

We note that the inverse relationship between $\S$ and $\PsI$ is the same as the one found in Sect.~\ref{sec:S_maps} over the full sky (Fig.~\ref{fig:S_p_GNILC_160arcmin_delta80}), which is dominated by the high-latitude diffuse ISM, once the difference in resolution, and therefore in the lag $\delta$ used to compute $\S$, is accounted for in the framework of our analytical model (see Appendix~\ref{subsec:appendix:fm}). Indeed, for this analysis towards selected molecular regions, we work at a finer 40\arcmin\ FWHM resolution, instead of 160\arcmin, and numerical results of the model show that the product $\StimesPsI$ scales as FWHM$^{0.18}$ (Eq.~\eqref{Eq-Sp_lag}). The prediction of this model is shown as the dashed lines in Figs.~\ref{PsI_S_NH_regions} and \ref{S_PsI_regions}. 

The bottom panel of Fig.~\ref{S_PsI_regions} shows the result of the same procedure applied to the phenomenological model described in Sect.~\ref{sec:rmodels} and smoothed to the same resolution of 40\arcmin. Our model, which was able to reproduce the trend $\S \propto 1/p$ observed at large scale (Figs.~\ref{fig:S_p_GNILC_160arcmin_delta80} and \ref{fig:S-vs-p-Andrea}), can also reproduce it at the smaller scales probed here, in regions of $12\deg\times12\deg$. The downward shift of the correlation observed in the data (compare Fig.~\ref{S_PsI_regions} with Figs.~\ref{fig:S_p_GNILC_160arcmin_delta80} and \ref{fig:S-vs-p-Andrea}), that is due to the change in resolution and is already integrated in our expression for $\S\times p$, is also observed in the simulation.

As quantified in Appendix~\ref{sec:appendix:Sp} within the framework of the phenomenological model of~\cite{planck2016-XLIV}, the mean value of the $\StimesPsI$ product depends on $f_{\rm M}$, the ratio of the dispersion of the turbulent component of the field to the mean field strength (see Eq.~\eqref{Eq-fS}, where $f_{\rm m}$ scales linearly with $f_{\rm M}$). Thus, the alignment of the data lines in the top panel of Fig.~\ref{S_PsI_regions} is a remarkable result, which suggests that the strength of the turbulent component of the magnetic field, relative to the mean field strength, is comparable among Gould Belt clouds, and between clouds and the diffuse ISM, despite differences in the local star-formation rate. This interpretation is illustrated by the correspondence between the top and bottom (data and model) plots of Fig.~\ref{S_PsI_regions}.

In the cold neutral medium, the magnetic and turbulent kinetic energies are known from {\sc Hi} Zeeman observations to be in approximate equipartition~\citep{Heiles05}. Our analysis of the \Planck\ data suggests that this energy equipartition also applies to the Gould Belt clouds. This result is consistent with the much earlier results derived from the modelling of stellar polarization data by~\cite{Myers91} and~\cite{JKD92}.

\begin{figure}[htbp]
\includegraphics[width=\hhsize]{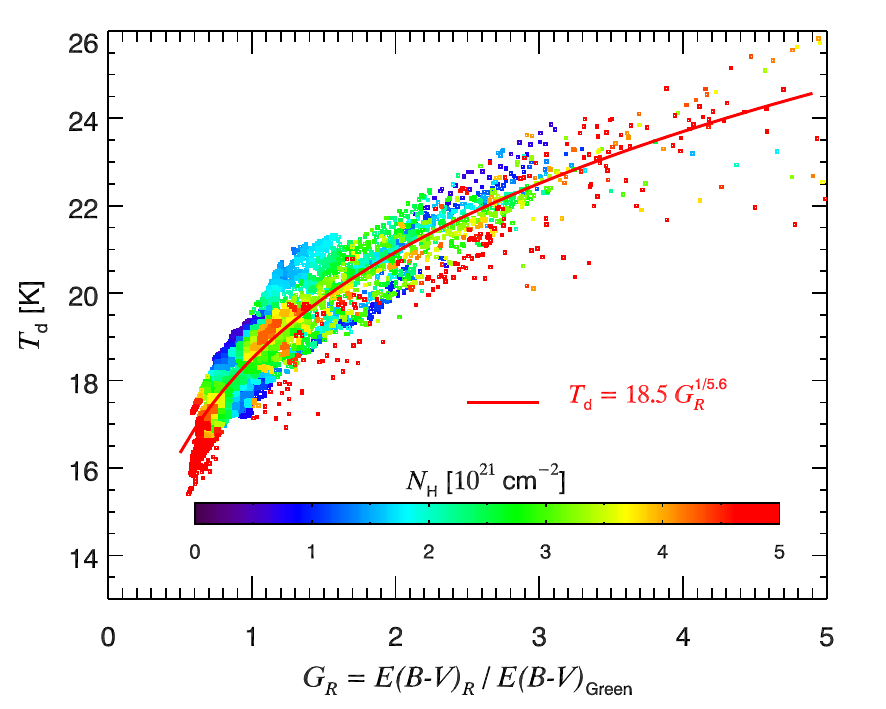}
\caption{Correlation between dust temperature $\Td$ and our estimate $\G$ for the radiation field intensity, in the selected regions, coloured by $\NH$, and for pixels with $\NH < 5\times10^{21}$\,\cmsq. The red curve is a prediction for a simple model of dust (see text). 
}
\label{T_G0_molec}
\end{figure}

\begin{figure}[htbp]
\includegraphics[width=\hhsize]{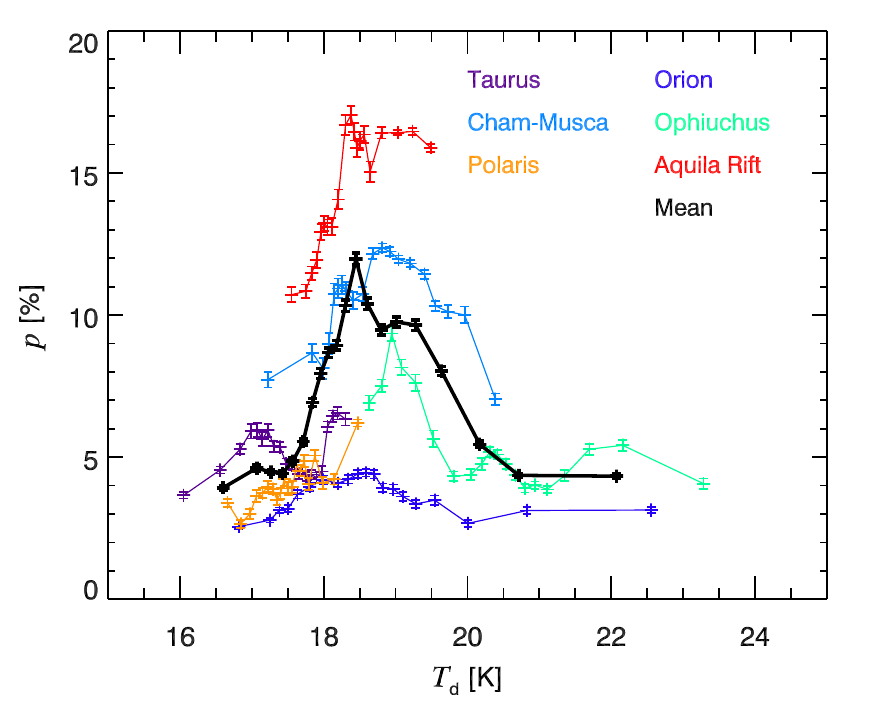}
\includegraphics[width=\hhsize]{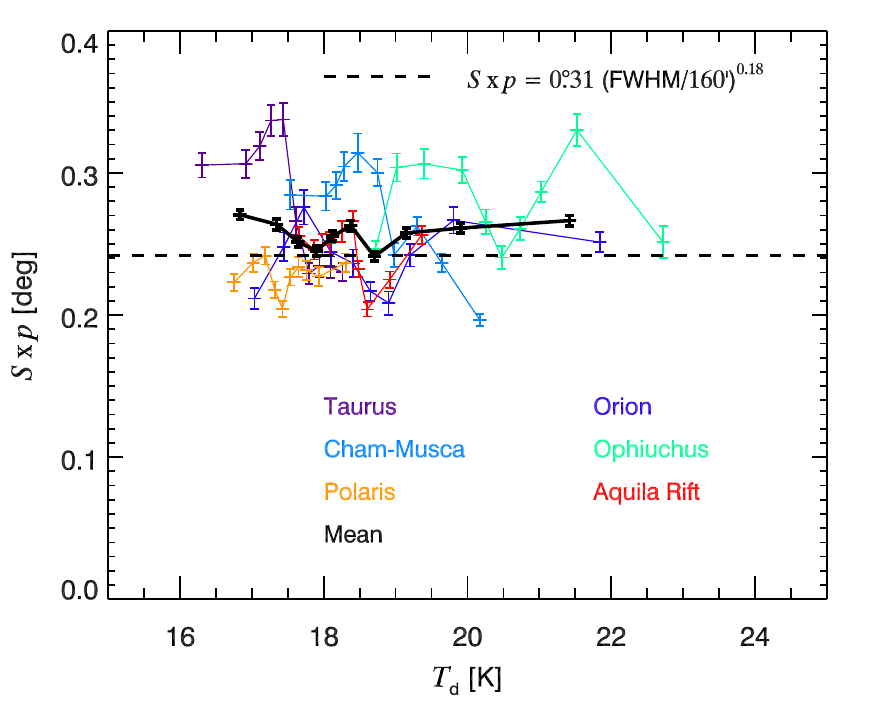}
\caption{Polarization fraction at 353\GHz\ (top) and product $\StimesPsI$ (bottom) as a function of dust temperature, at a resolution of 40\arcmin. The black curve indicates the mean trend averaged over all regions. The dashed horizontal line is the mean value of $\PsIxS$ at 160\arcmin\ (Fig.~\ref{fig:S_p_GNILC_160arcmin_delta80}), corrected for its dependence on the lag (Eq.~\eqref{Eq-Sp_lag}).}
\label{RAT_molec}
\end{figure}

To test the RAT theory, we need to estimate the relative intensity of the radiation field, $G_0$, in these regions and then look for a possible correlation between this value and the polarization fraction, once the latter is adjusted for the variations related to $\S$, i.e., look for a correlation between $G_0$ and $\StimesPsI$. To this end, we use an estimator $\G$ of the radiation field intensity \citep{guillet-et-al-2018,fanciullo-et-al-2015} that is based on the assumption of thermal equilibrium for large dust grains, which dominate the emission at this frequency. Under this hypothesis, the dust radiance $\mathcal{R}$, which is the integrated intensity of the dust emission~\citep{planck2013-p06b}, is balanced by the heating of dust by absorption of the ambient radiation field. The relative intensity $G_0$ of this ambient field is therefore estimated using the radiance per unit visual extinction $A_V$, and in practice, the estimate $\G$ is computed through
\begin{equation}
\G=\frac{\ebvR}{\ebvGreen} \, ,
\end{equation}
where $\ebvR$ stands for the dust radiance converted to a colour excess using a correlation with extinction to quasars \citep{planck2013-p06b}, while $\ebvGreen$ is from a colour excess map \citep{Green2018} based on Pan-STARRS1 (PS1).

Figure~\ref{T_G0_molec} shows the correlation between this estimate $\G$ of the radiation field intensity and the dust temperature $\Td$ from the MBB fit, in the molecular regions selected, with data points coloured according to the dust optical depth $\tau$ converted to a column density $\NH$. This is the equivalent of figure~2 in~\cite{guillet-et-al-2018}, limited to the Gould Belt regions selected. We see that the correlation is quite tight and follows a scaling $\Td\approx 18.5\,\G^{1/5.6}$, corresponding to an average temperature of $18.5\,\mathrm{K}$ for a standard radiation field $G_0=1$ and a spectral index $\beta=1.6$. We note that red points at low dust temperatures do not follow this trend perfectly because the reddening map of \cite{Green2018} tends to underestimate the true column density at high optical depths, so that $\G$ is overestimated.  We also note that at high optical depths the spectral shape of the ambient radiation field is altered by the frequency dependence of the extinction, which would also impact the amount of energy absorbed and thus the dust temperature.  Overall, over a wide range of $\G$ this plot demonstrates that a change in the dust temperature is a good indicator of a change in the ambient radiation field intensity $G_0$ in these molecular regions.

According to the RAT theory, we would expect grains in a more intense radiation field to be better aligned, and therefore that $\StimesPsI$ would tend to increase with increasing $\Td$. However, Fig.~\ref{RAT_molec} does not show any correlation between the polarization fraction and the dust temperature, and the product $\StimesPsI$, which we use as a proxy for dust alignment, does not show any trend with $\Td$ either. This analysis for molecular clouds at a resolution of 40\arcmin\ confirms our conclusion drawn in Sect.~\ref{subsec:originpvar} for the diffuse ISM that there is no strong indication of a link between the polarization fraction $\PsI$ and the dust temperature $\Td$.

\subsection{Multi-resolution view of the variations of $\PsI$, $\S$, and $\PsIxS$ across the ISM}
\label{view}

\begin{figure}[htbp!]
\includegraphics[width=\hhsize]{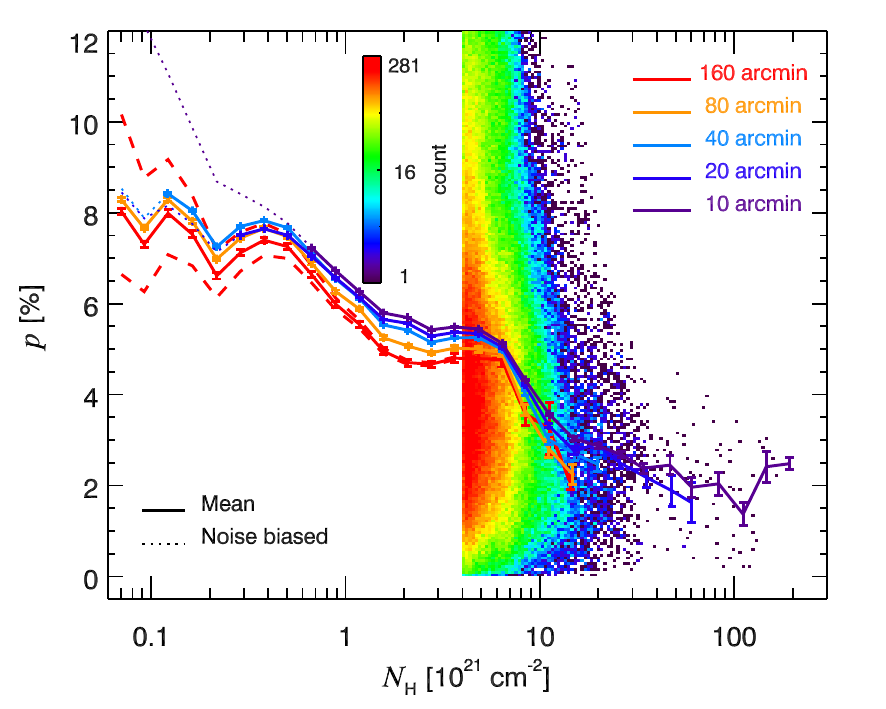}
\includegraphics[width=\hhsize]{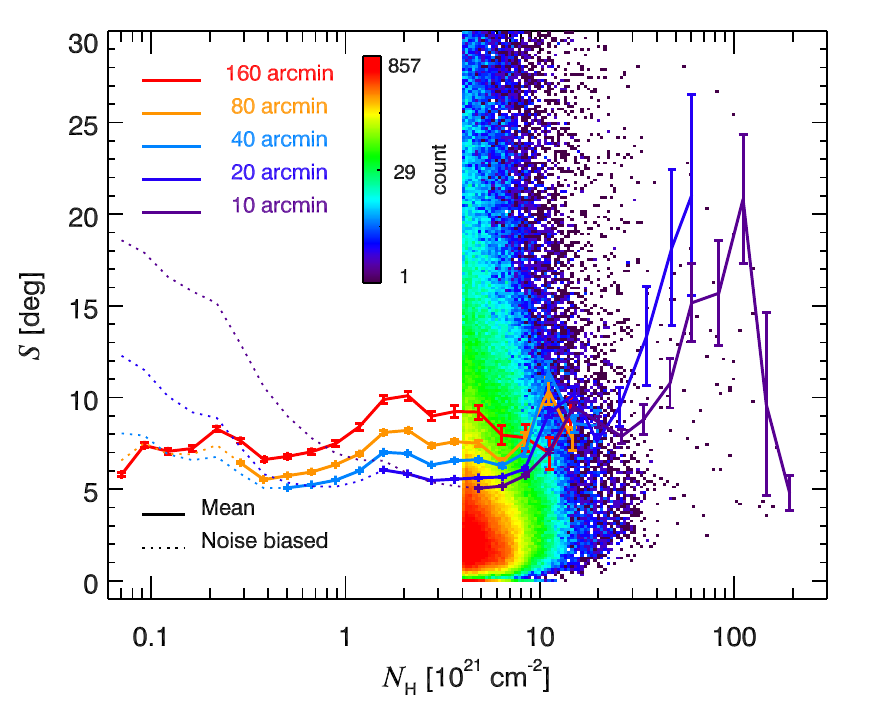}
\includegraphics[width=\hhsize]{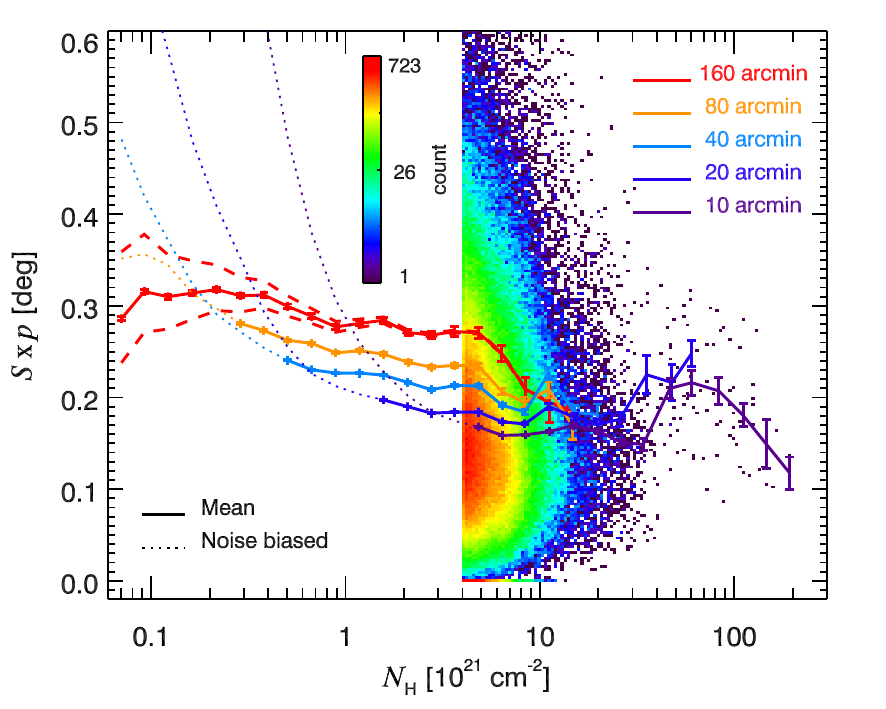}
\caption{Mean of $\PsI$ (top), $\S$ (middle), and $\PsIxS$ (bottom) as a function of $\NH$, for various resolutions, over the full sky (excluding the Galactic plane, $|b| > 5\deg$). Dotted lines correspond to trends affected by noise bias. Dashed lines correspond to the uncertainty on the total intensity offset, shown only for 160\arcmin\ data. The background colour represents the density of points at a resolution of 10\arcmin, shown only for $\NH>4\times10^{21}\,\mathrm{cm}^{-2}$.
}
\label{mean_reso}
\end{figure}

\begin{figure}[htbp!]
\includegraphics[width=\hhsize]{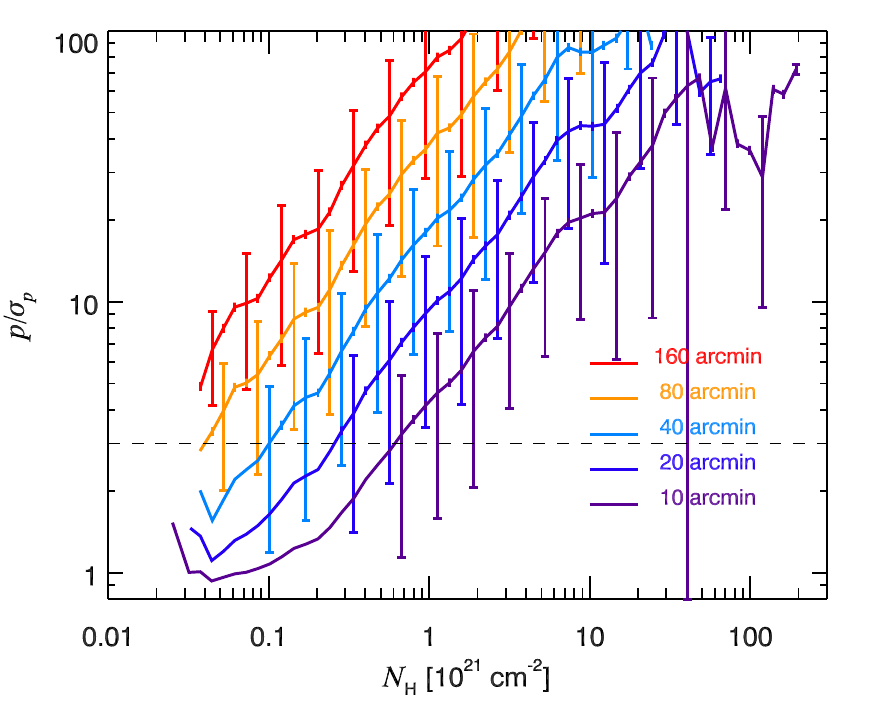}
\includegraphics[width=\hhsize]{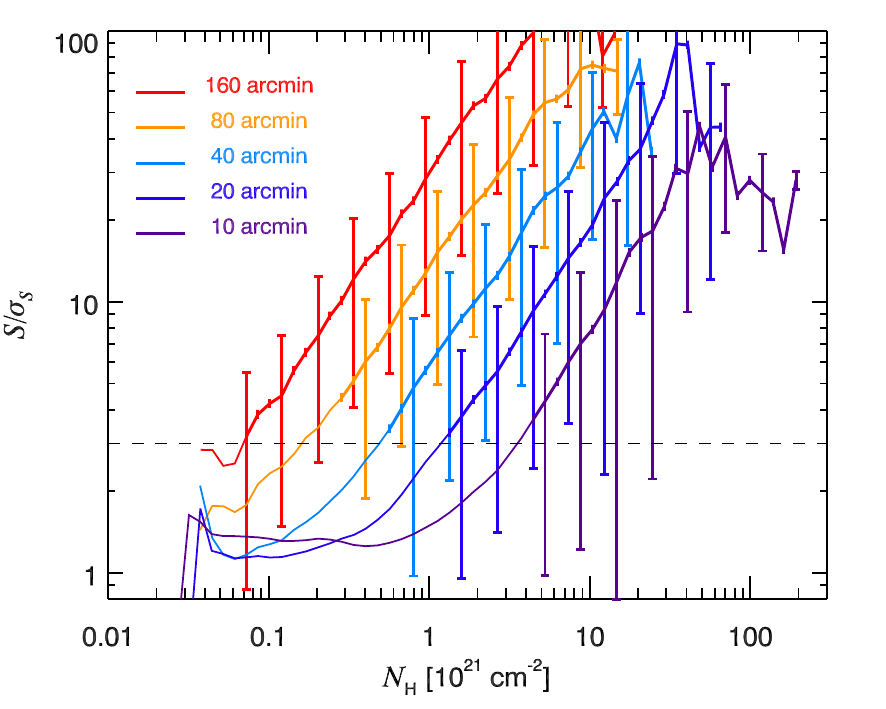}
\caption{Mean S/N of $\PsI$ (top), and $\S$ (bottom) as a function of $\NH$, for various resolutions, over the full sky (excluding the Galactic plane, $|b| > 5\deg$). Error bars correspond to the scatter in each bin, not to the uncertainty on the mean. The dashed line indicates the minimal S/N that ensures reliable mean values for debiased quantities.
}
\label{mean_reso_SNR}
\end{figure}

The discussion of the variations of $\PsI$, $\S$, and $\PsIxS$ in the diffuse ISM at 160\arcmin\ (Sect.~\ref{subsec:stimesp}) and in molecular clouds at 40\arcmin\ (Sect.~\ref{subsec:gbregions}) suggests a multi-resolution exploration of these trends. In Fig.~\ref{mean_reso}, we present such a view by compiling mean trends at all resolutions, from 10\arcmin\ to 160\arcmin. The impact of noise bias at low column densities in $\PsI$, and even more so in $\S$, is clearly seen as a rising deviation (dotted line) from the bundle of curves that otherwise roughly match - except for a global shift - at higher column densities, despite the fact that both $\PsI$ and $\S$ have been debiased. All debiasing methods indeed fail when the S/N becomes lower than 1. This occurs below a threshold in $\NH$ that is different for $\PsI$ and $\S$ and increases with decreasing FWHM. Our debiasing methods are known to be statistically efficient when the S/N is higher than about 3 \citep{SS85,PMA1}. 

Figure~\ref{mean_reso_SNR} presents the evolution of the mean S/N in $\PsI$ and $\S$ as a function of the column density for various resolutions of the map (still considering only $|b| > 5\deg$). Obviously, the S/N tends to increase with $\NH$ at a given resolution, and to decrease with increasing resolution at a given $\NH$. With these figures we can estimate, at each resolution, the column density threshold above which the mean S/N for $\PsI$ and $\S$ is larger than 3, i.e., above which debiased values of $\PsI$ and $\S$ are robust. The S/N is smaller for $\S$ than it is for $p$, and we note the large scatter in S/N for any given bin in $\NH$. Users of \Planck\ data in polarization should take into account these thresholds in column density to estimate the reliability of the debiasing. 

In Fig.~\ref{mean_reso}, the data points below these thresholds  are excluded from our analysis (dotted curves). The spread of values for $\PsI$ and $\PsIxS$ at low column densities, induced by the uncertainty on the offset in total intensity $I$, is indicated by dashed lines for 160\arcmin. For reference, the density of points at a resolution of 10\arcmin\ is plotted as a background. It is only plotted for $\NH> 4 \times 10^{21}$\,\cmsq, which is the column density threshold for $\S$ at this resolution.

\begin{figure}
\includegraphics[width=\hhsize]{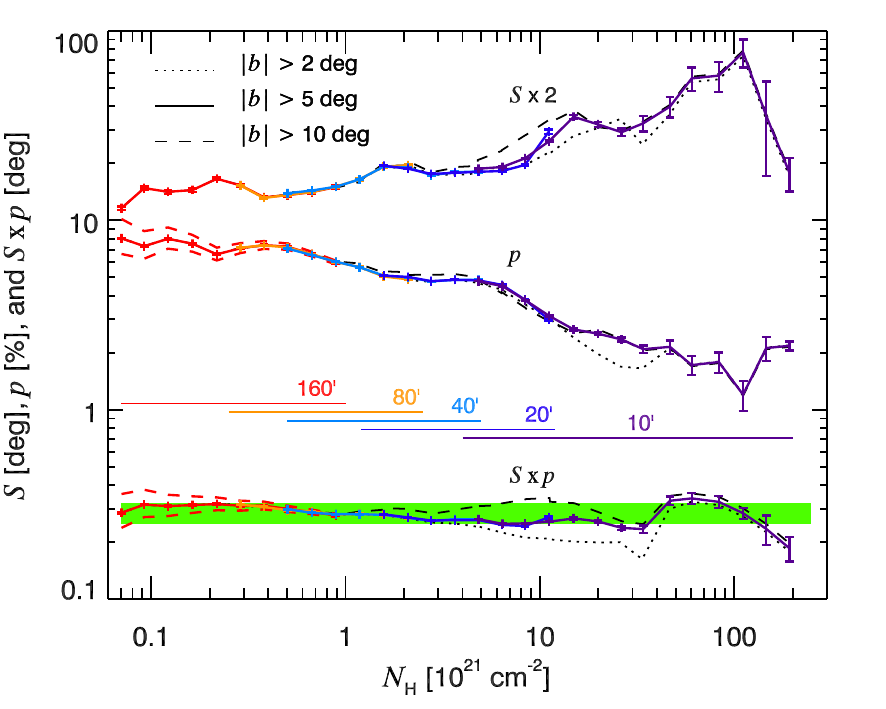}
\caption{Mean $\S$, $\PsI$, and $\PsIxS$ as a function of $\NH$, combining results from \Planck\ maps at optimal resolutions for all lines of sight above $|b| > 5\deg$ (solid curves). For clarity, $\S$ has been raised vertically by a factor of 2. Upper and lower dashed red curves show the corresponding values using the low and high total intensity offsets, respectively. In contrast to other plots, the running means are computed here for bins of equal logarithmic size, which therefore do not contain the same number of pixels. Error bars correspond to the uncertainty on the mean, $\sigma/\sqrt{n}$, where $\sigma$ is the statistical dispersion and $n$ is the number of lines of sight in the corresponding bin.
Results of the same analysis with different selection criteria on Galactic latitude are shown by thin black dashed ($|b| > 10 \deg$) and dotted ($|b| > 2 \deg$) curves.
Horizontal coloured bars indicate for each resolution $\ellres$ the column density interval $\mathcal{I}(\ellres)$ used in the renormalization procedure (see text). The green band highlights a $\maxdropSp$\,\% decrease in $\PsIxS$ with column density up to $\uptoNH$\,\cmsq.
}
\label{mean_reso_norm}
\end{figure}

Inspection of Fig.~\ref{mean_reso} shows that there always exists a range in column density where the curves at two consecutive resolutions are parallel to each other, i.e., they probe the same variations. It is therefore possible to obtain a unique, smooth and continuous trend for each quantity as a function of $\NH$ through a renormalization of the profiles at each resolution, leading to Fig.~\ref{mean_reso_norm}. 
We proceed as follows, for $\PsI$, $\S$, and $\PsIxS$. At each resolution $\ellres$ (expressed in arcmin), using Fig.~\ref{mean_reso_SNR} we determine the minimal column density $\NH^{\rm min}(\ellres)$ above which $\S$ is correctly debiased, and define a reference interval $\mathcal{I}(\ellres)=[\NH^{\rm min}(\ellres),10\times\NH^{\rm min}(\ellres)]$ (indicated by the horizontal colour bars in Fig.~\ref{mean_reso_norm}). For two consecutive resolutions $\ellres$ and $\ellres/2$, we compute the mean values of $\PsI$, $\S$, and $\PsIxS$ at both resolutions on the common interval $\mathcal{I}(\ellres)\cap\mathcal{I}(\ellres/2)$ and then the ratio of these two values, $r_{\ellres, \ellres/2}$. 
Finally, we compute the factor by which each curve at resolution $\ellres$ from Fig.~\ref{mean_reso} must be multiplied to be normalized to the curve at the coarsest resolution, $\ellres_{\rm max}=160\arcmin$, i.e.,
$r(\ellres) = r_{\ellres_{\rm max},\ellres_{\rm max}/2}\times ... \times r_{4\ellres,2\ellres} \times r_{2\ellres,\ellres}$. This renormalization removes the depolarization induced in $\PsI$ by the smoothing of the data, as well as the change of the lag $\delta$ with the resolution, as far as $\S$ in concerned.\footnote{For $\StimesPsI$, this renormalization is consistent with the scaling with the resolution, $\StimesPsI\propto \omega^{0.18}$ (Eq.~\eqref{Eq-Sp_lag}).} 

The mirrored similarity of each detailed variation in the logarithmic representation of $\PsI$ and $\S$ in Fig.~\ref{mean_reso_norm} clearly emphasizes the inverse relationship between these two quantities. In our multi-resolution normalized representation of the variations of $\PsI$ with the column density, the mean value of $\PsI$ decreases by a factor \maxdropp\ from the lowest column densities at high latitudes and a resolution of $160\arcmin$, to the highest column densities probed here at a resolution of 10\arcmin. This strong decrease is almost entirely mirrored as an increase in $\S$ by the same factor, demonstrating again that the decrease in $\PsI$ with the column density is mainly a consequence of the depolarization by the structure of the magnetic field. 

The residual trend in $\PsIxS$ is small, about a $\maxdropSp\,\%$ decrease with column density from $10^{20}$\,\cmsq\ to $\uptoNH$\,\cmsq.
For the case $|b| > 10 \deg$, the profile of $\PsIxS$ over this same range of $\NH$ is quite flat. For the case including $|b| > 2\deg$, $\PsIxS$ decreases systematically with $\NH$. In our phenomenological model, a systematic decrease in $\PsIxS$ is expected at low Galactic latitudes from an increased number of independent layers $N$ along the line of sight (see Eq.~\eqref{Eq-Sp}), related to the increased dust opacity and/or length probed \citep{JKD92}, that is not compensated by the inverse effect of increased
$\S$ due to an increased distance to the observed dust material (recall that $\S$ depends on the physical scale probed in the cloud, therefore on its distance). There remains therefore little room for a systematic variation of grain alignment for column densities up to $\uptoNH$\,\cmsq.

At slightly higher column densities ($\NH>3\times10^{22}$\,\cmsq), we observe a decrease in $\PsI$, together with an increase in $\S$ and $\PsIxS$. 
Such a combination cannot be explained by a decrease in grain alignment, which would not affect $\S$. These lines of sight, part of the Orion and Ophiuchus regions, are situated at intermediate latitudes ($10\deg< |b| < 20\deg$) and probably do not suffer from depolarization by the Galactic background, unlike other lines of sight at lower column densities situated at lower latitude. As can be seen from the density of points in Fig.~\ref{mean_reso}, this departure from the mean trend has a low statistical significance, which prevents us from commenting further. 

To conclude, most variations of the polarization fraction $\PsI$ with $\NH$ are inversely related to those of $\S$, a tracer of the depolarization by the turbulent magnetic field. The multi-resolution study of the variation of $\PsIxS$ with $\NH$ does not indicate any systematic variation of the grain alignment efficiency beyond around $\maxdropSp\,\%$, up to a column density of $\uptoNH$\,\cmsq.

\subsection{Grain alignment efficiency in the ISM}\label{discuss-alig}

In this section, we discuss the impact of our results on the question of grain alignment.

Since the pioneering work of \cite{Myers91} and \cite{JKD92}, it has been clear that the structure of the magnetic field along the line of sight plays a major role in the build-up of polarization observables. Nevertheless, the decrease in the polarization fraction with increasing column density is widely considered as evidence for a systematic decrease in the degree of grain alignment efficiency with increasing exinction~\citep{W08,Cashman2014,Alves2014}.

In this paper, we have shown that most (if not all) variations observed in the polarization fraction $\PsI$ are mirrored in the dispersion of polarization angles, $\S$, a quantity that is independent of the grain alignment efficiency and of dust optical properties. Quantitatively, the near constancy of $\PsIxS$ with increasing column density indicates that the variations of the polarization fraction are dominated by the structure of the magnetic field, not only in the diffuse ISM, but also in molecular clouds, at least up to a column density of $\NH\approx\uptoNH$\,\cmsq. The decrease in grain alignment efficiency with column density cannot exceed about $\maxdropSp\,\%$, from the most diffuse ISM up to this same column density, $\NH\approx \uptoNH$\,\cmsq. These results are significant constraints for theories of grain alignment.

Dust alignment in the ISM is widely thought to be associated with radiative torques (RATs). As mentioned, in the classical framework of RATs \citep{Lazarian2007}, the grain alignment efficiency depends on the radiation field intensity and on the angle between the radiation field anisotropy and the magnetic field. During the last decade, there have been a few claims for evidence of such effects \citep[e.g.,][]{AP10,VD15,Andersson15}. Analysing \Planck\ full-sky data, we could not find, either in the diffuse ISM or in molecular clouds, any signature in polarization observables that could point to a significant variation of grain alignment related to a variation in the grain temperature. 

However, the low resolution of \Planck\ data (5\arcmin), combined with the smoothing of the maps necessary to guarantee that $\PsI$ and $\S$ are not biased by noise (160\arcmin\ in the high latitude diffuse ISM, 40\arcmin\ in molecular clouds), does not allow us to probe the same physical conditions as, for example, NIR polarimetry through dense clouds \citep{JBK15}. A detailed analysis of \Planck\ polarization without the additional smoothing, and hence at higher resolution (7\arcmin) and higher column densities, will be pursued in a future paper dedicated to cold cores.

\section{Comparison with starlight polarization at high Galactic latitudes}
\label{sec:stars}

In this section, we correlate \Planck\ polarization with starlight polarization in the diffuse ISM at high Galactic latitudes, and derive new constraints on dust models. Following the approach in \cite{planck2014-XXI} for translucent lines of sight ($0.15 < \ebvsub < 0.8$)
at low Galactic latitudes ($|b| < 30$\deg), we can now derive emission-to-extinction polarization ratios for the diffuse high-latitude Galactic ISM $(\ebvsub < 0.15$, corresponding to column densities $\NH < 10^{21}$\,\cmsq). The ratios are
\begin{eqnarray}
\RPp & =& \frac{\Psub}{\pv} \, ; \\ 
\Rsv & =& \frac{\Psub/\Isub}{\pv/\tauv} \, .
\end{eqnarray}
Here $P$ and $I$ are what \Planck\ has measured in the submillimetre. In the optical \Vband\ band, $\pv$ is the degree of polarization for a star to which the optical depth is $\tauv$.  The latter is estimated from the colour excess of the star, $\ebvstar$, using $\Av=\Rv\times\ebvstar$ with the ratio of total to selective extinction $\Rv = 3.1$ \citep{FM07}, and $\tauv =\Av/1.086$.

These polarization ratios quantify the amount of polarized emission per unit polarized extinction.
Because they measure the effects of the same grains at different wavelengths, they remove the first-order dependencies of the polarization observables on the magnetic field structure and grain alignment efficiency \citep{planck2014-XXI}. As such, they directly provide observational constraints on the optical properties of the dust population that is aligned with the magnetic field, and strongly constrain dust models \citep{guillet-et-al-2018}.

The second of these ratios, $\Rsv$, being inversely proportional to $\Isub$, is sensitive to the total intensity offset. We comment on the derived values of $\Rsv$ for the low, fiducial, and high values of the offset in Sects.~\ref{sec:determination-polar-ratios} and \ref{sec:variations-polar-ratios}.

We are interested in examining how the amount of submillimetre polarization is related to the amount of optical polarization from the same dust. If the dust probed in the submillimetre and the optical is the same, then the polarization orientations should be orthogonal.  As discussed in Appendix~\ref{sec:stars_polang}, this is the case for the lines of sight used in our analysis of the ratios.  This is quantified by the polarization angle difference $\Delta\psi_{\rm S/V}$ which takes into account the 90\deg\ difference (see Eq.~\eqref{eq:deltapsi}).
 
In Appendix~\ref{sec:appendix:nands} we describe in detail our estimates of the noise and systematic uncertainties that affect the submillimetre and optical data as used in this new analysis. We highlight the relevant results in the discussion below.

\subsection{Estimates for starlight reddening}
\label{sec:reddening}

To enable appropriate comparison of polarization properties in the optical and in the submillimetre on lines of sight to stars, it is necessary to obtain an estimate of the reddening to the star. To this end, we obtain the distance $\dist$ to each star by extracting the parallax $\plx$ from the Gaia DR2 release \citep{GaiaMission,GaiaDR2Cat} or from the polarization catalogues when Gaia data are not available. Then we derive an estimate of the reddening to the star, $\ebvstar$, by interpolating the PS1-based 3-dimensional reddening data cube \citep{Green2018} at distance $\dist$. This data cube is composed of 31 maps, each representing a range in distance modulus. To limit the impact of noise in our analysis, the 31 maps were separately smoothed to a resolution of 40\arcmin\ and downgraded to $\nside= \nsidestars$.  We also converted the PS1-based reddening to the Johnson $\ebv$ scale using the relation $\ebv_{\rm Johnson} = \ebv_{\rm PS1} / 1.0735$ \citep[table 6 of][]{SF11}.  The Johnson scale is used hereafter without explicit subscripting. From the same maps we also obtain the total reddening along the line of sight, $\ebvtotal$. Uncertainties related to the reddening maps are estimated in Appendix~\ref{sec:uncertainreddening}.

\subsection{Polarization data} 
\label{sec:stars_pol}

For this analysis, we use the alternative \Planck\ 353-GHz $\Isub$, $\Qsub$, and $\Usub$ maps from the {\DRThree} data release (\asmaps, Sect.~\ref{sec:asm}), smoothed to a resolution of \reso\arcmin\ to limit the noise in $\Qsub$ and $\Usub$. 

From a series of optical polarization catalogues of high-latitude stars \citep{BerdIV2001,BerdV2001,BerdVI2002,BerdVII2004,BerdVIII2014}, we extract data for \nstarsuniq\ lines of sight to stars with measured degree of polarization $\pv$, uncertainty $\dpv$, and polarization angle $\Gv$ (in the IAU convention, consistently with our definition of $\psi$ for the \Planck\ data). These catalogues cover the northern Galactic hemisphere at high latitudes (\nstarsnorth\ stars with $b > 30$\deg), and part of the southern cap (\nstarssouth\ stars with $b < -59$\deg). After removing \nstarsoutofmask\ stars falling outside the region covered by PS1 (mainly in the southern hemisphere) and 3 stars without a distance estimate, there remain \nstarsebv\ stars for which we have both reddening estimates and optical polarization data.

As with the \Planck\ submillimetre data, we use the MAS estimator \citep{plaszczynski14} to debias the degree of polarization $\pv$ in the optical. Using these values of $\pv$ and $\Gv$, we then define Stokes parameters in extinction, $\qv$ and $\uv$, in the same \healpix\ convention as our \Planck\ data:
\begin{eqnarray}
\qv & = & \phantom{-}\pv \,\cos{(2\Gv)}\, ; \\
\uv & = & -\pv \,\sin{(2\Gv)} \, .
\end{eqnarray}

\begin{figure}
\includegraphics[width=\hsize]{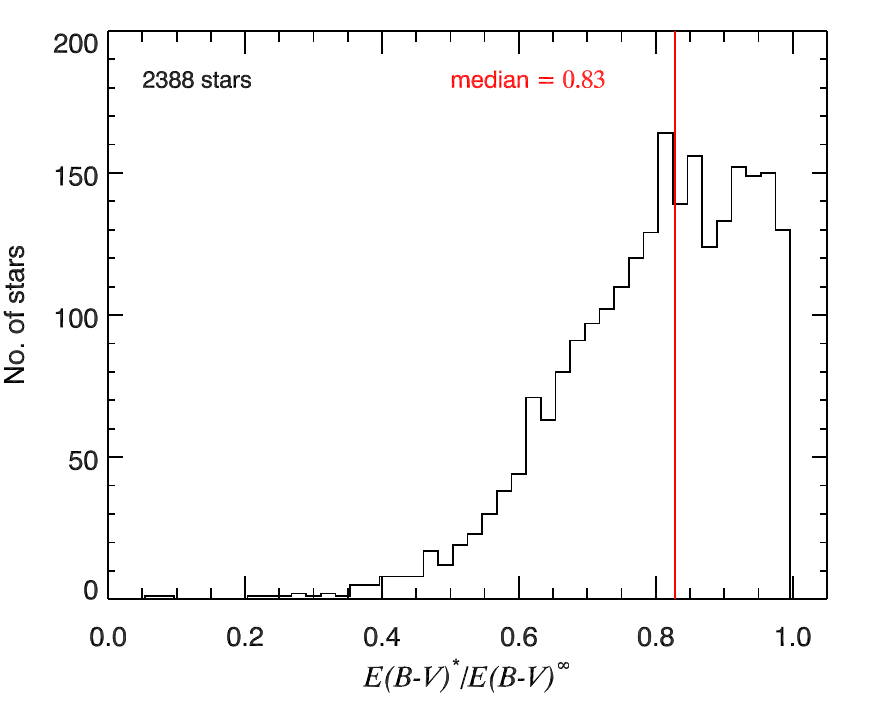}
\caption{Histogram of the ratio of the reddening to the star to the total reddening on the same line of sight, $\ebvstar/\ebvtotal$, as derived from the Pan-STARRS1 3D cube \citep{Green2018}. The red line indicates the median ratio. In practice we retain lines of sight for which $\ebvstar/\ebvtotal> 0.75$.
}
\label{Fig_EBVratio}
\end{figure}

Emission-to-extinction ratios are subject to systematic errors because extinction probes the ISM in the foreground to the star, while emission probes the entire line of sight \citep{planck2014-XXI}.  Figure~\ref{Fig_EBVratio} presents the histogram of the ratio $\ebvstar/\ebvtotal$ of the reddening to the star to the total reddening, i.e., the fraction of ISM material that is in front of each star. The median ratio for our full sample is $\medEBVratio$, illustrating the potential for systematic effects on the polarization ratios.

If we assumed for simplicity that the ISM along the line of sight were uniform (in density, magnetic-field orientation, and dust properties), then the polarization ratio $\RPp=\Psub/\pv$ would artificially increase linearly with decreasing $\ebvstar/\ebvtotal$. Consequently, by neglecting the presence of background material we would typically overestimate the polarization ratio $\RPp$ by 17\,\%. Given this contamination, to debias our estimate of $\RPp$ we replace $\pv$ by a linear estimate of what its value would be if the star were at infinity: 
\begin{equation}
\label{eq:pvinfty_uniform}
\pv^\infty = \pv \frac{\ebvtotal}{\ebvstar} \, ,
\end{equation}
with an associated uncertainty
\begin{equation}
\label{eq:dpvinfty_uniform}
\dpv^\infty = \dpv\frac{\ebvtotal}{\ebvstar}\, .
\end{equation}
We use similar expressions to estimate $\qv^\infty$ and $\uv^\infty$ and their uncertainties $\dqv^\infty$ and $\duv^\infty$.
On the other hand, $\Rsv$, as a ratio of non-dimensional quantities, would be unaffected by this uniform background.

\subsection{Selection of the lines of sight}
\label{sec:selectlines}

However, the ISM is not uniform and as a consequence our estimates of both $\Rsv$ and $\RPp$ could be biased by the presence of a background whose properties are different from those of the foreground to the star. The magnitude of this uncertainty is evaluated in Appendix~\ref{sec:bkgd}.
We minimize the contribution of this uncertainty by excluding those lines of sight with an important background, as inferred from the ratio $\ebvstar/\ebvtotal$ shown in Fig.~\ref{Fig_EBVratio}. We explicitly choose to keep only stars for which $\ebvstar/\ebvtotal> 0.75$.

We also exclude those lines of sight where $\Delta\psi_{\rm S/V}$ is significantly different than the expected 0\deg, having found that for these lines of sight the rms scatter about the best fit correlations yielding $\RPp$ and $\Rsv$ (Sect.~\ref{sec:determination-polar-ratios}) is indeed much larger.  To be conservative and retain enough lines of sight for our subsequent statistical analysis, we excluded only lines of sight for which $\Delta\psi_{\rm S/V} > \diffGmin$\deg, i.e., about 9\,\% of the sample.  Our results are not sensitive to this particular choice.

Our final sample contains $\nstarsselected$ stars.  The ISM towards these stars in emission is representative of diffuse dust at high Galactic latitudes, with MBB fit parameters $\Td=19.7\pm1.3$\,K, $\beta = 1.60\pm0.15$, and $\ebvsub \in [0.01, 0.24]$, with a mean reddening $\left\langle\ebvsub\right\rangle=0.03$.

\subsection{Determination of the polarization ratios}
\label{sec:determination-polar-ratios}

\begin{figure*}[htbp!]
\includegraphics[width=\hhsize]{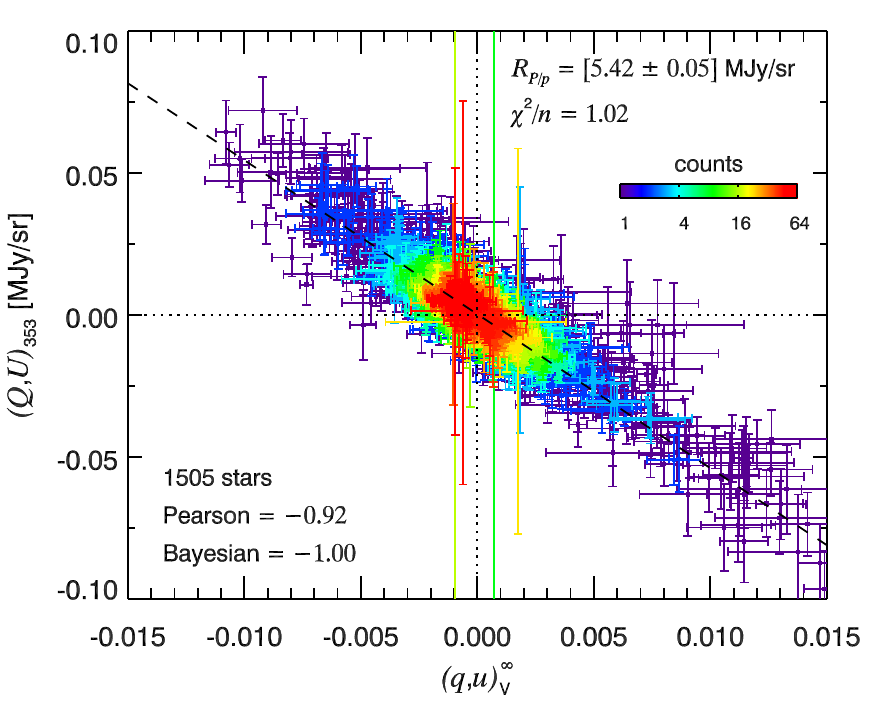}
\includegraphics[width=\hhsize]{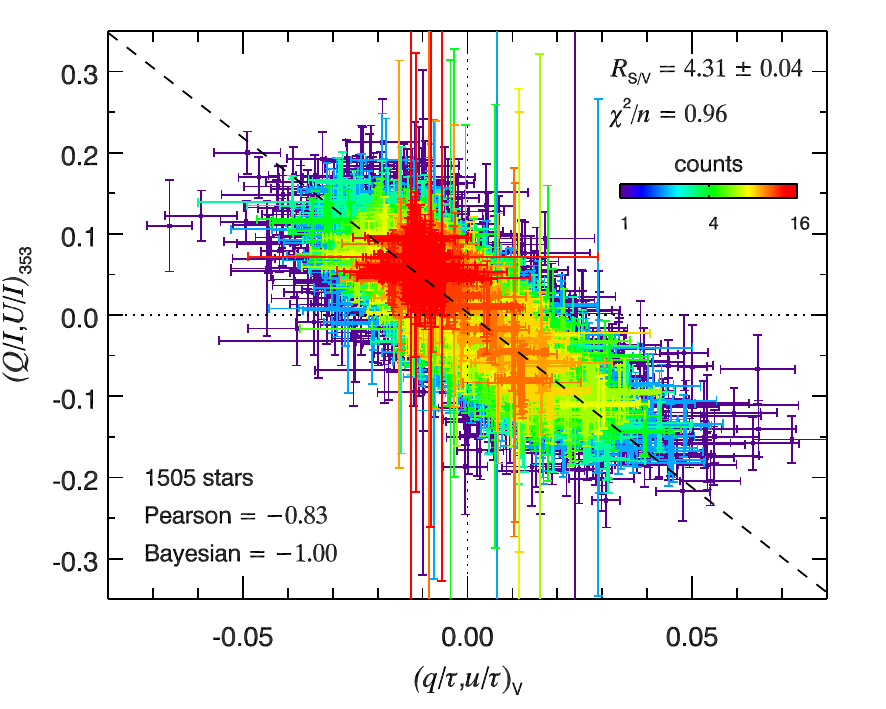}
\caption{Correlation between Stokes polarization parameters in emission at 353\GHz\ and in optical extinction, with the colour in the 2-dimensional histogram representing the density of points. {\it Left}: Stokes parameters $(\Qsub,\Usub)$ versus $(\qv^\infty,\uv^\infty)$, yielding an estimate of $\RPp$. {\it Right}: Normalized Stokes parameters $(\Qsub/\Isub,\Usub/\Isub)$ versus $(\qv/\tauv,\uv/\tauv)$, yielding an estimate of $\Rsv$. 
The slopes of the correlations are obtained using the Bayesian fitting method of \cite{Kelly07}. The reduced $\chi^2$, the Pearson correlation coefficient, and the correlation coefficient inferred from the Bayesian method \citep{Kelly07} are listed.}
\label{Fig_correl}
\end{figure*}

Following the approach in \cite{planck2014-XXI}, we derive $\RPp$ through a joint correlation of the pair ($\Qsub,\Usub$) with ($\qv^\infty,\uv^\infty$), and derive $\Rsv$ through a joint correlation of ($\Qsub/\Isub,\Usub/\Isub$) with ($\qv/\tauv,\uv/\tauv$).
In Fig.~\ref{Fig_correl} we present the two correlation scatterplots, that for $\RPp$ on the left, and that for $\Rsv$ on the right.
For the fitting, we settle on the Bayesian method of \cite{Kelly07}, but we obtain the same results with other fitting methods making use of uncertainties on both axes. Both for determining the value of the ratio (the slope) and for calculating the reduced $\chi^2$ to assess the quality of the fit, it is important to assess all sources of uncertainty, as addressed in this section and in more detail in Appendix~\ref{sec:appendix:nands}.

As a first test, we fit the data for the 206 translucent lines of sight from \cite{planck2014-XXI} with this estimator and find no change, even though we smooth \Planck\ Stokes parameter maps to 40\arcmin\ FWHM. We are therefore confident that it is legitimate to compare the polarization ratios that we derive here at 40\arcmin\ resolution with those measured at 7\arcmin\ resolution in \cite{planck2014-XXI}.

The correlation of ($\Qsub,\Usub$) with ($\qv^\infty,\uv^\infty$) shown in Fig.~\ref{Fig_correl} (left) is tight, with a Pearson correlation coefficient $-0.92$. For lines of sight where $p$ is low, error bars have been greatly increased by the correction factor $\sqrt{1+\delta p^2_{\rm beam}/p^2}$ for beam depolarization (Appendix~\ref{sec:beamdepol}).
When systematic uncertainties are taken into account, the reduced $\chi^2$ is good ($\chiRPpsys$, compared to $\chiRPpnosys$ when they are ignored). The fit yields a polarization ratio $\RPp = (\meanRPpfitted\pm \sigmeanRPpfitted)$\,\MJysr, similar to the value found for translucent lines of sight, $(5.4\pm0.5)$\,\MJysr \citep{planck2014-XXI}.

We find a somewhat larger scatter in the correlation plot ($\Qsub/\Isub,\Usub/\Isub$) with ($\qv/\tauv,\uv/\tauv$) in Fig.~\ref{Fig_correl} (right), as quantified by the slightly lower absolute value of the correlation coefficient. 
Here also, the reduced $\chi^2$ is good when systematic uncertainties are included ($\chiRsvsys$), and larger when they are not ($\chiRsvnosys$).
The fitted slope $\Rsv = \meanRsvfitted \pm \sigmeanRsvfitted$ is also compatible with the value $4.2\pm0.5$ found for translucent lines of sight at 7\arcmin\ resolution~\citep{planck2014-XXI}. We find similar values at other resolutions: $\Rsv=\Rsvfitlowres$ at \resolowres\arcmin\ resolution; and $\Rsv=\Rsvfithighres$ at \resohighres\arcmin\ resolution. These results are for the fiducial intensity offset. With the low and high offsets at 40\arcmin\ resolution, we obtain $\Rsv=\RsvfitlowCIB$ and $\Rsv=\RsvfithighCIB$, respectively, which makes the \Planck\ intensity offset the main source of uncertainty on $\Rsv$.

\begin{figure*}[htbp!]
\includegraphics[width=\hhsize]{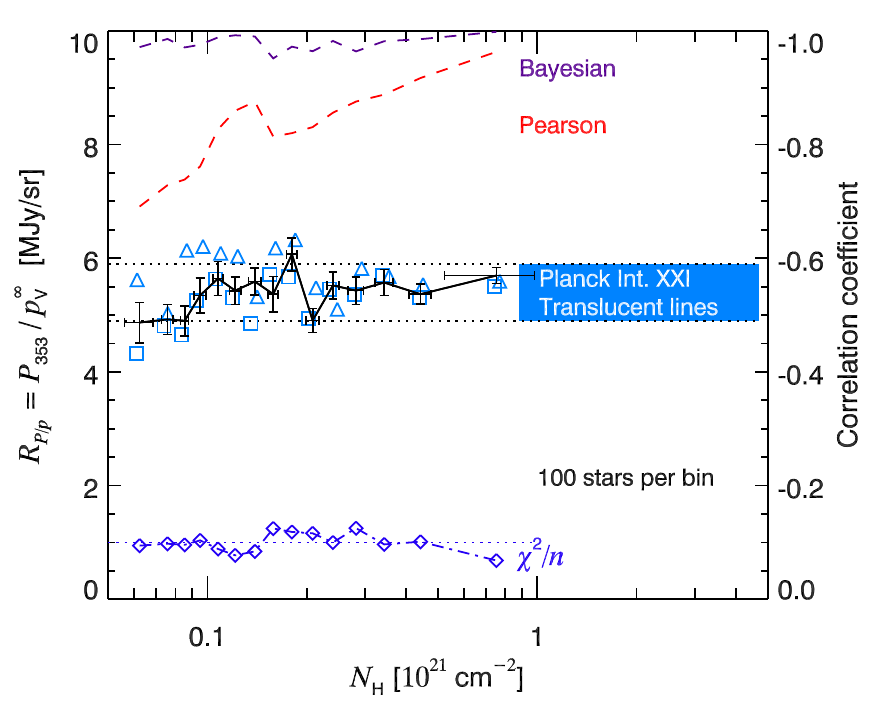}
\includegraphics[width=\hhsize]{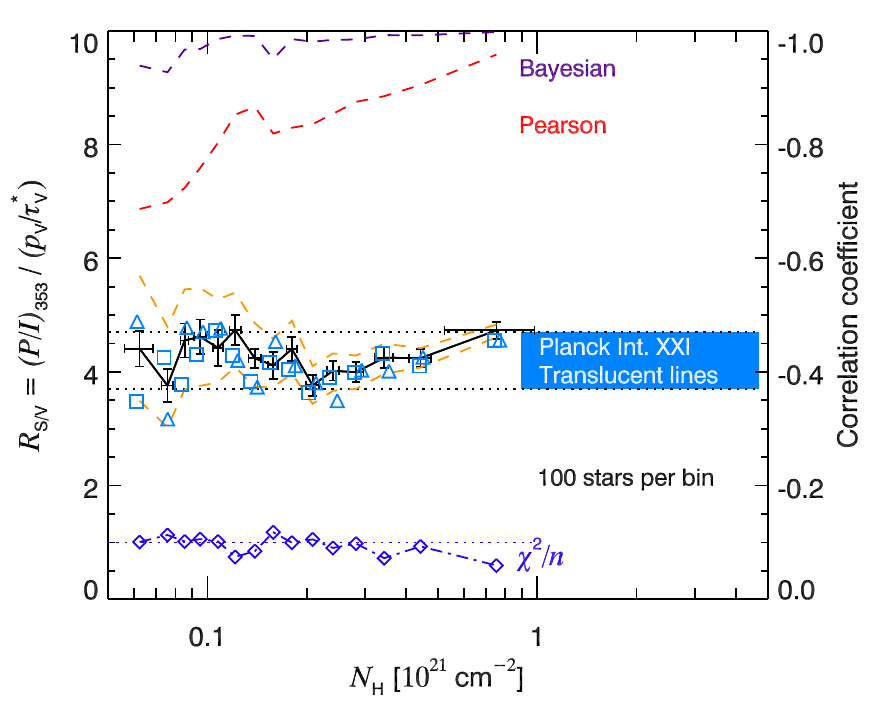}
\caption{Emission-to-extinction polarization ratios.  The black curves show the ratios $\RPp$ (left) and $\Rsv$ for the fiducial offset in $I$ (right), at 40\arcmin\ resolution, as a function of the column density, $\NH$. For the running mean each bin contains the same number (\nstarsperbin) of lines of sight. The lower dark blue dotted-dashed lines indicate the reduced $\chi^2$ of the fits (with the scale on the left axis).  Dashed red and purple curves represent the Pearson and Bayesian \citep{Kelly07} correlation coefficients, respectively (with the scale on the right axis). The results at a resolution of $\resolowres\arcmin$ (squares) and $\resohighres\arcmin$ (triangles) are also shown. On the right panel, the upper and lower dashed orange curves represent the trend for the low and high offsets in $I$, respectively, at the reference resolution of 40\arcmin. The blue band shows in each panel the mean value, together with its uncertainty domain, found for the range of column densities considered in \cite{planck2014-XXI}. 
}
\label{Fig_ratio}
\end{figure*}

\subsection{Variations of $\RPp$ and $\Rsv$ with column density}
\label{sec:variations-polar-ratios}

Our sample contains enough lines of sight to study variations of the polarization ratios with column density  $\NH$, which is determined from the dust optical depth at 353\,GHz as explained at the beginning of Sect.~\ref{sec:NewResults}. This is presented in Fig.~\ref{Fig_ratio}. The sample is divided into \nwin\ independent bins in $\NH$, each bin containing \nstarsperbin stars. For the polarization ratio $\RPp$ at low $\NH$, we observe a roughly 10\,\% increase with increasing $\NH$, from about 5.0\,\MJysr\ at $\NH \approx 6\times10^{19}$\,\cmsq\ to 5.4\,\MJysr\ at $\NH \approx 1.5\times 10^{20}$\,\cmsq. This is followed by a plateau at higher $\NH$. The normalized polarization ratio, $\Rsv$, is found to decrease with column density for the low total intensity offset, to be rather flat for the fiducial one, and to slightly increase for the high offset. The values obtained for $\RPp$ and $\Rsv$ at higher ($\resohighres\arcmin$) and lower ($\resolowres\arcmin$) resolutions are close to the 40\arcmin\ values. Both the Pearson correlation coefficients and the correlation coefficients provided by the Bayesian method \citep{Kelly07} are high enough at all column densities to bring confidence in our results.  For comparison, the values of the polarization ratios found for translucent lines of sight in \cite{planck2014-XXI}, together with their ranges of uncertainty, are also displayed in Fig.~\ref{Fig_ratio}. Altogether, $\RPp$ and $\Rsv$ are remarkably constant with column density and consistent with the values determined for translucent lines of sight. We note that there are some small variations of potential significance, such as the 10\,\% increase in  $\RPp$ over the range $6\times10^{19}-1.5\times10^{20}$\,\cmsq, but it is beyond the scope of this paper to discuss them.

\subsection{Maximum $\pv/\ebv$ at low column densities}
\label{sec:maxpsebv}

\begin{figure*}[htbp!]
\includegraphics[width=\hhsize]{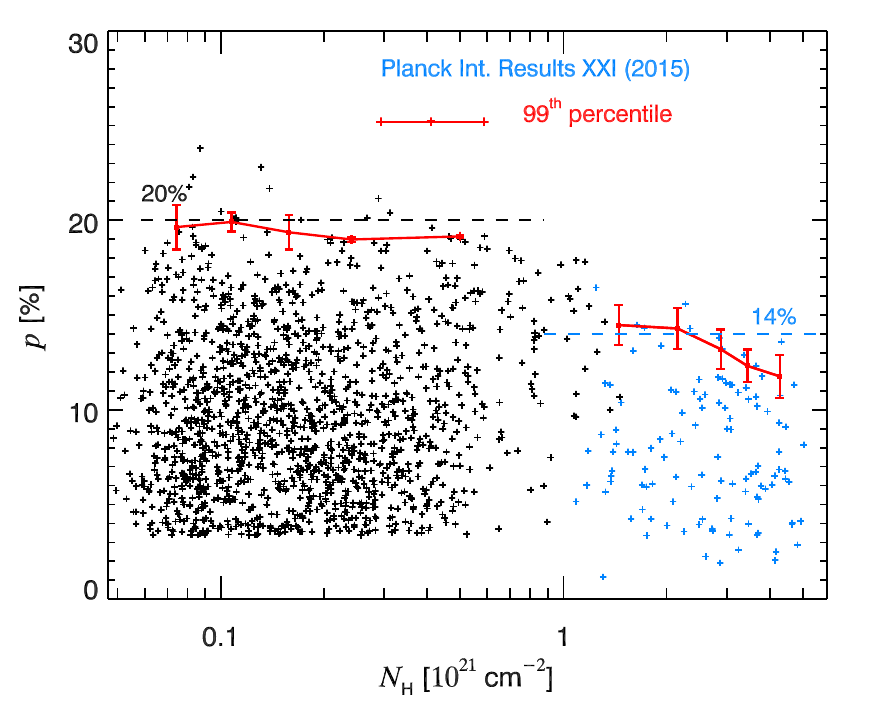}
\includegraphics[width=\hhsize]{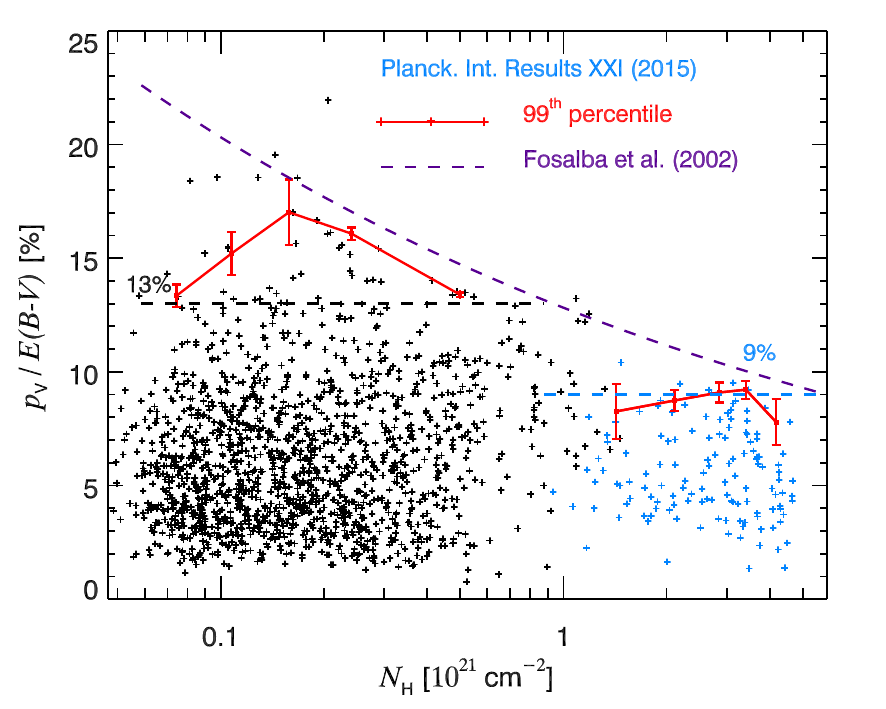}
\caption{
Polarization fraction in emission at 353\GHz, $\PsI=\Psub/\Isub$ (left), and in optical extinction, $\pv/\ebv$ (right), as a function of the column density, $\NH$. The sample in blue shows the 206 translucent lines of sight from \cite{planck2014-XXI}, along with the estimates of maximum polarization. The sample in black is the one in our current study (\nstarsselected\ stars), where data have been corrected for systematic effects such as beam depolarization (see Appendix~\ref{sec:appendix:nands}). For each sample, we plot the 99th percentile in $\PsI$ and $\pv/\ebv$, along with the uncertainty on the derivation of this percentile (see text). The fit from \cite{Fo02}, corresponding to $\pv/\ebv \propto \ebv^{-0.2}$, is shown for comparison.
}
\label{Fig_maxp}
\end{figure*}

Regarding the observed starlight polarization fraction per unit reddening in the optical, $\pv/\ebv$, its maximum value was first estimated to be at most 9\,\% by \cite{SMF75}. This maximum value is often considered as providing an upper limit on the dust alignment efficiency, although it is based on less than 300 stars at moderate extinction ($\ebv > 0.15$), characteristic of translucent lines of sight.
Several attempts have been made to constrain this maximum value at low reddening \citep{Fo02,FR15,SK18}, suggesting larger values, but with poor precision. Our present sample probes more diffuse lines of sight and, with more statistical significance, allows us to characterize the maximum polarization fraction at low reddening, extending to $\ebv \approx 0.01$. 

To study the dependence of the polarization fractions $\Psub/\Isub$ and $\pv/\ebv$ on column density, we combine data for the $\nstarsselected$ diffuse lines of sight to stars from this study (high latitude, low $\ebv$, 40\arcmin\ resolution for \Planck\ data and $\ebv$ map) and the 206 stars on translucent lines of sight (low latitude, moderate $\ebv$, 7\arcmin\ resolution for \Planck\ data) from \cite{planck2014-XXI}. 

Figure~\ref{Fig_maxp} shows how these polarization fractions in emission and extinction vary with column density. 
Polarization fractions at 353\GHz\ never reach low values because \Planck\ polarization data have been corrected for beam depolarization through Eq.~\eqref{Eq-pencil}. However, this does not affect our results for the high percentiles because high values of $p$ suffer very little depolarization. We overplot the upper limit $\pv \le 9\%\ebv$ of \cite{SMF75}, the non-linear fit $\pv/\ebv \propto \ebv^{-0.2}$ proposed by \cite{Fo02} for polarization in extinction (Fig.~\ref{Fig_maxp}, right), and the estimate for the maximum value of the polarization fraction in emission observed for translucent lines of sight on the left \citep[approximately $\PsImaxPIPXXI\,\%$, ][]{planck2014-XXI}. 

Assuming a maximum polarization in emission of $\pmax~=~20\,\%$ for our sample at a resolution of 40\arcmin\ (close to its 99th percentile\footnote{The percentile curves in Fig.~\ref{Fig_maxp} have uncertainties that are computed in the following way. For each $\NH$ bin, the values are sorted and the one closest to the 99th percentile of the distribution is taken as the value for the 99th percentile. The uncertainty interval then spans the range between the value just preceding the 99th percentile value and that just following it.}), and a polarization ratio $\Rsv = \Rsvfidround$ (Fig.~\ref{Fig_correl}), we would expect a maximum polarization fraction in extinction of $\pv/\ebv_{\rm max} = (3.1\times\pmax) / (\Rsv\times1.086) \approx \pmaxv\,\%$. As seen in Fig.~\ref{Fig_maxp} (right), this upper limit is somewhat smaller than the measured 99th percentile of our data in extinction. We would reach a similar conclusion using the value $\pmax = 22\,\%$ that we obtained for the diffuse ISM in Sect.~\ref{subsec:polfrac}, with $\pv/\ebv_{\rm max} \simeq 14.5\,\%$. This upper limit is also smaller than the upper limit proposed by \citet{Fo02} based on a study of the dependence of the mean starlight polarization fraction $\pv/\ebv$ with $\ebv$. 

However, the distribution in the density of points for polarization in extinction (right panel) compared to that for polarization in emission (left panel) suggests that lines of sight with high $\pv/\ebv$ might be outliers. 
One should indeed be aware that the direct derivation of the maximum polarization fraction in extinction is much more subject to noise and systematics than our derivation, which is based on the measurement of the polarization ratio $\Rsv$ and the much better characterized maximum polarization fraction $\PsI$ in emission. We therefore consider the value of \pmaxv\,\% as a well-constrained maximum value for $\pv/\ebv$ at very low $\NH$ ($< 5\times10^{20}$\,\cmsq). This is a strong new constraint on the grain optical properties used in dust models.

The observed maximum polarization fractions drop from the diffuse ISM at high Galactic latitudes to the translucent lines of sight at low Galactic latitudes. In emission, $\pmax$ decreases from 20\,\% to \PsImaxPIPXXI\,\%, whereas in extinction $(\pv/\ebv)_{\rm max}$ decreases from \pmaxv\,\% to 9\,\%. Such a decrease is usually attributed to a loss of grain alignment 
(see \citealp{Andersson15} and references therein).
However, inspection of Fig.~\ref{mean_reso_norm} for Galactic latitudes higher than $10\deg$ shows that the product $\PsIxS$ remains constant over the range of column densities probed here. Following our analysis in Sect.~\ref{sec:NewResults}, we therefore attribute most of this decrease in the maximum polarization fraction when the column density increases to the increasing depolarizing effect from the structure of the magnetic field along the line of sight, with little room for a systematic decrease in the grain alignment efficiency.

Dust models should therefore be able to reproduce the maximum observed polarization fractions, $\pmax=20\,\%$ in emission and $(\pv/\ebv)_{\rm max} = \pmaxv\,\%$ in extinction, even when applied to regimes in column densities where such values are actually never directly observed.

\section{Conclusions}
\label{sec:conclusions}

In this paper, we have analysed the {\DRThree} thermal dust polarization data at 353\,\GHz. Starting from full-sky maps of Stokes $I$, $Q$, and $U$ at a uniform 80\arcmin\ resolution, processed with the Generalized Needlet Internal Linear Combination (\GNILC) algorithm~\citep{Remazeilles2011b} to mitigate the contamination by CIB and CMB anisotropies as well as noise, we have presented the maps of polarization fraction $\PsI$, polarization angle $\psi$, and polarization angle dispersion function $\S$, with their associated uncertainties. The statistical analysis of these maps provides new insights into the astrophysics of dust polarization. 

We have been able to determine the maximum observed polarization fraction, $\pmax=22.0^{+3.5}_{-1.4}\pm0.1\,\%$, at this resolution and frequency, where the second uncertainty is statistical, underscoring the excellent quality of the data, and the first reflects the principal systematic uncertainty affecting this determination, which is linked to the uncertainty on the Galactic emission zero level in total intensity~\citep{planck2013-p06b}. This maximum polarization fraction provides strong constraints for models of dust properties and alignment in the Galactic magnetic field~\citep{guillet-et-al-2018}. Owing to the strong effect of the magnetic field morphology, a low value for the maximum polarization fraction in a given region is not an indication that grain alignment in that region is ineffective, but rather that polarization is strongly affected by depolarization because the direction of the large-scale field is closer to the line of sight in that region.

We confirmed that the statistical properties of $\PsI$, $\psi$, and $\S$ essentially reflect the structure of the Galactic magnetic field~\citep{planck2014-XIX}, with a clear peak of the polarization angle near $0\deg$, corresponding to the field being parallel to the Galactic plane, and an inverse proportionality between the polarization fraction $\PsI$ and the polarization angle dispersion function $\S$. 
We showed analytically, and using numerical models of the polarized sky, that this relationship can be reproduced statistically to first order by an interstellar magnetic field incorporating a turbulent component superposed on a small number of layers with a simple uniform field configuration. Looking for evidence in the diffuse ISM ($\NH < \maxNHSpT$\,\cmsq) of a correlation of the polarization fraction with the dust temperature, as one could expect from the classical radiative torque theory \citep{Lazarian2007}, we could not find any: all variations of $\PsI$ are here again mirrored with those of $\S$, which does not depend on the dust physics.

Based on this analysis, we introduced the product $\StimesPsI$ as a means of exploring the non-geometric elements of the polarization maps, such as variations in grain properties, in alignment physics, or in ISM phase contributions.
We showed that $\StimesPsI$ exhibits smaller and smoother variations than either $\PsI$ or $\S$ when considered as a function of the Galactic latitude $b$, the Galactic longitude $l$, or the column density
(which is simply scaled from the dust optical depth $\tau$ at 353\,\GHz).

We provided an analysis at a finer angular resolution of 40\arcmin\, using the {\DRThree} data, towards a sample of six molecular regions in the Gould Belt. This confirmed the trends observed at coarser resolution over the full sky, most notably that the polarization angle dispersion function is inversely proportional to the polarization fraction, $\S\propto 1/p$. Strikingly, the $\S$ versus $\PsI$ curves for the different regions all fall on the same line, demonstrating that most of the variations of $\PsI$ with column density are driven by variations of $\S$, i.e., by the structure of the magnetic field along the line of sight~\citep{planck2014-XX}. Considering then the product $\StimesPsI$ and how it varies with dust temperature $\Td$ in these regions, we found no evidence for a link between the polarization properties and the intensity of the radiation field. Comparing these properties with column density in a multi-resolution view, we found that the product $\StimesPsI$ does decrease, but only by about \maxdropSp\,\% between $\NH\approx10^{20}\,\mathrm{cm}^{-2}$ and $\NH\approx\uptoNH$\,\cmsq, while the polarization fraction $\PsI$ decreases by a factor of \maxdropp\ over the same interval.

We also compared the \Planck\ polarization data with optical stellar polarization data in the high Galactic latitude sky, expanding on the analysis done in~\cite{planck2014-XXI} for translucent lines of sight. We constrained the polarization properties of the dust at low column densities, quantified by the polarization ratios $\RPp = P/\pv$ and $\Rsv = (P/I)/(p/\tauv)$ defined in~\cite{planck2014-XXI}. The larger statistical sample ($\nstarsselected$ stars selected) allowed us to study the dependence of these polarization ratios on column density. We found $\RPp$ to increase from 5.0\,\MJysr\ at $\NH=6\times10^{19}\,\mathrm{cm}^{-2}$ to 5.4\,\MJysr\ for $\NH>1.5\times10^{20}\,\mathrm{cm}^{-2}$, the same value as for translucent lines of sight. The polarization ratio $\Rsv$ was found to be compatible on average (around \Rsvfidround) with that for translucent lines of sight ($4.2\pm0.5$), having a decreasing, flat, or slightly increasing trend with the column density, depending on the offset for the \Planck\ intensity map at 353\GHz, which is here again the dominant systematic effect. Combining emission and extinction measurements, we derived an estimate for the maximum value of the polarization fraction in the visible at high Galactic latitude, $\pv/\ebv \le 13\,\%$, significantly higher than the value of 9\,\% characterizing translucent lines of sight at low latitudes \citep{SMF75}. This is a strong new constraint that dust models must satisfy.

\begin{acknowledgements}
The \Planck\ Collaboration acknowledges the support of ESA; CNES, and CNRS/INSU-IN2P3-INP (France); ASI, CNR, and INAF (Italy); NASA and DoE (USA); STFC and UKSA (UK); CSIC, MINECO, JA, and RES (Spain); Tekes, AoF, and CSC (Finland); DLR and MPG (Germany); CSA (Canada); DTU Space (Denmark); SER/SSO (Switzerland); RCN (Norway); SFI (Ireland); FCT/MCTES (Portugal); ERC and PRACE (EU). A description of the \Planck\ Collaboration and a list of its members, indicating which technical or scientific activities they have been involved in, can be found at \href{http://www.cosmos.esa.int/web/planck/planck-collaboration}{\texttt{http://www.cosmos.esa.int/web/planck/planck-collaboration}}. 
This work has made use of data from the European Space Agency (ESA) mission Gaia (\url{https://www.cosmos.esa.int/gaia}), processed by the Gaia Data Processing and Analysis Consortium (DPAC, \url{https://www.cosmos.esa.int/web/gaia/dpac/consortium}). Funding for the DPAC has been provided by national institutions, in particular the institutions participating in the Gaia Multilateral Agreement. 
The research leading to these results has received funding from the European Research Council under the European Union's Horizon 2020 Research \& Innovation Framework Programme / ERC grant agreement ERC-2016-ADG-742719. 
This research has received funding from the Agence Nationale de la Recherche (ANR-17-CE31-0022).
We thank Pekka Teerikorpi and Andrei Berdyugin for kindly providing stellar polarization data and providing insights on stellar polarization references, Gina Panopoulou for discussions of the zero point calibration of the polarization angle, and Ralf Siebenmorgen and Nikolai Voshchinnikov for statistical discussions.
We gratefully acknowledge the help of Rosine Lallement regarding the handling of the Gaia data.
\end{acknowledgements}

\bibliographystyle{aat} 
\bibliography{Planck_bib,CPP_L11b,biblio}


\appendix
\section{A guide to {\Planck} papers on Galactic astrophysics using polarized thermal emission from dust}
\label{sec:appendix:PlanckPapers}

In this appendix, we give a summary of the contents and main results of the {\Planck} papers dealing with Galactic astrophysics using polarized thermal emission from dust, in the hope that it will provide a useful reference for many readers.

In \citet{planck2014-XIX}, we presented the first analysis of the 353-GHz polarized sky at $1^\circ$ resolution, focusing on the statistics of the polarization fraction $p$ and polarization angle $\psi$, at low and intermediate Galactic latitudes. We found a high maximum polarization fraction ($p_\mathrm{max}>19.8\,\%$) in the most diffuse regions probed. This maximum polarization fraction was found to decrease as total gas column density $\NH$ increases, which might be interpreted as changes of grain alignment properties or as the effect of magnetic-field structure along the line of sight. We also characterized the structure of the polarization angle map by computing its local dispersion function $\mathcal{S}$, which was found to be inversely related to the polarization fraction, lending support to the second explanation.

In \citet{planck2014-XX}, we presented an analysis of \Planck\ 353-GHz polarized thermal dust emission towards a set of molecular clouds and other nearby fields, at 15\arcmin\ resolution, and compared their statistics to those computed on synthetic maps derived from simulations of anisotropic magnetohydrodynamical (MHD) turbulence. We showed that, at these angular scales, the turbulent structure of the Galactic magnetic field (GMF) is able to reproduce the main statistical properties of polarized thermal dust emission in nearby molecular clouds, with no necessity to introduce spatial changes of dust alignment properties.

In~\citet{planck2014-XXI}, we compared the \Planck\ polarized emission at 353\,GHz to surveys of starlight polarization in extinction in the visible, selecting those stars for which the submillimetre and optical estimates of column densities and polarization angles match. For these lines of sight, we computed the ratio $R_{\rm S/V}$ of submillimetre to visible polarization fractions, and the ratio $R_{P/p}$ of the polarized intensity in the submillimetre to the degree of polarization in the visible. We found these to be $R_{\rm S/V}=4.2\pm0.2\pm0.3$ and $R_{P/p}=[5.4\pm0.2\pm0.3]\,\mathrm{MJy\,sr^{-1}}$, where the first uncertainty is statistical and the second is systematic. The value of $R_{P/p}$ provides strong constraints for models of dust polarized emission. The {\tt DustEM} model~\citep{compiegne-et-al-2011} has been updated by~\cite{guillet-et-al-2018} to take these constraints into account.

In~\cite{planck2014-XXXII}, we studied the correlation between the magnetic-field orientation inferred from polarization angles at 353\,GHz and the filamentary structures of matter, at 15\arcmin\ resolution, for intermediate and high Galactic latitudes, covering a range of total gas column densities from $10^{20}\,\mathrm{cm^{-2}}$ to $10^{22}\,\mathrm{cm^{-2}}$. The filaments were extracted using the Hessian matrix of the dust total intensity map. We found that the filaments are preferentially parallel to the magnetic orientation, in particular when the polarization fractions are high and the filaments are more diffuse. Conversely, some of the densest, molecular filaments are perpendicular to the magnetic orientation. This analysis also provided a first estimate for the ratio of the turbulent to mean components of the GMF, i.e., $f_\mathrm{M}=0.8\pm0.2$.

In~\cite{planck2014-XXXIII}, we further studied the signature of the magnetic-field geometry of interstellar filaments in \Planck\ 353-GHz dust polarization maps, at the native $4{\parcm}8$ resolution, focusing on three nearby, dense, star-forming filaments ($\NH\approx 10^{22}\,\mathrm{cm^{-2}}$), and interpreting the \Planck\ Stokes $I$, $Q$, and $U$ maps as the superposition of contributions from the filaments themselves and their respective backgrounds and foregrounds. In this way we found
differences in polarization angles between the filaments and their environments, reaching values up to $54^\circ$, and a decrease in polarization fraction within the filaments, although this could be due not only to the effect of magnetic field tangling along the line of sight, but also in part to changes of grain alignment properties.

In~\cite{planck2015-XXXIV}, we combined the polarization data from \Planck\ at 353 GHz with rotation measure (RM) observations from~\cite{savage-et-al-2013} towards a massive star-forming region, the Rosette Nebula in the Monoceros molecular cloud, to study the impact of an expanding {\sc Hii} region on the morphology of the Galactic magnetic field. We introduced an analytical model that describes the magnetic field structure in a spherical shell, following the expansion of an ionized nebula in a medium with uniform density and magnetic field, and fitted it to the data. This work was subsequently extended to non-spherical bubbles to model the structure of the large-scale magnetic field in the Local Bubble~\citep{alves-et-al-2018}. The \Planck\ polarization data towards the Orion-Eridanus superbubble provide additional evidence for the impact of massive stars on the magnetic field structure~\citep{soler-et-al-2018}.

In~\cite{planck2015-XXXV}, we studied the relative orientation between filamentary structures of matter and the magnetic field towards molecular clouds of the Gould Belt, probing lines of sight with total gas column densities $\NH$ from around $10^{21}$ to $10^{23}\,\mathrm{cm^{-2}}$, at 10\arcmin\ resolution, using the histogram of relative orientations (HRO) technique~\citep{Soler13}. We found that this relative orientation changes progressively from preferentially parallel in regions with the lowest gas column densities to preferentially perpendicular in the regions with the highest $\NH$, with a crossover at $\NH$ of a few $ 10^{21}\,\mathrm{cm^{-2}}$. This change in relative orientation was found to be compatible with simulations of sub- or trans-Alfv\'enic MHD turbulence, underlining the important dynamical role played by the magnetic field in shaping the structure of molecular clouds.

In~\cite{planck2015-XXXVIII}, we studied the $E$ and $B$ modes~\citep{kamionkowski-et-al-97,zaldarriaga-seljak-97} of dust polarization from the magnetized filamentary structure of the interstellar medium at high Galactic latitudes, isolating Hessian-extracted filaments at angular scales where the $E/B$ power-asymmetry and $TE$ correlation are observed~\citep{planck2014-XXX}. The preferred orientation of these filaments parallel to the magnetic field is able to account for both of these observations and was quantified by an oriented stacking of the maps of $I$, $Q$, $U$, $E$, and $B$. From these stacked maps and the histogram of relative orientations, we derived an estimate of the mean polarization fraction in the filaments, $\langle{p}\rangle\approx11\,\%$. 

In~\cite{planck2016-XLII}, we provided a comparison of three different models of the large-scale GMF to \Planck\ polarization data at low and high frequencies, respectively taken as templates for the polarized synchrotron and thermal dust emission. We found in particular that the models underpredict the dust polarization out of the Galactic plane, calling for an increased ordering of the GMF near the observer. 

In ~\cite{planck2016-XLIV}, we provided a phenomenological model of the polarized dust sky. Polarized emission is assumed to arise from a small number of layers in which the GMF is taken to be a superposition of a uniform component $\boldsymbol{B}_0$ and a turbulent component $\boldsymbol{B}_{\rm t}$. Applying this model to the \Planck\ maps of the southern Galactic cap at 353\,GHz and $1^\circ$ resolution, and using the 1-point statistics of $p$ and $\psi$, we could constrain the orientation of the large-scale field in the Solar neighbourhood, the number of layers ($N\approx 7$), the effective polarization fraction of dust emission ($p_0=26\pm3\,\%$), and the ratio of the strengths of the turbulent to mean components of the GMF ($f_\mathrm{M}=0.9\pm0.1$). This phenomenological framework was further improved by~\cite{ghosh-et-al-2017} and~\cite{vansyngel-et-al-2017}. 

\section{\texttt{GNILC} Stokes and covariance maps}
\label{sec:appendix:GNILC}

\subsection{Variable resolution \GNILC\ maps}

\begin{figure*}
\includegraphics[width=0.5\textwidth]{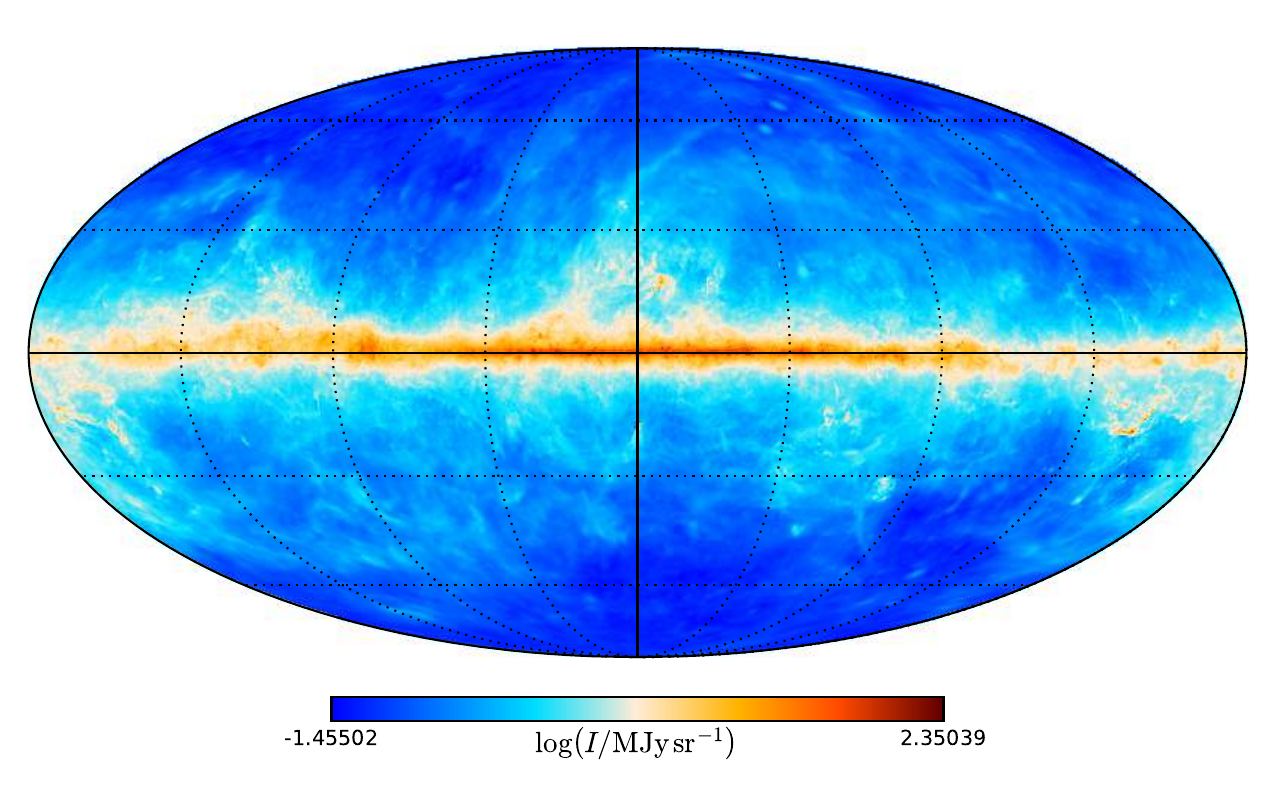}
\includegraphics[width=0.5\textwidth]{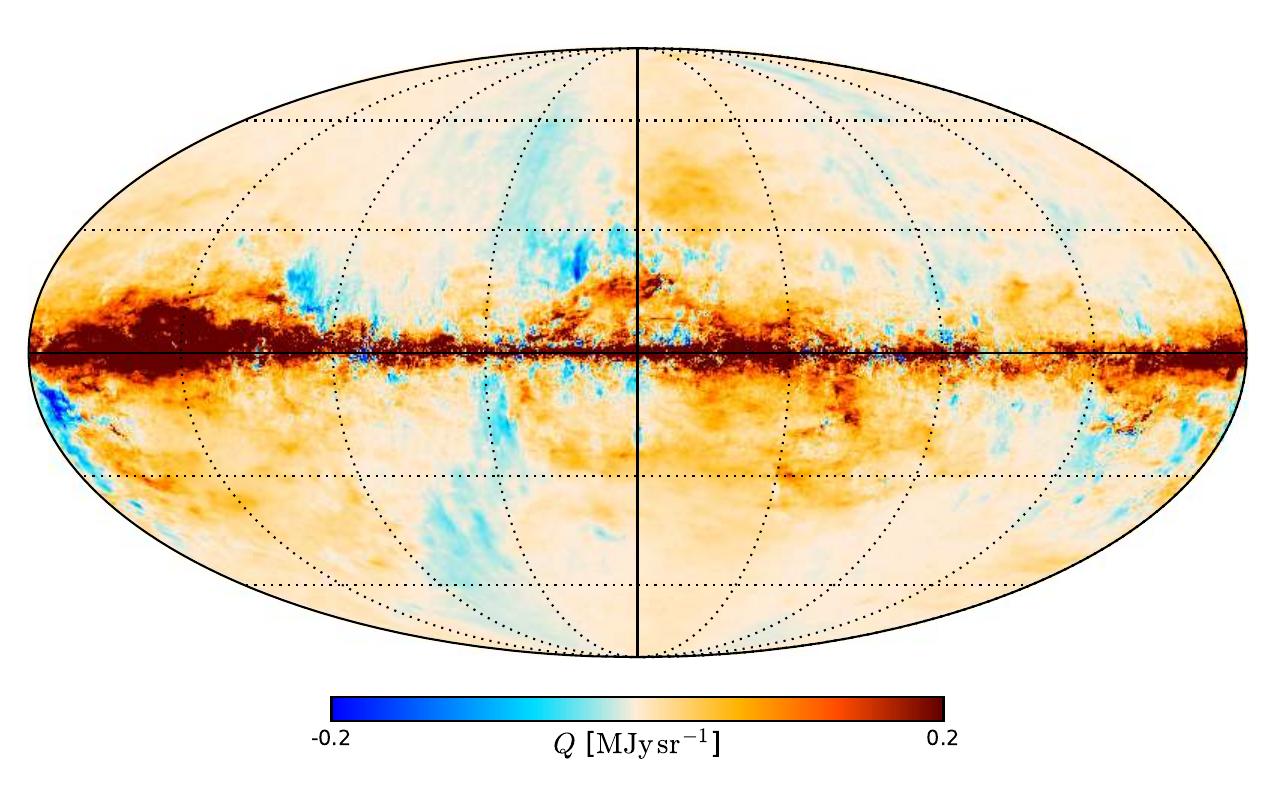}
\includegraphics[width=0.5\textwidth]{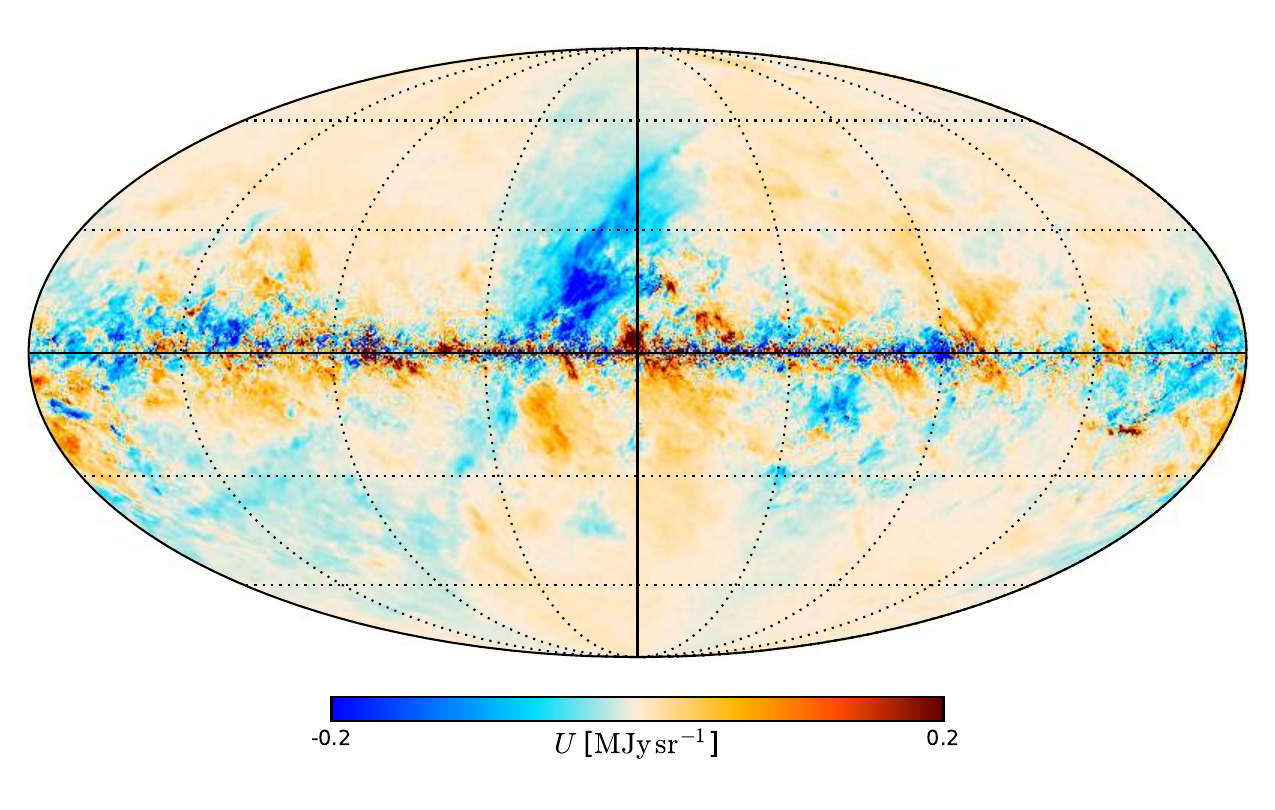}
\includegraphics[width=0.5\textwidth]{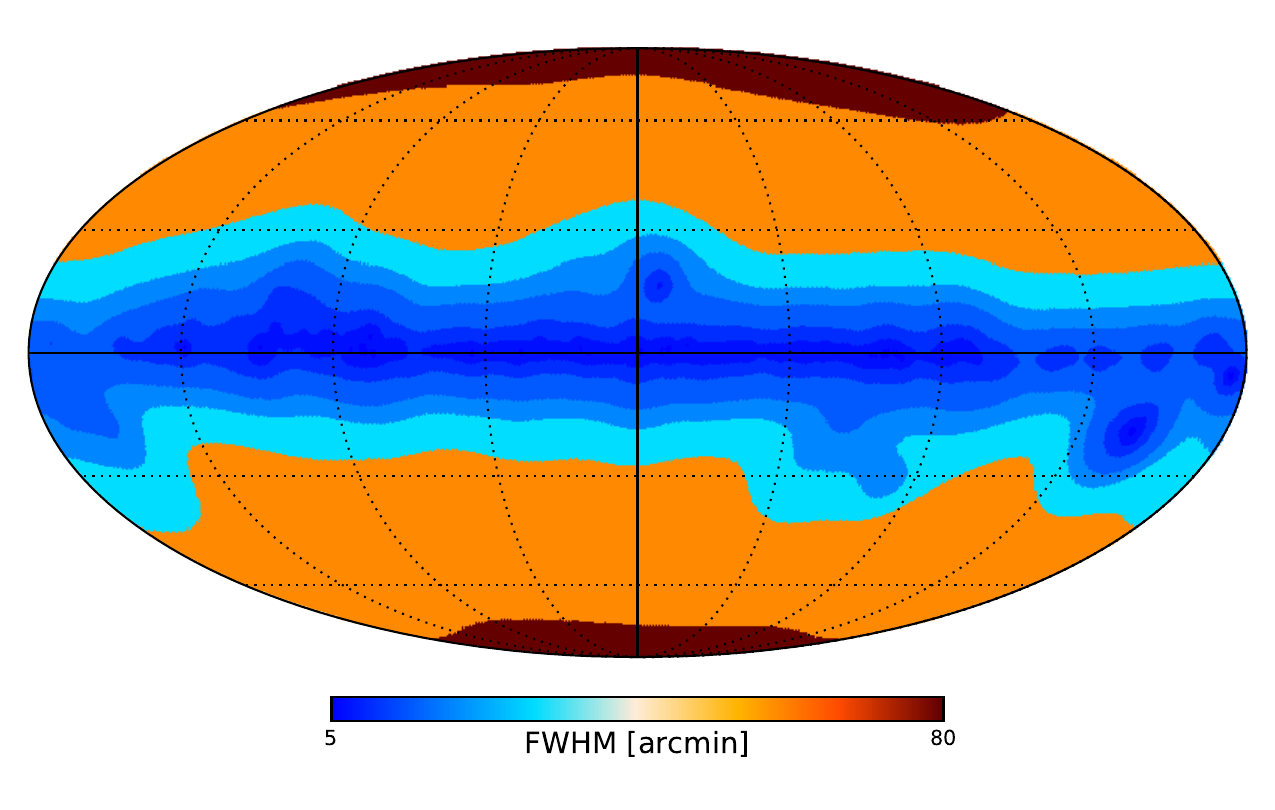}
\caption[]{\GNILC\ maps of Stokes $I$ (top left), $Q$ (top right), $U$ (bottom left), and effective FWHM (bottom right) at 353\,GHz and varying resolution.
The discrete values of the effective FWHM are 5\arcmin, 7\arcmin, 10\arcmin, 15\arcmin, 20\arcmin, 30\arcmin, 60\arcmin, and 80\arcmin. The scale of the $I$ map is logarithmic, while the rest are linear.}
\label{fig:IQU_353_GNILC}
\end{figure*}

For reference, in Fig.~\ref{fig:IQU_353_GNILC} we show the \GNILC\ Stokes maps at 353\,GHz and variable resolution over the sky, alongside the map of the effective FWHM, whose discrete values are $5\arcmin$, $7\arcmin$, $10\arcmin$, $15\arcmin$, $20\arcmin$, $30\arcmin$, $60\arcmin$, and $80\arcmin$. The total intensity map corresponds to the fiducial offset value.

\subsection{\GNILC-processed covariance maps}

To assess the statistical uncertainties affecting the dust polarization properties studied in this paper, the \GNILC\ algorithm provides the maps of covariances between the Stokes parameters at 353\,GHz, $\sII$, $\sIQ$, $\sIU$, $\sQQ$, $\sQU$, and $\sUU$. We now describe how these covariance maps are produced.

The \GNILC\ dust map, $D_{\tt GNILC}$, is a mimimum-variance weighted linear combination of the \Planck\ frequency maps $X_i$:
\begin{equation}
D_{\tt GNILC}=\sum_i\, w_i\, X_i \,,
\end{equation}
where the sum extends over the seven \Planck\ polarization-sensitive frequency channels. The weights $w_i$ are estimated by the \GNILC\ algorithm in order to extract the dust emission while filtering out the instrumental noise and the CMB.\footnote{For total intensity, the CIB anisotropies are also filtered out.} The residual noise rms fluctuations, $N_{\tt GNILC}$, of the \GNILC\ dust map are thus related to the instrumental noise rms fluctuations, $N_i$, in each \Planck\ frequency map $X_i$, according to
\begin{equation}
N_{\tt GNILC} = \sum_i\, w_i\, N_i \,.
\end{equation}
That residual noise is minimized by the \GNILC\ weights. An estimate of the instrumental noise rms fluctuations $N_i$ in each frequency channel $i$ can be obtained by computing the half-difference of the \Planck\ half-mission maps $X_{i,\mathrm{HM}_1}$ and $X_{i,\mathrm{HM}_2}$:
\begin{equation}
\widehat{N}_i = \frac{1}{2}\, \left(X_{i,\mathrm{HM}_2}\,-\,X_{i,\mathrm{HM}_1}\right) \, ,
\end{equation}
because the sky emission cancels out in the difference while the noise does not. We can thus estimate the residual noise in the \GNILC\ dust map as
\begin{equation}
\widehat{N}_{\tt GNILC}=\sum_i\, w_i\, \widehat{N}_i \, ,
\end{equation}
where the $\widehat{N}_i$ maps have first been smoothed to the actual resolution of the \GNILC\ maps, i.e., either 80\arcmin\ for the uniform resolution case or to the local resolution of the specific regions of the sky for the multi-resolution case. The estimate $\widehat{N}_{\tt GNILC}$ has the variance of the actual residual noise in the \GNILC\ dust maps, $D_{\tt GNILC}$. 

The \GNILC\ noise covariance maps are then estimated as follows. We first smooth the native \Planck\ noise covariance maps at $353$\,GHz, $\sigma_{XY}$,\footnote{Despite the symbol $\sigma$, these are covariances, not dispersions.} where $X$ and $Y$ stand for any one of the three Stokes $I$, $Q$, or $U$, to the resolution of the \GNILC\ maps, by following the procedure employed in~\citet{planck2014-XIX}. For the multi-resolution case, the value of the smoothing scale adopted in each region of the sky depends on the local resolution of the \GNILC\ maps in that region. Because a covariance is derived from the product of two Stokes parameter maps, the smoothing scale of a covariance map is $\sqrt{2}$ times the resolution of the Stokes maps.  We then compute the local (co)variance value in each region of a given resolution of the \Planck\ and \GNILC\ noise maps, $\widehat{N}_{\rm 353\, GHz}$ and $\widehat{N}_{\tt GNILC}$, for instance:
\begin{eqnarray}
{\rm cov}^{(j)}\left(\widehat{N}^Q_{\rm 353},\widehat{N}^U_{\rm 353}\right) &=& \frac{1}{n_j}\, \sum_{p \in R_j}\, \widehat{N}^Q_{\rm 353} (p)\, \widehat{N}^U_{\rm 353} (p)\, ,\cr
{\rm cov}^{(j)}\left(\widehat{N}^Q_{\tt GNILC},\widehat{N}^U_{\tt GNILC}\right) &=& \frac{1}{n_j}\, \sum_{p \in R_j}\, \widehat{N}^Q_{\tt GNILC}(p)\, \widehat{N}^U_{\tt GNILC}(p) \, ,
\end{eqnarray}
where $R_j$ is the set of sky pixels at the given \GNILC\ effective resolution, indexed by $j$, and $n_j$ is the number of such pixels. In each region $R_j$, the \Planck\ noise covariance maps, $\sigma_{XY}(p)$, are then scaled according to
\begin{equation}
\sigma_{XY,\,{\tt GNILC}}(p) = \frac{{\rm cov}^{(j)}\left(\widehat{N}^X_{\tt GNILC},\widehat{N}^Y_{\tt GNILC}\right)}{{\rm cov}^{(j)}\left(\widehat{N}^X_{\rm 353},\widehat{N}^Y_{\rm 353}\right)}\,\sigma_{XY}(p)\, .
\end{equation}
The resulting covariance maps $\sigma_{XY,\,{\tt GNILC}}(p)$ are what we refer to as the \GNILC-processed covariance maps in the rest of this paper. 

We note that they are built using the PSB-only data for polarization at 353\,GHz, both for the uniform 80$\arcmin$ resolution case and for the $B$-mode-driven, varying resolution case (5\arcmin--80\arcmin). In both cases, they are computed at a {\healpix} resolution $N_\mathrm{side}=2048$, but in the uniform 80\arcmin\ resolution case, they are downgraded to $N_\mathrm{side}=128$ to avoid oversampling. For the varying resolution case, we keep the original $N_\mathrm{side}=2048$. The maps are also converted from $\mathrm{K_{CMB}^2}$ to $\mathrm{MJy^2\,sr^{-2}}$ using the conversion factor $287.5\,\mathrm{MJy\,sr^{-1}\,K_{CMB}^{-1}}$ at 353\,GHz~\citep{planck2016-l03}. 

\begin{figure*}[htbp]
\includegraphics[width=0.5\textwidth]{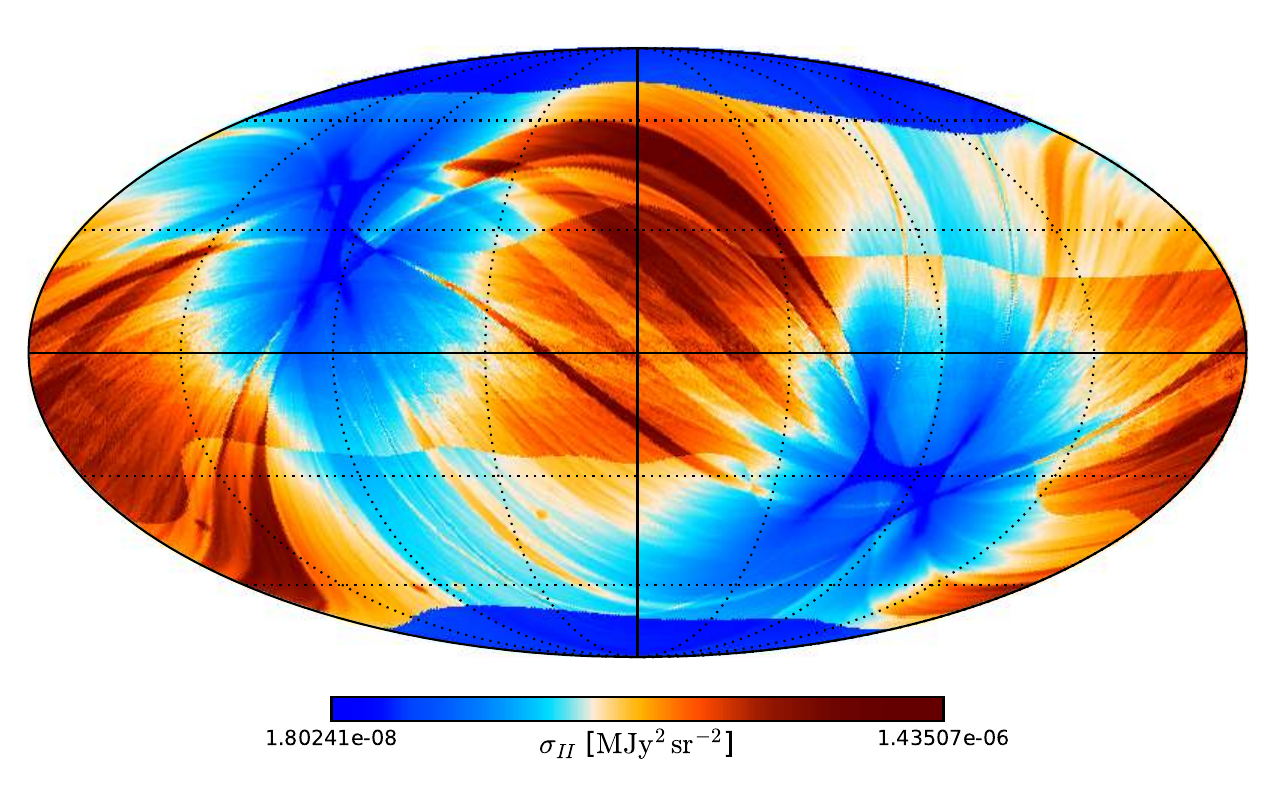}
\includegraphics[width=0.5\textwidth]{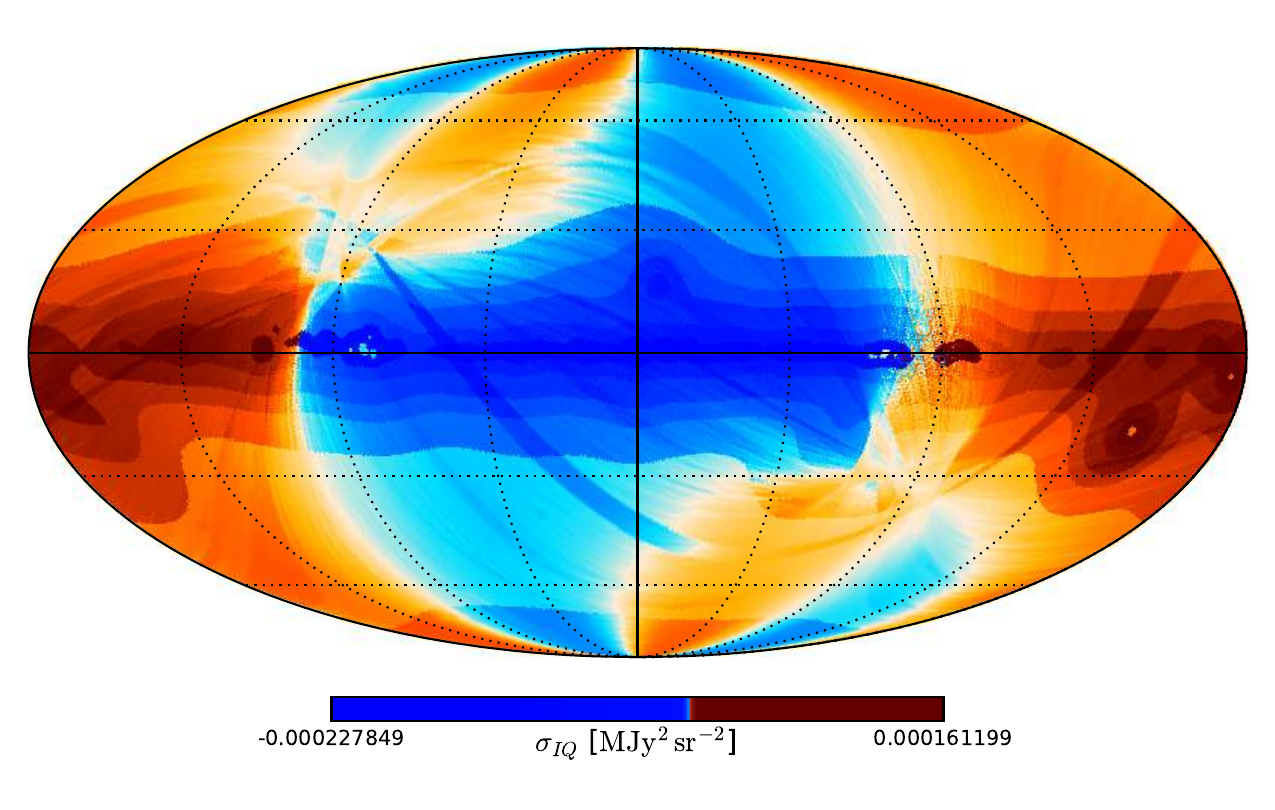}
\includegraphics[width=0.5\textwidth]{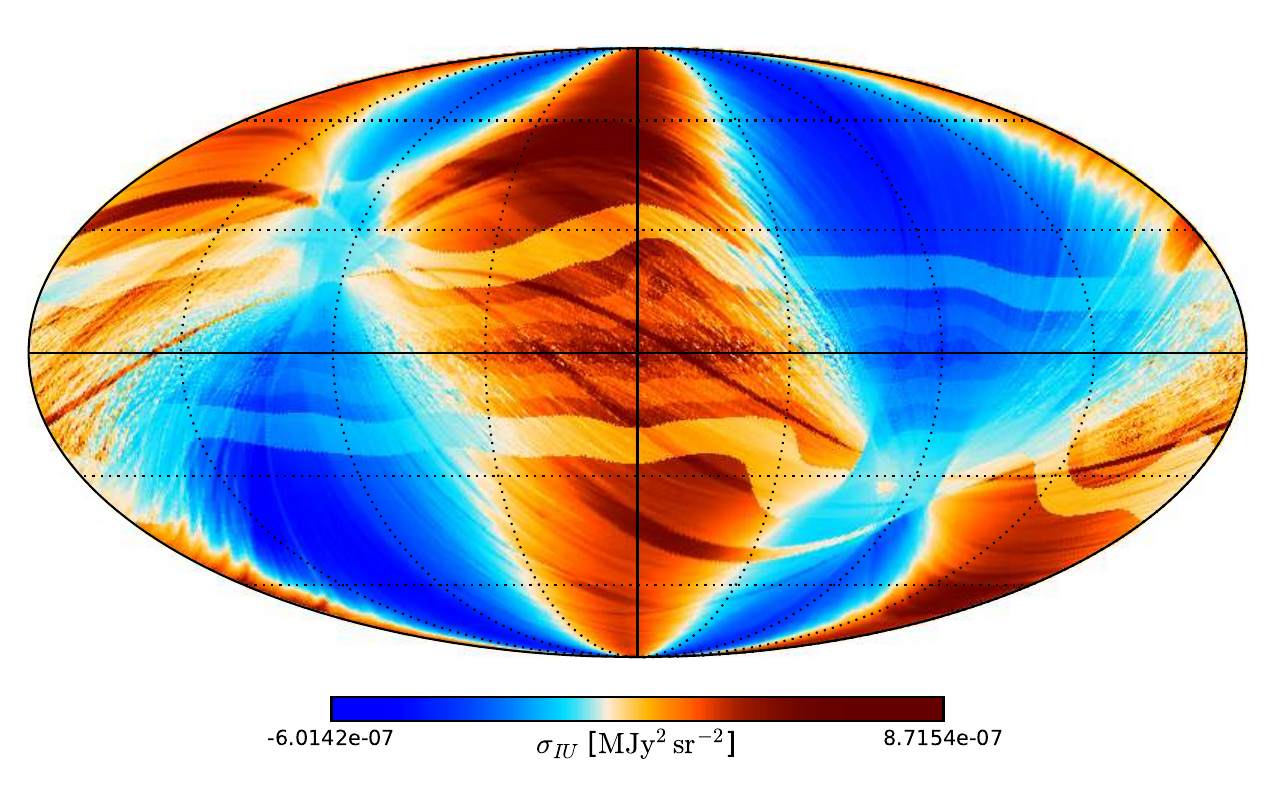}
\includegraphics[width=0.5\textwidth]{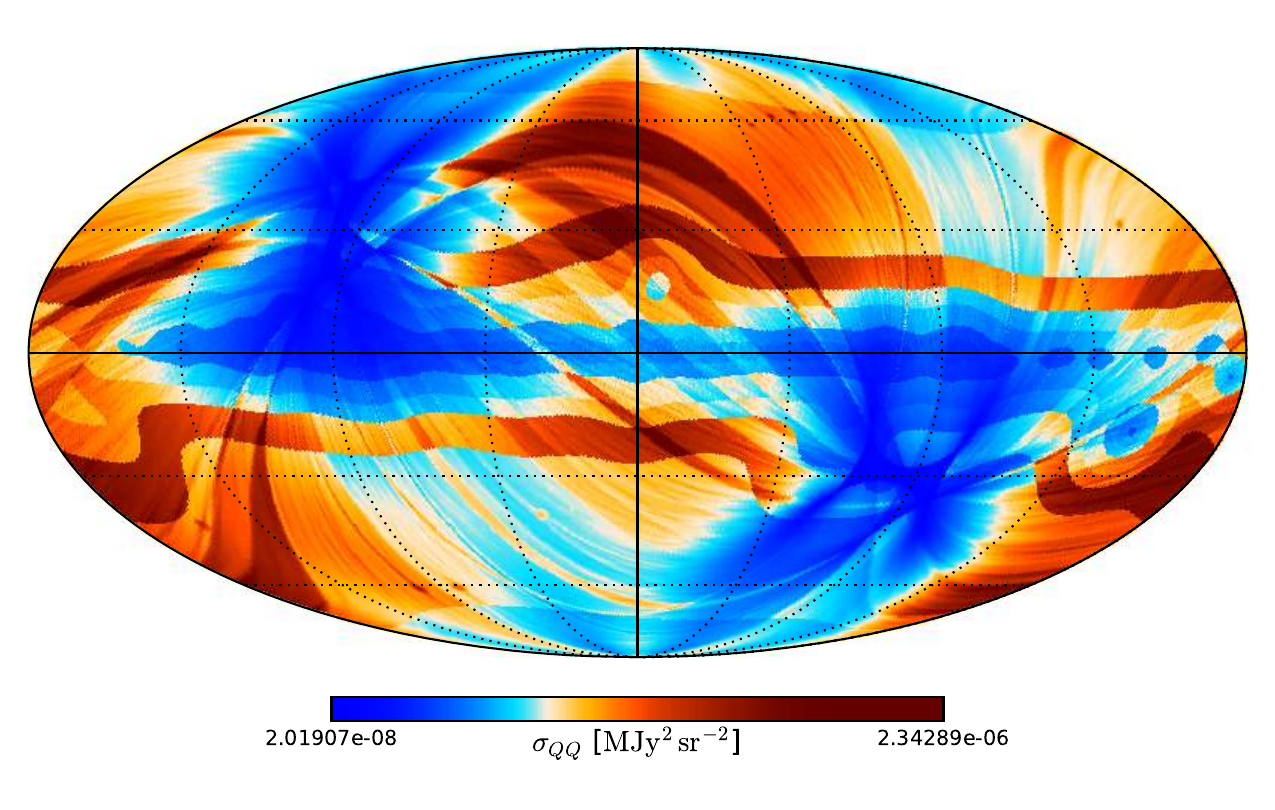}
\includegraphics[width=0.5\textwidth]{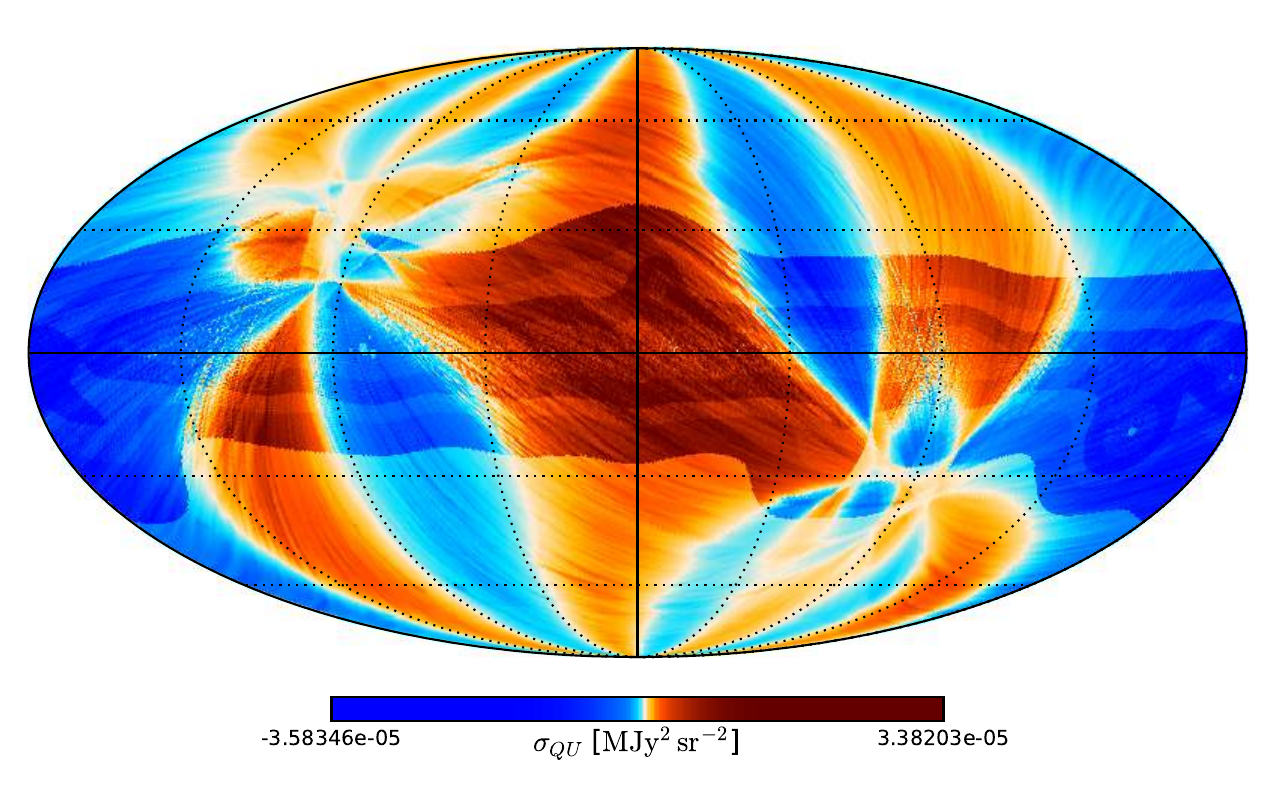}
\includegraphics[width=0.5\textwidth]{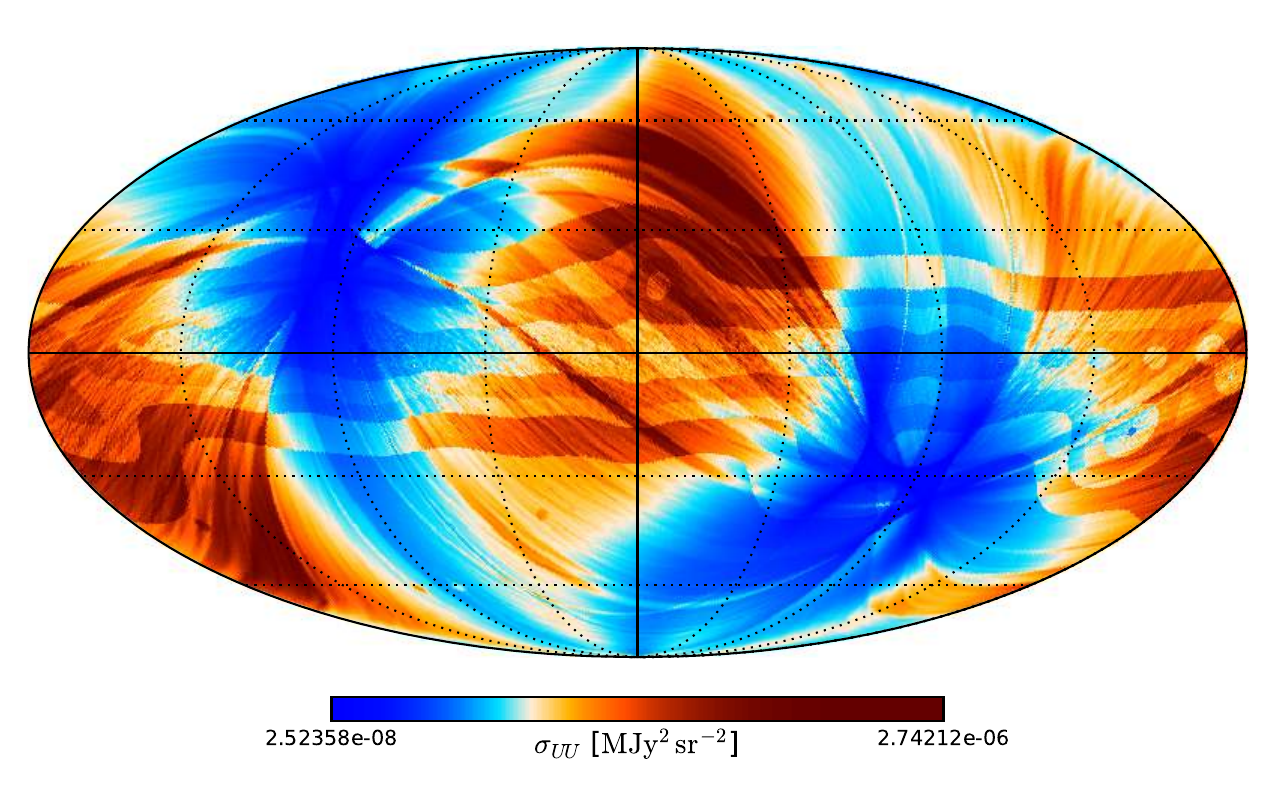}
\caption[]{\GNILC-processed covariance maps at 353\,GHz and varying resolution. From top to bottom and left to right, they are $\sII$, $\sIQ$, $\sIU$, $\sQQ$, $\sQU$, and $\sUU$.}
\label{fig:covariance_GNILC}
\end{figure*}

\begin{figure*}[htbp]
\includegraphics[width=0.5\textwidth]{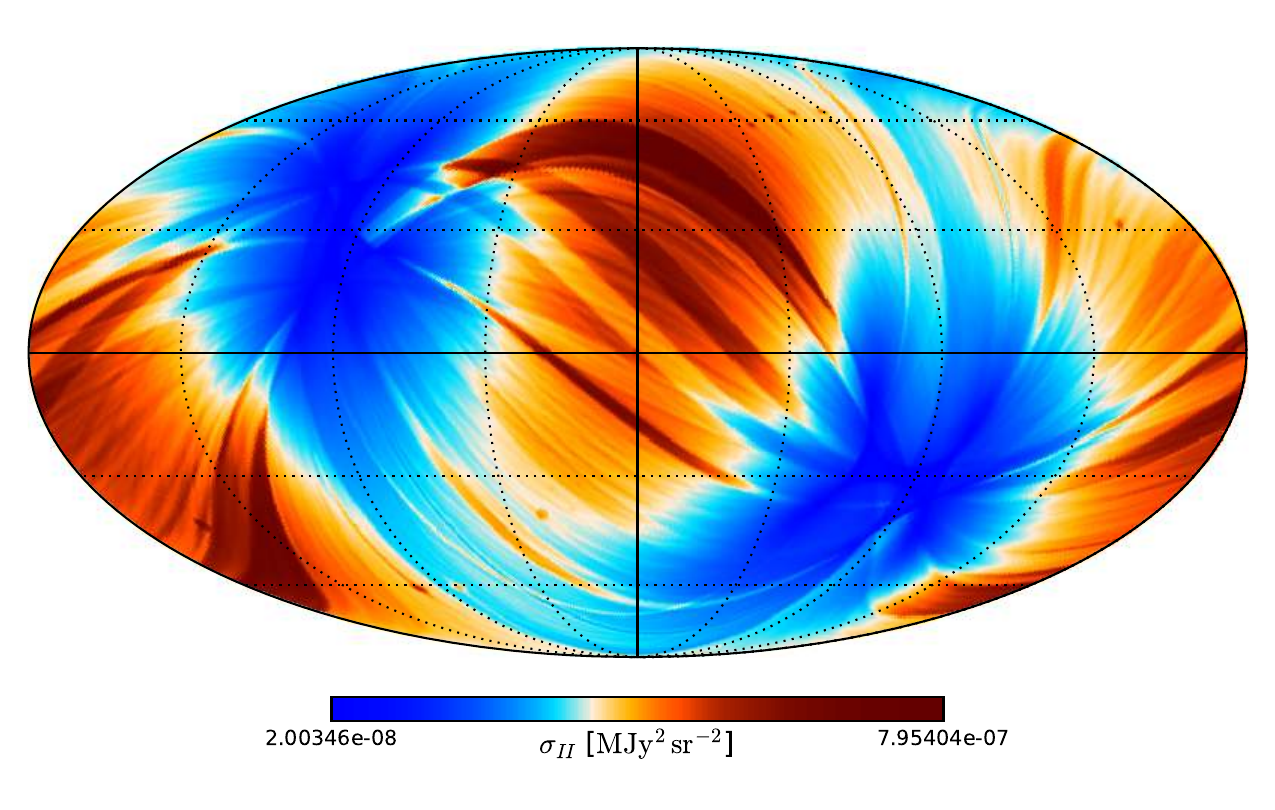}
\includegraphics[width=0.5\textwidth]{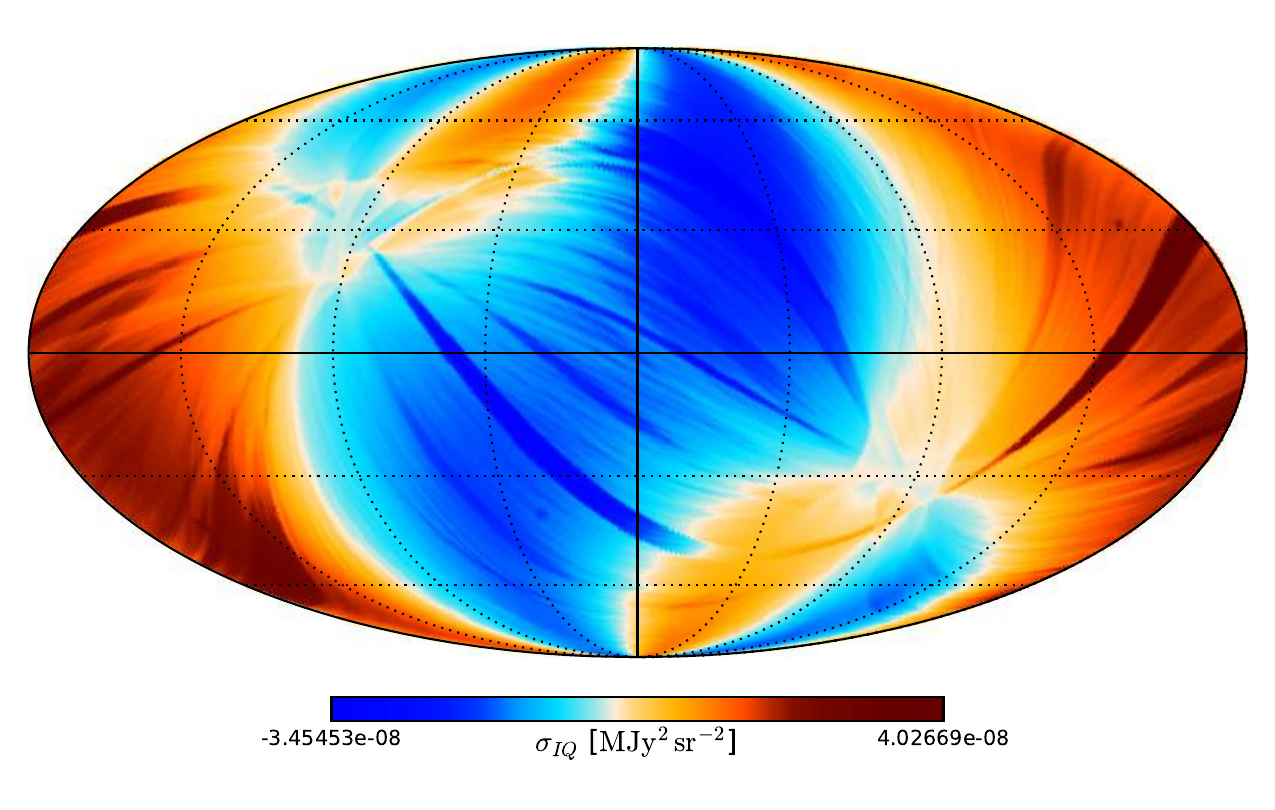}
\includegraphics[width=0.5\textwidth]{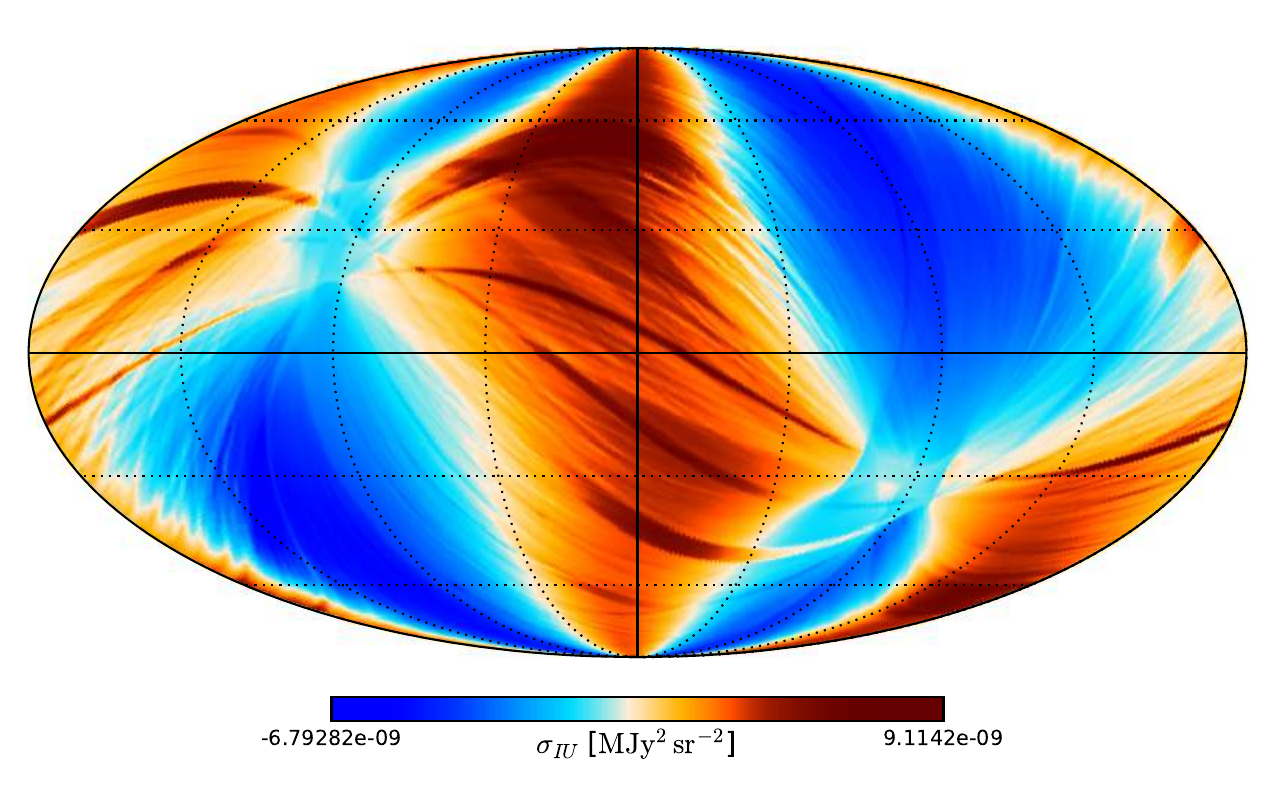}
\includegraphics[width=0.5\textwidth]{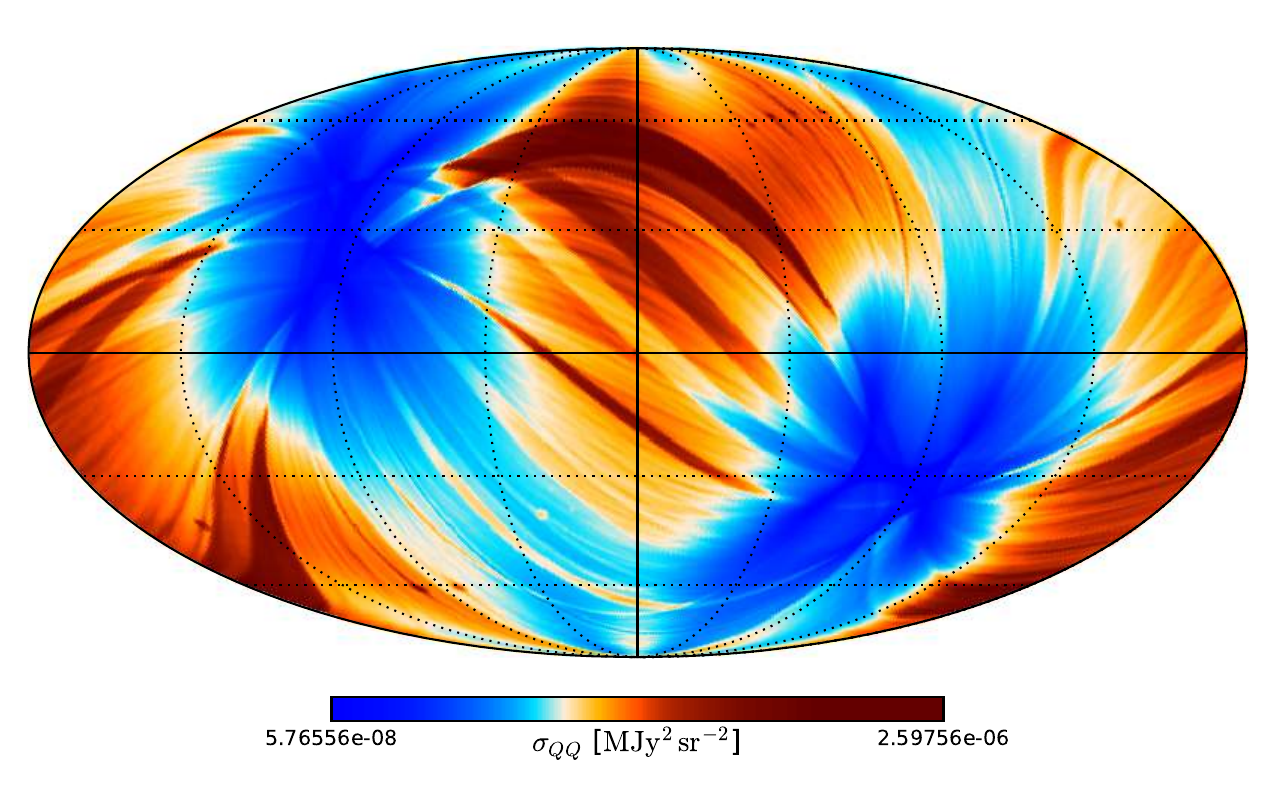}
\includegraphics[width=0.5\textwidth]{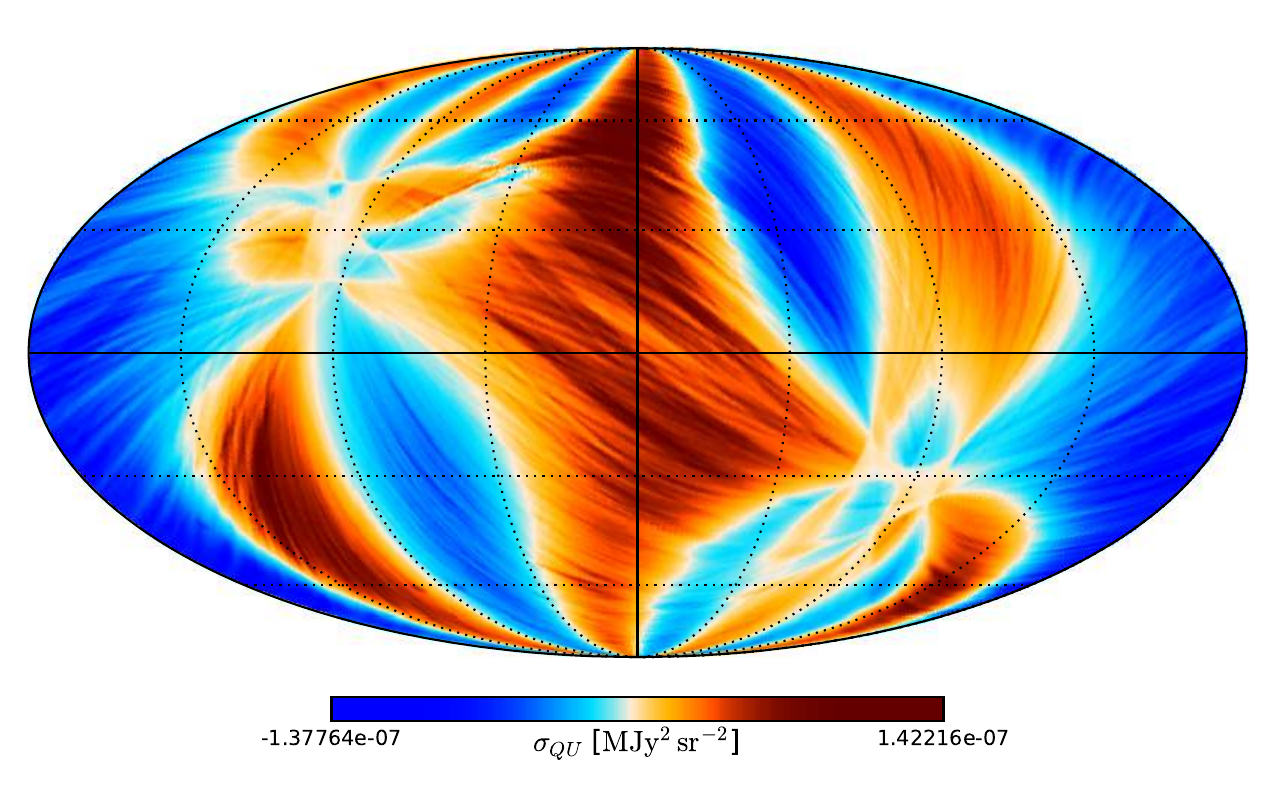}
\includegraphics[width=0.5\textwidth]{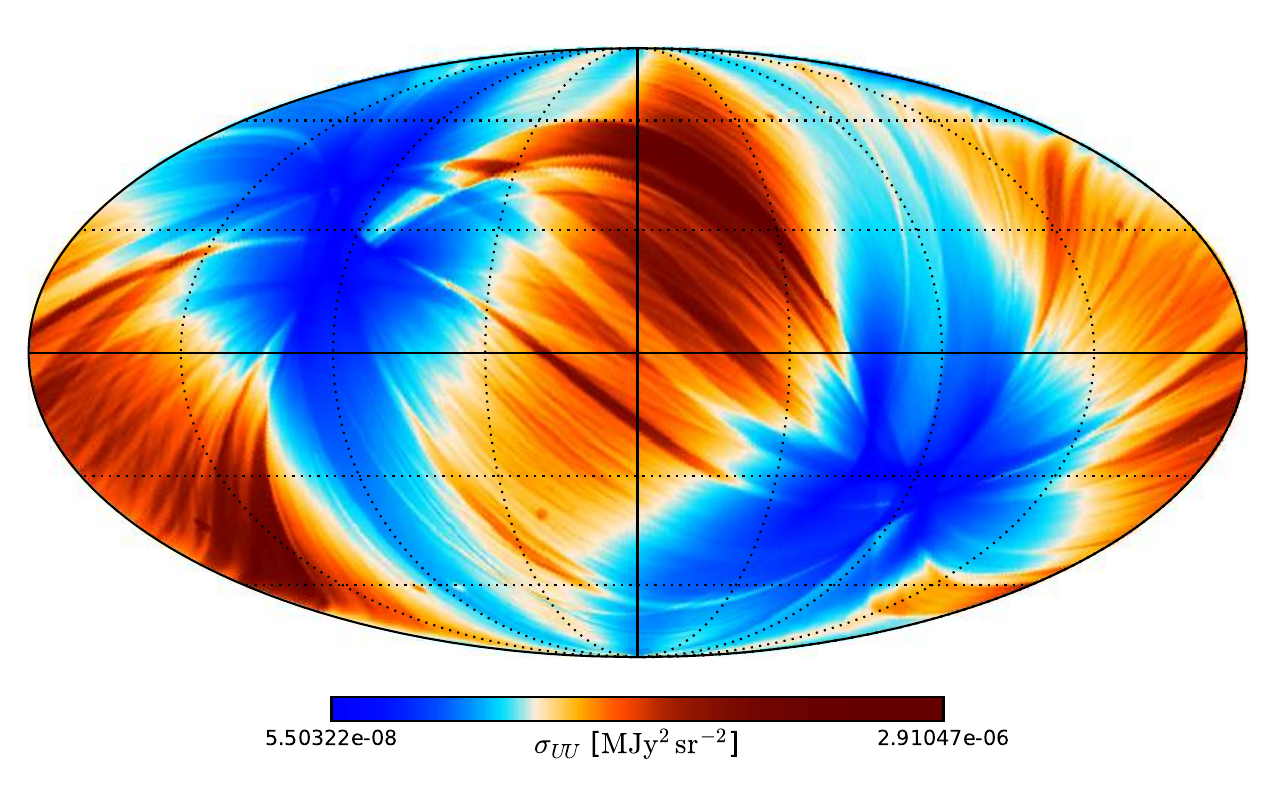}
\caption[]{\GNILC-processed covariance maps at 353\,GHz and unifom 80\arcmin\ resolution. From top to bottom and left to right, they are $\sII$, $\sIQ$, $\sIU$, $\sQQ$, $\sQU$, and $\sUU$. }
\label{fig:covariance_GNILC-80arcmin}
\end{figure*}

Figure~\ref{fig:covariance_GNILC} shows these covariance maps for the variable resolution case, while Fig.~\ref{fig:covariance_GNILC-80arcmin} shows the covariance maps at the common, uniform 80\arcmin\ resolution. The sky patterns of these uniform resolution covariance maps are by construction extremely similar to those directly taken from the {\DRThree} data, but with a significant improvement in the amplitudes, as shown in Table~\ref{table-GNILC-covariance-improvements}, which gives the mean ratios $\sigma_{XY,\mathrm{GNILC}}/\sigma_{XY}$ between the \GNILC\ covariance maps $\sigma_{XY,{\tt GNILC}}$ at 80\arcmin\ resolution and $N_\mathrm{side}=128$ to the corresponding maps $\sigma_{XY}$ in the {\DRThree} data release. Consequently, we use these \GNILC-processed covariance maps to assess statistical uncertainties in our analysis.

\begin{table}[htbp!]
\begingroup
\newdimen\tblskip \tblskip=5pt
\caption{
\GNILC\ versus {\DRThree} data release covariances.}
\label{table-GNILC-covariance-improvements}                            
\nointerlineskip
\footnotesize
\setbox\tablebox=\vbox{
   \newdimen\digitwidth 
   \setbox0=\hbox{\rm 0} 
   \digitwidth=\wd0 
   \catcode`*=\active 
   \def*{\kern\digitwidth}
   \newdimen\signwidth 
   \setbox0=\hbox{+} 
   \signwidth=\wd0 
   \catcode`!=\active 
   \def!{\kern\signwidth}
\halign{\hbox to 1.15in{#\leaderfil}\tabskip 1.0em&
\hfil#\hfil&
\hfil#\hfil\tabskip 0pt\cr
\noalign{\doubleline}
\omit\hfil $\sigma_{XY,{\tt GNILC}}/\sigma_{XY}$ \hfil& Mean*\cr
\noalign{\vskip 4pt\hrule\vskip 5pt}
$II$& 0.72\phantom{0}\cr
$IQ$& 0.22\phantom{0}\cr
$IU$& 0.065\cr
$QQ$& 0.93\phantom{0}\cr
$QU$& 0.26\phantom{0}\cr
$UU$& 0.88\phantom{0}\cr
\noalign{\vskip 4pt\hrule\vskip 5pt}
}}
\endPlancktable                    
\endgroup
\end{table}

\section{End-to-end simulations}
\label{sec:appendix:E2E}

The quality of the data presented here is assessed through a series of end-to-end (E2E) simulations that take into account all the known systematics and noise properties of the data. The process begins with a model of the sky (including foregrounds, in both total intensity and polarization), from which timelines (including all known effects) are simulated. These timelines are then processed through the \Planck\ mapmaking pipeline~\citep{planck2016-l03}. These E2E simulations are 
the same ones used in~\cite{planck2016-l11A} and we refer the reader to appendix A of that paper for a detailed description. The dust component of the model used in these simulations is a combination of a realization of the~\cite{vansyngel-et-al-2017} statistical model for $\ell\geqslant 20$ and the actual \DRThree\ $Q$ and $U$ maps at 353\,GHz for $\ell\leqslant 10$, with a smooth transition between the two in the $10\leqslant\ell\leqslant20$ range. Because of the latter, the large-scale component of the field varies over the simulated sky, which is essential to reproduce the statistics of $p$. The other input sky components are taken from the latest version of the Planck Sky Model~\citep{planck2014-a14}. These model component maps are then combined with the first 100 realizations of the systematic effects and noise~\citep{planck2016-l03}. This results in 100 E2E $I$, $Q$, and $U$ maps at 353\,GHz, which we smooth to 60\arcmin\ resolution, and from which we derive polarization quantities and compare them to the input dust maps, after subtraction of the CMB and the CIB monopole, as is done for the \asmaps\ (Sect.~\ref{sec:asm}). This allows us to assess the effects of residual systematics and data noise on the statistics presented in this paper.

\begin{figure}
\includegraphics[width=0.5\textwidth]{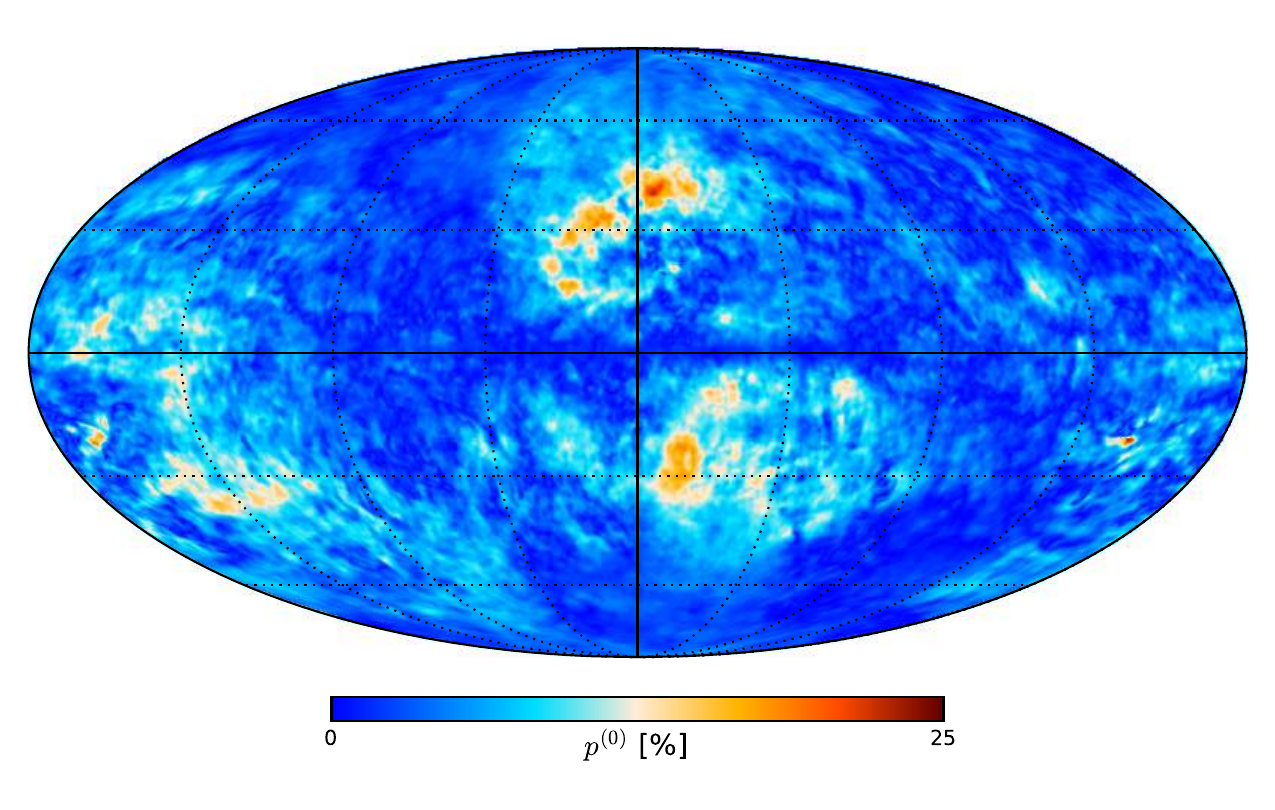}
\includegraphics[width=0.5\textwidth]{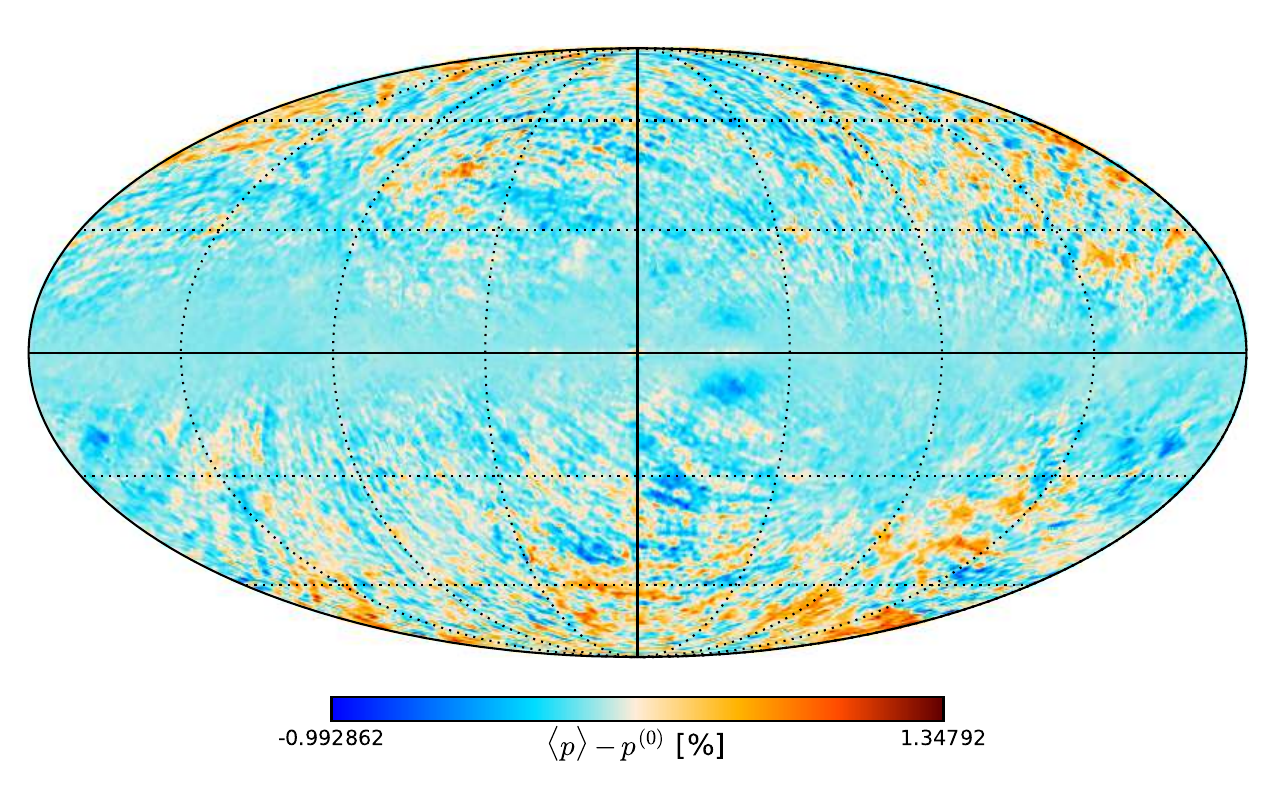}
\includegraphics[width=0.5\textwidth]{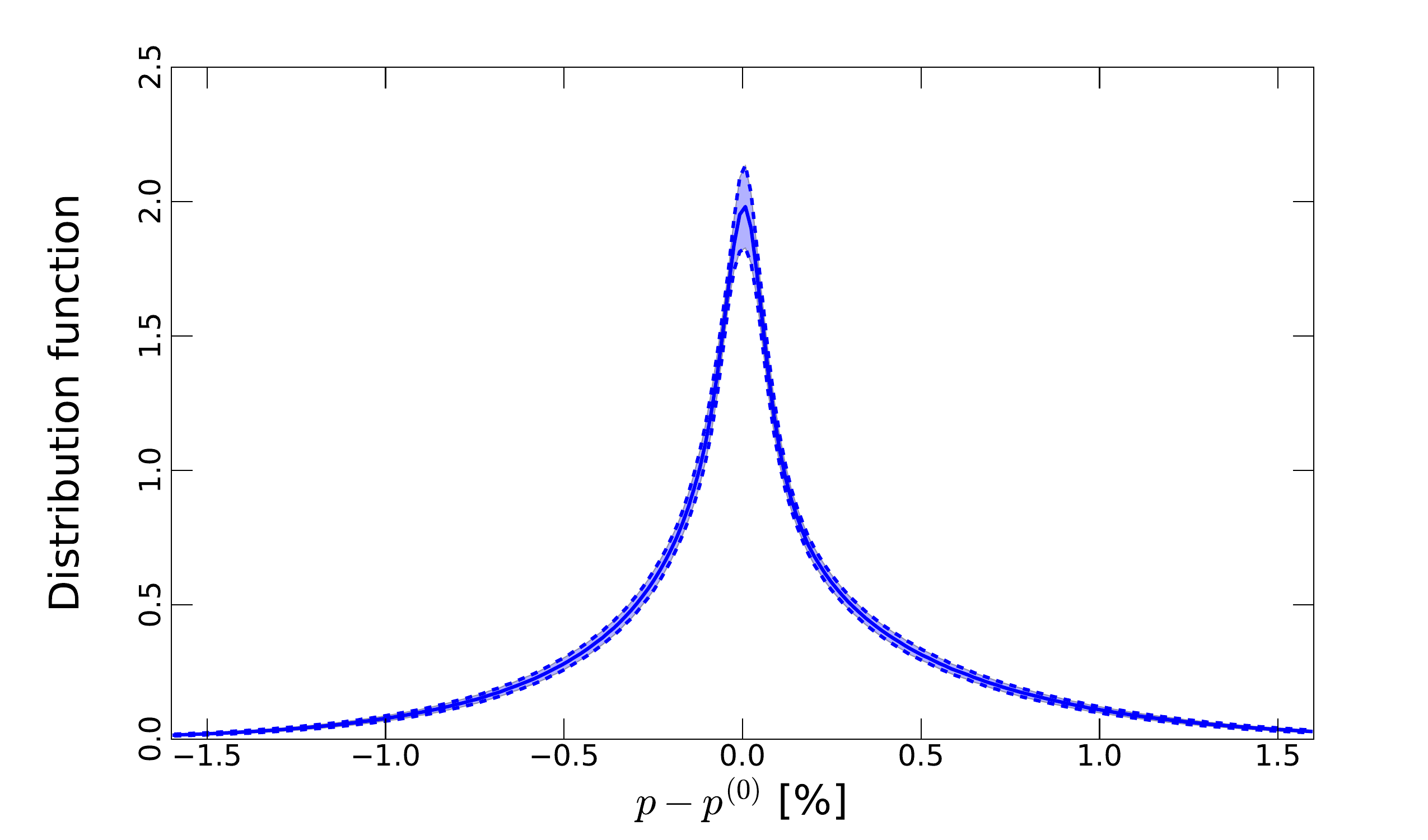}
\caption[]{{\it Top}: Map of polarization fractions $p^{(0)}$ for the input sky of the E2E simulations, using the MAS estimator at 60\arcmin\ resolution. 
{\it Middle}: Map of the difference between the polarization fraction averaged over the 100 realizations of the E2E simulations, $\langle{p}\rangle$, and the input polarization fraction $p^{(0)}$, at 60\arcmin\ resolution.
{\it Bottom}: Distribution function over the sky of the difference between the output and input polarization fractions. 
The solid blue curve is the average of 100 histograms of $p-p^{(0)}$ from the 100 realizations, while the dashed lines with blue shading between show the $\pm1\,\sigma$ dispersion.
}
\label{fig:PI_maps_E2E}
\end{figure}
 
Figure~\ref{fig:PI_maps_E2E} shows that the difference between the input dust polarization fraction $p^{(0)}$ and the average $\langle{p}\rangle$ over the 100 realizations of the E2E simulations is at most around 1\,\%. The distribution function of the difference $p-p^{(0)}$ is peaked around zero for each simulation.  The average of these distributions is shown in the bottom panel of Fig.~\ref{fig:PI_maps_E2E}, along with the $1\,\sigma$ dispersion around the average distribution, which has a mean of 0.03\,\% and a standard deviation of 0.47\,\%. 

\begin{figure}
\includegraphics[width=0.5\textwidth]{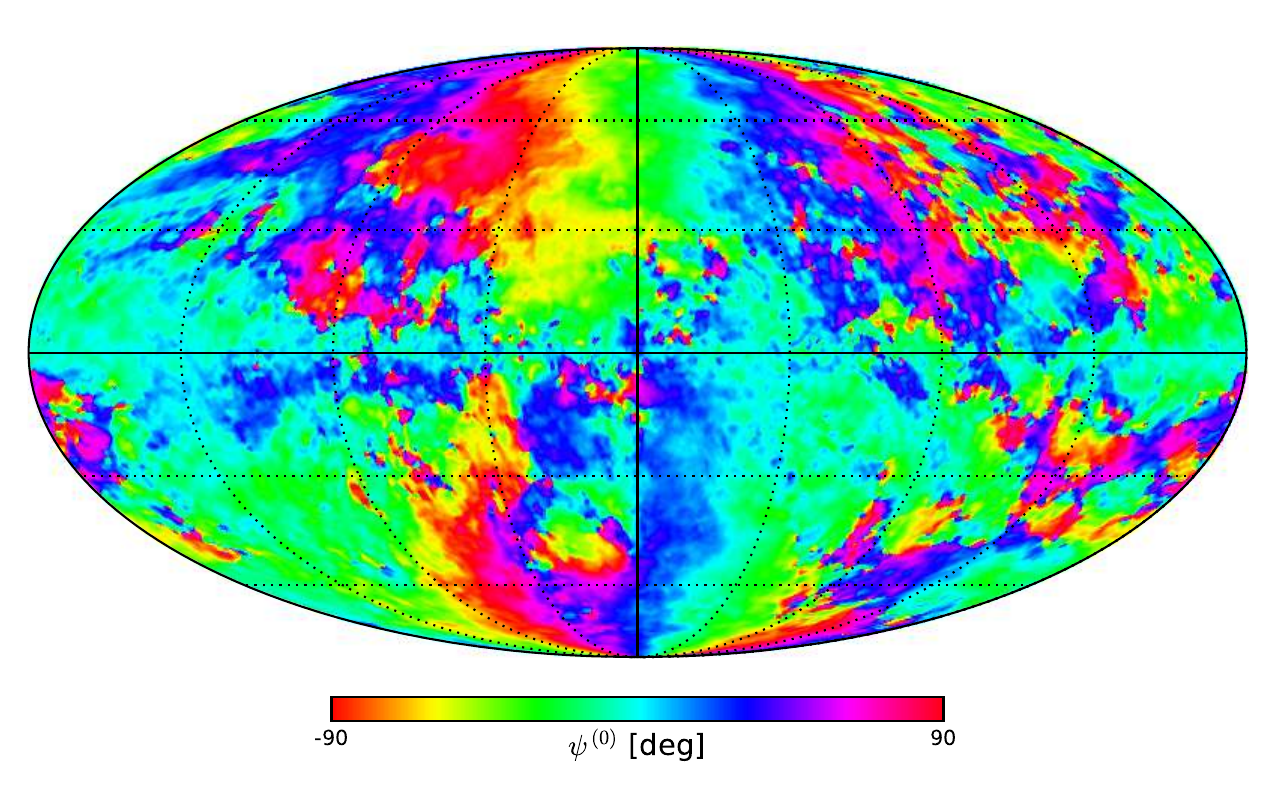}
\includegraphics[width=0.5\textwidth]{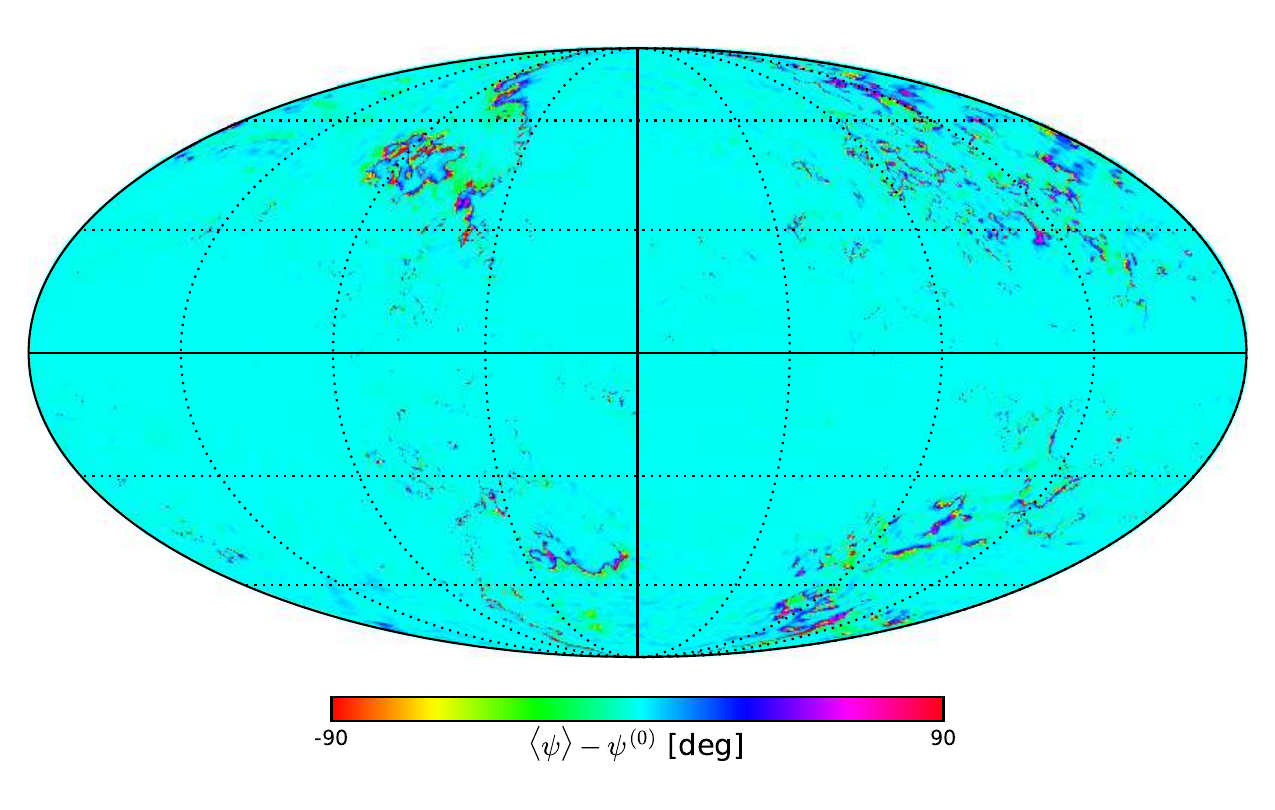}
\includegraphics[width=0.5\textwidth]{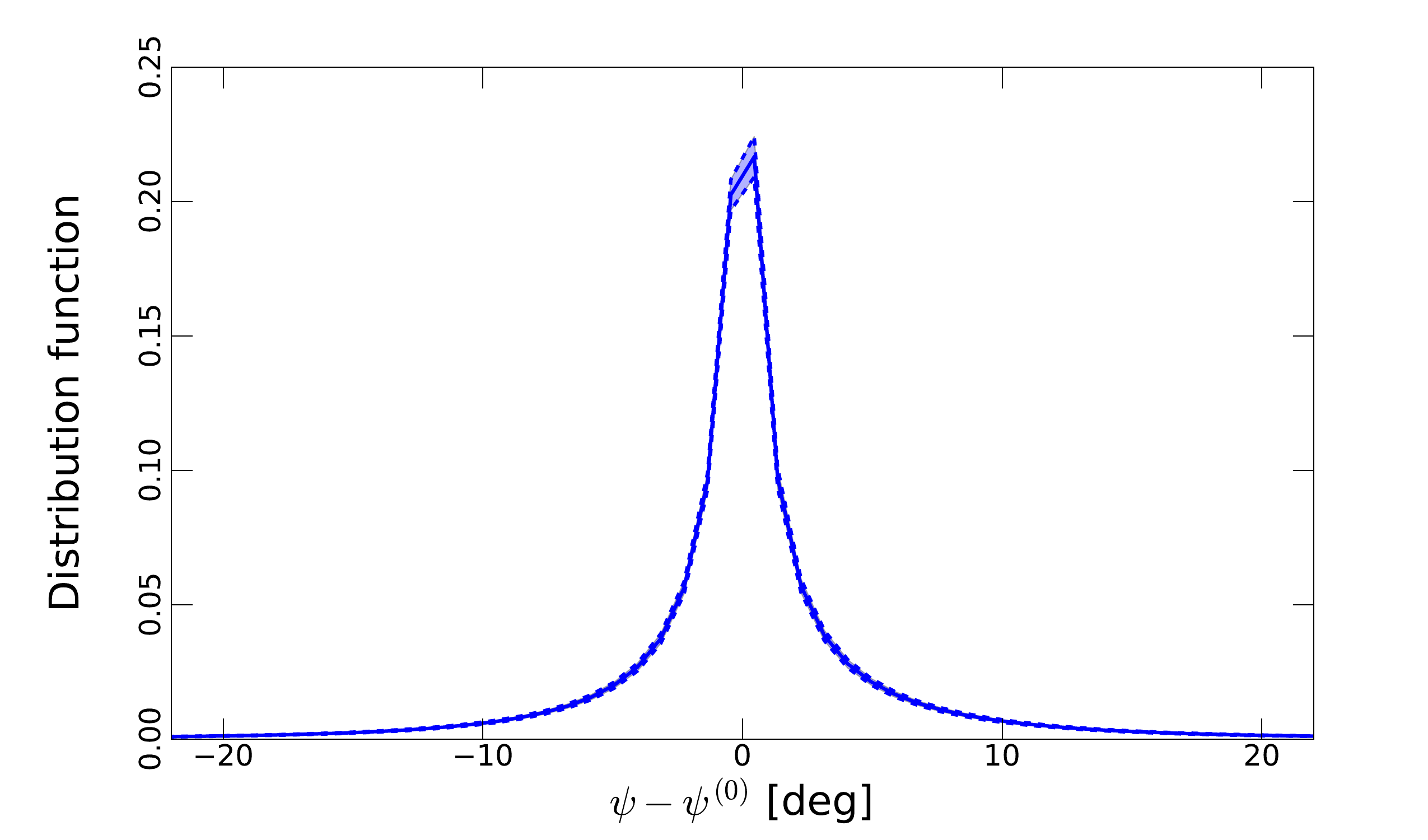}
\caption[]{{\it Top}: Map of the polarization angle $\psi^{(0)}$ for the input sky of the E2E simulations, at 60\arcmin\ resolution. 
{\it Middle}: Map of the difference between the polarization angle averaged over the 100 realizations of the E2E simulations, $\langle{\psi}\rangle$, and the input polarization angle $\psi^{(0)}$, at 60\arcmin\ resolution.
{\it Bottom}: Distribution function over the sky of the difference between the output and input polarization angle. The solid blue curve is the average of 100 histograms of $\psi-\psi^{(0)}$ from the 100 realizations, while the dashed lines with blue shading between show the $\pm1\,\sigma$ dispersion.
}
\label{fig:psi_maps_E2E}
\end{figure}
 
For the polarization angle data, the diagnostic of the E2E simulations is shown in Fig.~\ref{fig:psi_maps_E2E}. To compute the average difference between the output and input polarization angles, $\langle\psi\rangle-\psi^{(0)}$, account is taken of the circularity of the difference for each sky pixel and each realization independently. One can see that the regions of the sky where this difference is the largest are those where $\psi^{(0)}$ crosses the $\pm90\deg$ boundary. The average distribution function of these differences over the 100 simulations and the $1\,\sigma$ dispersion about the average are shown in the bottom panel of the figure. The average distribution has a mean of 0\pdeg3 and a standard deviation of 8{\pdeg}3.

\begin{figure}
\includegraphics[width=0.5\textwidth]{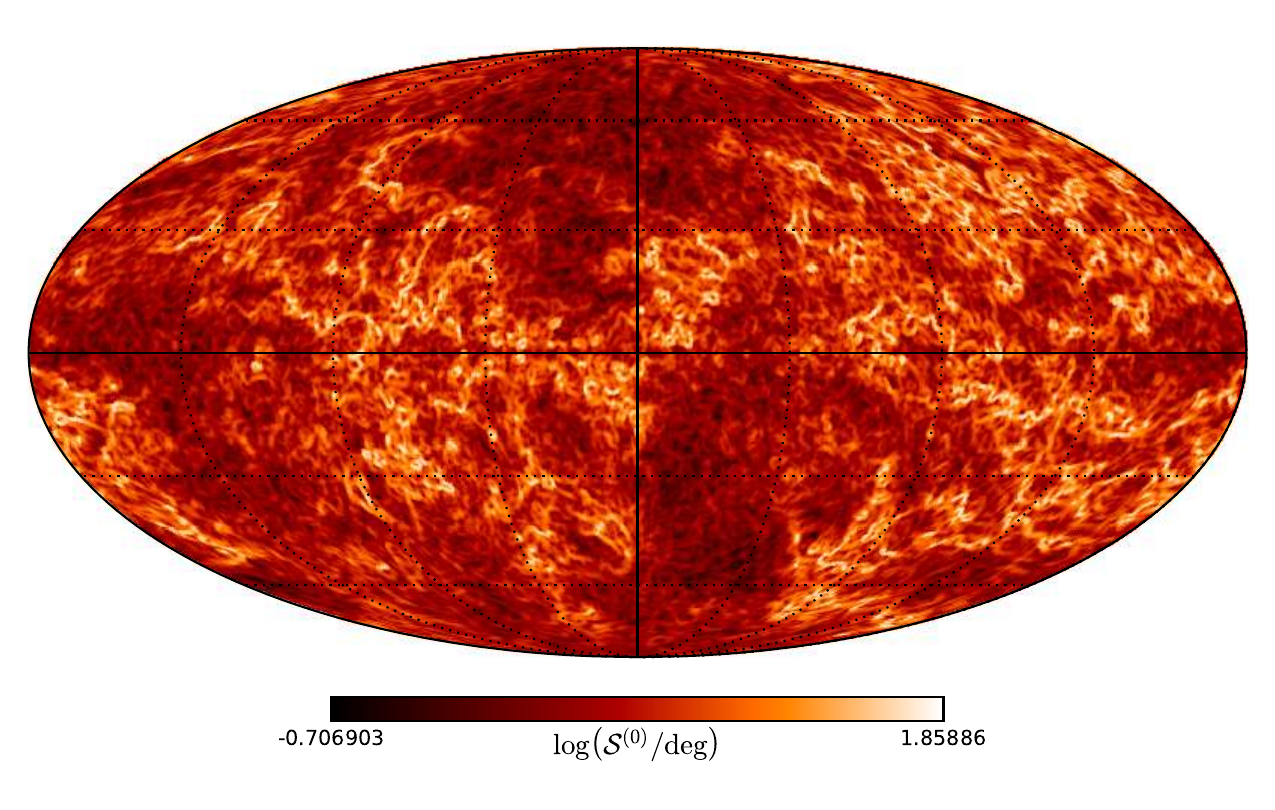}
\includegraphics[width=0.5\textwidth]{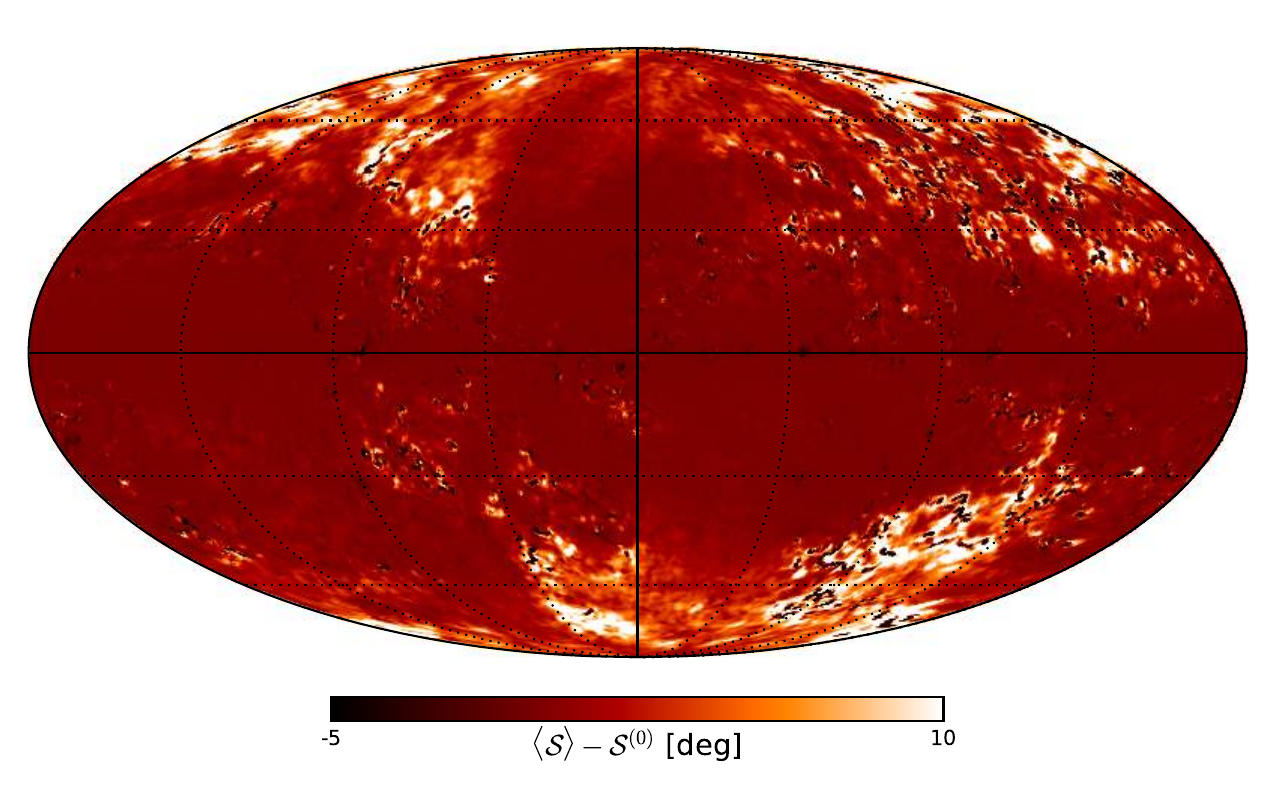}
\includegraphics[width=0.5\textwidth]{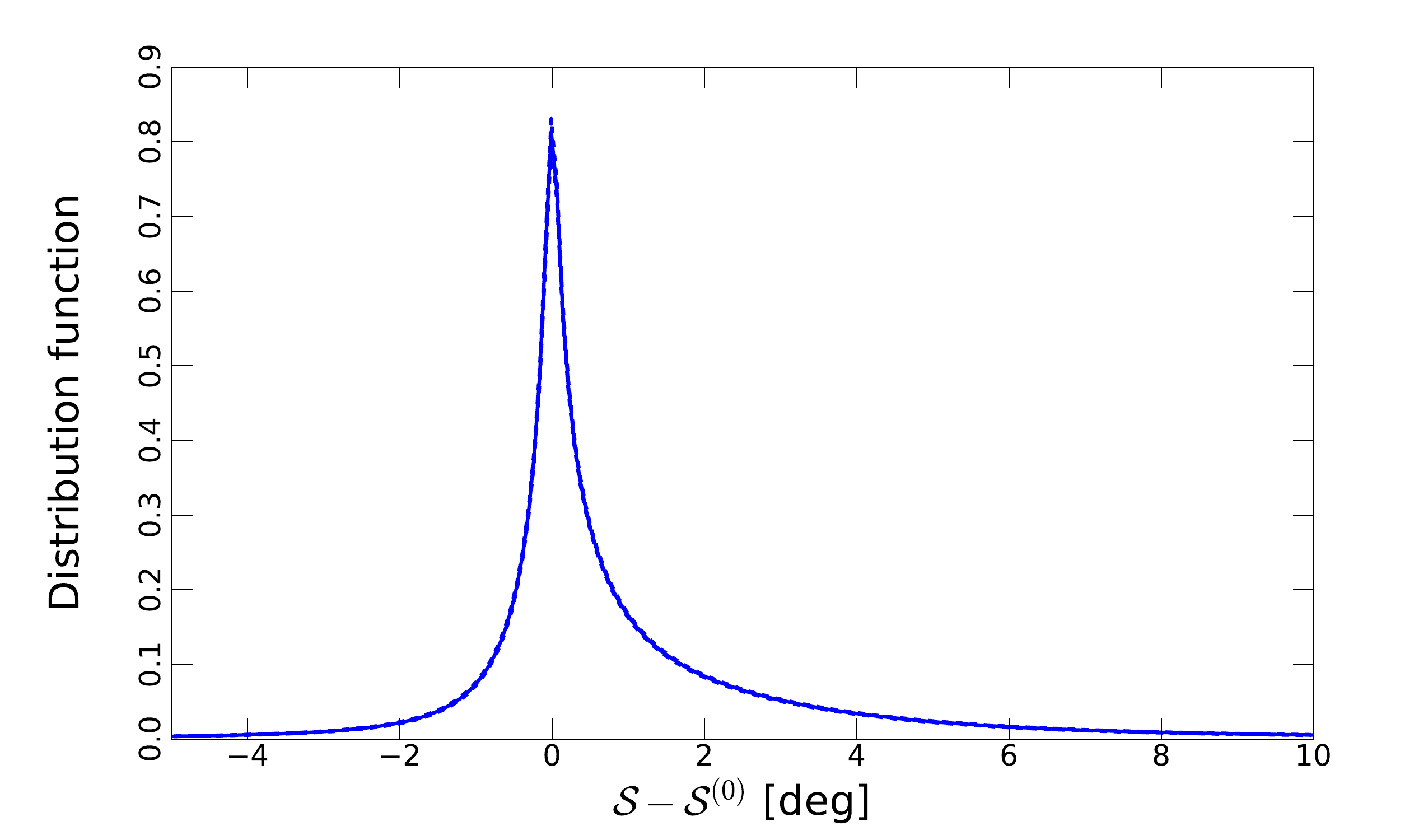}
\caption[]{{\it Top}: Map of the polarization angle dispersion function $\mathcal{S}^{(0)}$ for the input sky of the E2E simulations, at 60\arcmin\ resolution and a lag of $\delta=30\arcmin$. 
{\it Middle}: Map of the difference between the average $\langle\mathcal{S}\rangle$ and the input $\mathcal{S}^{(0)}$ in the E2E simulations.
{\it Bottom}: Distribution function of the difference between the output and input polarization angle dispersion function. The solid blue curve shows the average of 100 histograms of $\mathcal{S}-\mathcal{S}^{(0)}$ from the 100 simulations, with the $1\,\sigma$ dispersion shown as the blue area between dashed lines (barely visible).}
\label{fig:S_maps_E2E}
\end{figure}
 
The same diagnostics are run on the polarization angle dispersion function $\mathcal{S}$, which at 60\arcmin\ resolution we compute with a lag $\delta=30\arcmin$. Results are shown in Fig.~\ref{fig:S_maps_E2E}. We note that the average $\langle\mathcal{S}\rangle$ of the output $\mathcal{S}$ maps exhibits a significant positive bias with respect to the input $\mathcal{S}^{(0)}$ map, especially towards the Galactic poles. The distributions of pixel values in difference maps $\mathcal{S}-\mathcal{S}^{(0)}$ from the 100 simulations have a positive skewness. This shows that the polarization angle dispersion function is still affected by residual bias at this resolution, even though it is barely affecting the polarization angle map itself.

\begin{figure}
\includegraphics[width=0.5\textwidth]{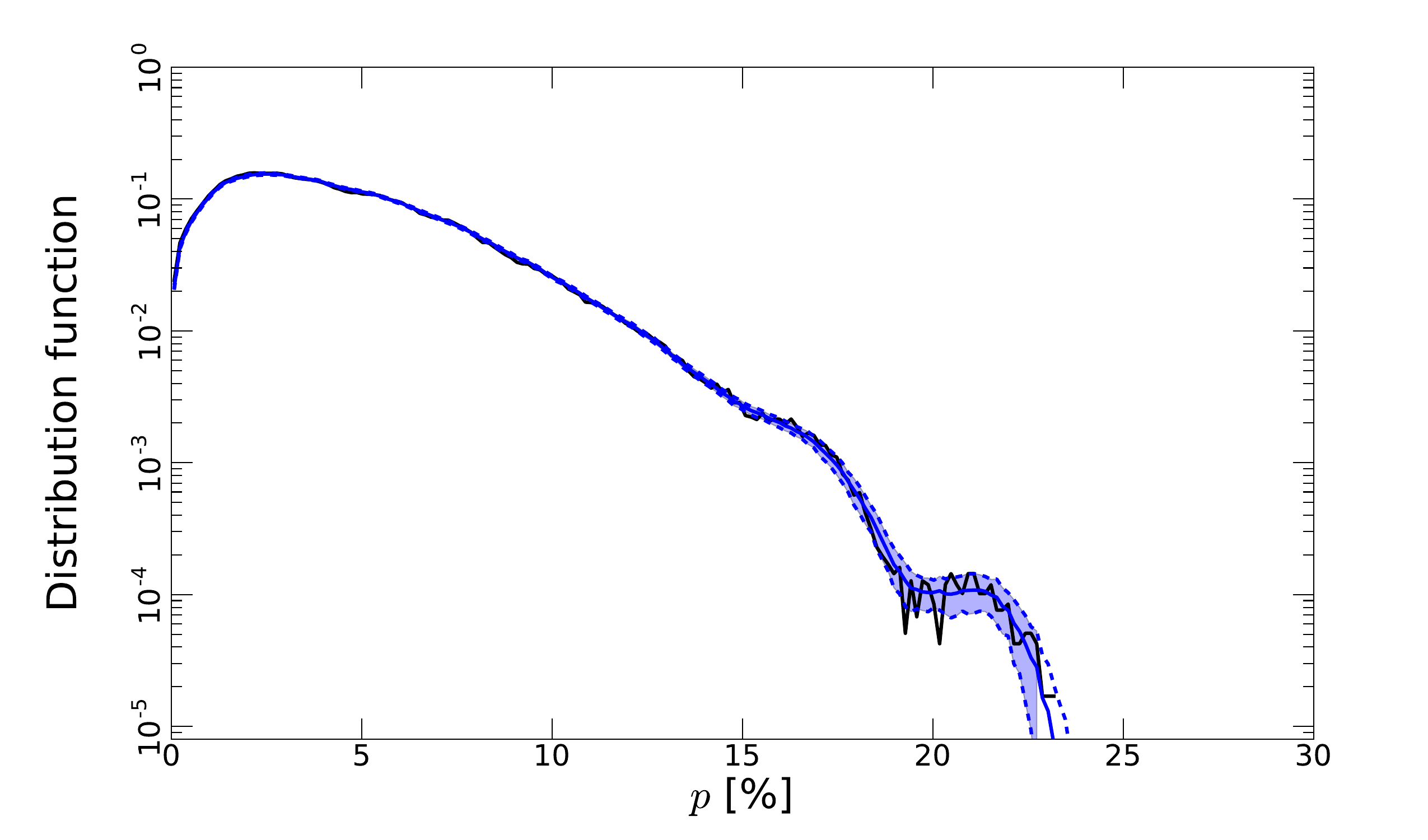}
\includegraphics[width=0.5\textwidth]{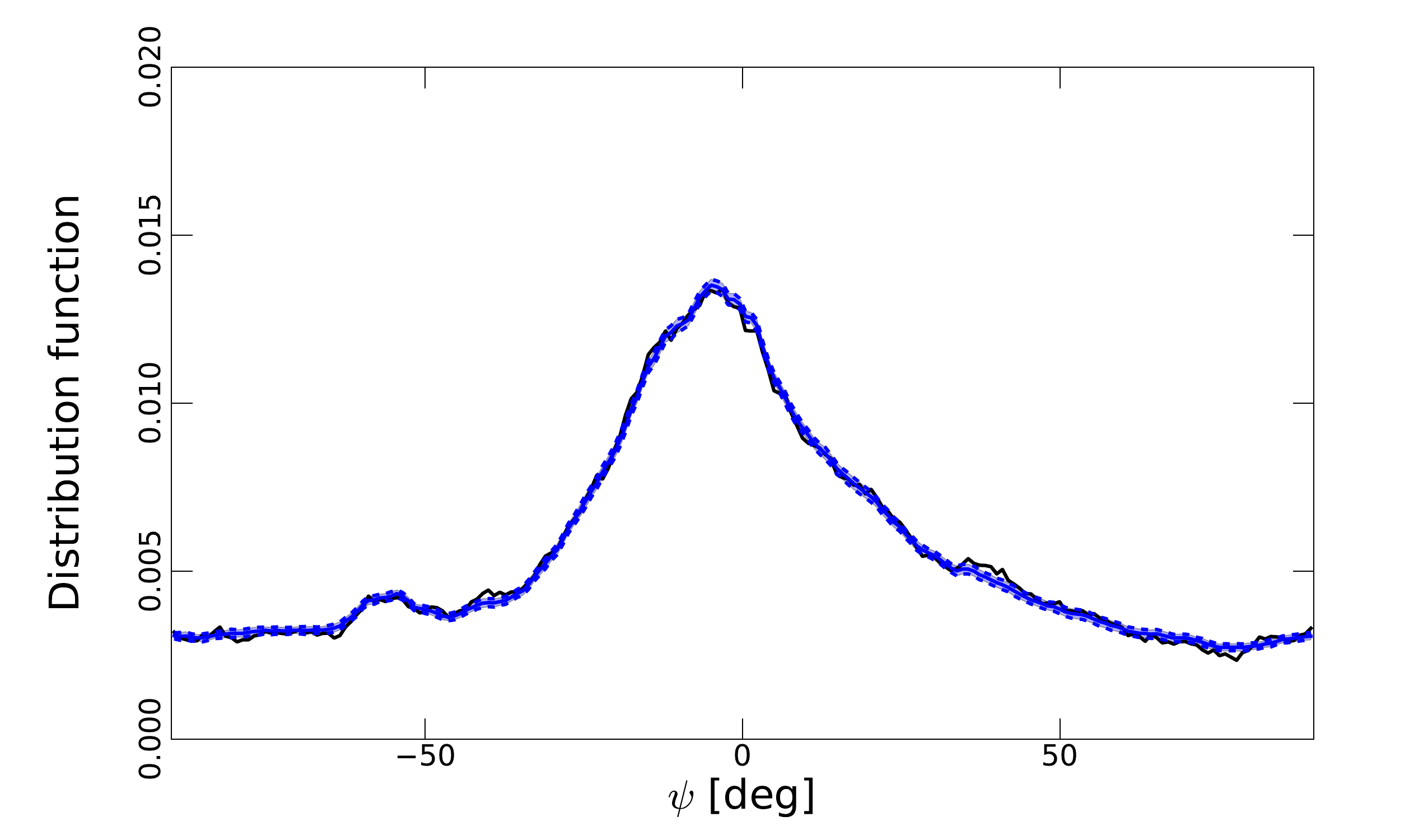}
\includegraphics[width=0.5\textwidth]{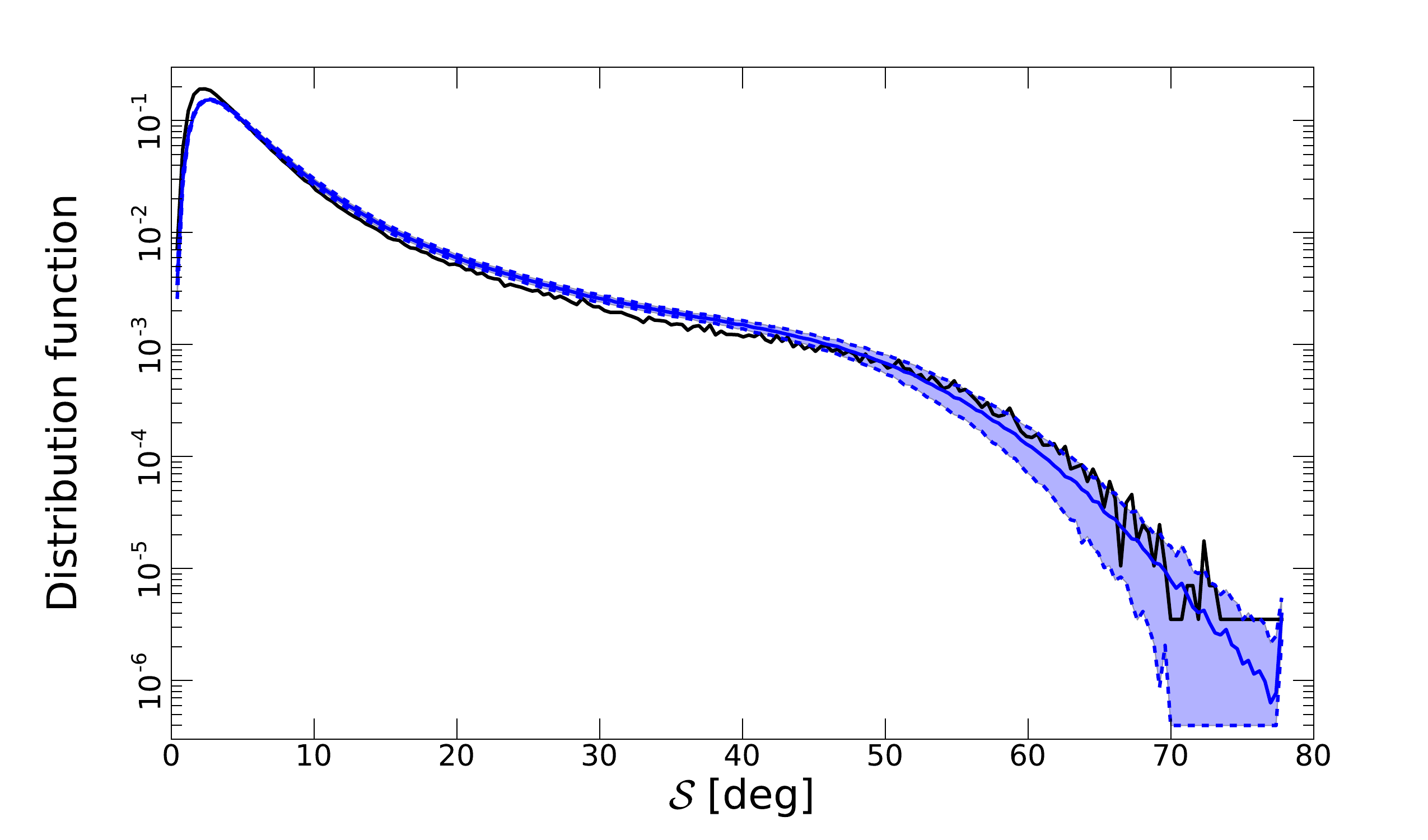}
\caption[]{Histogram of the polarization fractions (top), polarization angles (middle), and polarization angle dispersion functions (bottom).  The input data are shown by the black curves and the output of the E2E simulations by the solid blue curves (which are the average of 100 histograms from the 100 simulations), while the dashed lines with blue shading between show the $\pm1\,\sigma$ dispersion.
}
\label{fig:PI_psi_S_histograms_E2E}
\end{figure} 

For completeness, Fig.~\ref{fig:PI_psi_S_histograms_E2E} shows the histograms of the polarization fractions, polarization angles, and polarization angle dispersions for the input (black curve), and the outputs of the E2E simulations. For the latter, the blue curve on each panel shows the average histogram over the 100 simulations, with the $\pm1\,\sigma$ dispersion among histograms shown as the blue area between dashed lines. The agreement is excellent for both quantities $p$ and $\psi$, but for $\mathcal{S}$ we note that at intermediate values the positive bias already mentioned appears clearly. Finally, we stress that although the polarization fractions rarely go above 20\,\% for these simulated dust maps, this does not mean that the same range is expected in the \Planck\ data. 

\section{Link between $\mathcal{S}$ and the polarization gradients}
\label{sec:appendix:SdP}

\subsection{Analytical derivation}

The link between the polarization angle dispersion function $\S$ and the polarization angle gradient $|\nabla{\psi}|$ can be established analytically via a Taylor expansion of the polarization angle difference appearing in the definition of $\S$:
\begin{equation}
\S^2(\boldsymbol{r},\delta)=\left<\left[\psi\left(\boldsymbol{r}+\boldsymbol{\delta}\right)-\psi\left(\boldsymbol{r}\right)\right]^2\right>\approx\left<\left[\boldsymbol{\delta}.\boldsymbol{\nabla}\psi\right]^2\right> \, ,
\end{equation}
where the average is computed over the annulus centred on $\boldsymbol{r}$ having inner and outer radii $\delta/2$ and $3\delta/2$, respectively (Eq.~\eqref{eq:defS}), and $\boldsymbol{\nabla}\psi$ is the vector gradient of the polarization angle at the centre $\boldsymbol{r}$. Using a local reference frame with axes $y$ and $z$ in the plane of the sky, we can write the displacement vector as 
\begin{equation}
\boldsymbol{\delta}=l\left(\cos\theta\,\boldsymbol{e}_y+\sin\theta\,\boldsymbol{e}_z\right) \, ,
\end{equation} 
with $\delta/2\leqslant l\leqslant 3\delta/2$. The expression of $\mathcal{S}^2$ therefore becomes
\begin{equation}
\S^2(\boldsymbol{r},\delta)\approx\left<l^2\right>\left<\left(\cos\theta\,\frac{\partial\psi}{\partial y}+\sin\theta\,\frac{\partial\psi}{\partial z}\right)^2\right> \, ,
\end{equation}
where the spatial average takes into account that $l$ and $\theta$ are independent variables. The former simply yields
\begin{equation}
\left<l^2\right>=\frac{13}{12}\delta^2\approx\delta^2
\end{equation}
and the latter average is over $\theta\in[0,2\pi]$. In that average, taking the square and averaging over $\theta$ cancels the cross product, so that
\begin{equation}
\S^2(\boldsymbol{r},\delta)\approx\frac{\delta^2}{2}\left[\left(\frac{\partial\psi}{\partial y}\right)^2+\left(\frac{\partial\psi}{\partial z}\right)^2\right] \, .
\end{equation}
On the other hand, by defining the angular polarization gradient (Eq.~\eqref{eq:unitpolgrad}) for a unit polarization vector $Q/P = \cos(2\psi)$ and $U/P = \sin(2\psi)$, we have 
\begin{equation}
|\nabla{\psi}|=2\sqrt{\left(\frac{\partial\psi}{\partial y}\right)^2+\left(\frac{\partial\psi}{\partial z}\right)^2} \, ,
\end{equation}
which leads to the relation
\begin{equation}
\S(\boldsymbol{r},\delta)\approx\frac{\delta}{2\sqrt{2}}|\nabla{\psi}| \, .
\label{eq:sapprox}
\end{equation}

\subsection{The case of \Planck\ data}

\begin{figure*}
\includegraphics[width=0.5\textwidth]{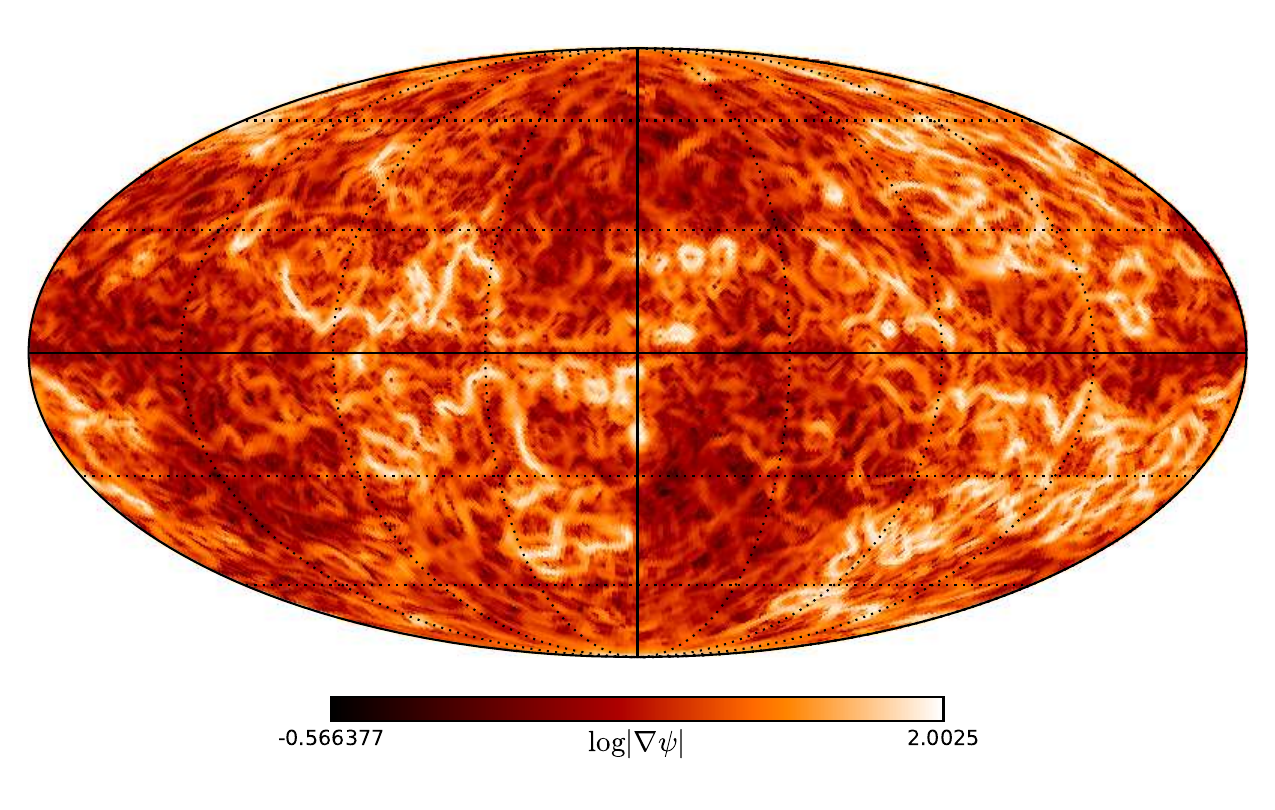}
\includegraphics[width=0.5\textwidth]{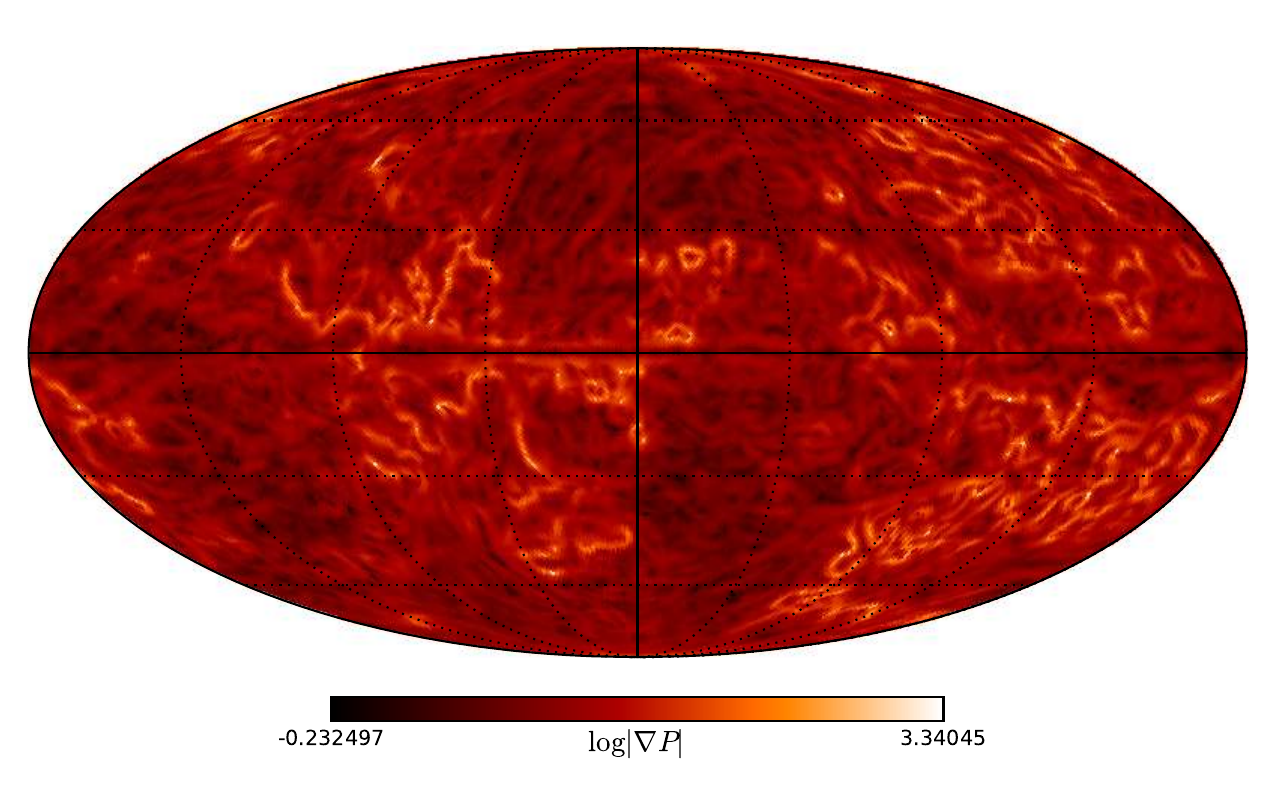}
\caption[]{Maps of the angular polarization gradient $|\nabla{\psi}|$ (left) and of the polarization gradient $|\nabla{P}|$ (right), built from the \GNILC-processed \Planck\ data at 353\,GHz and 160\arcmin\ resolution.
}
\label{fig:gradP_GNILC_80acm}
\end{figure*}

\begin{figure*}
\includegraphics[width=0.5\textwidth]{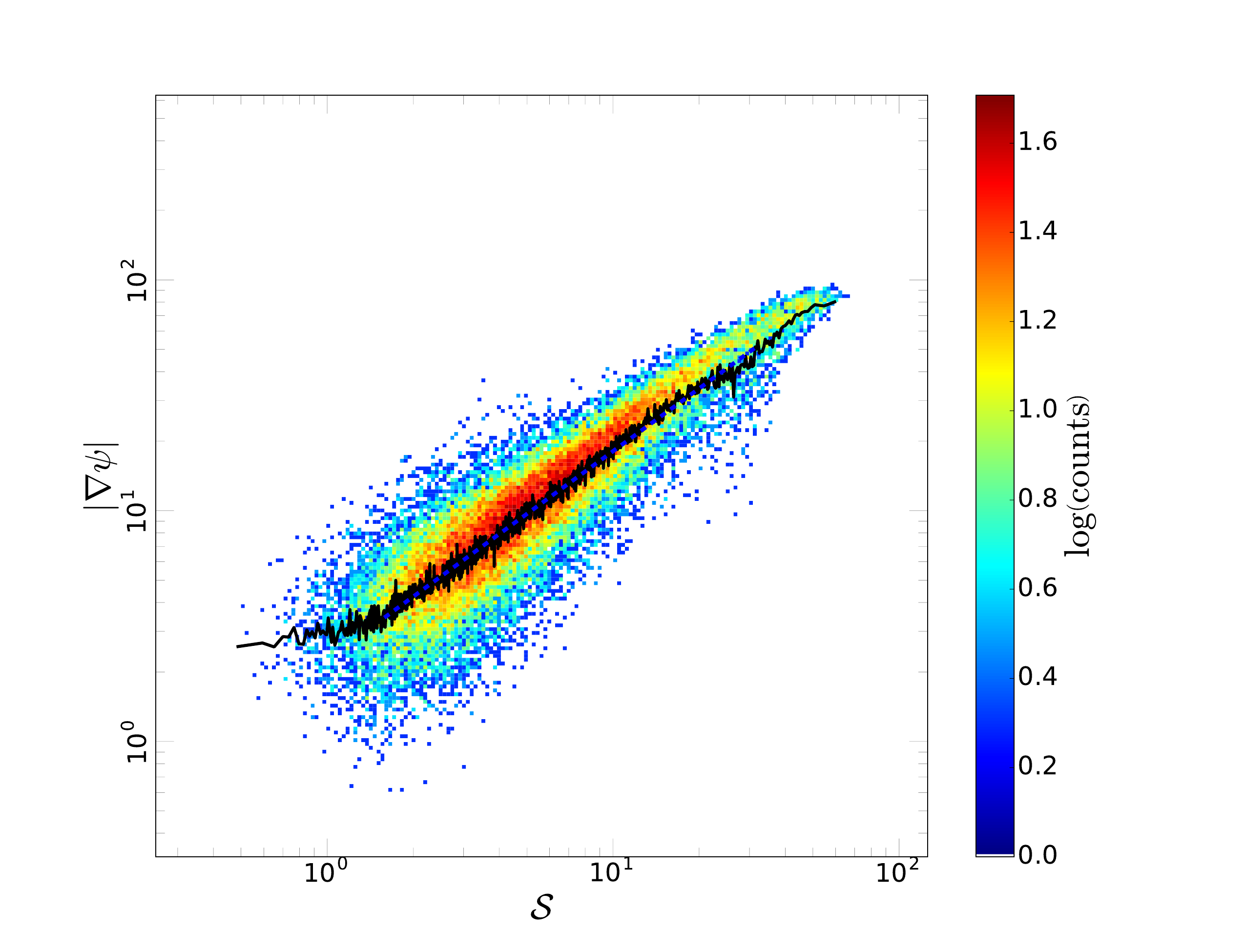}
\includegraphics[width=0.5\textwidth]{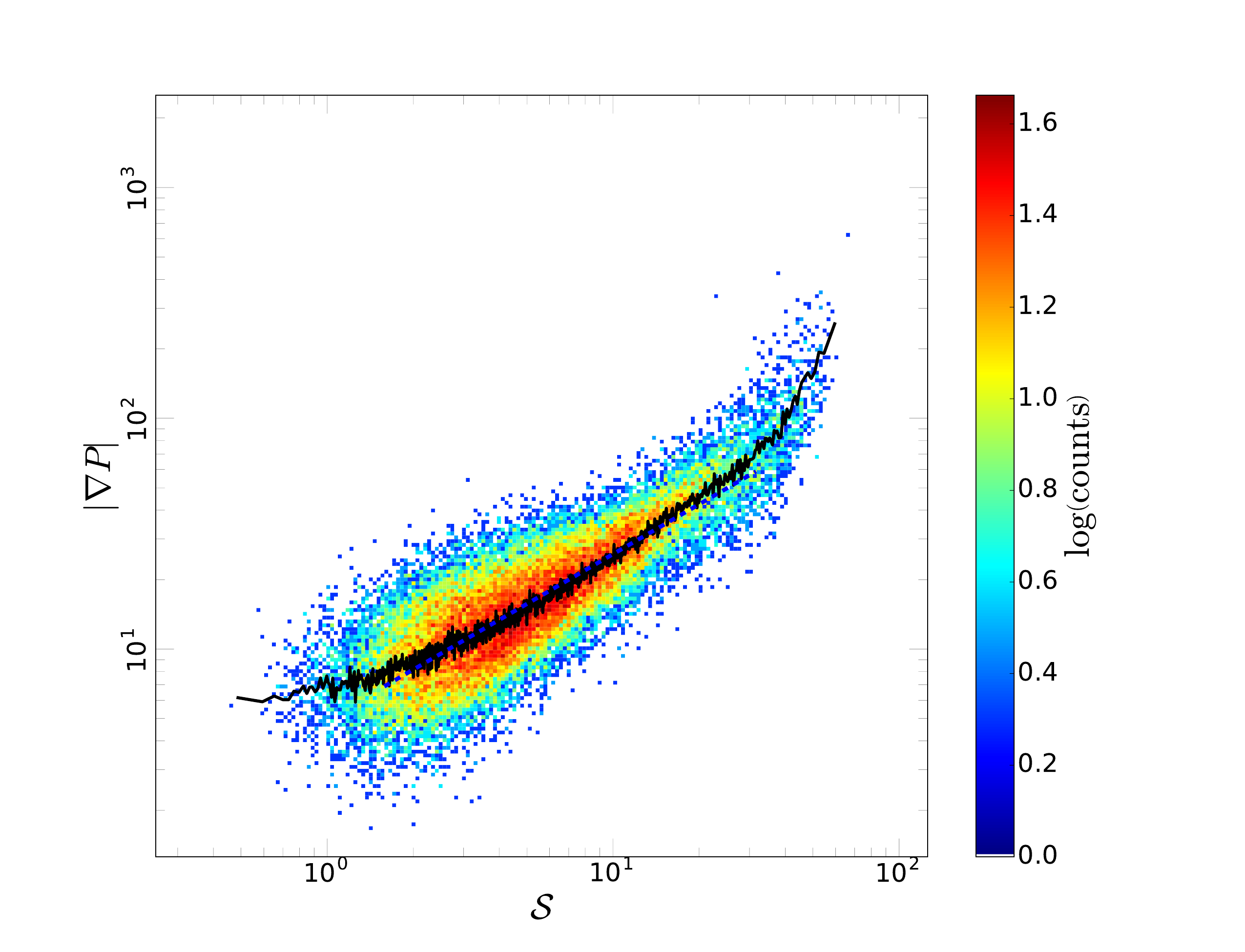}
\caption[]{
{\it Left}: Two-dimensional histogram representation of the correlation plot between the angular polarization gradient $|\nabla{\psi}|$ and the angular dispersion function $\mathcal{S}$ from
the \GNILC-processed \Planck\ data at 353\,GHz and 160\arcmin\ resolution, with a lag of 80\arcmin\ for $\mathcal{S}$.
{\it Right}: Correlation plot between the polarization gradient $|\nabla{P}|$ and the angular dispersion function $\mathcal{S}$. In both plots, the solid black curve shows the mean $|\nabla{\psi}|$ or $|\nabla{P}|$ in a given bin of $\mathcal{S}$.
}
\label{fig:S-gradP-GNILC-80arcmin}
\end{figure*}

Figure~\ref{fig:gradP_GNILC_80acm} shows the maps of both polarization gradients, $|\nabla{\psi}|$ and $|\nabla{P}|$ from Eqs.~\eqref{eq:unitpolgrad} and \eqref{eq:polgrad}, respectively,
for the \GNILC-processed \Planck\ data at 353\,GHz and 160\arcmin\ resolution. The correlations between $|\nabla{\psi}|$ and $\mathcal{S}$ on the one hand, and between $|\nabla{P}|$ and $\mathcal{S}$ on the other, are shown in Fig.~\ref{fig:S-gradP-GNILC-80arcmin} (for $\mathcal{S}$ a lag of 80\arcmin\ is used). These plots show that $\mathcal{S}$ correlates well with $|\nabla{\psi}|$, but not as well with $|\nabla{P}|$ and that $|\nabla{\psi}|$ is a very good proxy for the angular dispersion function $\mathcal{S}$, as would be expected from Eq.~\eqref{eq:sapprox} (and much faster to compute in practice). 

\section{Polarization fraction versus total gas column density for low and high offsets}
\label{sec:appendix:PsI-other-offsets}

\begin{figure}[htbp]
\includegraphics[width=0.55\textwidth,trim=50 0 0 50,clip=true]{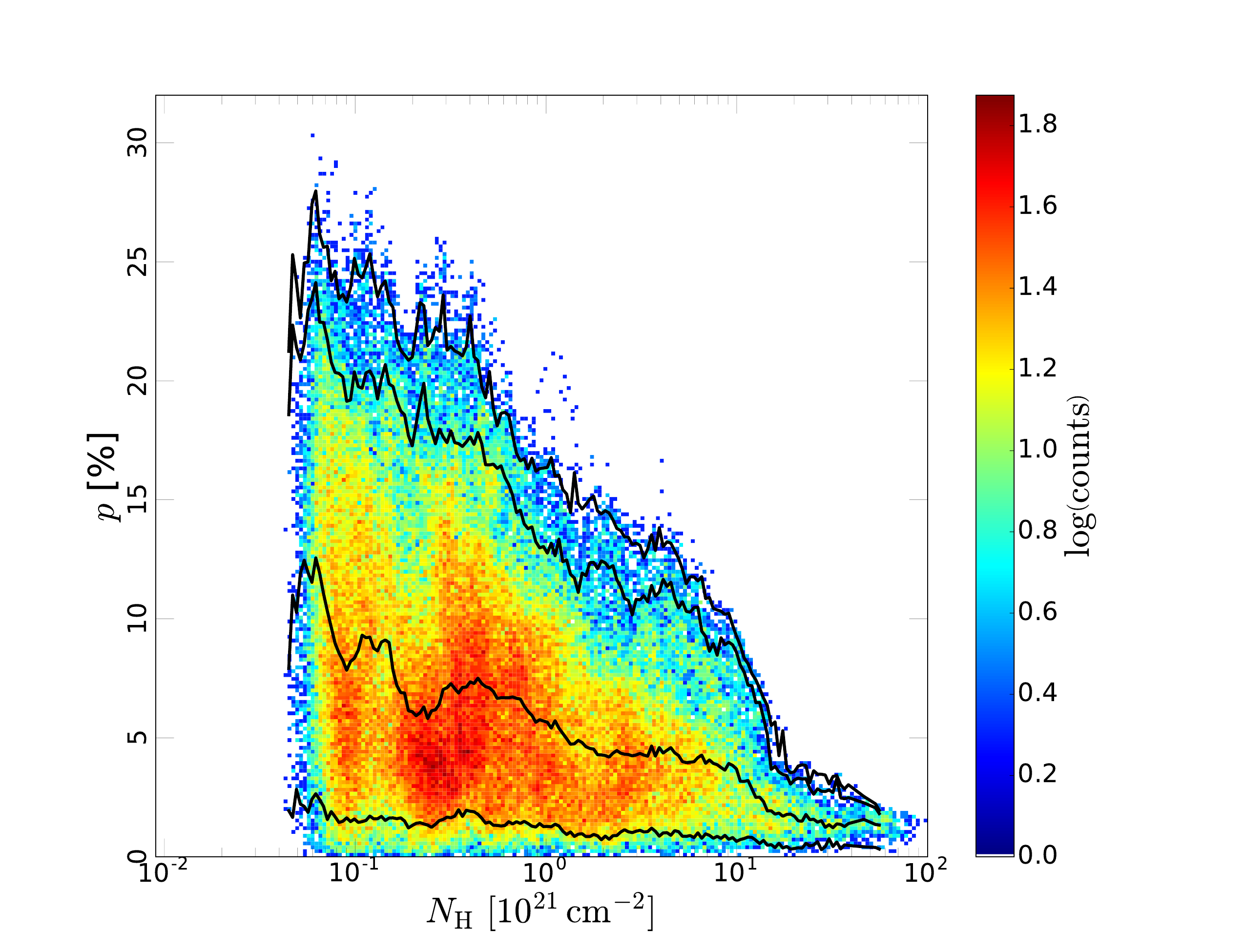}
\includegraphics[width=0.55\textwidth,trim=50 0 0 50,clip=true]{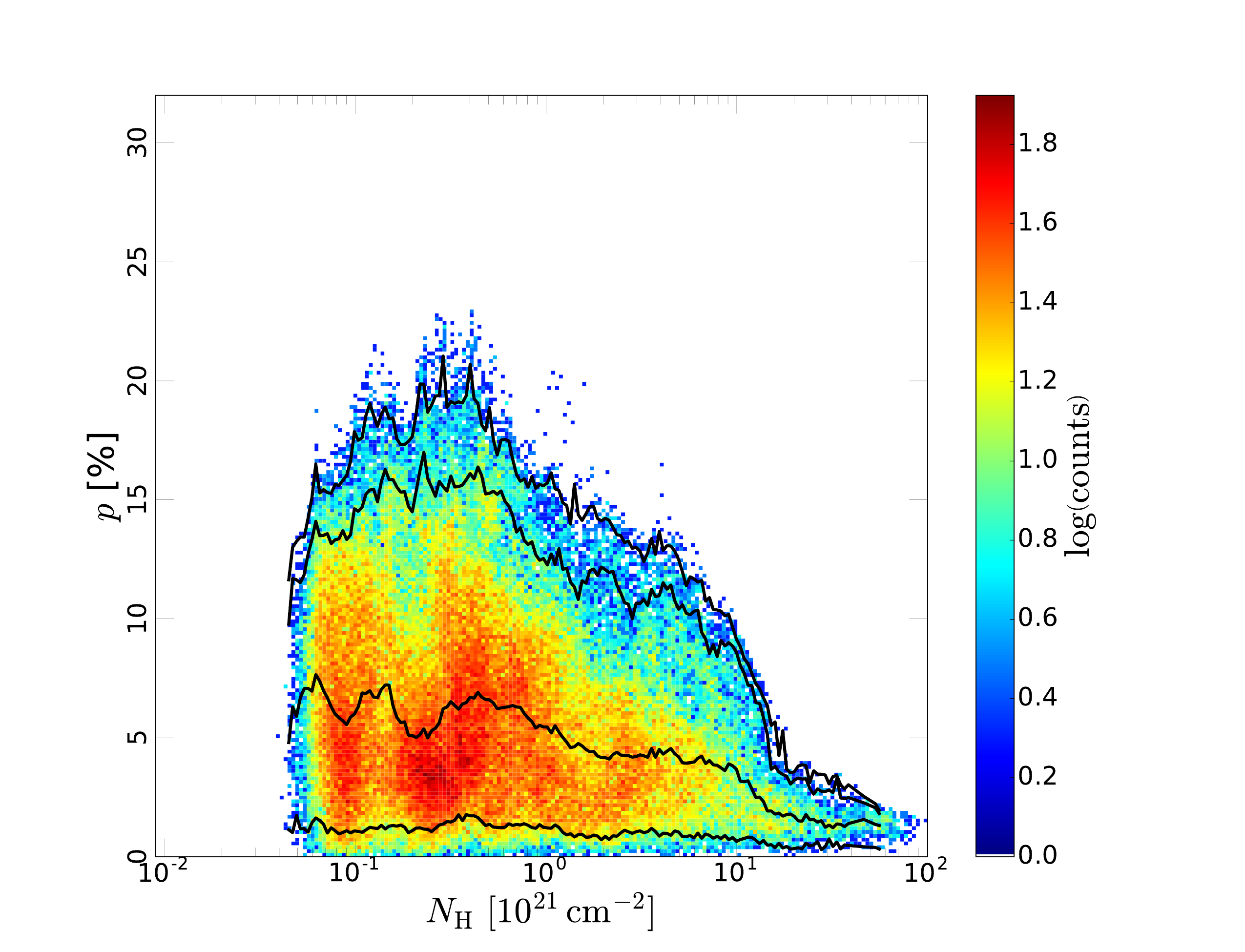}
\caption[]{
Two-dimensional histograms showing the joint distribution function of the polarization fraction $p$ from the \GNILC\ data (at 353\,GHz and uniform 80\arcmin\ resolution) and the total gas column density $\NH$. The top plot corresponds to the low total intensity offset, while the bottom plot corresponds to the high total intensity offset. The black lines show the 5th, 95th, and 99th percentiles of the $p$ distribution in each $\NH$ bin, as well as the median $p$ in each $\NH$ bin.
}
\label{fig:PI_I_GNILC_otheroffsets}
\end{figure}

In this appendix, we show in Fig.~\ref{fig:PI_I_GNILC_otheroffsets} plots similar to Fig.~\ref{fig:PI_I_GNILC}, but for the low and high total intensity offsets (Sect.~\ref{sec:I-offset}). The effects of the offset are discussed in Sect.~\ref{sec:pandI}.

\section{The inverse relationship between $\mathcal{S}$ and $p$}
\label{sec:appendix:Sp}

\begin{figure}
\includegraphics[width=\hhsize]{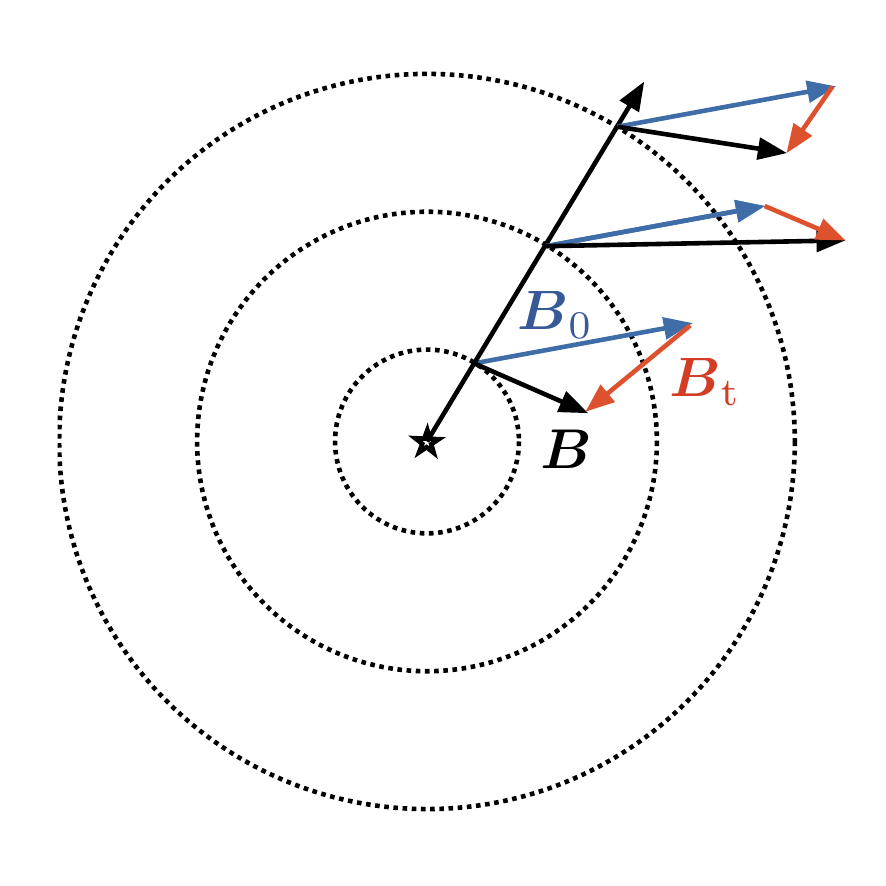}
\caption{Sketch of the phenomenological model of the dust polarized emission. The observer is represented by the central star, and the polarized emission is assumed to arise from a small number of layers (here $N=3$) in which the total magnetic field $\boldsymbol{B}=\boldsymbol{B}_0+\boldsymbol{B}_{\rm t}$ is the sum of a uniform field $\boldsymbol{B}_0$ and an isotropic turbulent field $\boldsymbol{B}_{\rm t}$ that is taken, in each layer, as a different realization of a Gaussian random field in three dimensions.}
\label{B-screens}
\end{figure}

In this appendix, we use a phenomenological model of the submillimetre polarized thermal dust emission, developed in~\citet{planck2016-XLIV}, \citet{ghosh-et-al-2017}, and \citet{vansyngel-et-al-2017}, to derive the relationship between the polarization fraction $p$ and the polarization angle dispersion function $\S$. In its most basic form presented in Fig.~\ref{B-screens}, this model assumes the polarized sky to result from a small set of $N$ concentric layers, each emitting a fraction $1/N$ of the total intensity,\footnote{The total intensity used in these models is the one observed by \Planck, because the focus is on modelling the polarization maps.} and harbouring a magnetic field $\boldsymbol{B}=\boldsymbol{B}_0+\boldsymbol{B}_{\rm t}$, where $\boldsymbol{B}_0$ is a uniform field (the same in each layer) and $\boldsymbol{B}_{\rm t}$ is an isotropic turbulent field that is taken, in each layer, as a different realization of a Gaussian random field in three dimensions. No effects of dust evolution or changes of intrinsic polarization properties of the dust grains are included in the model. By design, this model is able to reproduce the 1-point statistics of polarized thermal Galactic dust emission maps observed by \Planck, but it turns out that it is also able to reproduce the trend $\mathcal{S}\propto 1/p$ and the probability density function of $\PsIxS$, as we demonstrate below. 

\subsection{Reference frame and notations}

\begin{figure}
\includegraphics[width=\hhsize]{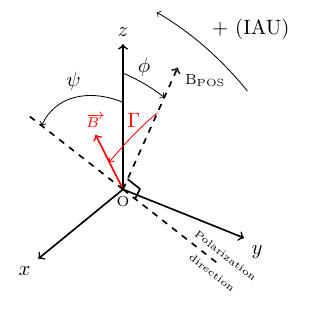}
\caption{Reference frame for our problem. $\Gamma \in [-\pi/2,\pi/2]$ is the inclination angle of the magnetic field vector $\boldsymbol{B}$ with respect to the plane of the sky $(yz)$, and $\phi \in [0,2\pi]$ is the angle, counted positively clockwise from the north, between the $z$ axis and the projection of the magnetic field vector onto the plane of the sky.
The polarization direction is also in the plane of the sky and perpendicular to that projection, making with the $z$ axis an angle $\psi = \phi - \pi/2~ [\pi] \in [-\pi/2,\pi/2]$. All constructions except $\boldsymbol{B}$ and $\Gamma$ are in the plane of the sky.}
\label{ref-frame}
\end{figure}

We use a reference frame defined in Fig.~\ref{ref-frame}. The $x$ axis is the line of sight, oriented towards the observer, and $Oyz$ is the plane of the sky. In that frame, the components of the large-scale magnetic field are ($\Bzerox, \Bzeroy,\Bzeroz$).

\subsection{Magnetic field in a layer at a given line of sight}

We begin by noting that, from one layer to the next, the different Gaussian realizations of the turbulent magnetic field can be taken to be independent. Therefore, in each layer $i$ (with $1\leqslant i\leqslant N$), we can write the components of the magnetic field $\Btot = \Bzero + \Bt$, at the position considered as the central pixel in the definition of $\S$ (Eq.~\eqref{eq:defS}), as 
\begin{eqnarray}
\Bx & = &\Bzerox + \fM\,B_0\,\Gx\,, \\
\By & = & \Bzeroy + \fM\,B_0\,\Gy\,, \\
\Bz & = & \Bzeroz+ \fM\,B_0\,\Gz\,. 
\end{eqnarray}
Here $\Gx$, $\Gy$, and $\Gz$ are Gaussian random variables with zero mean and variance $1/3$. The parameter 
\begin{equation}
\label{eq:appendix:fM}
\fM = \frac{\sigma_{\rm B}}{B_0}
\end{equation} 
is the ratio of the standard deviation $\sigma_B=\sqrt{\langle\Bt^2\rangle}$ of the turbulent magnetic field to the magnitude $B_0=||\Bzero||$ of the ordered field. The orientation of the magnetic field at the central pixel in layer $i$ is given by a set of two angles $\Gamma_i \in [-\pi/2,\pi/2]$ and $\phi_i \in [0,2\pi]$:
\begin{eqnarray}
\Bx & = & B_i\,\sin{\Gamma_i}\,; \\
\By & = & B_i\,\cos{\Gamma_i}\,\sin{\phi_i}\,; \\
\Bz & = & B_i\,\cos{\Gamma_i}\,\cos{\phi_i}\,.
\end{eqnarray}
As presented in Fig.~\ref{ref-frame}, the angle $\Gamma_i$ is the inclination angle of the vector $\Btot$ with respect to the plane of the sky, 
while $\phi_i$ is the angle, counted positively clockwise from the north, between the $z$ axis and the projection $\BPOS$ of $\Btot$ onto the plane of the sky.

\subsection{Fluctuations within each layer over the scale $\delta$}

When computing the polarization angle dispersion function $\S$, we introduce a specific scale, the lag $\delta$, which we always take as half the FWHM $\ellres$, so that $\delta=\ellres/2$. Presumably, the orientation of the magnetic field in each layer, i.e., the angles $\Gamma_i$ and $\phi_i$, vary little over these scales. Let us therefore consider a small Gaussian fluctuation $\dB$ around the direction of $\Btot$. The rms $\sigma_{B_i}(\delta)$ of this fluctuation can be cast into a parameter similar in form to Eq.~\eqref{eq:appendix:fM}, 
\begin{equation}\label{Eq-fmell}
\fm = \frac{\sigma_{B_i}(\delta)}{B_0}\ll 1\,,
\end{equation}
which depends on the lag $\delta$ considered, and is related to the overall turbulence parameter $f_{\rm M}$ and to the spectral index $\alpha_{\rm M}$ of the magnetic field.\footnote{Because the polarization angle dispersion function $\S$ involves an average over lags in $[\delta/2,3\delta/2]$, we note that $\fm$ actually stands for an average of the fluctuation ratio of the magnetic field over this range (see Appendix~\ref{subsec:appendix:fm}.)} This fluctuation $\dB$ corresponds to small variations for angles $\delta\Gamma_i$ and $\delta\phi_i$:
\begin{eqnarray}
\frac{\dBx}{B_0} & = & \phantom{-}\cos{\Gamma_i}\,\delta\Gamma_i= \fm\,\gx\,; \\
\frac{\dBy}{B_0} & = & -\sin{\Gamma_i}\sin{\phi_i}\,\delta\Gamma_i + \cos{\Gamma_i}\cos{\phi_i}\,\delta\phi_i = \fm\,\gy\,; \\
\frac{\dBz}{B_0} & = & -\sin{\Gamma_i}\cos{\phi_i}\,\delta\Gamma_i - \cos{\Gamma_i}\sin{\phi_i}\,\delta\phi_i = \fm\,\gz\,.
\end{eqnarray}
where $\gx$, $\gy$, and $\gz$ are Gaussian random variables with zero mean and variance 1/3. This allows us to compute the small variations of the angles as
\begin{eqnarray}
\label{eq:appendix:deltaGamma}
\delta\Gamma_i & = & \frac{\fm}{\cos{\Gamma_i}} \,\gx\,, \\
\label{eq:appendix:deltaPhi}
\delta\phi_i & = & \frac{\fm}{\cos{\Gamma_i}} \,\left(\gy\,\cos{\phi_i}-\gz\,\sin{\phi_i}\right)\,,
\end{eqnarray}
provided that the ratio $\fm/\cos{\Gamma_i}$ is still small, which only fails if the direction of the mean magnetic field is very close to the line of sight. 

\subsection{Polarization angle and Stokes parameters}\label{sec:PolStokes}

The angle $\phi_i$ is related to the polarization angle $\psi_i$, appearing in the definition of the Stokes parameters below, by a $\pi/2$ rotation, i.e., $\psi_i=\phi_i-\pi/2 \left[\pi\right]$. The $\pi$-ambiguity arises because the Stokes parameters are unchanged in the transformation $\BPOS \mapsto -\BPOS$. The polarization angle thus lies in the range $[-\pi/2,\pi/2]$, and the Stokes parameters ($I_i,Q_i,U_i)$ at the central pixel for each layer $i$ are then\footnote{In this appendix, for simplicity, we use a consistent convention (IAU or {\healpix}) for Stokes $U$ and polarization angles. The results do not depend on that choice.}
\begin{eqnarray}
\label{eq:appendix:Qi}
Q_i & = & \pmax \,\I_i\, \cos^2\Gamma_i\,\cos{2\psi_i} = P_i\,\cos{2\psi_i}\,,\\
\label{eq:appendix:Ui}
U_i & = & \pmax \,\I_i\, \cos^2\Gamma_i\,\sin{2\psi_i} = P_i\,\sin{2\psi_i}\,,
\end{eqnarray}
where $I_i$ and $P_i$ are the total and polarized intensity at the central pixel in layer $i$, respectively, and $\pmax$ is the polarization fraction of thermal dust emission that would be observed in the case of a uniform magnetic field parallel to the plane of the sky ($\Gamma_i=0$). A fluctuation $\dB$ of the magnetic field at the scale $\delta$ therefore produces a small variation of these Stokes parameters that can be written as
\begin{eqnarray}
\delta Q_i & = & -2\left(Q_i\,\tan{\Gamma_i}\,\delta\Gamma_i+U_i\,\delta\psi_i\right)\,,\\
\delta U_i & = & -2\left(U_i\,\tan{\Gamma_i}\,\delta\Gamma_i-Q_i\,\delta\psi_i\right)\,,
\end{eqnarray}
where it is assumed that the total intensity varies little on the scale $\delta$. Because we work with a lag smaller than the FWHM, $\delta=\ellres/2$, this is a reasonable assumption. The fluctuation of the polarization angle is $\delta\psi_i=\delta \phi_i$, and so by inserting Eqs.~\eqref{eq:appendix:deltaGamma} and \eqref{eq:appendix:deltaPhi} for the fluctuations of the angles we obtain
\begin{eqnarray}
\label{eq:appendix:deltaQi}
\delta Q_i & = & -\frac{2\fm}{\cos{\Gamma_i}}\left[Q_i\,\gx\tan{\Gamma_i}+U_i\,\left(\gy\cos{\phi_i} - \gz\sin{\phi_i}\right)\right], \\
\delta U_i & = &- \frac{2\fm}{\cos{\Gamma_i}}\left[U_i\,\gx\tan{\Gamma_i}-Q_i\,\left(\gy\cos{\phi_i} - \gz\sin{\phi_i}\right)\right].\label{eq:appendix:deltaUi}
\end{eqnarray}
These expressions will be helpful in determining the fluctuations of the Stokes parameters over which to average when computing the polarization angle dispersion function in the next section.

\subsection{Polarization angle dispersion function}

The polarization angle dispersion function $\S$ is computed for a central pixel $c$, and consists of an average over the $n$ pixels, indexed by $j$ (with $1\leqslant j\leqslant n$), in an annulus of mean radius $\delta=||\boldsymbol{\delta}||$ and width $\delta$ around the central pixel, as defined in Eq.~\eqref{eq:defS}. This can also be written in terms of the Stokes parameters $Q$ and $U$ at the central pixel, and $Q(j)$ and $U(j)$ at a pixel $j$ in the annulus~\citep{planck2014-XIX}:
\begin{eqnarray}
\S(\delta) & = & \sqrt{\frac{1}{n}\sum_{j=1}^n \left[\frac{1}{2}\arctan{\frac{Q(j)\,U- U(j)\,Q}{Q(j)\,Q + U(j)\,U}}\right]^2}\,.
\end{eqnarray}
Because we are interested in the average behaviour of $\S$, we will ultimately consider the mean of this expression over the position of the central pixel as well. 

The distribution function of $\S$ (Fig.~\ref{fig:PDF_S_GNILC_160arcmin}) shows that most pixels have a small dispersion of polarization angles, $\S\lesssim 10\deg$. For these values, it is safe to approximate the arctangent by its argument, so that we may write
\begin{eqnarray}
4\,\S^2(\delta) & = & \left\langle \left[\frac{Q(j) U - U(j)\,Q}{Q(j)\,Q + U(j)\,U}\right]^2\right\rangle_j\,.
\end{eqnarray}

The Stokes parameters at pixels $c$ and $j$ can be written as sums over the $N$ layers. More precisely, for the central pixel we have, by definition,
\begin{eqnarray}
Q=\sum_{i=1}^N Q_i,\\
U=\sum_{i=1}^N U_i,
\end{eqnarray}
while for the displaced pixel $j$ we can write
\begin{eqnarray}
\label{eq:appendix:Qj}
Q(j)=\sum_{i=1}^N \left[Q_i+\delta Q_i(j)\right]=Q+\delta Q(j)\,,\\
U(j)=\sum_{i=1}^N \left[U_i+\delta U_i(j)\right]=U+\delta U(j),\label{eq:appendix:Uj}
\end{eqnarray}
exhibiting the fluctuations of the Stokes parameters given in Eqs.~\eqref{eq:appendix:deltaQi} and \eqref{eq:appendix:deltaUi}. We use this decomposition to write the numerator and denominator that appear in the squared quantity above as
\begin{eqnarray}
Q(j)\,U- U(j)\,Q &=& U\,\delta Q(j)-Q\,\delta U(j)\,,\\
Q(j)\,Q+U(j)\,U & =& P^2+Q\,\delta Q(j)+U\,\delta U(j)\,.
\end{eqnarray}
In the latter expression, the second and third terms are most likely negligible compared to the polarized emission $P^2$ at the central pixel, especially when averaged over index $j$, and so they can be ignored. We therefore have
\begin{eqnarray}
\label{eq:appendix:A25}
4\,\S^2(\delta) & \approx & \frac{\left\langle\left[U\,\delta Q(j)-Q\,\delta U(j)\right]^2\right\rangle_j}{P^4}\,,
\end{eqnarray}
because $P$ is independent of the pixel $j$. Appearing in the numerator are the averages $\langle\delta Q(j)^2\rangle_j$, $\langle\delta U(j)^2\rangle_j$, and $\langle\delta Q(j)\delta U(j)\rangle_j$, which can be computed using the fact that for the above Gaussian random variables $\gx$, $\gy$, and $\gz$
\begin{eqnarray}
\left<\gx^2\right>_j=\left<\gy^2\right>_j=\left<\gz^2\right>_j=\frac{1}{3}\,,\\
\left<\gx\gy\right>_j=\left<\gx\gz\right>_j=\left<\gy\gz\right>_j=0\,.
 \end{eqnarray}
Because the random variables $g_x$, $g_y$, and $g_z$ are also uncorrelated from one layer to the next, we have
\begin{eqnarray}
\left<\delta Q(j)^2\right>_j&=& \sum_{i=1}^N\left<\delta Q_i(j)^2\right>_j\,,\\
\left<\delta U(j)^2\right>_j& =&\sum_{i=1}^N\left<\delta U_i(j)^2\right>_j\,,\\
\left<\delta Q(j)\,\delta U(j)\right>_j& =& \sum_{i=1}^N \left<\delta Q_i(j)\,\delta U_i(j)\right>_j\,, 
\end{eqnarray}
which yields, using the expressions of Eqs.~\eqref{eq:appendix:deltaQi} and \eqref{eq:appendix:deltaUi},
\begin{equation}
\left<\delta Q(j)^2\right>_j=\frac{4}{3}\fmsq\sum_{i=1}^N\frac{Q_i^2\tan^2{\Gamma_i}+U_i^2}{\cos^2{\Gamma_i}}\,,\label{Eq-deltaQ2}
\end{equation}
\begin{equation}
\left<\delta U(j)^2\right>_j=\frac{4}{3}\fmsq\sum_{i=1}^N\frac{U_i^2\tan^2{\Gamma_i}+Q_i^2}{\cos^2{\Gamma_i}}\,,\label{Eq-deltaU2}
\end{equation}
\begin{equation}
\left<\delta Q(j)\,\delta U(j)\right>_j=\frac{4}{3}\fmsq\sum_{i=1}^N\frac{Q_i\,U_i\left(\tan^2{\Gamma_i}-1\right)}{\cos^2{\Gamma_i}} \,.
\end{equation}
Combining the above expressions, we then have
\begin{align}
\S^2(\delta) = \frac{\fmsq}{3P^4}\sum_{i=1}^{N}\frac{\left(Q\,Q_i + U\,U_i\right)^2+\left(Q\,U_i-U\,Q_i\right)^2\tan^2{\Gamma_i}}{\cos^2{\Gamma_i}}\,. 
\end{align}
The combinations of $Q$, $U$, $Q_i$, and $U_i$ appearing in this expression can be cast into another form by introducing the angular shift $\Delta\psi_i = \psi_i-\psi$ between the polarization angle in each layer $\psi_i$ and the observed polarization angle $\psi$, both considered at the central pixel:
\begin{eqnarray}
Q\,Q_i + U\,U_i & = & P_i\,P\,\cos{2\Delta\psi_i}\,, \\
Q\,U_i - U\,Q_i & = & P_i\,P\,\sin{2\Delta\psi_i}\,.
\end{eqnarray}
We then have
\begin{eqnarray}
\S^2(\delta)= \frac{\fmsq}{3P^2}\sum_{i=1}^N\frac{P_i^2\left[\sin^2{2\Delta\psi_i}\tan^2{\Gamma_i}+\cos^2{2\Delta\psi_i}\right]}{\cos^2\Gamma_i}\,,
\end{eqnarray}
which can be simplified further, using $P_i^2=p_\mathrm{max}^2I_i^2\cos^4{\Gamma_i}$, to
\begin{align}
\S^2(\delta) = \frac{\fmsq\,\pmax^2}{3P^2}\sum_{i=1}^NI_i^2\Bigg(\sin^2{2\Delta\psi_i}\sin^2{\Gamma_i} 
+ \cos^2{2\Delta\psi_i}\cos^2\Gamma_i\Bigg)\,.
\end{align}
Furthermore, in our phenomenological model the total intensity is split equally among the $N$ layers, so that $I_i=I/N$.  Therefore,
\begin{align}
\S(\delta) = \frac{\fm}{\sqrt{3N}}\,\frac{\pmax}{P/I}\,\mathcal{A}\,,
\end{align}
where $\mathcal{A}$ is a geometrical factor that depends on the magnetic-field structure in the layers, with
\begin{equation}
\mathcal{A}^2 = \frac{1}{N}\sum_{i=1}^N \Bigg(\sin^2{2\Delta\psi_i}\sin^2{\Gamma_i} + \cos^2{2\Delta\psi_i}\cos^2\Gamma_i\Bigg)\,.
\end{equation}

\subsection{Application to {\Planck} data: the case for strong turbulence}

In line with our analysis of the data, we compute the mean of $\S(\delta)$ over those pixels that have the same polarization fraction $\PsI$. This gives
\begin{equation}
\left<\S(\delta)\right>_{p} = \frac{\fm}{\sqrt{3N}}\,\frac{\pmax}{p}\left<\mathcal{A}\right>_{p}\,.
\end{equation}

Application of the phenomenological model to the {\Planck} data in~\cite{planck2016-XLIV} shows that good fits are obtained for the parameters $p_\mathrm{max}=0.26$, $f_{\rm M}=0.9$, and $N=7$. This value of $f_{\rm M}$ implies rather strong turbulence, and therefore the angles $\Gamma_i$ and $\Delta\psi_i$ are uncorrelated, yielding $\left<\mathcal{A}\right>_p\approx1/\sqrt{2}$. In that case, the polarization angle dispersion function simply reads
\begin{equation}
\left<\S(\delta)\right>_p\approx \frac{\fm}{\sqrt{6N}}\,\frac{\,\pmax}{p}\,.
\label{Eq-Sp}
\end{equation}

We reach the important conclusion that the trend $\S\propto 1/p$ observed in the {\Planck} data can be reproduced as a generic behaviour that depends only on the statistical properties of the turbulent magnetic field, without invoking changes in properties of the dust or in its alignment with respect to the magnetic field.

We note that the typical value and dispersion of the product $\StimesPsI$ depend not only on the properties of the turbulence at the scale of the lag, via the $\fm$ parameter, and on the number of layers $N$, but also on the maximum polarization fraction $\pmax$ that the dust can produce. Estimates of the latter are quite sensitive to the uncertainty on the zero level of the total intensity, as discussed in the main text.

\begin{figure}
\includegraphics[width=\hhsize]{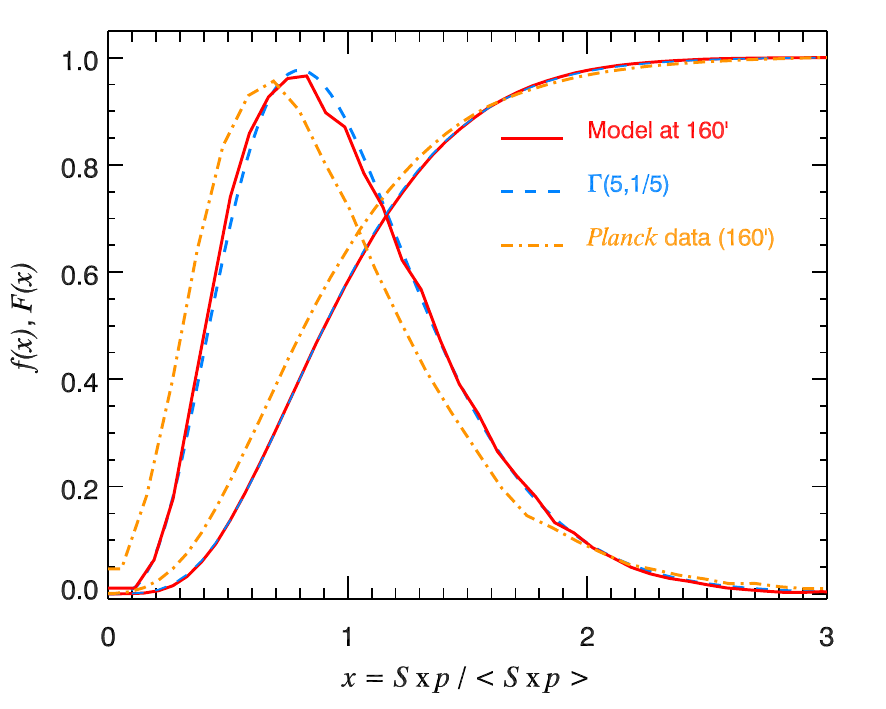}
\caption{
Probability density function (PDF) and cumulative density function (CDF) 
of various distribution functions. Shown are the mean-normalized product $\StimesPsI/\left\langle \StimesPsI \right\rangle$ for our model taken at a resolution of 160\arcmin\ (solid red curves), the same for \Planck\ data at 160\arcmin\ resolution (dot dashed orange curves), and a Gamma distribution with shape parameter $k=5$ and scale parameter $\theta=1/5$ (dashed blue curves).}
\label{Fig-pdf_Sxp}
\end{figure}

For completeness, Fig.~\ref{Fig-pdf_Sxp} presents the probability density function (PDF) and cumulative density function (CDF) 
of various distribution functions.
The solid curves correspond to the distribution of
the mean-normalized product $\StimesPsI/\left\langle \StimesPsI \right\rangle$
for our model taken at a resolution of 160\arcmin\ (models with up to 20\arcmin\ resolution are similar). Empirically, the corresponding density functions for a Gamma distribution~\citep{hazewinkel} with shape parameter $k=5$ and scale parameter $\theta=1/5$ reproduce these curves well, i.e., this Gamma distribution has similar statistics. These model and empirical density functions are also in reasonable agreement with the PDF and CDF of $\StimesPsI/\left\langle \StimesPsI \right\rangle$ for \Planck\ data at the same 160\arcmin\ resolution.

\subsection{Derivation of the expression for $f_{\rm m}(\delta)$}
\label{subsec:appendix:fm}
In our model, each component $B_{ix}$, $B_{iy}$, and $B_{iz}$ of the  magnetic field vector in layer $i$ is the sum of the uniform field and a realization of a Gaussian random variable on the sphere, with a power-spectrum $C_\ell = C\,\ell^{\,\alpha_{\rm M}}$, where $\ell$ is the multipole. As in \cite{planck2016-XLIV}, we consider that the non-vanishing modes of the turbulent component start at $\ell = 2$.

We normalize the turbulent component by imposing that
\begin{equation}
\sigma_B^2=\langle \delta B_{ix}^2\rangle+\langle \delta B_{iy}^2\rangle+\langle \delta B_{iz}^2\rangle=3\langle \delta B_{ix}^2\rangle=1\,,
\end{equation}
where $\delta B_{ix}=B_{ix}-B_{0x}$, and similarly for the other components. This in turn imposes that for the uniform magnetic field, from the definition of $\fM$ (Eq.~\eqref{eq:appendix:fM}),
\begin{equation}
B_0^2=\frac{1}{\fM^2}\,.
\end{equation}
Parseval's theorem then relates this normalization to the power spectrum by
\begin{equation}
\frac{1}{2\pi}\sum_{\ell=2}^{\infty} \left(\ell+\frac{1}{2}\right) C_\ell=\langle \delta B_{ix}^2\rangle=\frac{1}{3}\,.
\end{equation}

\begin{figure}
\includegraphics[width=\hhsize]{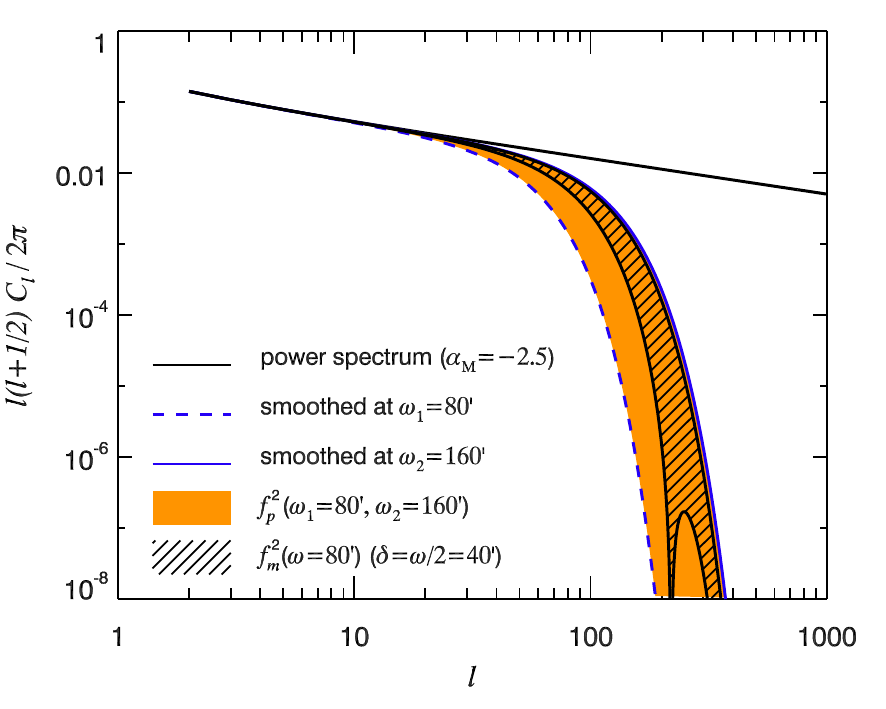}
\caption{Power spectrum $\ell(\ell + 1/2)\,C_\ell/(2\pi)$ of a turbulent component of index $\alpha_{\rm M}=-2.5$, as a function of the multipole $\ell$. The differential energy lost by smoothing the maps from an initial resolution $\ellres_1=80\arcmin$ to $\ellres_2=160\arcmin$ is filled in orange, representing a fraction $f_p^2(\ellres_1=80\arcmin, \ellres_2=160\arcmin)$ of the original power in the turbulent component (see text in Appendix~\ref{A-depolar} and Eq.~\eqref{Eq-fsqp}). Shown as hatched is the turbulent energy implied in the calculation of $\S$ at a resolution of $\ellres=80\arcmin$ (with $\delta=\ellres/2=40\arcmin$), which is a fraction $f^2_{\rm m}(\ellres=80\arcmin)$ of the original power in the turbulent component (see text in Appendix~\ref{subsec:appendix:fm} and Eq.~\eqref{Eq-fmdelta}). Both coloured and hatched regions scale with the resolution $\ellres$ as $\ellres^{-2-\alpha_{\rm M}}$.}
\label{Fig-depolar}
\end{figure}

The maps of $B_{ix}$, $B_{iy}$, and $B_{iz}$ are smoothed to a FWHM resolution $\ellres= 2 \sqrt{2 \log{2}} \,\sigma$, where $\sigma$ is the standard deviation of the smoothing circular Gaussian beam. This results in smoothed maps denoted $B_{ix,\ellres}$, $B_{iy,\ellres}$, and $B_{iz,\ellres}$. Through the Fourier transform, the total power in the turbulent part of each of these smoothed maps is:
\begin{equation}
\left\langle \delta B_{ix,\ellres}^2\right\rangle = \left\langle \delta B_{iy,\ellres}^2\right\rangle = \left\langle \delta B_{iz,\ellres}^2\right\rangle =
\frac{1}{2\pi}\sum_{\ell=2}^{\infty} \left(\ell+\frac{1}{2}\right) C_\ell \exp^{-\sigma^2 \ell^2}\, ,
\end{equation}
where $\delta B_{ix,\ellres}=B_{ix,\ellres}-B_{0x}$, and similarly for the other components.  The loss of power at large $\ell$ associated with the smoothing is clearly seen in Fig.~\ref{Fig-depolar}.

The factor $f_{\rm m}(\delta)$ appearing in the expression of $\StimesPsI$ (Eq.~\eqref{Eq-Sp}) is by definition (Eq.~\eqref{Eq-fmell}) the typical relative fluctuation of the magnetic field at those scales comprised in the annulus between $\delta/2$ and $3\delta/2$ (with $\delta=\ellres/2$), i.e., 
\begin{equation}
\fmsq = \frac{1}{B_0^2}\sum_{k=x,y,z}\left\langle\left(\delta B_{ik,\ellres}(\boldsymbol{r}+\boldsymbol{\delta'})-\delta B_{ik,\ellres}(\boldsymbol{r})\right)^2\right\rangle_{\delta/2 \le ||\boldsymbol{\delta'}|| \le 3\delta/2} \, .
\end{equation}
Through Parseval's theorem, and using the Fourier transform of this annulus of mean radius $\delta=||\boldsymbol{\delta}||$ and width $\delta$~\citep{Gautier92}, we find (Fig.~\ref{Fig-depolar})
\begin{equation}\label{Eq-fmdelta}
\fmsq = \frac{3}{2\pi}\ \fM^2\sum_{\ell=2}^{\infty} \left(\ell+\frac{1}{2}\right) C_\ell \exp^{-\sigma^2 \ell^2}\left[H^2\left(\ell\frac{\delta}{2}\right)-H^2\left(\ell\frac{3\delta}{2}\right)\right] \, ,
\end{equation}
where $H(x) = 2J_1(x)/x$, with $J_1$ the Bessel function of the first kind of order one, and $\delta$ expressed in radians. Using Eq.~\eqref{Eq-fmdelta}, we compute $\fmsq$ at a resolution $\ellres=80\arcmin$ (corresponding to $\delta=40\arcmin$) for $\alpha_{\rm M}=-2.5$. 
We find $\fmsq=0.0192\,\fM^2$, which corresponds graphically to the hatched region in Fig.~\ref{Fig-depolar}.

To determine the dependence of $f_m^2$ on the lag $\delta$, we simplify Eq.~\eqref{Eq-fmdelta} by considering a unique $\ell \sim 1/\delta$ and a constant $C_\ell = C_{1/\delta}$ in the sum, \ie\ we replace the smooth, Gaussian, cutoff by a step in $\ell$. This simplification, while being numerically incorrect, conserves the scaling of the integral with $\delta$ as long as $\ell \gg 1$.

Thus
\begin{equation}\label{Eq-scalingfm}
\fmsq \propto  \fM^2\,\left(\frac{2}{\delta}-\frac{2}{3\delta}\right)\left(\frac{1}{\delta}+\frac{1}{2}\right)C_{1/\delta} \,.\\
\end{equation}
We note that working with resolutions of 160\arcmin\ and less gives $1/\delta>40$. The $1/2$ term can therefore be neglected compared to $1/\delta$, yielding $\fmsq \propto \fM^2\,\delta^{-2-\alpha_{\rm M}}$. 

Recalling $\delta=\ellres/2$ to convert to $\ellres$ and renormalizing to 160\arcmin, this analysis yields the following scaling with resolution:
\begin{equation}\label{Eq-fS}
f_{\rm m}(\ellres) = 0.164 \,f_{\rm M}\, \left(\frac{\ellres}{160\arcmin}\right)^{-1-\alpha_{\rm M}/2} \, ,
\end{equation}
valid as long as $\alpha_{\rm M}$ does not depart too much from $-2.5$.

\subsection{Beam depolarization}\label{A-depolar}

In this section, we estimate the effect of the resolution on the polarization fraction by quantifying the depolarization that occurs within the beam. This is important not only for comparing our results at 80\arcmin\ and 160\arcmin\ but also for taking into account the effects of the difference in resolution between the \Planck\ polarization data and the starlight polarization that occurs within a pencil beam.

Following our approach in the previous section, we compute the difference in the total energy of the turbulent component, $f_p^2(\ellres_1,\ellres_2)$, between two given resolutions $\ellres_1$ and $\ellres_2>\ellres_1$, for a given line of sight. We have
\begin{equation}\label{Eq-fsqp}
f^2_p(\ellres_1,\ellres_2)=\frac{3}{2\pi}\ \fM^2 \sum_{\ell=2}^{\infty} \left(\ell+\frac{1}{2}\right) C_\ell \left(\exp^{-\sigma_1^2 \ell^2}-\exp^{-\sigma_2^2 \ell^2}\right) \,,
\end{equation}
where $\sigma_1$ and $\sigma_2$ are related to $\omega_1$ and $\omega_2$ by $\omega=2\sqrt{2\log{2}}\,\sigma$, as already mentioned.
We compute $f^2_p(\ellres_1,\ellres_2)$ for $\ellres_1=80\arcmin$, $\ellres_2=2\ellres_1 =160\arcmin$ and $\alpha_{\rm M}=-2.5$. We find $f^2_p(80\arcmin,160\arcmin) =0.058\,\fM^2$ (see also Fig.~\ref{Fig-depolar}). Following the same approach as for 
Eq.~\eqref{Eq-scalingfm}, 
this yields the following scaling with the resolution $\ellres$:
\begin{equation}
\label{Eq-fdepol}
f_p(\ellres,2\ellres)  =  0.285 \,f_{\rm M}\,\left(\frac{\ellres}{160\arcmin}\right)^{-1-\alpha_{\rm M}/2} \, .
\end{equation}
From Eqs.~\eqref{Eq-fS} and \eqref{Eq-fdepol}, we conclude that the factor $f^2_p(\ellres,2\ellres)/\fm^2(\ellres)$ is independent of $\fM$ and $\ellres$, and only depends on $\alpha_{\rm M}$. 
For $\alpha_{\rm M} = -2.5$, it is equal to 
$3.02$. 

We now study the effect of smoothing on the polarization fraction map. For that we note that Stokes $Q$ and $U$ exhibit a power spectrum similar to that of the turbulent component of the magnetic field at $\ell\gg 1$ (see Appendix~\ref{sec:cmp_analytic_numeric}). From Parseval's theorem, we therefore have, for each layer $i$
\begin{equation}
\left\langle\PsI^2_i\right\rangle_\ellres-\left\langle\PsI^2_i\right\rangle_{2\ellres}=k\pmax^2f^2_p(\ellres,2\ellres)\,,
\end{equation}
where the factor $f^2_p(\ellres,2\ellres)$ comes from the loss of power in the turbulent component of the field between the two resolutions (Fig.~\ref{Fig-depolar}) and $k$ is a constant to be determined numerically. The different random realizations $\delta \boldsymbol{B}_i$ are independent from one another for $\ell\geqslant 2$. At small enough spatial scales ($\ell\gg 1$), this ensures that the various realizations of $Q_i$ and $U_i$ are also independent from one another. We therefore have
\begin{equation}
\left\langle\PsI^2_\ellres\right\rangle-\left\langle\PsI^2_{2\ellres}\right\rangle=k\frac{\pmax^2}{N}f^2_p(\ellres,2\ellres)\,.
\end{equation}
A comparison with our numerical model shows that $k\simeq 1$ (Fig.~\ref{Fig-cmp_model_analytic} presents the agreement with the model for $\PsI$ when taking $k=1$), and therefore
\begin{equation}
\left\langle p^2_{\ellres} \right\rangle - \left\langle p^2_{2\ellres} \right\rangle=\frac{\pmax^2}{N}\,f^2_p(\ellres,2\ellres)\,.
\end{equation}
Using Eq.~\eqref{Eq-Sp}, this yields
\begin{equation}
\label{Eq-deltap2}
\left\langle p^2_{\ellres} \right\rangle - \left\langle p^2_{2\ellres}\right\rangle  = 
6 \frac{f^2_p(\ellres,2\ellres)}{\fm^2(\ellres)} \left\langle \StimesPsI\right \rangle^2_{\ellres} \, .
\end{equation}
Eq.~\eqref{Eq-deltap2} quantifies by how much the polarization fraction decreases when smoothing from resolution $\ellres$ to $2\ellres$, because on average (\textit{only}) $\left\langle p^2_{2\ellres}\right\rangle <\left\langle p^2_{\ellres} \right\rangle$. 

We can now generalize to the case of smoothing data from a finer resolution $\ellres/2^n$ to a resolution $\ellres$. In that case, we can compute the beam depolarization by a chain sum, 
\begin{equation}
\left\langle p^2_{\ellres/2^n} \right\rangle - \left\langle p^2_{\ellres} \right\rangle =  \sum_{i=0}^{n-1} \left[ \left\langle p^2_{\ellres/2^{i+1}} \right\rangle -\left\langle p^2_{\ellres/2^i} \right\rangle\right] \, ,
\end{equation}
where all averages are taken over the entire map.
This yields
\begin{eqnarray}
\left\langle p^2_{\ellres/2^n} \right\rangle - \left\langle p^2_{\ellres} \right\rangle & \approx & 6 \times 3.02 \sum_{i=0}^{n-1} \left\langle \StimesPsI\right \rangle^2_{\ellres/2^{i+1}} \,,\nonumber\\
 &\approx& 18.1 \left\langle \StimesPsI\right \rangle^2_{\ellres} \sum_{i=0}^{n-1} \left(2^{i+1}\right)^{2+\alpha_{\rm M}}\,, \nonumber\\
 &\approx& 18.1 \times \frac{1-2^{n\left(2+\alpha_{\rm M}\right)}}{2^{-2-\alpha_{\rm M}}-1} \left\langle \StimesPsI\right \rangle^2_{\ellres} \,.
\end{eqnarray}

In summary, the change in squared polarization fraction from the coarser scale $\ellres$ to the finer resolution $\ellres/2^n$ is
\begin{equation}\label{Eq-depol_omega}
\left\langle p^2_{\ellres/2^n} \right\rangle \approx  \left\langle p^2_{\ellres} \right\rangle + 18.1\times \frac{1-2^{n\left(2+\alpha_{\rm M}\right)}}{2^{-2-\alpha_{\rm M}}-1} \left\langle \StimesPsI\right \rangle^2_{\ellres} \, .
\end{equation}
In the pencil-beam limit ($n=\infty$), we have
\begin{equation}
\delta p^2_{\rm beam} \equiv \left\langle p^2_{\rm pencil}\right\rangle - \left\langle p^2_{\ellres} \right\rangle \simeq \frac{18.1}{2^{-2-\alpha_{\rm M}}-1} \left\langle \StimesPsI\right \rangle^2_{\ellres} 
\label{Eq-pencil}
\end{equation}
From symmetry arguments, we also have 
\begin{equation}
\delta \left(Q/I\right)^2_{\rm beam} \equiv   \left\langle \left[\left(Q/I\right)_{\rm pencil} - (Q/I)_{\ellres} \right]^2 \right\rangle_{\omega} \simeq 0.5\,\delta p^2_{\rm beam} 
\end{equation}
and the same for $U/I$.

\subsection{Comparison of the analytical expressions with numerical results and application to pencil beams}
\label{sec:cmp_analytic_numeric}

\begin{figure}
\includegraphics[width=\hhsize]{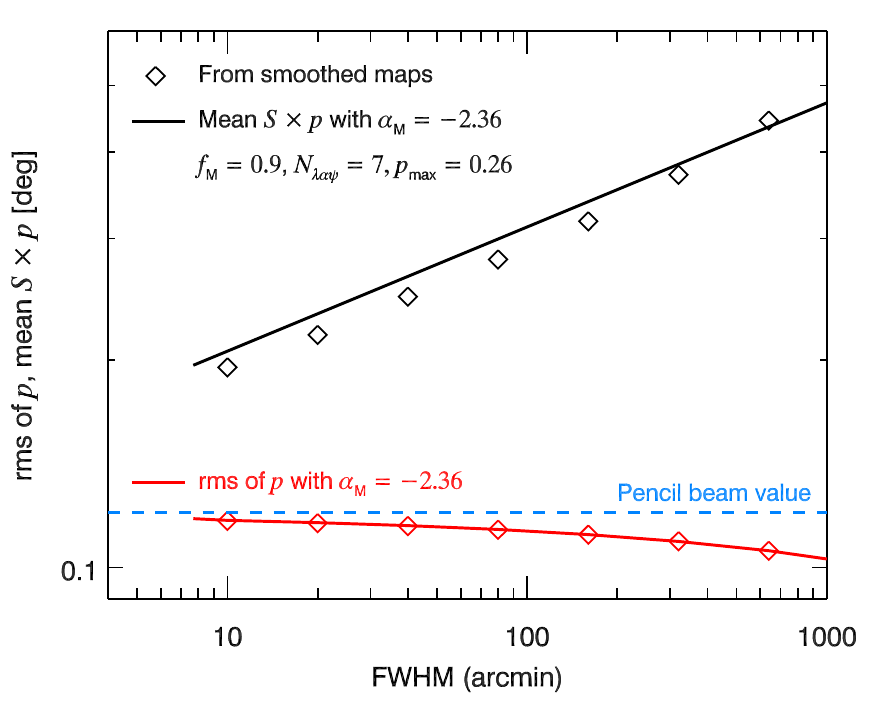}
\caption{Comparison between numerical results based on smoothed maps of our model of the turbulent magnetic field (diamonds) and the application of our analytical expressions for the decrease in the rms of $\PsI$ (red) by depolarization (Eq.~\eqref{Eq-depol_omega} with $\omega=160\arcmin$) and the increase in $\PsIxS$ (black) with the resolution (Eqs.~\eqref{Eq-fS} and \eqref{Eq-Sp}).
The fractional difference is less than 10\,\% for $\PsIxS$. 
The dashed blue line represents the pencil beam value for the rms of $\PsI$, as calculated from Eq.~\eqref{Eq-pencil} based on the model taken at $\omega=160\arcmin$.}
\label{Fig-cmp_model_analytic}
\end{figure}

In Fig.~\ref{Fig-cmp_model_analytic}, we compare our analytical expressions for the mean $\PsIxS$ (Eqs.~\eqref{Eq-fS} and \eqref{Eq-Sp}) and the rms of $p$ (Eq.~\eqref{Eq-depol_omega}) as a function of the resolution, with numerical results directly computed from the smoothed Stokes maps of our simulated model from~\cite{planck2016-XLIV}, i.e., $\fM=0.9$, $\alpha_{\rm M}=-2.5$, $N=7$ and $p_0=26\,\%$.  There are two aspects of the comparison for the analytical model, the normalization and the dependence on resolution.

For $\PsIxS$, we observe a slight normalization difference (7\,\%) between the analytical and numerical results. More precisely, at 160\arcmin, we find $\PsIxS = 0\textrm{\pdeg}34$ for the analytical expression and $0\textrm{\pdeg}32$ for the simulation. Both are nevertheless very close to the observational value of $0\textrm{\pdeg}31$. A deviation from the trend predicted by Eq.~\eqref{Eq-fS} is observed at low resolutions, as expected from our demonstration in Appendix~\ref{subsec:appendix:fm}.
Concerning the beam depolarization, the decrease in the polarization fraction with a decreasing resolution (larger $\ellres$) is well reproduced over two orders of magnitude in resolution: the example shown corresponds to a mean rms of $\PsI$ of 11\,\% over the full sky. Nevertheless, 
\cite{vansyngel-et-al-2017} already noted a small (approximately 0.1) difference between the index $\alpha_{\rm M}$ characterizing the power spectrum of the turbulent component of the magnetic field in the simulation, and the index $\alpha_{EE}$ and $\alpha_{BB}$ recovered from the analysis of the $EE$ and $BB$ power spectra. 
This is also what we find here: the scaling of $f_\mathrm{m}$ and $f_p$ with $\ellres$ is actually closer to $\ellres^{0.18}$, which would correspond to $\alpha_{\rm M}=-2.36$, when the model is produced with $\alpha_{\rm M}=-2.5$. We show this scaling as the solid lines in Fig.~\ref{Fig-cmp_model_analytic} and this is why in the rest of the paper we consider a scaling 
\begin{equation}
\fm \propto  \delta^{0.18} \propto  \ellres^{0.18}\, .
\end{equation}

The case of a pencil beam is of interest. Applying Eq.~\eqref{Eq-pencil} to the highest polarization fraction observed at 160\arcmin, $\pmax(\ellres$=160$\arcmin) \approx 20\,\%$, 
we can estimate the corresponding rms polarization over that scale in pencil-beam data, which corresponds to $n=\infty$ and for which the pre-factor in Eq.~\eqref{Eq-pencil} is 63.9 for $\alpha_{\rm M} = -2.36$. Expressing the observed $\langle \StimesPsI\rangle_{160\arcmin} = 0\textrm{\pdeg}31$ in radians, we find an rms $\pmax(\ellres$=0$\arcmin)=20.5\,\%$. In that particular case, depolarization is expected to be very small because when $p=\pmax$, the magnetic field is already ordered and within the plane of the sky and therefore little subject to depolarization. 
However, the effect would be stronger for lines of sight characterized by a low polarization fraction.  For example, if $p(\ellres$=160$\arcmin) = 6.0\,\%$ at 160\arcmin\ resolution, Eq.~\eqref{Eq-pencil} gives an rms polarization fraction $p(\ellres$=0$\arcmin)= 7.4\,\%$ over that scale for pencil-beam data.

\section{Noise and systematics in the comparison of submillimetre and optical polarization data}
\label{sec:appendix:nands}

For our estimation of the emission-to-extinction polarization ratios $\RPp$ and $\Rsv$ (Sect.~\ref{sec:stars}), all observations should ideally be done in a pencil beam and probe the full line of sight through the Galactic dust. Unavoidable departures from this ideal situation therefore introduce systematic effects on the quantities appearing in these ratios. In this appendix, we estimate these various systematic effects.

\subsection{Beam depolarization} 
\label{sec:beamdepol}

Systematic distortions of the submillimetre polarization signal occur due to the averaging of Stokes $\Qsub$ and $\Usub$ in the telescope beam. This does not happen with the pencil beam of optical observations. In Appendix~\ref{A-depolar} we demonstrate that beam depolarization produces a negative bias in $\PsI$ with respect to the pencil-beam value, with a scatter around this biased value (Eq.~\eqref{Eq-pencil}). For $\alpha_M=-2.36$, we find $\delta p^2_{\rm beam}\simeq63.9  \left\langle \StimesPsI\right \rangle^2_{\ellres}$ (Appendix~\ref{sec:cmp_analytic_numeric}).
As a consequence, to compare optical and submillimetre polarization data, we compensate for this systematic beam depolarization by multiplying all \Planck\ Stokes parameters and uncertainties ($P$, $\PsI$, $\sigma_P$, $\sigma_{\PsI}$) by the factor $\sqrt{1+\delta p^2_{\rm beam}/p^2}$. Because $Q$ and $U$ play a symmetric role in $P^2~=~ Q^2+U^2$, the same correction factor is applied to $\Qsub$, $\Usub$, $\Qsub/\Isub$, and $\Usub/\Isub$.

\subsection{Background distortion}
\label{sec:bkgd}

As mentioned in Sect.~\ref{sec:stars_pol}, the optical polarization degree $\pv$ is potentially biased by the difference in length probed along the line of sight, compared to the polarized emission in the submillimetre. Under the assumption of a uniform ISM, this bias due to the background may be corrected using Eq.~\eqref{eq:pvinfty_uniform}. Recognizing that the background is not uniform, we estimate that the uncertainty on $\pv^\infty$ (the ideal measure from the infinity) is proportional to the amount of reddening behind the star $\Delta  E(B-V)^{\infty,\star} = \ebvtotal - \ebvstar$ and to the rms of the polarization degree per unit reddening in the background, i.e.,
\begin{equation}
\sigma_{\pv}^\infty \simeq \Delta  E(B-V)^{\infty,\star}\times{\rm rms} \left(\frac{\pv}{\ebv}\right)_{\rm bkgd}\,.
\label{Eq:sigp_bkgd}
\end{equation} 
This equation replaces Eq.~\eqref{eq:dpvinfty_uniform} when the background is not uniform. We discuss the appropriate value of the last factor in Appendix~\ref{sec:Stddev}.

For the polarization fraction $\pv/\tauv$, no renormalization is needed as both $\pv$ and $\tauv$ are measured from the star. The systematic uncertainty on $\pv/\tauv$ from background distortion in then simply taken as:
\begin{equation}
\sigma\left(\frac{\pv}{\tauv}\right) = \frac{\sigma_{\pv}^\infty}{\tauv}\,.
\label{Eq:sigpst_bkgd}
\end{equation}
These equations also apply respectively to uncertainties on $\qv^\infty$ and $\uv^\infty$~(Eq.~\eqref{Eq:sigp_bkgd}), and $\qv/\tauv$ and $\uv/\tauv$~(Eq.~\eqref{Eq:sigpst_bkgd}).

\subsection{Uncertainties related to the reddening maps}
\label{sec:uncertainreddening}

\begin{figure}
\includegraphics[width=\hsize]{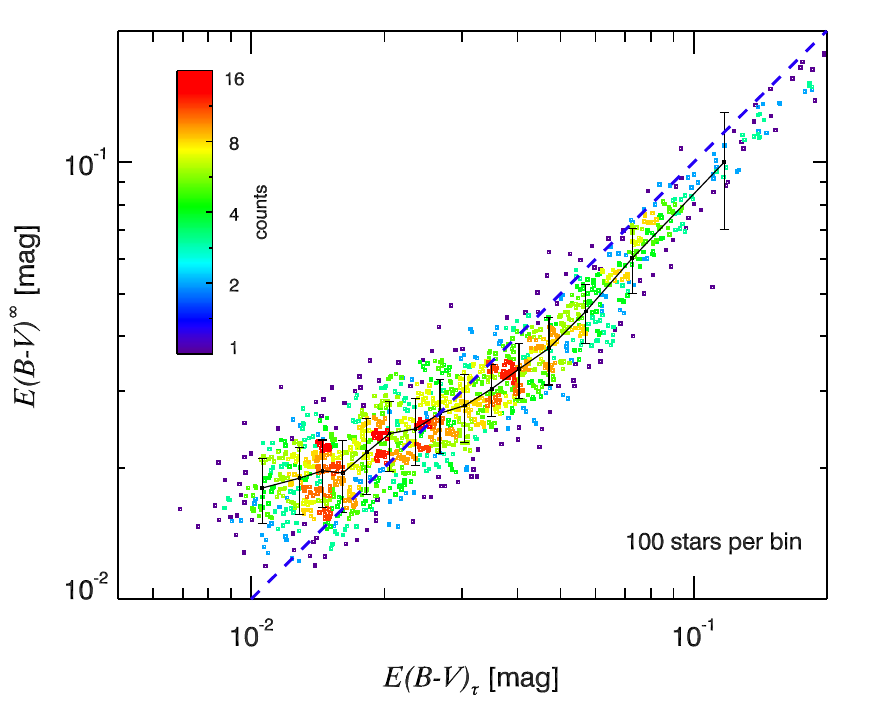}
\caption{Total reddening observed in the optical, $\ebvtotal$, as a function of $\ebvsub$, the dust optical depth at 353\GHz\ converted to a reddening, for the \nstarsselected\ selected stars (Sect. \ref{sec:selectlines}). Each bin of the running mean contains the same number of lines of sight. Error bars represent the standard deviation in each bin. The dashed line corresponds to a one-to-one correlation.
}
\label{ebvtot_ebvsub_stars}
\end{figure}

The uncertainty on $\ebvstar$ stems from the uncertainty on the PS1-based reddening data and the uncertainty $\sigma_{\rm \plx}$ on the stellar parallax. We estimate the former by correlating the PS1 total reddening, $\ebvtotal$, with the \Planck\ optical depth at 353\GHz\ converted to a reddening, $\ebvsub$. Figure~\ref{ebvtot_ebvsub_stars} shows that these quantities are remarkably well correlated, with a Pearson correlation coefficient $r=0.98$ and little scatter (around $15\,\%$ of the running mean on average). 
Lacking more precise information, we assume equal contributions from instrinsic scatter between $\ebvtotal$ and $\ebvsub$ on the one hand and from noise in $\ebvtotal$ on the other hand, so that we take $\sigma_{\ebv}/\ebv = 0.1$ for the PS1-based estimates.
The small departure from linearity in this correlation has been interpreted as evidence for dust evolution in the diffuse ISM \citep{planck2013-p06b,planck2014-XXIX}, i.e., for the modification of the dust optical properties in its lifecycle through the ISM. This aspect will be investigated in its relation to dust polarization properties in a future paper.

The uncertainty $\sigma_{\rm \plx}$ on the parallax leads to an additional uncertainty on $\ebvstar$ that can be estimated roughly by considering the variations of $\ebvstar$ when the parallax varies from $\plx-\sigma_{\rm \plx}$ to $\plx+\sigma_{\rm \plx}$, i.e., 
\begin{equation}
\sigma_{\ebv}^{\plx} = \frac{\ebvstar_{\plx-\sigma_{\rm \plx}}-\ebvstar_{\plx+\sigma_{\rm \plx}}}{2} \, ,
\end{equation}
where $\ebvstar_{\plx-\sigma_{\rm \plx}}$ and $\ebvstar_{\plx+\sigma_{\rm \plx}}$ are the reddenings to the star obtained for the altered parallaxes $\plx-\sigma_{\rm \plx}$ and $\plx+\sigma_{\rm \plx}$, respectively. Gathering the two sources of uncertainty, the total uncertainty $\sigebvstar$ on $\ebvstar$ is then
\begin{equation}
\sigebvstar = \sqrt{\left[0.1\, \ebvstar\right]^2+\left[\sigma_{\ebv}^{\plx}\right]^2} \, .
\label{Eq:sigebvstar}
\end{equation}
This uncertainty then propagates to the quantities used in determining $\Rsv$ (see Sect.~\ref{sec:determination-polar-ratios}), e.g.,
\begin{equation}
\sigma\left(\frac{\pv}{\tauv}\right) \simeq \frac{1}{\tauv}\,\sqrt{\left(\sigma_{\pv}^{\infty}\right)^2+\left(\pv\,\frac{\sigebvstar}{\ebvstar}\right)^2}\,,
\end{equation}
and similarly for $\qv/\tauv$ and $\uv/\tauv$.

\begin{figure}
\includegraphics[width=\qhsize]{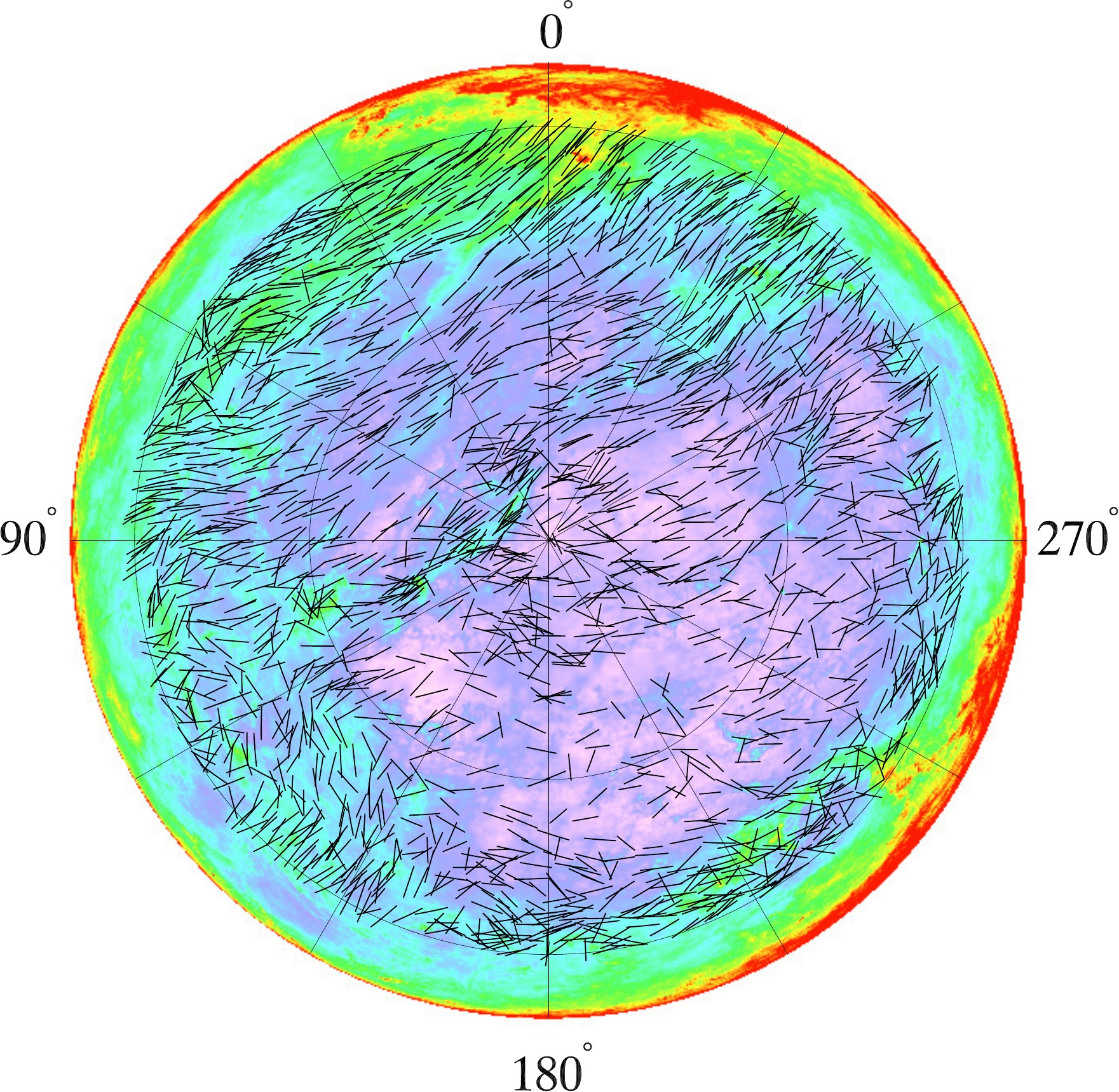}
\includegraphics[width=\qhsize]{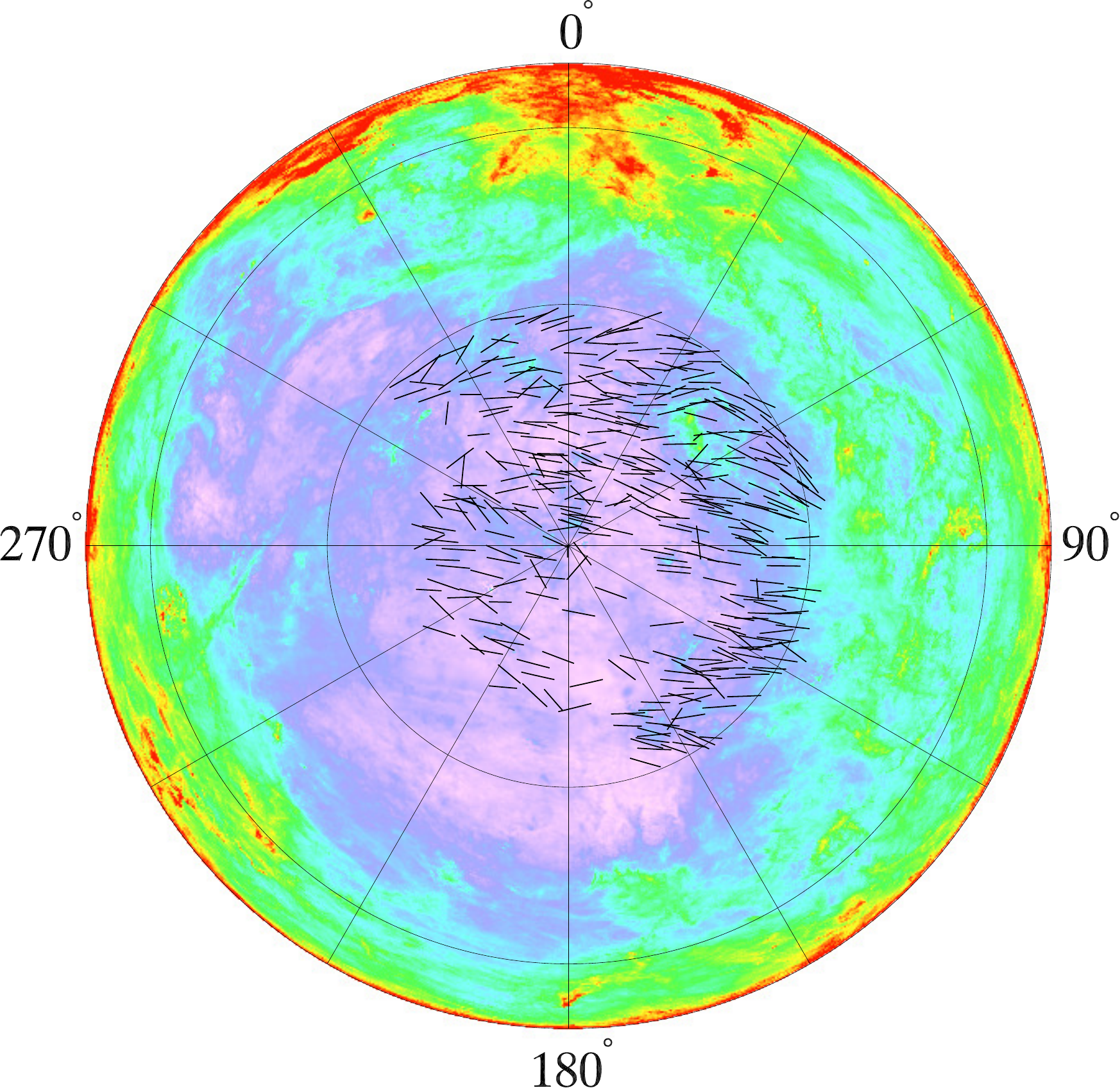}
\includegraphics[width=\qhsize]{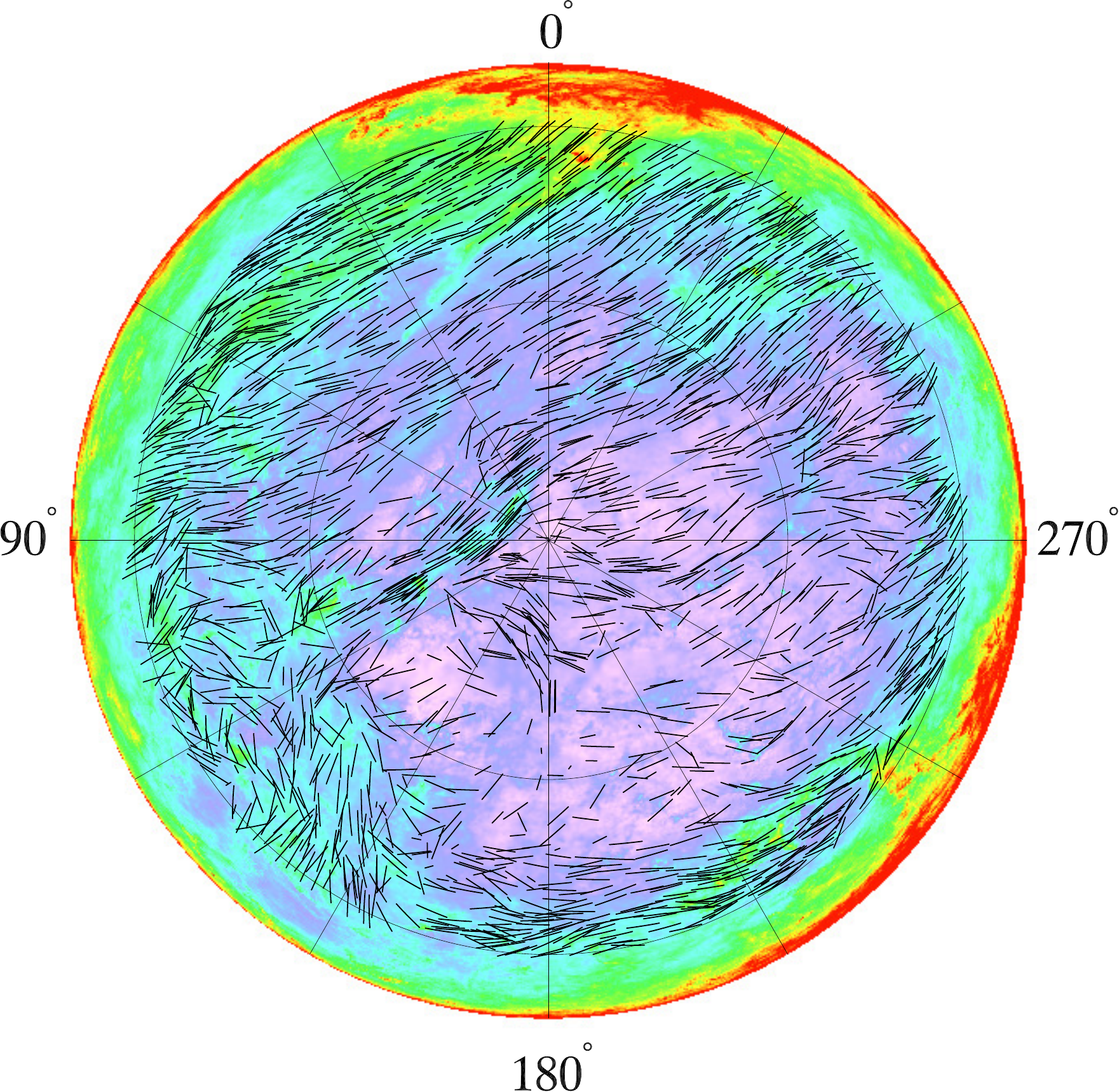}
\includegraphics[width=\qhsize]{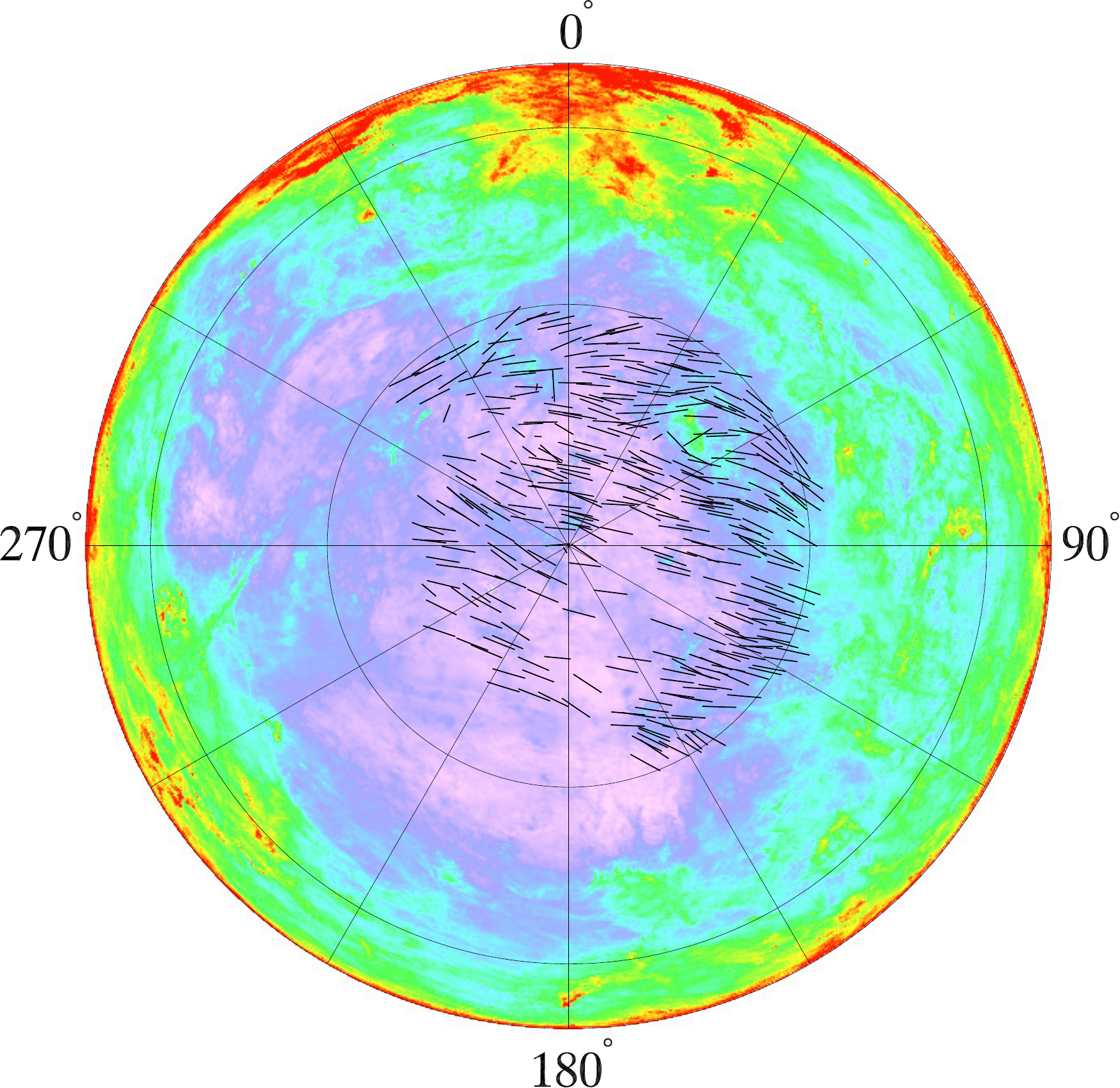}
\caption{Comparison of the orientation of the projection of the magnetic field on the plane of the sky, in orthographic projection with the dust optical depth at 353\GHz\ as the coloured background, from optical data (top panels) and from \Planck\ data at 353\GHz\ (bottom panels).
The line length is proportional to the S/N on the polarization angle.
Northern (left panels) and southern (right panels) Galactic hemispheres are shown, with the Galactic centre situated at the top of each map. 
}
\label{Fig_mapangle}
\end{figure}

\begin{figure}
\includegraphics[width=\hsize]{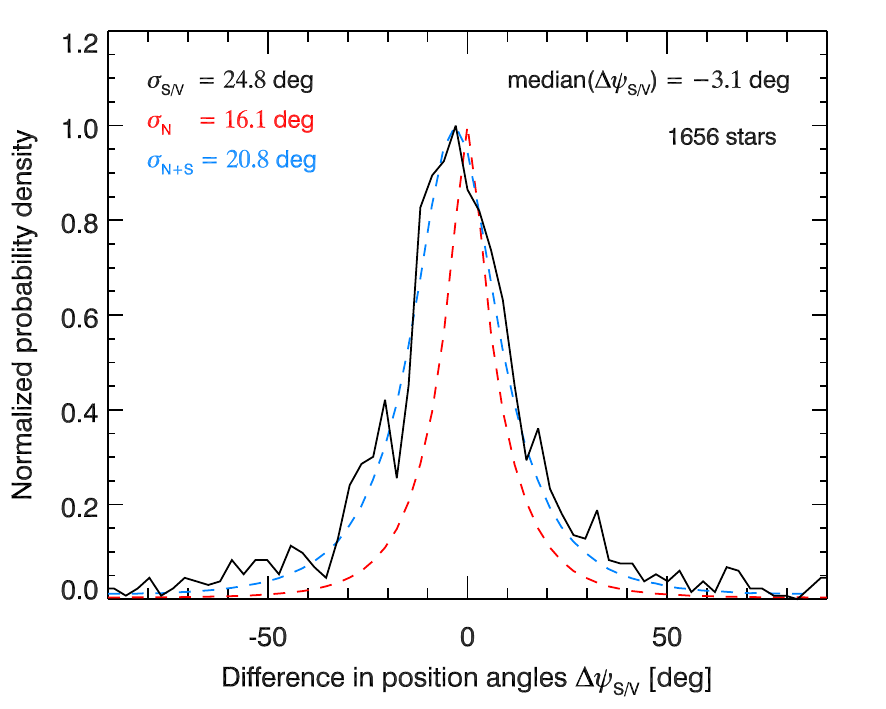}
\caption{Histogram of the difference in polarization angles between \Planck-derived angles and optical-polarization-derived angles, $\Delta\psi_{\rm S/V}=\left(\psi+90\deg\right)-\psi_V$, with its standard deviation $\sigma_\mathrm{S/V}$ and median value indicated. 
Histograms are overplotted for simulations based on noise only (dashed red curve), and noise plus systematics (dashed blue curve) -- see text.  For easier comparison, the latter histogram has been shifted horizontally by $-3\pdeg1$ to peak at the same position as the data.
}
\label{Fig_hist}
\end{figure}

\subsection{Polarization angle difference} 
\label{sec:stars_polang}

The unbiased comparison between submillimetre and optical measurements also requires an agreement in polarization angles. We recall that for interstellar polarization of starlight, the direction of the projection of the magnetic field on the plane of the sky ($\BPOS$) can be inferred directly from the polarization angle. For polarization in emission at 353\,GHz, a rotation by 90\deg\ is required. In Fig.~\ref{Fig_mapangle}, we compare the direction of $\BPOS$ inferred from the two tracers.  
The length of each line is proportional to the S/N value for the corresponding polarization angle.
If the same dust is probed in the optical and the submillimetre, the directions should be identical, and we do find that the agreement is quite remarkable, although not perfect.

To quantify this agreement, we define the difference in orientation angles between the \Planck\ (or submillimetre, `S') and optical (or visual, `V')
polarization data as $\Delta\psi_{\rm S/V}~=~\left(\psi+90\deg\right)~-~\psi_V$.  In terms of Stokes parameters this can be written as \citep{planck2014-XXI}
\begin{equation}
\label{eq:deltapsi}
\Delta\psi_{\rm S/V} = \frac{1}{2}\mathrm{atan2}{\left[\left(\Usub\,\qv-\Qsub\,\uv\right)\, , \, -\left(\Qsub\,\qv+\Usub\,\uv\right)\right]} \, .
\end{equation}
Fig.~\ref{Fig_hist} presents the histogram of $\Delta\psi_{\rm S/V}$, of which two important aspects should be understood, i.e., its width and any shift relative to the expected centring on zero.

\subsubsection{Standard deviation}
\label{sec:Stddev}

We attempt to explain the width of the histogram $\Delta\psi_{\rm S/V}$ (Fig.~\ref{Fig_hist}) by simulating what is expected from noise and systematic effects. We start from \Planck\ data at a resolution of 40\arcmin\ and starlight data, rotated to assume perfect orthogonality. Then we add fluctuations drawn from Gaussian distributions having the estimated variances for each random and systematic uncertainty considered among the following: noise in the submillimetre  and optical, beam depolarization (Appendix \ref{sec:beamdepol}), and background distortion (Appendix \ref{sec:bkgd}). Finally, with the resulting Stokes parameters, $\Qsub$ and $\Usub$ at 353\GHz, and $\qv$ and $uv$ in the optical, we compute $\Delta\psi_{\rm S/V}$ for these simulated data.

The width of the observed histogram is too large to be explained by noise only, as shown by the dashed red curve that results from simulating only effects of the noise in the submillimetre and optical data (Fig.~\ref{Fig_hist}). Part of this discrepancy may be accounted for by the dispersion of polarization angles in the optical within a \Planck\ beam, i.e., at scales that cannot be probed by \Planck. To estimate this contribution, we analyse the starlight polarization data alone, ignoring \Planck\ data. We compute the standard deviation $\sigma_{\rm V/V}$ of the difference in polarization angles in the optical, for those stars at an angular distance $\delta$ smaller than 40\arcmin\ (one FWHM~of the \Planck\ beam). We find $\sigma_{\rm V/V} = 26\deg$. We note that by its nature this dispersion incorporates twice the variance of optical polarization angles compared to that which enters into the $\Delta\psi_{\rm S/V}$ histogram. Thus the expected standard deviation from this effect would be about \sigmaVtoref. Three random and systematic effects affecting starlight polarization measures could explain that star-to-star dispersion : noise in the optical, turbulence at scales smaller than the \Planck\ beam (Appendix~\ref{sec:beamdepol}); and background distortion (Appendix \ref{sec:bkgd}). For these three, we consider Gaussian fluctuations having variances $\dpv^2$, $\delta p^2_{\rm beam}$\footnote{Whether we add the Gaussian random realizations to the submillimetre or to the optical Stokes parameters does not affect the resulting histogram of differences in polarization angles and so we do not show the conversion of the equations for the submillimetre beam depolarization into their analogous form in the optical.} and $(\sigma_{\pv}^\infty)^2$, respectively, the latter depending on an unknown parameter, ${\rm rms}(\pv/\ebv)_{\rm bkgd}$. We find that simulations including both optical noise and turbulence within the \Planck\ beam produce a star-to-star dispersion of 13\deg, well below the observed value of $\sigmaVtoref$.  
This points to the need for a contribution from background distortion, and thus a need to estimate ${\rm rms}(\pv/\ebv)_{\rm bkgd}$. The present data sample provides lower and upper bounds to the rms of $\pv/\ebv$, respectively 7\% for the rms of $\pv/\ebv$ in our sample of $\nstarsselected$ stars, and 13\% for the maximum observed polarization fraction (see Sect. \ref{sec:maxpsebv}). Assuming a value of ${\rm rms}(\pv/\ebv)_{\rm bkgd}= \rmspsebv$ close to the rms over our sample of stars, we obtain a dispersion of $\sigmabkgd$ arising from the background distortion $\sigma_{\pv}^\infty$. The total dispersion including optical noise, turbulence within the \Planck\ beam, and background distortion is then \sigmasimuVtoref, still significantly below our estimate of \sigmaVtoref. We note that the individual contributions to the angular dispersion do not quite add up in quadrature to the one obtained when taking into account all sources of errors, because even though the fluctuations of Stokes parameters are Gaussian-distributed, the polarization angles are not~\citep{naghizadeh-khouei-clarke-1993}. A total dispersion of \sigmaVtoref\ would require ${\rm rms}(\pv/\ebv_{\rm bkgd}) = 11\%$, a choice that is probably too extreme given the observational constraints, the uncertainties on our estimated value for $\sigma_{\rm V/V}$, and other physical effects not included in this analysis  (see Sect. \ref{sec:meandiff} below). Therefore, to be conservative, we will adopt a value of \rmspsebv\ for the rms of $\pv/\ebv_{\rm bkgd}$ to compute the background distortion, which then by itself contributes a dispersion of \sigmabkgd.

Our simulated histogram of $\Delta\psi_{\rm S/V}$ based on optical and submillimetre noise, beam depolarization at 40\arcmin\ and background distortion is presented in Fig.~\ref{Fig_hist} (dashed blue line). 
It is close in shape and width (standard deviation $\sigma_{\rm N+S}=\sigmasimu$) to the observed histogram and by construction it is centred near zero. The contributions to the standard deviation are 
\sigmasubnoise\ from submillimetre noise, 
\sigmaoptnoise\ from optical noise, 
\sigmabeam\ from beam depolarization, and  
\sigmabkgd\ from background distortion.
Here again, we warn against the simple quadratic addition of individual sources of uncertainty. We note that this model does not include a contribution from possible \Planck\ systematics, which can be assessed through the E2E simulations presented in Appendix~\ref{sec:appendix:E2E}. For the lines of sight to stars, we built the histogram of the difference in polarization angles between the input maps at 5\arcmin\ resolution and those at 40{\arcmin} that went through the simulation pipeline and therefore include estimates of \Planck\ noise and known systematics. The standard deviation of these histograms is $14\deg \pm1 \deg$, depending on the number of simulations used, a value close to that found from \Planck\ noise alone, underlining the low level of remaining systematics in the \planck\ data.

This analysis of the standard deviation of $\Delta\psi_{\rm S/V}$ provides some assurance that the subset of uncertainties needed for the quantification of the emission-to-extinction polarization ratios in Sect.~\ref{sec:determination-polar-ratios} have been quantified adequately.

\subsubsection{Mean difference}
\label{sec:meandiff}

The histogram of $\Delta\psi_{\rm S/V}$ in the data peaks at $-3\pdeg1$, revealing a systematic angle difference (shift) between polarization angle measurements in emission relative to extinction. Given the large number of lines of sight, this shift cannot be explained by a random process, thus pointing to a systematic effect. Although the shift is small and unimportant for evaluating the polarization ratios discussed in Sect.~\ref{sec:stars}, it is potentially important for other investigations.  For example, accurate absolute calibration of the polarization angle is critical for future CMB $B$-mode experiments to avoid systematic effects that could compromise reaching the precision required \citep{aumont2018}. We have explored the possible origins of such a shift.

There is some evidence for the possibility of background distortion, a systematic difference arising because in the submillimetre dust is observed along a longer path length than probed in the optical. First, within the full \nstarsebv-star sample the shift appears to depend on Galactic longitude. Second, when we apply a more stringent criterion on the reddening ratio $\ebvstar/\ebvtotal$ (see Sect.~\ref{sec:selectlines}), thus reducing the chance of a significant background contamination, we find that the shift is smaller,  $-2\deg$, and a longitude dependence is no longer evident. 

In the above-mentioned \Planck\ E2E simulations (Appendix~\ref{sec:Stddev}), the mean and median of the difference in polarization angles are consistently shifted from zero by less than $0\pdeg25$, which shows that the observed difference is not ascribable to any known systematic in the \Planck\ data. Concerning the calibration of the zero point of the polarization angle in the submillimetre, the uncertainties on the orientation of the HFI PSBs at 100, 143, and 217\,GHz are below $0\pdeg3$~\citep{planck2014-a10,planck2016-l11A}, and it seems likely that the uncertainty at 353\,GHz is of the same magnitude. This is consistent with multifrequency measurements of the polarization angle of the bright synchrotron emission of the Crab Nebula from 23 to 353\,GHz \citep{aumont2018}.

The calibration of the zero point in the optical is less clear. For the high-latitude polarization surveys considered here, three highly polarized stars were used \citep{BerdVIII2014}. We have examined the extensive historical record of the polarization angle of these stars and find good evidence that each varies with an rms of typically 1\pdeg5, but with excursions as large as 5\deg. Without confidence in sub-degree accuracy of the optical zero point, the possibility remains of a significant contribution to the shift due to this uncertain calibration. There is not a lot of overlap between the optical polarization compilation of \citet{H00} and the surveys of \citet{BerdVIII2014} used here.  However, by a statistical comparison between the polarization angles of stars in one catalogue with those of stars in the other, binned as a function of angular distance, it is possible to investigate any systematic shift.  This analysis can be extended by replacing the optical polarization angle measurements of one or both catalogues with the \Planck\ polarization angle at the catalogue positions. This reveals small systematic shifts up to a few degrees that could arise from different zero point calibrations and/or different path lengths probed. From these investigations, a residual contribution to the shift of order 1\deg\ seems possible.

A final possibility is actual decorrelation between the submillimetre and optical polarization, e.g., due to a temperature-weighting effect in the submillimetre coupled with a correlation of variations in the heating of aligned grains (or in the dust properties themselves) along the line of sight with variations of the magnetic field orientation \citep{Tassis2015}. 

We cannot pinpoint a single cause for a shift of the magnitude seen.  Instead, it seems that several smaller contributions might have conspired to produce the effect observed.

\end{document}

%% file: Planck.tex
\def\setsymbol#1#2{\expandafter\def\csname #1\endcsname{#2}}
\def\getsymbol#1{\csname #1\endcsname}

\def\Planck{\textit{Planck}}

\newbox\tablebox    \newdimen\tablewidth
\def\leaderfil{\leaders\hbox to 5pt{\hss.\hss}\hfil}
\def\endPlancktable{\tablewidth=\columnwidth 
    $$\hss\copy\tablebox\hss$$
    \vskip-\lastskip\vskip -2pt}

\def\tablenote#1 #2\par{\begingroup \parindent=0.8em
    \abovedisplayshortskip=0pt\belowdisplayshortskip=0pt
    \noindent
    $$\hss\vbox{\hsize\tablewidth \hangindent=\parindent \hangafter=1 \noindent
    \hbox to \parindent{$^#1$\hss}\strut#2\strut\par}\hss$$
    \endgroup}
\def\doubleline{\vskip 3pt\hrule \vskip 1.5pt \hrule \vskip 5pt}

\def\L2{\ifmmode L_2\else $L_2$\fi}

\def\DeltaT{\ifmmode \Delta T\else $\Delta T$\fi}
\def\deltat{\ifmmode \Delta t\else $\Delta t$\fi}
\def\fknee{\ifmmode f_{\rm knee}\else $f_{\rm knee}$\fi}
\def\Fmax{\ifmmode F_{\rm max}\else $F_{\rm max}$\fi}
\def\solar{\ifmmode{\rm M}_{\mathord\odot}\else${\rm M}_{\mathord\odot}$\fi}
\def\Msolar{\ifmmode{\rm M}_{\mathord\odot}\else${\rm M}_{\mathord\odot}$\fi}
\def\Lsolar{\ifmmode{\rm L}_{\mathord\odot}\else${\rm L}_{\mathord\odot}$\fi}
\def\inv{\ifmmode^{-1}\else$^{-1}$\fi}
\def\mo{\ifmmode^{-1}\else$^{-1}$\fi}
\def\sup#1{\ifmmode ^{\rm #1}\else $^{\rm #1}$\fi}
\def\expo#1{\ifmmode \times 10^{#1}\else $\times 10^{#1}$\fi}
\def\,{\thinspace}
\def\lsim{\mathrel{\raise .4ex\hbox{\rlap{$<$}\lower 1.2ex\hbox{$\sim$}}}}
\def\gsim{\mathrel{\raise .4ex\hbox{\rlap{$>$}\lower 1.2ex\hbox{$\sim$}}}}

\def\simprop{\mathrel{\raise .4ex\hbox{\rlap{$\propto$}\lower 1.2ex\hbox{$\sim$}}}}
\def\deg{\ifmmode^\circ\else$^\circ$\fi}
\def\pdeg{\ifmmode $\setbox0=\hbox{$^{\circ}$}\rlap{\hskip.11\wd0 .}$^{\circ}
          \else \setbox0=\hbox{$^{\circ}$}\rlap{\hskip.11\wd0 .}$^{\circ}$\fi}
\def\arcs{\ifmmode {^{\scriptstyle\prime\prime}}
          \else $^{\scriptstyle\prime\prime}$\fi}
\def\arcm{\ifmmode {^{\scriptstyle\prime}}
          \else $^{\scriptstyle\prime}$\fi}
\newdimen\sa  \newdimen\sb
\def\parcs{\sa=.07em \sb=.03em
     \ifmmode \hbox{\rlap{.}}^{\scriptstyle\prime\kern -\sb\prime}\hbox{\kern -\sa}
     \else \rlap{.}$^{\scriptstyle\prime\kern -\sb\prime}$\kern -\sa\fi}
\def\parcm{\sa=.08em \sb=.03em
     \ifmmode \hbox{\rlap{.}\kern\sa}^{\scriptstyle\prime}\hbox{\kern-\sb}
     \else \rlap{.}\kern\sa$^{\scriptstyle\prime}$\kern-\sb\fi}
\def\ra[#1 #2 #3.#4]{#1\sup{h}#2\sup{m}#3\sup{s}\llap.#4}
\def\dec[#1 #2 #3.#4]{#1\deg#2\arcm#3\arcs\llap.#4}
\def\deco[#1 #2 #3]{#1\deg#2\arcm#3\arcs}
\def\rra[#1 #2]{#1\sup{h}#2\sup{m}}

\def\dots{\relax\ifmmode \ldots\else $\ldots$\fi}
\def\WHzsr{\ifmmode $W\,Hz\mo\,sr\mo$\else W\,Hz\mo\,sr\mo\fi}
\def\mHz{\ifmmode $\,mHz$\else \,mHz\fi}
\def\GHz{\ifmmode $\,GHz$\else \,GHz\fi}
\def\mKs{\ifmmode $\,mK\,s$^{1/2}\else \,mK\,s$^{1/2}$\fi}
\def\muKs{\ifmmode \,\mu$K\,s$^{1/2}\else \,$\mu$K\,s$^{1/2}$\fi}
\def\muKRJs{\ifmmode \,\mu$K$_{\rm RJ}$\,s$^{1/2}\else \,$\mu$K$_{\rm RJ}$\,s$^{1/2}$\fi}
\def\muKHz{\ifmmode \,\mu$K\,Hz$^{-1/2}\else \,$\mu$K\,Hz$^{-1/2}$\fi}
\def\MJysr{\ifmmode \,$MJy\,sr\mo$\else \,MJy\,sr\mo\fi}
\def\MJysrmK{\ifmmode \,$MJy\,sr\mo$\,mK$_{\rm CMB}\mo\else \,MJy\,sr\mo\,mK$_{\rm CMB}\mo$\fi}
\def\microns{\ifmmode \,\mu$m$\else \,$\mu$m\fi}

\def\muK{\ifmmode \,\mu$K$\else \,$\mu$\hbox{K}\fi}
\def\microK{\ifmmode \,\mu$K$\else \,$\mu$\hbox{K}\fi}
\def\muW{\ifmmode \,\mu$W$\else \,$\mu$\hbox{W}\fi}
\def\kms{\ifmmode $\,km\,s$^{-1}\else \,km\,s$^{-1}$\fi}
\def\kmsMpc{\ifmmode $\,\kms\,Mpc\mo$\else \,\kms\,Mpc\mo\fi}
\providecommand{\sorthelp}[1]{}

%% file: Macros.tex
\def\arXivFIG{}

\newcommand{\planck}{\Planck}
\newcommand{\healpix}{{\tt HEALPix}}

\def\GNILC{{\tt GNILC}}

\def\aap{{A\&A}}

\def\apjs{{ApJ Supp.}}

\def\Vband{$V$}
\def\hi{\sc Hi}

\newcommand{\NH}{N_{\rm H}} 
\newcommand{\Bfield}{\boldsymbol{B}}
\newcommand{\BPOS}{\Bfield_\mathrm{POS}}

\newcommand{\Td}{T_\mathrm{d}}
\def\plx{\theta^\star}
\def\G{G_{\mathcal{R}}}
\def\ellres{\omega}
\def\ebv{E(B-V)}
\def\ebvR{\ebv_\mathcal{R}}
\def\ebvGreen{\ebv_{\rm Green}}
\def\ebvstar{\ebv^\star}
\def\sigebvstar{\sigma_{E(B-V)}^{\star}}
\def\ebvtotal{\ebv^\infty}
\def\ebvsub{\ebv_{\tau}}

\def\Rv{R_V}
\def\Av{A_V}

\def\Rsv{R_{\rm S/V}}
\def\RPp{R_{P/p}}
\def\pv{p_V}

\def\dpv{\sigma_{p_V}}
\def\dqv{\sigma_{u_V}}
\def\duv{\sigma_{q_V}}
\def\Gv{\psi_V}
\def\qv{q_V}
\def\uv{u_V}
\def\dist{D^\star}
\def\ie{\emph{i.e.\/}}
\def\tauv{\tau_V}

\def\maxNHSpT{8\times10^{21}}
\def\maxdropp{{3--4}}
\def\maxdropSp{25}
\def\uptoNH{2\times10^{22}}
\def\pmaxv{13}
\def\PsImaxPIPXXI{14}

\def\Rsvfidround{4.3}

\def\Isub{I}
\def\Psub{P}

\def\Qsub{Q}
\def\Usub{U}

\def\sigmasimu{20\pdeg8}
\def\sigmasubnoise{13\pdeg9}
\def\sigmaoptnoise{8\pdeg8}
\def\sigmabeam{9\pdeg6}
\def\sigmabkgd{12\pdeg6}

\def\sigmaVtoref{18\deg}

\def\rmspsebv{8\%}
\def\sigmasimuVtoref{16\deg}

\def\PsI{P/I}

\def\PsI{p}
\def\S{\mathcal{S}}

\def\sII{\sigma_{II}}
\def\sIQ{\sigma_{IQ}}
\def\sIU{\sigma_{IU}}
\def\sQQ{\sigma_{QQ}}
\def\sQU{\sigma_{QU}}
\def\sUU{\sigma_{UU}}
\def\StimesPsI{\S\times\PsI}
\def\PsIxS{\StimesPsI}

\def\MJysr{MJy\,sr$^{-1}$}

\def\cmsq{cm$^{-2}$}
\def\kms{{\rm km\,s}$^{-1}$}

\def\nside{N_{\rm side}}

\def\nsidePsIxS{64-bgt5_v8}

\def\L{\emph{(left)}}

\newlength{\thsize}
\newlength{\hhsize}
\newlength{\qhsize}
\def\ie{i.e.,}

\def\nstarsuniq{2461}

\def\resolowres{80}
\def\Rsvfitlowres{4.2}
\def\reso{40}
\def\nwin{15}
\def\nstarsperbin{100}
\def\nsidestars{256}
\def\nstarsebv{2388}
\def\nstarsoutofmask{70}
\def\nstarsselected{1505}
\def\nstarsnorth{2092}
\def\nstarssouth{369}

\def\RsvfitlowCIB{4.8}
\def\RsvfithighCIB{3.9}

\def\medEBVratio{0.83}
\def\resohighres{20}
\def\Rsvfithighres{4.2}

\def\diffGmin{45}
\def\chiRPpnosys{3.9}
\def\chiRsvnosys{3.0}
\def\chiRPpsys{1.02}
\def\chiRsvsys{0.96}

\def\meanRPpfitted{5.42}
\def\sigmeanRPpfitted{0.05}
\def\meanRsvfitted{4.31}
\def\sigmeanRsvfitted{0.04}

\def\revisedVG{18DEC2018}

\newlength{\full}
\newlength{\half}
\newlength{\third}
\newlength{\fourth}

\def\pmax{p_{\rm max}}

\def\I{I}

\def\Bzero{\boldsymbol{B}_0}
\def\Bzerox{B_{0x}}
\def\Bzeroy{B_{0y}}
\def\Bzeroz{B_{0z}}
\def\Btot{\boldsymbol{B}_i}
\def\Bt{\boldsymbol{B}_{\rm t}}
\def\Bx{B_{ix}}
\def\By{B_{iy}}
\def\Bz{B_{iz}}
\def\dB{\delta\boldsymbol{B}_{i}}
\def\dBx{\delta B_{ix}}
\def\dBy{\delta B_{iy}}
\def\dBz{\delta B_{iz}}
\def\gx{g_{x}}
\def\gy{g_{y}}
\def\gz{g_{z}}

\def\fm{f_{\rm m}(\delta)}
\def\fmsq{f^2_{\rm m}(\delta)}
\def\fM{f_{\rm M}}
\def\Gx{G_{x}}
\def\Gy{G_{y}}
\def\Gz{G_{z}}

\def\goffasm{68} 

\def\goff{63} 
\def\goffl{23} 
\def\goffh{103} 
\def\goffu{40} 

\newcommand{\DRThree}{\Planck\ 2018}

\newcommand{\asmap}{ASM}
\newcommand{\asmaps}{ASMs}

%% file: L11_Dust_B_authors_and_institutes.tex
\author{\small
Planck Collaboration: N.~Aghanim\inst{48}
\and
Y.~Akrami\inst{13, 51, 53}
\and
M.~I.~R.~Alves\inst{90, 8, 48}
\and
M.~Ashdown\inst{60, 5}
\and
J.~Aumont\inst{90}
\and
C.~Baccigalupi\inst{74}
\and
M.~Ballardini\inst{19, 35}
\and
A.~J.~Banday\inst{90, 8}
\and
R.~B.~Barreiro\inst{55}
\and
N.~Bartolo\inst{25, 56}
% ...
}
\institute{\small
AIM, CEA, CNRS, Universit\'{e} Paris-Saclay, Universit\'{e} Paris-Diderot, Sorbonne Paris Cit\'{e}, F-91191 Gif-sur-Yvette, France\goodbreak
\and
APC, AstroParticule et Cosmologie, Universit\'{e} Paris Diderot, CNRS/IN2P3, CEA/lrfu, Observatoire de Paris, Sorbonne Paris Cit\'{e}, 10, rue Alice Domon et L\'{e}onie Duquet, 75205 Paris Cedex 13, France\goodbreak
\and
African Institute for Mathematical Sciences, 6-8 Melrose Road, Muizenberg, Cape Town, South Africa\goodbreak
\and
Aix Marseille Univ, CNRS, CNES, LAM, Marseille, France\goodbreak
\and
Astrophysics Group, Cavendish Laboratory, University of Cambridge, J J Thomson Avenue, Cambridge CB3 0HE, U.K.\goodbreak
\and
Astrophysics \& Cosmology Research Unit, School of Mathematics, Statistics \& Computer Science, University of KwaZulu-Natal, Westville Campus, Private Bag X54001, Durban 4000, South Africa\goodbreak
\and
CITA, University of Toronto, 60 St. George St., Toronto, ON M5S 3H8, Canada\goodbreak
\and
CNRS, IRAP, 9 Av. colonel Roche, BP 44346, F-31028 Toulouse cedex 4, France\goodbreak
\and
Cahill Center for Astronomy and Astrophysics, California Institute of Technology, Pasadena CA,  91125, USA\goodbreak
\and
California Institute of Technology, Pasadena, California, U.S.A.\goodbreak
\and
Computational Cosmology Center, Lawrence Berkeley National Laboratory, Berkeley, California, U.S.A.\goodbreak
\and
D\'{e}partement de Physique Th\'{e}orique, Universit\'{e} de Gen\`{e}ve, 24, Quai E. Ansermet,1211 Gen\`{e}ve 4, Switzerland\goodbreak
\and
D\'{e}partement de Physique, \'{E}cole normale sup\'{e}rieure, PSL Research University, CNRS, 24 rue Lhomond, 75005 Paris, France\goodbreak
\and
Departamento de Astrof\'{i}sica, Universidad de La Laguna (ULL), E-38206 La Laguna, Tenerife, Spain\goodbreak
\and
Departamento de F\'{\i}sica, Universidad de Oviedo, C/ Federico Garc\'{\i}a Lorca, 18 , Oviedo, Spain\goodbreak
\and
Department of Astrophysics/IMAPP, Radboud University, P.O. Box 9010, 6500 GL Nijmegen, The Netherlands\goodbreak
\and
Department of Mathematics, University of Stellenbosch, Stellenbosch 7602, South Africa\goodbreak
\and
Department of Physics \& Astronomy, University of British Columbia, 6224 Agricultural Road, Vancouver, British Columbia, Canada\goodbreak
\and
Department of Physics \& Astronomy, University of the Western Cape, Cape Town 7535, South Africa\goodbreak
\and
Department of Physics and Astronomy, University of Sussex, Brighton BN1 9QH, U.K.\goodbreak
\and
Department of Physics, Gustaf H\"{a}llstr\"{o}min katu 2a, University of Helsinki, Helsinki, Finland\goodbreak
\and
Department of Physics, Princeton University, Princeton, New Jersey, U.S.A.\goodbreak
\and
Department of Physics, University of California, Santa Barbara, California, U.S.A.\goodbreak
\and
Department of Physics, University of Illinois at Urbana-Champaign, 1110 West Green Street, Urbana, Illinois, U.S.A.\goodbreak
\and
Dipartimento di Fisica e Astronomia G. Galilei, Universit\`{a} degli Studi di Padova, via Marzolo 8, 35131 Padova, Italy\goodbreak
\and
Dipartimento di Fisica e Scienze della Terra, Universit\`{a} di Ferrara, Via Saragat 1, 44122 Ferrara, Italy\goodbreak
\and
Dipartimento di Fisica, Universit\`{a} La Sapienza, P. le A. Moro 2, Roma, Italy\goodbreak
\and
Dipartimento di Fisica, Universit\`{a} degli Studi di Milano, Via Celoria, 16, Milano, Italy\goodbreak
\and
Dipartimento di Fisica, Universit\`{a} degli Studi di Trieste, via A. Valerio 2, Trieste, Italy\goodbreak
\and
Dipartimento di Fisica, Universit\`{a} di Roma Tor Vergata, Via della Ricerca Scientifica, 1, Roma, Italy\goodbreak
\and
European Space Agency, ESAC, Planck Science Office, Camino bajo del Castillo, s/n, Urbanizaci\'{o}n Villafranca del Castillo, Villanueva de la Ca\~{n}ada, Madrid, Spain\goodbreak
\and
European Space Agency, ESTEC, Keplerlaan 1, 2201 AZ Noordwijk, The Netherlands\goodbreak
\and
Gran Sasso Science Institute, INFN, viale F. Crispi 7, 67100 L'Aquila, Italy\goodbreak
\and
Helsinki Institute of Physics, Gustaf H\"{a}llstr\"{o}min katu 2, University of Helsinki, Helsinki, Finland\goodbreak
\and
INAF - OAS Bologna, Istituto Nazionale di Astrofisica - Osservatorio di Astrofisica e Scienza dello Spazio di Bologna, Area della Ricerca del CNR, Via Gobetti 101, 40129, Bologna, Italy\goodbreak
\and
INAF - Osservatorio Astronomico di Padova, Vicolo dell'Osservatorio 5, Padova, Italy\goodbreak
\and
INAF - Osservatorio Astronomico di Trieste, Via G.B. Tiepolo 11, Trieste, Italy\goodbreak
\and
INAF, Istituto di Radioastronomia, Via Piero Gobetti 101, I-40129 Bologna, Italy\goodbreak
\and
INAF/IASF Milano, Via E. Bassini 15, Milano, Italy\goodbreak
\and
INFN - CNAF, viale Berti Pichat 6/2, 40127 Bologna, Italy\goodbreak
\and
INFN, Sezione di Bologna, viale Berti Pichat 6/2, 40127 Bologna, Italy\goodbreak
\and
INFN, Sezione di Ferrara, Via Saragat 1, 44122 Ferrara, Italy\goodbreak
\and
INFN, Sezione di Milano, Via Celoria 16, Milano, Italy\goodbreak
\and
INFN, Sezione di Roma 1, Universit\`{a} di Roma Sapienza, Piazzale Aldo Moro 2, 00185, Roma, Italy\goodbreak
\and
INFN, Sezione di Roma 2, Universit\`{a} di Roma Tor Vergata, Via della Ricerca Scientifica, 1, Roma, Italy\goodbreak
\and
Imperial College London, Astrophysics group, Blackett Laboratory, Prince Consort Road, London, SW7 2AZ, U.K.\goodbreak
\and
Infrared Processing and Analysis Center, California Institute of Technology, Pasadena, CA 91125, U.S.A.\goodbreak
\and
Institut d'Astrophysique Spatiale, CNRS, Univ. Paris-Sud, Universit\'{e} Paris-Saclay, B\^{a}t. 121, 91405 Orsay cedex, France\goodbreak
\and
Institut d'Astrophysique de Paris, CNRS (UMR7095), 98 bis Boulevard Arago, F-75014, Paris, France\goodbreak
\and
Institut dÕAstrophysique Spatiale, CNRS, Univ. Paris-Sud, Universite Paris-Saclay, Bat. 121, 91405 Orsay cedex, France\goodbreak
\and
Institute Lorentz, Leiden University, PO Box 9506, Leiden 2300 RA, The Netherlands\goodbreak
\and
Institute of Astronomy, University of Cambridge, Madingley Road, Cambridge CB3 0HA, U.K.\goodbreak
\and
Institute of Theoretical Astrophysics, University of Oslo, Blindern, Oslo, Norway\goodbreak
\and
Instituto de Astrof\'{\i}sica de Canarias, C/V\'{\i}a L\'{a}ctea s/n, La Laguna, Tenerife, Spain\goodbreak
\and
Instituto de F\'{\i}sica de Cantabria (CSIC-Universidad de Cantabria), Avda. de los Castros s/n, Santander, Spain\goodbreak
\and
Istituto Nazionale di Fisica Nucleare, Sezione di Padova, via Marzolo 8, I-35131 Padova, Italy\goodbreak
\and
Jet Propulsion Laboratory, California Institute of Technology, 4800 Oak Grove Drive, Pasadena, California, U.S.A.\goodbreak
\and
Jodrell Bank Centre for Astrophysics, Alan Turing Building, School of Physics and Astronomy, The University of Manchester, Oxford Road, Manchester, M13 9PL, U.K.\goodbreak
\and
Kavli Institute for Cosmological Physics, University of Chicago, Chicago, IL 60637, USA\goodbreak
\and
Kavli Institute for Cosmology Cambridge, Madingley Road, Cambridge, CB3 0HA, U.K.\goodbreak
\and
Kavli Institute for Particle Astrophysics and Cosmology, Physics and Astrophysics Building, 452 Lomita Mall, Stanford, CA 94305, USA\goodbreak
\and
Kavli Institute for the Physics and Mathematics of the Universe (Kavli IPMU, WPI), UTIAS, The University of Tokyo, Chiba, 277- 8583, Japan\goodbreak
\and
Laboratoire Univers et Particules de Montpellier, Universit{\'e} de Montpellier, CNRS/IN2P3, CC 72, Place Eug{\`e}ne Bataillon, 34095 Montpellier Cedex 5, France\goodbreak
\and
Laboratoire d'Oc{\'e}anographie Physique et Spatiale (LOPS), Univ. Brest, CNRS, Ifremer, IRD, Brest, France\goodbreak
\and
Laboratoire de Physique Subatomique et Cosmologie, Universit\'{e} Grenoble-Alpes, CNRS/IN2P3, 53, rue des Martyrs, 38026 Grenoble Cedex, France\goodbreak
\and
Laboratoire de Physique Th\'{e}orique, Universit\'{e} Paris-Sud 11 \& CNRS, B\^{a}timent 210, 91405 Orsay, France\goodbreak
\and
Low Temperature Laboratory, Department of Applied Physics, Aalto University, Espoo, FI-00076 AALTO, Finland\goodbreak
\and
Max-Planck-Institut f\"{u}r Astrophysik, Karl-Schwarzschild-Str. 1, 85741 Garching, Germany\goodbreak
\and
Mullard Space Science Laboratory, University College London, Surrey RH5 6NT, U.K.\goodbreak
\and
NAOC-UKZN Computational Astrophysics Centre (NUCAC), University of KwaZulu-Natal, Durban 4000, South Africa\goodbreak
\and
National Centre for Nuclear Research, ul. A. Soltana 7, 05-400 Otwock, Poland\goodbreak
\and
Nicolaus Copernicus Astronomical Center, Polish Academy of Sciences, Bartycka 18, 00-716 Warsaw, Poland\goodbreak
\and
Nordita, KTH Royal Institute of Technology and Stockholm University, Roslagstullsbacken 23, 10691 Stockholm, Sweden\goodbreak
\and
SISSA, Astrophysics Sector, via Bonomea 265, 34136, Trieste, Italy\goodbreak
\and
San Diego Supercomputer Center, University of California, San Diego, 9500 Gilman Drive, La Jolla, CA 92093, USA\goodbreak
\and
School of Chemistry and Physics, University of KwaZulu-Natal, Westville Campus, Private Bag X54001, Durban, 4000, South Africa\goodbreak
\and
School of Physical Sciences, National Institute of Science Education and Research, HBNI, Jatni-752050, Odissa, India\goodbreak
\and
School of Physics and Astronomy, Cardiff University, Queens Buildings, The Parade, Cardiff, CF24 3AA, U.K.\goodbreak
\and
School of Physics and Astronomy, Sun Yat-sen University, 2 Daxue Rd, Tangjia, Zhuhai, China\goodbreak
\and
School of Physics and Astronomy, University of Nottingham, Nottingham NG7 2RD, U.K.\goodbreak
\and
School of Physics, Indian Institute of Science Education and Research Thiruvananthapuram, Maruthamala PO, Vithura, Thiruvananthapuram 695551, Kerala, India\goodbreak
\and
Simon Fraser University, Department of Physics, 8888 University Drive, Burnaby BC, Canada\goodbreak
\and
Sorbonne Universit\'{e}, Observatoire de Paris, Universit\'{e} PSL, \'{E}cole normale sup\'{e}rieure, CNRS, LERMA, F-75005, Paris, France\goodbreak
\and
Sorbonne Universit\'{e}-UPMC, UMR7095, Institut d'Astrophysique de Paris, 98 bis Boulevard Arago, F-75014, Paris, France\goodbreak
\and
Space Research Institute (IKI), Russian Academy of Sciences, Profsoyuznaya Str, 84/32, Moscow, 117997, Russia\goodbreak
\and
Space Science Data Center - Agenzia Spaziale Italiana, Via del Politecnico snc, 00133, Roma, Italy\goodbreak
\and
Space Sciences Laboratory, University of California, Berkeley, California, U.S.A.\goodbreak
\and
The Oskar Klein Centre for Cosmoparticle Physics, Department of Physics, Stockholm University, AlbaNova, SE-106 91 Stockholm, Sweden\goodbreak
\and
UPMC Univ Paris 06, UMR7095, 98 bis Boulevard Arago, F-75014, Paris, France\goodbreak
\and
Universit\'{e} de Toulouse, UPS-OMP, IRAP, F-31028 Toulouse cedex 4, France\goodbreak
% ...
}